# Molecular Modeling of
# Self-Assembling Hybrid Materials

# Molecular Modeling of
# Self-Assembling Hybrid Materials

Memòria presentada per optar al títol de
Doctor per la Universitat Rovira i Virgili

**ALESSANDRO PATTI**

Departament d'Enginyeria Química

Escola Tècnica Superior d'Enginyeria Química

Universtitat Rovira i Virgili

Av. Països Catalans 26

43007 Tarragona

Tesi dirigida per els Dr. Flor Siperstein i Dr. Allan Mackie

Universitat Rovira i Virgili

Tarragona, 2007

Els sota signants, Dr. Flor Siperstein i Dr. Allan Mackie

FAN CONSTAR

Que la tesi doctoral que porta per títol

MOLECULAR MODELING OF SELF-ASSEMBLING HYBRID MATERIALS

i que presenta el Sr. Alessandro Patti per optar al grau de Doctor per la Universitat Rovira i Virgili, ha estat realitzada sota la seva direcció i que tots els resultats presentats i la seva anàlisi són fruit de la investigació realitzada per l'esmentat doctorand.

I perquè se'n prengui coneixement i tingui els efects corresponents, signem aquest certificat.

Dr. Flor Siperstein                                  Dr. Allan Mackie

School of Chemical Engineering          Dep. D'Enginyeria Química

and Analytical Science                         Universitat Rovira I Virgili

The University of Manchester

Tarragona, 23 d'agost del 2007

*A mio papà e a mia mamma*

*A Eva*

# Ringraziamenti

disponibilità nel visionare la mia tesi come commissari esterni. In special modo ringrazio il Prof. Nigel Seaton per avermi dato l'opportunità di studiare durante alcuni mesi presso il suo gruppo di ricerca ad Edinburgo, nel quale ho avuto il privilegio di lavorare a fianco di Claudia Prosenjak, senza la quale parte di questa tesi non sarebbe stata scritta. A lei e a tutto il gruppo di Edinburgo, vanno i miei più sinceri ringraziamenti.

Un grazie va anche al Dr. Vladimir Zeleňák, che non solo mi ha dedicato il suo tempo per la sintesi e la caratterizzazione di alcuni materiali mesoporosi presso il suo dipartimento a Košice, Slovacchia, ma anche per lasciarmi conoscere le bellezze naturali del suo magnifico paese.

Ringrazio la mia famiglia per la loro costante presenza, nonostante la lontananza che in questi quattro anni ci ha diviso. Ho sentito il loro calore ed appoggio durante ogni momento della mia permanenza in Spagna, ed è grazie a loro se nei momenti più difficili sono riuscito a guardare avanti, rinfrancato dal loro amore. Ai miei genitori, Filippo ed Antonietta, dedico questa tesi, che è anche frutto del loro lavoro di educatori preziosi, stimolanti e pazienti. Soprattutto pazienti. Tutto quello che ho ottenuto in questi anni di studio lo devo specialmente a loro e ai sacrifici che hanno fatto con tanto affetto per me. Grazie papà, grazie mamma!

E un grazie dal profondo del mio cuore va ad Eva. A lei, alla sua infinita dolcezza e all'amore che ci lega dedico questa tesi. Gracias, señorita. Sin ti a mi lado, todo esto no tendria algún sentido.

# Contents

# List of Figures

# List of Tables

*Molecular Modeling of Self-Assembling Hybrid Materials.*

*Dietro ogni problema c'è un'opportunità.*
*Galileo Galilei*

# Chapter 1

## INTRODUCTION

Since the very beginning of our history, engineers have had to face with more and more difficult challenges, imposed by kings, popes, or simply by the needs of societies. The Egyptian pyramids, the Acropolis in Athens, Saint Peter's in Rome, and in general all the huge works of engineering have characterized our times and have given us the opportunity to discover old cultures, otherwise doomed to be forgotten.
Across the centuries, these enormous and wonderful works contributed to develop a popular image of the engineer and the tasks he should deal with. Colossal skyscrapers, boats, chemical plants, aircrafts are indeed the most classical examples which sculpted in our minds the stereotype of the modern engineer.
Nowadays, this image cannot be considered complete. In the last decades, the exponential interest towards the *nano-world* and its vast possibilities of applications, has changed it. An engineer builds ordered geometrical nanometer scale structures, and the building blocks are no longer bricks or pillars, they are atoms or molecules, and the rules to build these structures are based on different laws. The Nobel Prize in Physics in 1965, Richard Feynman, was the first who pointed out the problem of manipulating and controlling things on a small scale [*Feynman*, 1960]. He is generally considered as one of the fathers of nanotechnology, although the term itself was introduced in 1974 by Norio Taniguchi [*Taniguchi*, 1974], and it was only in the late 1980s that nanotechnology started giving the first important results leading to the discovery of fullerenes [*Kroto et al.*, 1985] and carbon nanotubes [*Iijima*, 1991], which are allotropes of carbon as well as diamond and graphite.

Since then, the fields based on technology involving the nano-world has been growing exponentially, and nowadays we can even say that nanotechnology is everywhere [*Hullmann*, 2007]. For instance, microelectronics and computers have become part of our everyday life, and the companies have succeeded in satisfying the increasing demands for faster and faster processors, by increasing the speed by a factor of 4 every three years [*Zhang et al.*, 2003]. This incredible growth is mainly due to the reduction of chips to a size belonging to the nanometer scale, as Richard Feynman predicted decades ago: "*If we wanted to make a computer that had all these marvelous extra qualitative abilities, we would have to make it, perhaps, the size of the Pentagon. This has several disadvantages. [...] Because of its large size, there is finite time required to get the information from one place to another. [...] But there is plenty of room to make them smaller. There is nothing that I can see in the physical laws that says the computer elements cannot be made enormously smaller than they are now. In fact, there may be certain advantages*"[*Feynman*, 1960].

In many commercial computers, the average length scale of the latest devices is around 150 nm, and it needs to reach 50 nm in the next generation to satisfy the demands of the market. Indeed, this aim imposes a very difficult and intriguing challenge to the scientific community, because at these length scales the rules governing the physical properties are still not completely understood. For instance, the material traditionally used as insulator between interconnecting wires (silica) is not insulating enough to separate the signals [*Zhang et al.*, 2003]. However, such difficulties do not seem to discourage too much public research in new nanotech devices [*Hullmann*, 2007].

In this research work, we are more interested in that area of nanotechnology dealing with nanostructured materials, and in particular with self-assembling nanostructures that lead to the formation of ordered porous structure. Nanomaterials have become one of the most important nanotech research area in a very short period of time, because of their tremendous economical, technological, and scientific impact [*Zhang et al.*, 2003]. As a matter of fact, one fourth of the total number of nanotech patents worldwide comes from the area of nanomaterials, which keeps on growing [*Hullmann*, 2007].

Self-assembling nanomaterials are the result of the spontaneous organization of atoms, molecules, particles, or, more generally, building blocks, into ordered and functional structures. Such a process, driven by the interaction energies established between the building blocks, represents the key factor for a bottom-up approach to nanotechnology. In this work, the amphiphilic molecules are the building blocks able to give rise to interesting complex structures, from spherical micelles to ordered liquid crystal phases.

Porous nanostructured materials are located in the mesoporous region, as their pore size ranges between 2 and 50 nm. Since 1992, the year of their discovery [*Beck et al.*, 1992], the self-assembling approach for the synthesis of mesoporous materials has been growing in importance, and today many different synthesis can be used for a wide range of purposes. The usefulness of self-assembled periodic structures is also due to the possibility to use them as structural frameworks to develop new materials (1) by incorporating functionalities into the pore walls; (2) by filling the mesopores with carbon or metals (nanocasting); or (3) by confining other materials with interesting properties inside the pores.

Self-assembly is the key factor to fabricate mesoporous materials, and a careful choice of the structure directing agent as well as of the inorganic precursor is fundamental to obtain the desired mesostructure. Many chemical and physical parameters play a very important role in the synthesis of mesoporous materials, such as the nature of the template, functional group of the inorganic precursor (if any), position of this organic group, *pH* of the solution, temperature, salt concentration, etc. The position of the functional organic group in the precursor molecules, for instance, deeply affects the final structure of the hybrid material, as well as its possible applications. The length of the surfactant affects the pore size and, in some cases, the thickness of the channel walls.

Our aim is not to consider every parameter, but rather those that are considered to have a strong affect on the phase and aggregation behavior of these hybrid systems and that can be studied by using simple models where amphiphiles, inorganic precursors and a model solvent are present [*Patti et al.*, 2007]. This choice gave us the opportunity to appreciate the macroscopic phase separation between a solvent-rich phase and a liquid crystal, and the periodicity of the self-assembled mesophases, and to analyze their structural properties. In particular, we aim to model the formation of self-assembling hybrid materials by considering how changing the inorganic precursors affect the formation of ordered liquid crystal phases, being the result of a phase separation driven by the strong interactions between the inorganic component and the surfactant heads. The analysis of the phase separation leads to the calculation of the associated ternary phase diagrams that can be of use in the design of ordered mesoporous materials.

In particular, we aim to understand why it is more difficult to observe the formation of ordered mesoporous materials when a terminal organosilica precursor, instead of a bridging organosilica precursor, is used during a co-condensation synthesis. The nature of the functionalizing organic group being more or less solvophobic can affect the formation of ordered structures and the organization of the precursor in the framework surrounding the pores. Therefore, we intend to understand to which extent the solubility of the precursor and the solvophobicity of its organic group affects its arrangement in the material framework.

The final ordered structure is affected by the surfactant chosen as structure directing agent. Therefore, we will also point out which consequences, in terms of phase separation and self-assembling behavior, can be detected when the architecture of the surfactant chains is modified.

The structure of the present thesis is organized as follows.

In Chapter 2, we introduce the amphiphilic systems by focusing our attention on their ability to self-assemble in ordered liquid crystal phases. Amphiphiles are molecules presenting a double behavior in a given solvent. If this solvent is water, the hydrophobic part (*tail*) of the molecule tries to lower the contact with the solvent as much as possible, and the hydrophilic part (*head*) plays an important role to achieve this aim. Amphiphiles are encountered in our everyday life as detergents, soaps, shampoos, and dispersants for paints, food processing aids, foaming agents, emulsifiers, cosmetic ingredients, etc. [*Zhang et al.*, 2003].

At low concentrations, when amphiphiles are present in solution as free unimers, it is usual to find them absorbed at the water surface with the tails pointing outwards and the heads towards the water solution. When the concentration becomes high enough and reaches the so called *critical micelle concentration*, the amphiphiles start assembling in spherical or elongated aggregates with a hydrophobic core shielded from the contact with water by the hydrophilic heads. In Chapter 2, we discuss the conditions to observe self-assembling into a given aggregate size and shape. A literature review of the most important achievements in theoretical, computational, and experimental works related with the aggregation and phase behavior of amphiphilic systems is given.

In Chapter 3, hybrid and mesoporous materials are presented. Hybrid organic-inorganic materials are synthesized through self-assembling of amphiphilic molecules acting as templates in water solutions containing an inorganic precursor, usually a silica precursor. Depending on the physical and chemical properties of the system, the surfactant concentration, and the nature of the inorganic precursor, the equilibrium phases can be different, such as cubic-ordered spherical aggregates, hexagonally ordered cylindrical aggregates, or lamellae. In particular, hybrid materials presenting rod-like micelles organized into a hexagonal order are the key step for the synthesis of mesoporous materials, obtained by removing the organic template from the inorganic framework [*Beck et al.*, 1992].

The so-synthesized porous structure shows a very large surface area and narrow pore size distribution (between 2 and 50 nm), being very attractive for several applications, such as catalysis, adsorption, molecular separation. We analyze the main features of such materials with particular interest given to the different ways to modify their pore walls with organic functionalities. Co-condensation and post-synthesis with pure silica or organosilica precursors are critically reviewed, by underlining the advantages and drawbacks of each [*Hatton et al.*, 2005]. The experimental techniques of characterization, such as X-ray diffraction or adsorption, are discussed as well as the most important applications. Finally, a brief review of the most important works in the modeling of mesoporous materials is reported.

In Chapter 4, the lattice model and the simulation methodology used in this study are presented. Our lattice is a three-dimensional network of fully occupied sites, each of them interacting with the nearest or diagonally nearest neighbors, according to the model commonly used in the study of the aggregation behavior of amphiphilic systems [*Larson et al.*, 1985]. The surfactant and inorganic precursor are modeled as chains of connected sites, representing coarse-grained groups of molecules. The inorganic precursor is modeled by considering the eventual presence of a functionalizing organic group, its solubility with the solvent, and its position with respect to the inorganic source, being terminal or bridging.

The interactions are qualitatively assumed to reproduce the repulsion between surfactant tails and solvent, and the strong attraction between surfactant heads and inorganic precursor. Monte Carlo (MC) simulations in the *NVT* ensemble are performed in boxes of different lengths and shape: elongated boxes have been used to study the phase behavior and to locate the concentration range in which the system phase separates; cubic boxes have been used to analyze the structural properties of the liquid crystals formed through the calculation of cluster size distribution, radial distribution functions, and density profiles in the aggregates. We review the basics of the MC method with particular attention to the moves performed in our study to sample the configurational space.

The results of the simulations have been compared with a lattice-based mean field approximation, the quasi-chemical theory (QCT) [*Guggenheim*, 1952], sharing the same model with MC simulations. The fundamental equations governing such a theory are presented with regard to our ternary system. This theory was previously applied by other researchers in similar amphiphilic systems to compare with the results of their simulations [*Kim et al.*, 2002; *Larson*, 1988; *Mackie et al.*, 1996; *Mackie et al.*, 1995]. In these works, the phase behavior in different model amphiphilic systems was investigated and a very good agreement between theory and simulations was observed except when self-assembled aggregates were formed, since QCT does not distinguish between ordered and disordered phases. Our conclusions confirm such quantitative disagreement between MC simulations and QCT, when ordered structures are formed.

The macroscopic phase separation in self-assembling amphiphilic systems is discussed in Chapter 5, with a focus on the equilibrium ternary phase diagrams obtained by using different inorganic precursors. Our aim here is to underline the possibility to obtain ordered hybrid materials even at low surfactant concentrations, by taking benefit of a phase separation where the inorganic component plays an important role. After a general introduction on the phase behavior of binary and ternary amphiphilic systems, we report the phase diagrams obtained by performing MC simulations and by applying QCT. We discuss the results according to the driving force for the phase separation, whose strength is deeply affected by changing the nature of the inorganic precursors.

Systems with terminal and bridging organosilica precursors are presented separately, and their phase behavior compared. A biphasic region is generally observed where a concentrated phase at high content of surfactant is in equilibrium with a dilute solvent-rich phase. In some cases, the content of surfactant in the dense phase is high enough to observe the formation of ordered liquid crystal phases. Interestingly, the phase separation observed in systems containing a bridging precursor leads to a bigger immiscibility gap than that observed in systems with terminal precursors. As we will see, such a difference is due to the strong attraction between surfactant heads and the inorganic group of the precursor, and is of key importance to determine the eventual formation of ordered structures in the surfactant-rich phase.

In Chapter 6, the microscopic segregation in the phases obtained as a result of a macroscopic phase separation is analyzed. We use different techniques to gain information on the structural order of those phases containing spherical aggregates or more complex liquid crystal phases. In systems presenting spherical aggregates, the analysis of the cluster-size distributions help us in understanding to which extent a given inorganic precursor affects the quality of the solvent surrounding the aggregate, by promoting or inhibiting the aggregation. The shape of such aggregates has been evaluated by the calculation of the three principal radii of gyration, or the asphericity factor being a combination of the three radii of gyration.

The analysis of the density profiles of surfactant heads and tails, along with those of inorganic precursor, gave us the opportunity to study the details of the organization of the components inside and around the core of the micelle. In general, we observed a solvophobic core completely composed by surfactant tails, and a corona in which the presence and the organization of the inorganic precursor is strongly determined by its organic group and the interactions established with the surfactant and the solvent. If this group is solvophobic, it is not even possible to obtain ordered structures, regardless of the surfactant concentration, or the miscibility of the inorganic segment with the solvent. When hybrid precursors are used, the distribution of inorganic and organic segments in the framework is usually homogeneous. Only in some cases a slightly larger concentration of organic segments is obtained near the solvophobic core.

In the last chapter, we consider the main factors affecting the design of ordered mesoporous materials. The distance between neighboring pores, the wall thickness, the organization of the functional groups and the pore size distribution will be analyzed to estimate how the presence of the surfactant and the precursor, along with its organic groups, can change the morphology of the materials. Systems containing terminal or bridging precursors will be analyzed and compared to find the pros and cons in the use of one or another. Moreover, we present a brief analysis of the effect of the surfactant architecture on the equilibrium structures. In particular, a branched-head surfactant will be compared to a linear surfactant in order to appreciate the eventual modifications of the mesophase. A many-scale approach, finding a common point between atomistic and coarse-grained models, will be discussed and the preliminary results presented.

The conclusions and possible future developments close the thesis.

**References Chapter 1**

Beck, J. S., J. C. Vartuli, W. J. Roth, M. E. Leonowicz, C. T. Kresge, K. D. Schmitt, C. T. W. Chu, D. H. Olson, E. W. Sheppard, S. B. Mccullen, J. B. Higgins, and J. L. Schlenker, A New Family of Mesoporous Molecular-Sieves Prepared with Liquid-Crystal Templates, *Journal of the American Chemical Society*, 114, 10834-10843, 1992.

Feynman, R., There's Plenty of Room at the Bottom, *Engineering and Science Magazine*, 23, 1960.

Guggenheim, E. A., *Mixtures*, Clarendon Press, Oxford, 1952.

Hatton, B., K. Landskron, W. Whitnall, D. Perovic, and G. A. Ozin, Past, present, and future of periodic mesoporous organosilicas - The PMOs, *Accounts of Chemical Research*, 38, 305-312, 2005.

Hullmann, A., Measuring and assessing the development of nanotechnology, *Scientometrics*, 70, 739-758, 2007.

Iijima, S., Helical Microtubules of Graphitic Carbon, *Nature*, 354, 56-58, 1991.

Kim, S. Y., A. Z. Panagiotopoulos, and M. A. Floriano, Ternary oil-water-amphiphile systems: self-assembly and phase equilibria, *Molecular Physics*, 100, 2213-2220, 2002.

Kroto, H. W., J. R. Heath, S. C. Obrien, R. F. Curl, and R. E. Smalley, C-60 - Buckminsterfullerene, *Nature*, 318, 162-163, 1985.

Larson, R. G., Monte-Carlo Lattice Simulation of Amphiphilic Systems in 2 and 3 Dimensions, *Journal of Chemical Physics*, 89, 1642-1650, 1988.

Larson, R. G., L. E. Scriven, and H. T. Davis, Monte-Carlo Simulation of Model Amphiphilic Oil-Water Systems, *Journal of Chemical Physics*, 83, 2411-2420, 1985.

Mackie, A. D., K. Onur, and A. Z. Panagiotopoulos, Monte Carlo simulation of phase behavior and micellization for model amphiphile systems., *Abstracts of Papers of the American Chemical Society*, 211, 76-Iec, 1996.

Mackie, A. D., A. Z. Panagiotopoulos, and S. K. Kumar, Monte-Carlo Simulations of Phase-Equilibria for a Lattice Homopolymer Model, *Journal of Chemical Physics*, 102, 1014-1023, 1995.

Patti, A., A. D. Mackie, and F. R. Siperstein, Monte Carlo Simulation of Self-Assembled Ordered Hybrid Materials, *Langmuir*, 23, 6771-6780, 2007.

Schumacher, C., J. Gonzalez, P. A. Wright, and N. A. Seaton, Generation of atomistic models of periodic mesoporous silica by kinetic Monte Carlo simulation of the synthesis of the material, *Journal of Physical Chemistry B*, 110, 319-333, 2006.

Siperstein, F. R., and K. E. Gubbins, Phase separation and liquid crystal self-assembly in surfactant-inorganic-solvent systems, *Langmuir*, 19, 2049-2057, 2003.

Taniguchi, N., On the Basic Concept of 'Nano-Technology', *Proc. Intl. Conf. Prod. Eng. Tokyo, Part II*, 1974.

Zhang, J., Z. Wang, J. Liu, S. Chen, and G. Liu, *Self-Assembled Nanostructure*, Kluwer Academic/Plenum Publishers, Ottawa, 2003.

*È più facile resistere all'inizio che alla fine.*
*Leonardo da Vinci*

# Chapter 2

## AMPHIPHILES AND

## AMPHIPHILIC SYSTEMS

### Introduction

Soft matter is a huge area of research in which different fields converge. In Figure 2.1, a schematic picture with the most important areas of interest embraced by soft matter is given. There are at least two important features justifying the presence of these systems in the same family: the degree of order observed is between that of a crystalline solid and that of a liquid, and all the energetic phenomena are of the order of $kT$, where $k$ is the Boltzmann constant and $T$ the absolute temperature. The last definition is actually more precise: polymeric solutions or melts, for instance, do not show any orientational or translational order, but they are included in this group because their high viscosity or viscoelasticity make them very different from liquids [*Hamley*, 2000].

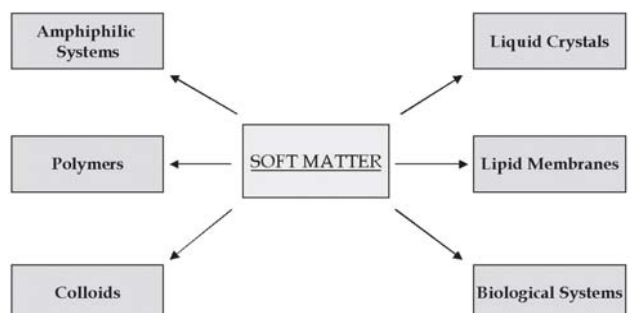

**Fig. 2.1**. Areas of interest in Soft Matter.

One of the groups belonging to the soft matter area is constituted by the amphiphilic and micellar systems. Amphiphiles are molecules with a double nature: one part of them is solvophilic and the other is solvophobic. This feature brings them to aggregate in order to shield the solvophobic part from the solvent. The ability to self-aggregate in ordered and complex structures, more commonly known as self-assembly, makes the amphiphilic systems of fundamental importance in nanotechnology.

Self-assembly is a bottom-up approach where simple nano-building blocks arrange together to form more complex structures, driven by the energetic constraints of the system. Self-assembly involves a huge range of length scales, from molecular to macroscopic scales, in Nature and technology, and different kinds of interactions. To celebrate its 125th year of life, the scientific magazine *Science* published the most interesting 125 questions that scientists will deal with for the next quarter century [*Service*, 2005]. "*How far can we push chemical self-assembly?*" is one of the 20 top questions in this special list. In the last thirty-forty years, discovering the rules governing self-assembly has contributed to opening new and more challenging research areas in those fields where nanostructures are implied. As a general trend, it is not at all practical to build nanostructures: it is much better to let Nature and its detailed laws work for us. One of the big challenges for the next generation material scientists will be to construct self-assembled complex hierarchical architectures, minimizing structural defects as much as possible.

In this chapter, the amphiphiles and their properties will be analyzed with particular attention to their ability to self-assemble in micelles or more complex structures, such as hexagonally ordered cylinders. The self-assembly of the amphiphilic molecules in such cylindrical micelles constitutes the necessary condition for the synthesis of hybrid organic-inorganic materials in which the amphiphilic molecules template the organization of the inorganic framework. After a brief introduction regarding the main properties of the amphiphilic molecules, a literature review of the most important publications embracing theoretical, computational, and experimental work is presented.

## 2.1 Amphiphiles and Amphiphilic Systems

The term *amphiphile* takes its origins from the ancient Greek *αμφί*, meaning *both*, and *φιλέω*, meaning *I like*. The term refers to the double nature of those molecules consisting of at least two parts, one being soluble in a given solvent, and one which is insoluble. When the solvent is water, one usually refers to the soluble part and to the insoluble part as the hydrophilic head and hydrophobic tail, respectively. In the following, we assume that the solvent is water unless otherwise specified.

The term *surfactants*[1] (a contraction for **surf**ace-**act**ive **age**nts) is commonly used as an alternative to *amphiphiles*, although it focuses on other properties of these molecules which are a consequence of their amphiphilic nature: when used at low concentrations in a given liquid-gas (liquid-liquid) system, surfactants reduce the surface (interfacial) tension[2] of the liquid by adsorbing at the interface between one phase and the other [*Rosen*, 2004]. This explains why such molecules are widely used in cleaning solutions, where the hydrophobic tails adhere to the surface of oil droplets and cover them completely [*Edler*, 2004].

In Figure 2.2, a schematic representation of a surfactant is reported.

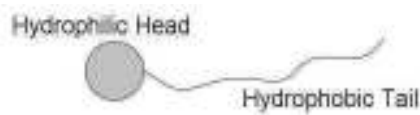

**Fig. 2.2**. Schematic representation of an amphiphilic molecule.

---

[1] "*A new word, surfactants, has been coined by Antara Products, General Aniline & Film Corporation, and has been presented to the chemical industry to cover all materials that have surface activity, including wetting agents, dispersants, emulsifiers, detergents and foaming agents*", American Dyestuff Reporter, 1950, *39*, 379/3.

[2] The surface tension of a given liquid is the free energy change when its surface area is increased by a unit area; the interfacial tension between two immiscible liquids is the free energy change to increase their interfacial area by a unit area. The term *interface* refers to a boundary between two immiscible phases, whereas the term *surface* is used when one of the two phases is a gas, most commonly air.

It is common to classify surfactants according to the polarity of the head group in anionic, cationic, non-ionic, or zwitterionic surfactants[3]. The anionic surfactants, mainly due to their low cost, are the largest surfactant group: of the 10 million tons per year of production of surfactants, the anionic ones represent around 60% [*Holmberg et al.*, 2002]. Another possible classification can be done by considering the fields of application of surfactants, such as detergency, painting, materials templating.

In Table 2.1, we report a very short list of surfactants used in different fields, according to the polarity of their group head.

**Table 2.1.** Surfactants and applications.

| Surf. | Molecular Formula | Group | Application | Reference |
|-------|-------------------|-------|-------------|-----------|
| CTAB | $C_{16}H_{33}(CH_3)_3NBr$ | Cationic | template | [Beck et al., 1992] |
| $C_{12}EO_8$ | $CH_3(CH_2)_{15}(OCH_2CH_2)_8OH$ | Non-ionic | template | [Attard et al., 1995] |
| P123 | $EO_{20}PO_{70}EO_{20}$ | Non-ionic | template | [Zhao et al., 1998] |
| NHM | $(OCH_2CH_3)_3Si(CH_2)_3NH_2$ | Cationic | template | [Burleigh et al., 2001] |
| SDS | $C_{12}H_{25}SO_4Na$ | Anionic | house products | [Rosen, 2004] |
| SLES | $CH_3(CH_2)_{10}CH_2(OCH_2CH_2)_2OSO_3Na$ | Anionic | personal care | [Stirton et al., 1952] |
| CAPB | $C_{19}H_{38}N_2O_2$ | Zwitterionic | cosmetics | [Pena and Peters, 1988] |

In bulk solutions, surfactants are able to aggregate into complex structures, such as micelles, rods, and bilayers, according to the interactions resulting from the inter- and intra-aggregate forces, and according to the thermodynamic laws governing their aggregation behavior [*Israelachvili*, 1991]. The first important condition is to reach a concentration, called the critical micelle concentration (*cmc*), at which surfactants start forming aggregates. The *cmc* is different for each surfactant and for the same surfactant can vary according to the physical and chemical conditions, such as *pH*, temperature, presence of a co-surfactant or salts.

There is no unique way to define the *cmc*. Some researchers define it as the concentration at which micelles appear in the system [*Mackie et al.*, 1997], others define it as the surfactant concentration at which the slope of the curve of the osmotic pressure plotted against the total surfactant concentration sharply changes [*Floriano et al.*, 1999], others as the surfactant concentration at which the curve of the equivalent conductivity shows a sharp break if plotted against the normality of solution [*Rosen*, 2004]. In this work, we assume that the *cmc* is the concentration at which micelles start to form in the solution.

---

[3] Zwitterionic surfactants are electrically neutral, but carry positive and negative charges on different atoms.

In particular, well below the *cmc*, most of the surfactant is present as free monomers in the bulk or at the interface of the system, and the concentration of micelles is practically negligible, although there is, in principle, a small probability of finding a cluster of any given size [*Mackie et al.*, 1997]. The surfactant molecules at the water interface organize themselves in such a way that the hydrophilic heads are oriented towards the water solution and the hydrophobic tails outwards. When the concentration reaches the *cmc*, the surfactant molecules already present in the bulk start aggregating into micelles, whose corona of hydrophilic heads shields the tails from contact with the water (see Figure 2.3.). Any further addition of surfactant in the system will not modify the monomer concentration, which stays roughly constant, but will increase the concentration of the aggregates.

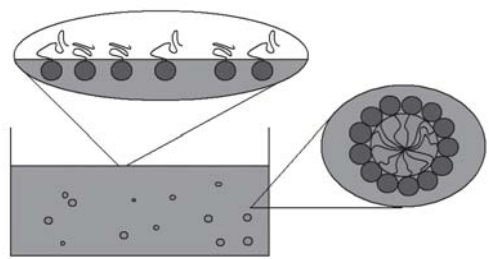

**Fig. 2.3**. Surfactants at the surface and in the bulk of a water solution.

The formation of micelles, also called micellization, leads to a reduction of the free energy of the system, and can be considered as a compromise between different competing factors: the attraction between hydrocarbon chains, the steric and/or electrostatic repulsion between the head groups, the entropic penalty for the deformation of the hydrocarbon chains and for curvature effects at the water interface. Micellization of surfactants is an example of the hydrophobic effect, referring to those processes where non-polar molecules or non-polar parts of them are spontaneously removed from contact with water [*Kronberg et al.*, 1995].

Since surfactants can aggregate in several shapes and structures, it makes sense to ask which micelles can be formed and which geometry do they show when some given conditions are satisfied. In the following three sections, we review the most interesting theories, computer simulations, and experimental techniques concerning the analysis of the aggregation behavior of amphiphilic systems.

## 2.2 Aggregation Behavior. Theories.

In this section, we review some of the most interesting theoretical contributions of the last two decades, whereas the computer simulation studies and the experimental techniques will be treated in sections 2.3 and 2.4, respectively.

The aggregation behavior of amphiphilic systems has been treated either by assuming the micelles at infinite dilution and then neglecting the interactions between aggregates [*Israelachvili*, 1991; *Nagarajan and Ruckenstein*, 1991], or by considering the deviations from ideal behavior [*Al-Anber et al.*, 2005; *Ben-Shaul et al.*, 1984; *Bhattacharya and Mahanti*, 2001].

In the theory proposed by Israelachvili [*Israelachvili et al.*, 1976] and Tanford [*Tanford*, 1980], the contributions to the repulsive interactions are not explicitly included because they are too complex to calculate and also because it is not necessary to know them explicitly. They are often modeled as $K/a$, where $K$ is a constant and $a$ the area per head-group. As for the contributions of the hydrocarbon chains, a common hypothesis is to look at the hydrophobic core of the micelle as a liquid hydrocarbon droplet. Such contributions are proportional to the head-group area and modeled as a surface tension term, $\gamma a$, being $\gamma$ the water-hydrocarbon surface tension. Therefore, the total interactions taken into account to describe the surfactant aggregation behavior are those concentrating at the hydrocarbon-water surface.

The balance of such interactions furnishes the interfacial free energy per molecule of surfactant in an aggregate [*Israelachvili*, 1995]:

$$\mu_M^0 = \gamma a + \frac{K}{a} = 2\gamma a_0 + \frac{\gamma}{a}\left(a - a_0\right)^2 \tag{2.2.1}$$

where $a_0$ is the optimal head-group area which minimizes $\mu_M^0$.

The amphiphiles tend to aggregate at this value of the head-group area; nevertheless, different geometries, such as spheres or cylinders, can exist for the same $a_0$. As a matter of fact, the optimal head-group area is not the only parameter necessary to define the packing geometry of a given aggregate: the length and the volume of the hydrocarbon chains also affect such a geometry, and are included in the definition of the so-called packing parameter, $v/a_0 l_c$, whose critical values are given in Table 2.2.

**Table 2.2**. Critical Packing Parameters.

| Structure formed | Critical packing shape | Critical packing parameter |
|---|---|---|
| Spherical micelles | Cone | <1/3 |
| Cylindrical micelles | Truncated cone | 1/3 – 1/2 |
| Vesicles | Truncated cone | 1/2 – 1 |
| Planar bilayers | Cylinder | ~1 |
| Inverted micelles | Inverted truncated cone | >1 |

For common surfactants, the ratio between the volume of the hydrocarbon chain and its length is constant, and its value is around 21 Å$^2$ for single tail surfactants and 42 Å$^2$ for double tail [*Tanford*, 1980]. Therefore, according to the above considerations, only the head-group area would affect the critical packing parameter, and the surfactant tail would not have any active role in determining the shape and size of the self-assembled structure.

Recently, Nagarajan showed that the surfactant tail directly or implicitly affects the head-group area [*Nagarajan*, 2002]. In particular, in spherical ionic micelles, the tails influence the ionic strength and hence the equilibrium head-group area; in cylindrical aggregates, the tail does not show a direct influence on the packing parameter, but affects the number of surfactant molecules in the endcaps of the rod-like micelles, and this explains why in some cases rod-like micelles do not form, although the packing parameter predicts their formation; in vesicles, the tails do not directly influence the packing parameter, but have an important effect on the volume of the aqueous cavity inside the vesicle, and if this volume cannot accommodate the surfactant head-groups, then vesicles cannot form, even when the packing parameter states the contrary [*Nagarajan*, 2002].

A very deep analysis has been dedicated to micelle formation by means of physicochemical models. Such models calculate the free energy of micellization by considering the transfer of a free surfactant monomer from the bulk of the solution into a micelle. Generally, the micelle is considered to be in an infinitely dilute solution and the interaggregate interactions are therefore neglected. From the free energy of micellization, the properties of interest can be calculated. Often these models make use of various phenomenological parameters based on data obtained from experiments. For instance, to compute the repulsive contributions in the free energy of micellization, an empirical constant has been estimated [*Nagarajan and Ruckenstein*, 1979], and geometrical packing constraints have been defined to calculate the optimal micellar shape [*Israelachvili et al.*, 1976].

More recently, the single contributions to the repulsive free-energy associated to the formation of micelles, such as steric, interfacial, and electrostatic terms, were explicitly estimated [*Nagarajan and Ruckenstein*, 1991; *Puvvada and Blankschtein*, 1990]. In particular, the work of Puvvada and Blankschtein included the following factors: (1) tail-water hydrophobic interactions; (2) conformational effects associated with tail packing; (3) curvature effects at the water interface; and (4) steric and electrostatic interactions between the surfactant heads.

As the surfactant concentration is increased, there are two important factors to deal with. One is the possible modification of micellar shape and size, and the other is the strength of the interaggregate interactions, becoming more and more important. The first works that tried to model micellar systems at higher surfactant concentrations, assumed an ideal solution behavior by neglecting the interactions between micelles and considering only the entropy of mixing [*Nagarajan and Ruckenstein*, 1979]. Nevertheless, if the aim is to analyze certain properties, such as phase separation, micellar size distribution, stability of some given micellar structures, or micellar diffusion coefficients, then these interaggregate interactions should be included [*Zoeller et al.*, 1997].

Zoeller and coworkers used the McMillan-Mayer theory of multicomponent solutions [*McMillan and Mayer*, 1945] to formulate a statistical-thermodynamic framework in order to calculate the Gibbs free energy of systems composed of micelles of nonionic surfactants [*Zoeller et al.*, 1997]. They describe the repulsive interactions by excluded volume considerations, whereas the attractive ones are described by a mean-field approximation. The chemical potentials of each component in the system are calculated and used to predict the micellar size distribution. Moreover, a comparison with the experimental data of a micellar solution of polyethylene oxide surfactants was performed, and the values of the *cmc*, the index of polydispersity, and the characteristics of phase separation, resulted to be in very good agreement with experimental results.

To avoid restrictions or assumptions on the molecular conformation for surfactant tails and heads, and also to avoid the use of empirical or semi-empirical parameters, two theories have been developed: the lattice self-consistent field (SCF) theory [*Scheutjens and Fleer*, 1979; *Scheutjens and Fleer*, 1980] and the single-chain mean-field (SCMF) theory [*Ben-Shaul and Szleifer*, 1985].

Single-chain theories consider the properties of a central chain whose intramolecular interactions are explicitly taken into account, whereas the intermolecular interactions are considered as a mean field, that is, an average approximation of neighboring chains. This is why single-chain theories are also called single-chain mean-field theories. Such theories were originally developed to study the chain packing in amphiphilic dry core aggregates [*Ben-Shaul and Szleifer*, 1985; *Szleifer et al.*, 1985], and later were extended to study polymers close or adsorbed to a surface [*Carignano and Szleifer*, 1993], and to micellar systems in which the positions occupied by the amphiphilic molecules were not restricted [*Mackie et al.*, 1997].

The properties of the system are calculated by considering the probability distribution function (*pdf*) of chain conformations, that is the probability to find any given chain with a conformation *c* in a micelle of geometry *G* [*Ben-Shaul and Szleifer*, 1985]. The *pdf* is calculated by minimization of the aggregate's free energy subject to the lattice single occupancy constraint [*Mackie et al.*, 1997]. The model chosen to represent the chain and the hypotheses done to derive the *pdf*, characterize a given SCMF theory. In particular, the *pdf* depends on the chain configuration, the temperature, the surface density of the chain, and in the case of mixtures, the composition of the sample [*Szleifer*, 1997].

The original version of the SCMF theory was published by Ben-Shaul and coworkers in 1985, to study the amphiphile organization in micellar aggregates of various geometries [*Ben-Shaul and Szleifer*, 1985; *Szleifer et al.*, 1985]. From the partition function of a single aggregate, they separate the translational-orientational factor and focus on the internal partition function, where the core and surface contributions are separated and the standard chemical potential per amphiphile is obtained. The *pdf* is derived by considering the following assumptions: (1) the density inside the hydrophobic core of the micelle is uniform and liquid-like; and (2) the most important interactions between the amphiphilic chains are the excluded volume repulsions. The first hypothesis was supported by contemporaneous experimental results [*Bendedouch et al.*, 1983], and molecular dynamics simulations [*Jonsson et al.*, 1986]. The theory was first applied to cubic chains and the results compared with the chain-packing theory proposed by Dill and Flory [*Dill and Flory*, 1980a; *Dill and Flory*, 1980b].

The Dill-Flory theory represented the chains on a cubic lattice, and the conformational part of the partition function was calculated by approximating it as an analytical function of the surface density of the chains. The agreement with the experimental data was good for bilayers, but poorer for curved aggregates. The problem was in the prediction of the bond order parameter: instead of decreasing from head to tail, as observed in experiments, it increased. Four years later, a modification was proposed for the calculation of the *pdf* [*Dill and Cantor*, 1984]: an energy factor for each chain kink was included to differentiate between a linear walk and a perpendicular walk. This new term solved the problem with the wrong estimate of the bond order parameter, but, according to Szleifer [*Szleifer et al.*, 1985], it had only a secondary role in determining the conformational behavior of the chains.

On the other hand, the agreement of the SCMF theory with the experiments was very good, including the behavior of the bond order parameter which was found to follow the experimental results without including any internal energy factor for the chain kinks [*Szleifer*, 1988]. The theory was also extended to the study of mixed aggregates and monolayers [*Ben-Shaul and Szleifer*, 1985].

Mackie *et al.* applied the SCMF theory to study the aggregation behavior of amphiphilic systems at low concentrations, and in particular to calculate the *cmc* and the micellar size distribution [*Mackie et al.*, 1997]. They modeled different diblock surfactants, and found a very good agreement between the *cmc* predicted by the theory and that obtained by performing lattice Monte Carlo (MC) simulations in the *NVT* ensemble. The slight differences could be attributed to the assumption of an ideal solution, namely no interaggregate interactions were considered, being a very good approximation near the *cmc*. However, the micellar distribution profiles appear quite different to the ones predicted by the SCMF theory being sharper and with its peak located at approximately half the value predicted by MC simulations.

The aggregation behavior of similar diblock surfactants was also studied by Guerin and Szleifer [*Guerin and Szleifer*, 1999]. They applied the SCMF theory to a continuous space model of spherical aggregates to study the *cmc* and the micellar size distribution as a function of the surfactant architecture and temperature. They found that the *cmc* shows a very slight dependence on the architecture of the surfactant head, but the micellar size distribution can be very different when linear and branched-head surfactants are compared. The core of the micelle does not show any significant trace of solvent, in agreement with the results of the self-consistent field theory applied to similar systems [*Wijmans and Linse*, 1995]. In particular, it was found that the hydrophobic region was more compact and larger for longer hydrophobic tails and more compact for higher temperatures.

The *cmc* was also calculated in binary surfactant solutions by Al-Anber and coworkers who applied the SCMF theory in the *NVT* ensemble for the calculation of the free energy of micellization [*Al-Anber et al.*, 2005]. Two symmetric non-ionic surfactants were studied and their *cmc* calculated, also including the non-ideal effects of mixing into the theoretical framework of the SCMF theory. By analyzing the dependence of *cmc* on the temperature, they showed that such effects become important when the *cmc* reaches the value of around 0.01 vol fraction. The same research group studied the sphere to rod transition in model non-ionic surfactants by applying the SCMF theory and grand-canonical MC simulations [*Al-Anber et al.*, 2003]. They observed that the surfactant modeled by four head segments and four tail segments prefers to form spherical micelles regardless of the surfactant concentration; on the other hand, by studying the asymmetric surfactant with three head segments and six tail segments, the SCMF theory predicted the presence of two peaks in the cluster size distribution, the first one representing the transition from free monomers to spherical micelles, and the other the transition from spherical to sphero-cylindrical micelles, which can be identified as a second *cmc*. The authors also studied the dependence of the second *cmc* on temperature and hydrophobic tail length, and found that it increases with increasing the temperature or decreasing the surfactant tail length.

The SCF theory was originally proposed by Scheutjens and Fleer to calculate the conformation of polymer chains adsorbed at interfaces in a lattice, filled by polymer and solvent molecules [*Scheutjens and Fleer*, 1979; *Scheutjens and Fleer*, 1980], and then generalized to study surfactant aggregates [*Leermakers and Scheutjens*, 1988]. The theory is based on mean-field approximations and calculates the statistical weight of a given conformation of the amphiphilic chains in specified self-consistent potentials [*de Bruijn et al.*, 2002]. Such potentials are computed from the knowledge of the distribution of the molecules in space, and, consequently, of the interactions they establish with each other. The different interactions between molecules are taken into account by considering the Flory-Huggins parameter $\chi$, and the possible chain conformations are generated by following the matrix method proposed by Di Marzio and Rubin [*Di Marzio and Rubin*, 1971], being very convenient to calculate the polymer segment density as a function of the distance from the interface. The theory presents the important advantage of considering explicitly the solvent molecules, so that it is possible to construct the phase diagrams by modifying the total concentration, but self-avoiding walks are not considered, namely different segments of the same chain can occupy the same lattice site unlike in the SCMF theory [*Szleifer*, 1988].

Bohmer *et al.* applied the SCF theory to the study the equilibrium structures formed by ionic surfactants in water solutions [*Bohmer et al.*, 1991]. They calculated the density profiles of the surfactant heads and tails in the micelles, and observed that the surfactant head groups have a rather wide distribution, whereas the boundary between the micelle core and the surrounding solution is quite sharp. By increasing the content of salt, the profile of this boundary is not significantly affected. The effects of the chain length and degree of branching on the *cmc* and on the average aggregation numbers are also evaluated, and the agreement with the experimental results is very good. It is interesting to note the dependence of the spherical to rod transition on the lattice geometry. The same research group applied the SCF theory to study the aggregation behavior of non-ionic surfactants and their adsorption in water solutions on hydrophilic or hydrophobic surfaces [*Bohmer and Koopal*, 1990].

Linse applied the SCF theory to study the phase behavior of triblock copolymers, and found regions containing free monomers, spherical micelles, or rod-like micelles [*Linse*, 1993b]. He observed a sudden radial reduction of around 10% at the transition from spherical to elongated micelles. The same author applied the SCF theory to evaluate the effect of temperature [*Linse*, 1993a], polymer impurities [*Linse*, 1994a], and polymer polydispersity [*Linse*, 1994b] on micellization in aqueous solutions for different Pluronic triblock copolymers. The diblock copolymers, considered as impurities and self-assembling more readily than the triblock copolymers, reduce the *cmc* and increases the average aggregation number of the system. Other impurities, such as homopolymers, also reduce the *cmc*, but the effect is less significant. The polymer polydispersity leads to a strong reduction of the *cmc*, whose exact value was found to strictly depend on the method to calculate it, and to an increase in micellar size. However, if the global polymer concentration is increased, the micellar size decreases and the effect of the polydispersity becomes softer. Finally, the temperature dependence of the *cmc* was found to be almost insensible to the polymer polydispersity.

To compare the results of their simulations, Larson and coworkers made use of a lattice-based mean-field approximation, the quasi-chemical theory (QCT) in a two-dimensional [*Larson et al.*, 1985], and then in a three-dimensional system [*Larson*, 1988]. The details of this theory will be presented in Chapter 4. Here, we briefly review the most important publications regarding the phase behavior of amphiphilic systems.

Larson *et al.* calculated the equilibrium ternary phase diagrams of systems composed of a linear model surfactant, water, and oil. The phase separation predicted by MC simulations was confirmed by the application of QCT, with good qualitative results. A three-phase region was observed for different surfactants, and, in particular for the surfactant presenting four head segments and four tail segments, the two-phase regions showed a partial miscibility between the solvents and the surfactant, at the reduced temperature $T^*$=6.5. Such an immiscibility was not observed in the simulations, as also reported a few years later by Mackie *et al.* [*Mackie et al.*, 1996]. This is probably due to the self-assembly of the surfactant in ordered aggregates, which cannot be predicted by the QCT, which reduces the formation of micelles to a macroscopic phase separation. At higher temperatures, the binary system solvent/surfactant does not form anymore ordered aggregates [*Larson*, 1992], and the agreement between MC simulations and QCT is qualitatively very good [*Kim et al.*, 2002].

Regardless of the temperature, in those systems presenting a three-phase region, namely a surfactant-rich phase at equilibrium with a water-rich phase and an oil-rich phase, the surfactant concentration in the surfactant-rich phase is overestimated by the QCT. Mackie and coworkers explained this discrepancy, by considering that this phase can absorb more water than that predicted by the QCT, because a microemulsion is formed [*Mackie et al.*, 1996].

## 2.3 Aggregation Behavior. Computer Simulations.

Computer simulations of self-assembling systems and, in particular, of surfactant solutions have been largely used during the last twenty years [*Shelley and Shelley*, 2000]. Simulations are not a panacea [*Rajagopalan*, 2001]: they cannot replace theories or experiments. Simulations give microscopic or macroscopic information not (easily) accessible in experiments. Even simple and short simulations are useful guidelines to test theories and to support experiments, especially when a given real system is affected by so many variables that understanding the effects of each would be very difficult, if we did not switch on only a few by computer experiments.

The development of complex structures in surfactant solutions has been studied by using atomistic, coarse-grained, and mesoscopic models, depending on the time and length scales of the physical properties of interest. In atomistic simulations, all the atoms or some groups of atoms are explicitly modeled, and their interactions are determined by different types of potentials[4]. The detailed description of the system by using an atomistic model makes the simulations computationally expensive. In fact, it is not feasible to model large systems involving thousands of molecules as it would take too long to obtain appreciable results. The typical time for the micelles to form and to reach the equilibrium is generally of the order of several microseconds or more [*Gelbart and BenShaul*, 1996; *Shelley and Shelley*, 2000], and this is still large compared to the time scales accessible with current atomistic simulations [*Cheong and Panagiotopoulos*, 2006]. This approach is more useful when local properties, such as the area per head group on the micelle surface or the orientations of the alkyl chains are analyzed.

---

[4] Chemical bonds potentials, electrostatic potentials, bond and dihedral angle potentials, non-bonded and non-electrostatic potentials.

Böcker *et al.* [*Böcker et al.*, 1994] used molecular dynamics (MD)[5] to simulate a water-like solution of 30 amphiphiles and around 2200 water molecules. Bandyopadhyay *et al.* have recently applied an atomistic model to a binary mixture of 32 chains of polyethylene oxide surfactant and 512 molecules of water in its liquid crystalline lamellar phase by using MD [*Bandyopadhyay et al.*, 2000]. Other works have included a higher number of surfactant chains and water molecules [*Maillet et al.*, 1999]. Generally, a fair choice of the initial configuration is needed, otherwise it would take too long to equilibrate the modeled systems whose properties of interest range over a time scale of picoseconds up to tens of nanoseconds, and in a length scale of Angstroms up to tens of nanometers. However, often one is not interested in a fully atomistic description of a phenomenon, but a coarser picture is satisfactory to get the essence of the physics behind it.

The choice of a less detailed model (and hence less computationally expensive) becomes mandatory if global structural properties implying many more molecules or longer periods of time have to be analyzed. In coarse-grained models, a number of atoms are arranged together in a simplified manner. Time and length scales involved are up to milliseconds and hundreds of nanometers, respectively. Lattice and off-lattice MC techniques represent the two main families of this group of simulations, even though MD coarse-grained simulations have also been used extensively [*Fodi and Hentschke*, 2000; *Palmer and Liu*, 1996]. Continuum models are more realistic, but lattice models are more computationally efficient, and have been largely used to study non-ionic surfactant systems.

The choice between MC or MD tools is obviously related to the system analyzed and to the physical properties of interest. MD is used to calculate time dependent properties, such as diffusion, residence time, etc.; MC generates random configurations where the particles are moved according to the probability of lowering the free energy of the system, and provides equilibrium properties of the system. In lattice MC simulations, the continuous space is replaced with a discretized lattice in which each site is occupied by only one particle, and the interactions are usually restricted to the nearest or diagonally-nearest neighboring sites, being *26* in the model proposed by Larson [*Larson et al.*, 1985].

The mesoscopic models try to connect some of the elements of the microscopic world to the macroscale behavior (such as fluid dynamics). Two interesting methods can be pointed out here: dissipative particle dynamics (DPD) and lattice gas calculations.

---

[5] In MD, each configuration of the system studied is generated by solving the Newton's equations of motions. The position and the velocity of any particle are therefore expressed as functions of time.

DPD has some common points with MD coarse-grained simulations, except for the fact that the force acting between the particles has conservative, dissipative[6] and fluctuating (random) parts. DPD has been used to study the self-assembly of a dense solution of amphiphilic species into micellar, hexagonal and lamellar phases [*Jury et al.*, 1999]. Lattice gas simulations are dynamic simulations on a lattice and have been used to study the behavior of binary immiscible and ternary microemulsion for an amphiphilic system [*Boghosian et al.*, 1996].

The idea to model surfactant solutions in a lattice box is due to Larson [*Larson et al.*, 1985] who studied the aggregation behavior of surfactants in systems where a water-like and an oil-like solvent were present. In this model, the surfactant chains were described as sequences of connected hydrophobic and hydrophilic segments, each unit occupying one single site in the lattice box. A site on the amphiphilic chain was connected to any of its $z=26$ nearest or diagonally-nearest neighbors, while solvent and oil molecules occupied one single site each. Larson's simulations showed the self-assembly of liquid crystal phases at high surfactant concentrations (>20% by volume for a short symmetric surfactant), including hexagonal (> 40%) and lamellar phases (> 75%). This simple lattice model for surfactant-oil-water mixtures illustrated very well the aggregation behavior of several real surfactant systems in which the hydrophilic to hydrophobic ratio is not too small to produce oil-water phase separation or too high to produce their miscibility. Therefore, this suggested that the phase behavior of these surfactants is dominated by physical phenomena that are captured by the lattice model [*Larson*, 1996]. This model will be discussed in detail in Chapter 4 as it has been applied to simulate the hybrid systems analyzed in this work.

The model proposed by Larson has been extensively applied to study amphiphilic systems at low surfactant concentrations, that is, when micellization occurs. However, Siperstein and Mackie applied this model to study the phase behavior of a binary surfactant-solvent system at intermediate and high densities [*Siperstein and Mackie*, 2005]. They performed lattice MC simulations in the grand-canonical ensemble and used parallel tempering techniques to observe the order-disordered transition and the formation of hexagonally ordered cylindrical aggregates. To locate the order-disorder transition temperatures at different surfactant concentrations, the authors calculated the heat capacity and the isothermal compressibility of the system as functions of the surfactant volume fraction. The peaks observed could be due to a phase transition, but, as the authors underlined, they may be a consequence of finite size effects. At high surfactant concentrations, these curves do not show any indication of the formation of hexagonal phases. However, by plotting the molar heat capacity versus the temperature at constant surfactant concentrations, it was possible to observe the peaks and to locate the transition temperatures.

---

[6] The dissipative force reduces the relative velocity of pairs of particles, dissipating their kinetic energy.

Glotzer *et al.* performed Brownian dynamic simulations at high surfactant concentrations to study the self-assembly of block copolymer tethered particles in hexagonal and lamellar structures [*Chan et al.*, 2006]. Three different architectures were studied: (1) a diblock copolymer with a cubic tethered block; (2) a diblock copolymer with a spherical tethered block; and (3) a triblock linear copolymer. They investigated how the architecture of the nanoparticles and the interactions established with the solvent affect the equilibrium structures. In general, lamellar and hexagonal structures were observed for all the geometries studied, with some differences related to the nature of the solvent. However, a couple of differences should be mentioned: the rigid and bulky nature of the cubic particles gives rise to an interfacial curvature not observed in triblock copolymers; and the shape of the nanoparticles, although it does not affect the morphology, has an interesting influence on the cross-sectional shape of the hexagonally ordered cylinders.

Studies at low surfactant concentrations with the Larson lattice model, were presented by Floriano and coworkers to analyze the micellization in model block surfactant systems [*Floriano et al.*, 1999]. They combined the data on free energy from a series of MC simulations performed in the grand-canonical ensemble by histogram reweighting, to obtain accurate values of the *cmc* over a broad range of temperatures. They found that the *cmc* increases with increasing the temperature, and the micellar aggregation numbers decrease at higher temperatures, but increases with the total surfactant concentration at temperatures near the upper limit of *cmc*. Analyses of the enthalpy of micellization revealed that it is proportional to the surfactant chain length.

Two years later, the same researchers published a study on micellization and phase separation of diblock and triblock surfactants [*Panagiotopoulos et al.*, 2002]. They observed that a system composed of a surfactant and a solvent either forms micelles or phase separates. In particular, high values of the hydrophilic/hydrophobic segment ratio favor phase separation and low values favor micellization. The *cmc* has been reported for several symmetric and asymmetric block surfactants and, for some of them, studied as a function of the reduced temperature.

An interesting study on the structural properties and phase behavior of triblock surfactants was published by Frenkel and coworkers [*Wijmans et al.*, 2004]. They use a coarse-grained model of polymeric chains and perform standard MC moves on a lattice, in order to investigate to which extent the micelles behavior can be considered as if it was of hard spheres. First they study the radial distribution functions of the aggregates in solution, the aggregates size and shape, by considering different interaction parameters and the effect of a different lattice coordination number. The chemical potential was calculated at low concentrations by applying the chain insertion method of Widom [*Frenkel and Smit*, 2001; *Widom*, 1963], which fails at high concentrations because of the low probability of accepting a trial move. They found that at high concentrations the structure factor calculated for the amphiphile-solvent system is very close to that of a mono-disperse hard-sphere fluid; at low concentrations, the hard-sphere model fails.

A modification of the model of Larson was proposed by Lisal and coworkers who studied the formation of micelles in supercritical carbon dioxide in the *NVT* ensemble [*Lisal et al.*, 2002]. The main difference with the standard Larson model is that in this case the lattice is not fully occupied, but some vacancies are included to study how the micellization is affected by changing the solvent density. The interactions established by these vacancies with all the other components in the box are set equal to zero by definition. Three different linear asymmetric diblock surfactants are modeled to study the effect of the head/tail ratio on the micellar behavior. The *cmc* was defined as the concentration at which the numbers of surfactants in aggregates matches the number of free surfactants in the solution, and it was shown that it increases by decreasing the number of head-groups, and that it is not significantly affected by the length of the hydrocarbon chain. This is in agreement with experimental results. An interesting difference with experiments concerns the dependence of the *cmc* on the solvent density: the *cmc* should increase with the solvent density, but the simulations show the opposite. The authors explained this apparent incongruence by considering that the solvent quality for the solvophilic and solvophobic blocks changes when the solvent density is modified. The pseudo-phase diagrams showed four different regions containing (1) free monomers, (2) spherical micelles, (3) spherical and elongated micelles at equilibrium, and (4) an unstable region at high surfactant concentration.

The aggregation and phase behavior of ternary surfactant-oil-water systems have been studied extensively by different researchers. Mackie and coworkers performed simulations in the *NVT* ensemble and in the Gibbs ensemble to calculate the ternary phase diagrams of symmetric and asymmetric diblock surfactants, and their results were in good qualitative agreement with the quasi-chemical theory, especially when no ordered structures were observed [*Mackie et al.*, 1996]. Kim and coworkers studied the effect of the oil chain length on the phase behavior of amphiphilic solutions and they obtained the phase diagrams by applying grand-canonical MC and the quasi-chemical theory [*Kim et al.*, 2002]. The dependence of the aggregation behavior on the oil volume fraction for different oil chain lengths shows that the *cmc* decreases by increasing the oil chain length, at the same reduced temperature. The density profiles calculated inside the spherical micelles are in agreement with the results obtained with the SCMF theory.

Ionic surfactants have been studied much less than the non-ionic ones. Bhattacharya *et al.* performed MC simulations to study the self-assembly of mesophases in a two-dimensional systems [*Bhattacharya and Mahanti*, 2001; *Bhattacharya et al.*, 1998]. The counterions were not explicitly considered and the interaction potentials were described as Lennard-Jones potential, a bond-bending potential, and a screened Coulombic potential. The presence of the Coulombic forces affects the micelle geometry in such a way that it results to be quite different from that of neutral surfactant at the same density. Since the model was not three-dimensional, it was not possible to do any quantitative estimate of the bulk micellization properties.

The first work focusing on the micellization of ionic surfactants in a three-dimensional system is very recent [*Cheong and Panagiotopoulos*, 2006]. In this work, a very simple lattice model was used to study the properties of various linear surfactants with ionic head-groups, by performing MC simulations in the grand-canonical ensemble. The surfactant is modeled as a chain of connected sites: the head-group is the biggest segment of the chain, it is charged, and is neutralized by a given number of monovalent counterions; the tail beads are as big as the counterions. The charged heads interact by hardcore repulsions and Coulombic forces. The authors determine the *cmc* by plotting the osmotic pressure as a function of the total surfactant concentration at different temperatures; the point where the curve undergoes a clear change in its slope corresponds to the critical micelle concentration. Their data of *cmc* result to be lower than those obtained experimentally, although of the same order of magnitude, and this is possibly caused by too many favorable interactions; maybe the inclusion of short-range repulsions between head-groups or between heads and tails, or a weaker interaction between tails could solve the problem, as the authors suggest.

## 2.4  Aggregation Behavior. Experiments.

The phase behavior and the structural and micellization properties of amphiphilic systems have also been studied through experimental techniques, and the values of the *cmc* and the average aggregation number for a very large range of surfactants are available in the literature [*Rosen*, 2004].

Hayashi and Ikeda studied the micelle size and shape of sodium dodecyl sulfate (SDS) in NaCl aqueous solutions at different temperatures by using light scattering measurements [*Hayashi and Ikeda*, 1980]. The authors analyzed the behavior of such solutions above and below the limit of solubility of SDS, and observed the formation of spherical micelles at low NaCl concentrations, and also rod-like micelles well above the *cmc* by increasing the content of salt. It was shown that at low salt concentrations, the spherical micelles are very stable and their size decreases by increasing the temperature, as generally observed for ionic surfactants [*Rosen*, 2004]. It was also pointed out that in the temperature range below the solubility limit or around it, before forming rod-like micelles, some of the spherical micelles associate into microgels, which can be metastable. The authors believe that this could be due to the big polar head-groups of SDS, which tend to link the micelles with each other to form microgel particles.

SDS aqueous solutions were also studied by using small-angle neutron scattering (SANS) over a wide range of surfactant concentrations and temperatures in order to measure the micellar size and the average aggregation number [*Bezzobotnov et al.*, 1988]. A slight dependence of the aggregation number on the temperature was observed; in particular, the aggregation number decreases, by increasing the temperature. The authors suggest that an increase in the temperature causes an "evaporation" of some surfactants attached to the micelle, and, if their concentration in the solution is high enough, such free surfactants will aggregate again to form new micelles, whose size will be lower than that of the micelles already present in the system. To support such considerations, they presented a simple theoretical model, being in very good quantitative agreement with their experimental results.

The critical micelle concentration is the most measured property of amphiphilic systems. Its determination has been obtained by different experimental techniques, such as conductimetric methods [*de Moraes and Rezende*, 2004; *Mysels and Mysels*, 1965], surface tension measurements [*Lyons et al.*, 1993], fluorescence methods [*Nakahara et al.*, 2005; *Priev et al.*, 2002], calorimetry [*Andersen and Christensen*, 2000; *Paula et al.*, 1995], capillary electrophoresis [*Lin*, 2004; *Lin et al.*, 1999; *Liu et al.*, 2007; *Nakamura et al.*, 1998] and other electrochemical means [*Chang et al.*, 1998a; *Chang et al.*, 1998b; *Nesmerak and Nemcova*, 2006]. The *cmc* depends slightly on the method used to determine it, and is affected by temperature, pressure, salts, and architecture of the surfactant.

The effect of temperature on the *cmc* of chloride and bromide cationic surfactants in water has been analyzed by Mehta *et al.* [*Mehta et al.*, 2005], who also estimated the thermodynamic properties associated with the formation of micelles, such as enthalpy, entropy, and Gibbs energy. As generally observed for ionic surfactants, the *cmc* decreases to a minimum value and then increases again; whereas for non-ionic surfactants the *cmc* decreases with increasing temperature [*Kim and Lim*, 2004]. Conductivity measurements have been done and it was observed that if the counterion is changed from chloride to bromide, the *cmc* showed a decrease. Moreover, it was reported that the entropic effects drive the micellization of the bromide cation surfactants only at lower temperatures, whereas the micellization of the chloride cationic surfactants is entropically driven regardless of the range of temperatures selected.

The effect of high pressure on the formation and the structure of micelles formed by the self-assembling of the non-ionic surfactant $C_8E_5$ in deuterium oxide ($D_2O$) was studied through SANS measurements by Lesemann *et al.* [*Lesemann et al.*, 2003]. The experiments were performed on a system containing 1% (wt) of surfactant at around 30°C and pressure up to 310 MPa. By increasing the pressure, the micelle radius decreases as well as the interaction between micelles, being practically negligible already at 150 MPa. The authors explain the effect of the pressure on the micelle size, by considering that it induces the dehydration of surfactant heads and the collapse of the corona at pressures between ambient and 150 MPa.

For ionic surfactants, the presence of charged heads can deeply affect the value of the *cmc*, and the micellization can even be prevented because of the strong repulsion between the head-groups [*Rosen*, 2004]. In these cases, counterions play an important role because they modify the head-group interactions, as analyzed by Bijma and Engberts [*Bijma and Engberts*, 1997], who observed that the *cmc* decreases by increasing the counterion size and the counterion hydrophobicity in systems containing alkylpiridinium surfactants.

The effect of salt on the *cmc* was reported by Florenzano and Dias for zwitterionic surfactants [*Florenzano and Dias*, 1997]. They proved that the head-group repulsions are shielded by electrolytes, leading to a decrease in the *cmc* and an increase of the average aggregation number. Such a behavior is generally observed for ionic surfactants, and the effect is stronger for long surfactant chains; for non-ionic surfactants, the addition of salt has a very slight effect on the *cmc* [*Hamley*, 2000]. Borbely and coworkers studied the effect of the hydrocarbon chain length on the *cmc* of SDS by SANS [*Borbely et al.*, 1989]. They observed that by increasing the hydrophobic part of the amphiphilic molecules, the *cmc* decreases.

The structural characteristics of self-assembling triblock copolymers, such as polyethylene/polypropylene/polyethylene oxides (PEO/PPO/PEO) of different compositions, have been studied by small-angle X-ray scattering techniques and by proton nuclear magnetic resonance ([1]H-NMR) spectroscopy [*Alexandridis et al.*, 1996]. These copolymers associate into cubic, hexagonal and lamellar liquid crystalline phases, and the number of such phases increases with the PEO content and the molecular weight of the polymers. The increase of temperature has been found to affect the stability of the structures in the order lamellar < hexagonal < cubic phases, as well as the interfacial area which decreases by heating up the system. However, in water solutions of the triblock copolymer $(EO)_{37}(PO)_{58}(EO)_{37}$ (EO=ethylene oxide; PO=propylene oxide), the phase regions containing hexagonal and lamellar structures are not much affected by the temperature. Finally, the lamellar periodicity and the interfacial area were studied as functions of the copolymer content, and it was shown that both of them decrease by increasing the copolymer content for all the block copolymers considered.

SANS was used to analyze how the structural properties of micelles formed in a solution of a more complex pentablock copolymer were affected by changing the *pH* [*Determan et al.*, 2006]. Between *pH* 3.0 and *pH* 7.4, spherical micelles have been observed; above *pH* 7.4, the size of the micelles was found to increase till the transition to cylindrical micelles in the *pH* range 8.1-10.5; for *pH*>11, the solution shows a macroscopic phase separation between a water-rich phase and a hexagonally ordered phase, as observed by performing small-angle X-ray scattering. The authors underline the reversibility of the process: by lowering the *pH* below 7.0, the solution becomes again homogenous.

Many researchers have focused their interests on the analysis of the transitions observed in amphiphilic systems, such as the sphere to rod transition [*Ganguly et al.*, 2005; *Heerklotz et al.*, 2004]; lamellar to hexagonal transition [*Epand et al.*, 1989; *Gutberlet et al.*, 1998]; or lamellar to gyroid transition [*Imai et al.*, 2000]. In particular, studies on the aggregation behavior of the triblock copolymer $(EO)_{20}(PO)_{70}(EO)_{20}$ showed that its spherical micelles do not undergo any significant growth up to the cloud point[7]. However, when small amounts of ethanol (5% - 10%) are added to this binary solution, then at high temperatures a sphere to rod phase transition can be observed. Such a transition can be even observed at ambient temperature if NaCl is added to the solution [*Ganguly et al.*, 2005].

Tiddy and coworkers studied the aggregation behavior of aqueous solutions of hexaethylene glycol n-hexadecyl ether, $C_{16}EO_6$, at moderate and high surfactant concentrations, in order to analyze the mesophases formed between the isotropic micellar and hexagonal phases, and the hexagonal and lamellar phases [*Funari et al.*, 1994]. The authors observed several *intermediate* phases which can be regarded as the result of a gradual modification of the aggregate curvature, being a balance between the steric repulsion and the solvation of the heads, and the interactions of the hydrocarbon chains. In particular, a nematic phase of rod-like micelles, a cubic phase, and a perforated lamellar phase, were observed and reported in the binary phase diagram. Their formation is made possible by two factors: (1) the increase of temperature which reduces the curvature and gives rise to phases with flatter surfaces; and (2) the increase of surfactant concentration which increases the strength of the interaggregate interaction. The authors compared the aggregation behavior of this system with that of systems containing longer [*Funari et al.*, 1992] or shorter [*Mitchell et al.*, 1983] hydrocarbon chains. In particular, three different micellar cubic phases have been detected in the binary system containing $C_{12}EO_{12}$, and the analysis of the transition from the body-centered cubic phase to the hexagonal phase suggested an *undulating* cylinder mechanism [*Sakya et al.*, 1997].

---

[7] The cloud point is the temperature at which a surfactant becomes insoluble in water by increasing the temperature.

## References Chapter 2

Al-Anber, Z. A., B. J. Avalos, M. A. Floriano, and A. D. Mackie, Sphere-to-rod transitions of micelles in model nonionic surfactant solutions, *Journal of Chemical Physics*, 118, 3816-3826, 2003.

Al-Anber, Z. A., J. B. Avalos, and A. D. Mackie, Prediction of the critical micelle concentration in a lattice model for amphiphiles using a single-chain mean-field theory, *Journal of Chemical Physics*, 122, -, 2005.

Alexandridis, P., D. Zhou, and A. Khan, Lyotropic Liquid Crystallinity in Amphiphilic Block Copolymers: Temperature Effects on Phase Behavior and Structure for Poly(ethylene oxide)-b-poly(propylene oxide)-b-poly(ethylene oxide) Copolymers of Different Composition, *Langmuir*, 12, 2690-2700, 1996.

Andersen, S. I., and S. D. Christensen, The critical micelle concentration of asphaltenes as measured by calorimetry, *Energy & Fuels*, 14, 38-42, 2000.

Attard, G. S., J. C. Glyde, and C. G. Goltner, Liquid-Crystalline Phases as Templates for the Synthesis of Mesoporous Silica, *Nature*, 378, 366-368, 1995.

Bandyopadhyay, S., M. Tarek, M. L. Lynch, and M. L. Klein, Molecular dynamics study of the poly(oxyethylene) surfactant C12E2 and water, *Langmuir*, 16, 942-946, 2000.

Beck, J. S., J. C. Vartuli, W. J. Roth, M. E. Leonowicz, C. T. Kresge, K. D. Schmitt, C. T.-W. Chu, D. H. Olson, E. W. Sheppard, S. B. McCullen, J. B. Higgins, and J. L. Schlenker, A new family of mesoporous molecular sieves prepared with liquid crystal templates, *J. Am. Chem. Soc*, 114, 10834-10843, 1992.

Bendedouch, D., S. H. Chen, and W. C. Koehler, Determination of interparticle structure factors in ionic micellar solutions by small angle neutron scattering, *J. Phys. Chem.*, 87, 2621-2628, 1983.

Ben-Shaul, A., and I. Szleifer, Chain Organization and Thermodynamics in Micelles and Bilayers .1. Theory, *Journal of Chemical Physics*, 83, 3597-3611, 1985.

Ben-Shaul, A., I. Szleifer, and W. M. Gelbart, Statistical Thermodynamics of Amphiphile Chains in Micelles, *Proceedings of the National Academy of Sciences of the United States of America-Physical Sciences*, 81, 4601-4605, 1984.

Bezzobotnov, V. Y., S. Borbely, L. Cser, B. Farago, I. A. Gladkih, Y. M. Ostanevich, and S. Vass, Temperature and Concentration Dependence of Properties of Sodium Dodecyl Sulfate Micelles Determined from Small Angle Neutron Scattering Experiments., *J. Phys. Chem.*, 92, 5738-5743, 1988.

Bhattacharya, A., and S. D. Mahanti, Self-assembly of ionic surfactants and formation of mesostructures, *Journal of Physics-Condensed Matter*, 13, 1413-1428, 2001.

Bhattacharya, A., S. D. Mahanti, and A. Chakrabarti, Self-assembly of neutral and ionic surfactants: An off-lattice Monte Carlo approach, *Journal of Chemical Physics*, 108, 10281-10293, 1998.

Bijma, K., and J. B. F. N. Engberts, Effect of Counterions on Properties of Micelles Formed by Alkylpyridinium Surfactants. 1. Conductometry and 1H-NMR Chemical Shifts, *Langmuir*, 13, 4843-4849, 1997.

Bocker, J., J. Brickmann, and P. Bopp, Molecular-Dynamics Simulation Study of an N-Decyltrimethylammonium Chloride Micelle in Water, *Journal of Physical Chemistry*, 98, 712-717, 1994.

Boghosian, B. M., P. V. Coveney, and A. N. Emerton, A lattice-gas model of microemulsions, *Proceedings of the Royal Society of London Series a-Mathematical Physical and Engineering Sciences*, 452, 1221-1250, 1996.

Bohmer, M. R., and L. K. Koopal, Association and adsorption of nonionic flexible chain surfactants, *Langmuir*, 6, 1478-1484, 1990.

Bohmer, M. R., L. K. Koopal, and J. Lyklema, Micellization of ionic surfactants: calculations based on a self-consistent field lattice model, *J. Phys. Chem.*, 95, 9569-9578, 1991.

Borbely, S., L. Cser, Y. M. Ostanevich, and S. Vass, Influence of Hydrocarbon Chain-Length on Micellar Size, *Journal of Physical Chemistry*, 93, 7967-7969, 1989.

Burleigh, M. C., M. A. Markowitz, M. S. Spector, and B. P. Gaber, Direct synthesis of periodic mesoporous organosilicas: Functional incorporation by co-condensation with organosilanes, *Journal of Physical Chemistry B*, 105, 9935-9942, 2001.

Carignano, M. A., and I. Szleifer, Statistical thermodynamic theory of grafted polymeric layers, *The Journal of Chemical Physics*, 98, 5006-5018, 1993.

Chan, E. R., L. C. Ho, and S. C. Glotzer, Computer simulations of block copolymer tethered nanoparticle self-assembly, *Journal of Chemical Physics*, 125, -, 2006.

Chang, H. C., B. J. Hwang, Y. Y. Lin, L. J. Chen, and S. Y. Lin, Measurement of critical micelle concentration of nonionic surfactant solutions using impedance spectroscopy technique, *Review of Scientific Instruments*, 69, 2514-2520, 1998a.

Chang, H. C., Y. Y. Lin, C. S. Chern, and S. Y. Lin, Determination of critical micelle concentration of macroemulsions and miniemulsions, *Langmuir*, 14, 6632-6638, 1998b.

Cheong, D. W., and A. Z. Panagiotopoulos, Monte Carlo simulations of micellization in model ionic surfactants: Application to sodium dodecyl sulfate, *Langmuir*, 22, 4076-4083, 2006.

de Bruijn, V. G., L. J. P. van den Broeke, F. A. M. Leermakers, and J. T. F. Keurentjes, Self-consistent-field analysis of poly(ethylene oxide)-poly(propylene oxide)-poly(ethylene oxide) surfactants: Micellar structure, critical micellization concentration, critical micellization temperature, and cloud point, *Langmuir*, 18, 10467-10474, 2002.

de Moraes, S. L., and M. O. O. Rezende, Determination of the critical micelle concentration of humic acids by spectroscopy and conductimetric measurements, *Quimica Nova*, 27, 701-705, 2004.

Determan, M. D., L. Guo, P. Thiyagarajan, and S. K. Mallapragada, Supramolecular Self-Assembly of Multiblock Copolymers in Aqueous Solution, *Langmuir*, 22, 1469-1473, 2006.

Di Marzio, E. A., and R. J. Rubin, Adsorption of a Chain Polymer between Two Plates, *The Journal of Chemical Physics*, 55, 4318-4336, 1971.

Dill, K. A., and R. S. Cantor, Statistical Thermodynamics of Short-Chain Molecule Interphases .1. Theory, *Macromolecules*, 17, 380-384, 1984.

Dill, K. A., and P. J. Flory, Interphases of Chain Molecules - Monolayers and Lipid Bilayer-Membranes, *Proceedings of the National Academy of Sciences of the United States of America-Physical Sciences*, 77, 3115-3119, 1980a.

Dill, K. A., and P. J. Flory, Statistical-Mechanics of the Disorder Gradient in Lipid Bilayer-Membranes, *Federation Proceedings*, 39, 1985-1985, 1980b.

Edler, K. J., Soap and sand: construction tools for nanotechnology, *Phil. Trans. R. Soc. Lond. A*, 362, 2635-2651, 2004.

Epand, R. M., K. S. Robinson, M. E. Andrews, and R. F. Epand, Dependence of the bilayer to hexagonal phase transition on amphiphile chain length, *Biochemistry*, 28, 9398-9402, 1989.

Florenzano, F. H., and L. G. Dias, Critical micelle concentration and average aggregation number estimate of zwitterionic amphiphiles: Salt effect, *Langmuir*, 13, 5756-5758, 1997.

Floriano, M. A., E. Caponetti, and A. Z. Panagiotopoulos, Micellization in model surfactant systems, *Langmuir*, 15, 3143-3151, 1999.

Fodi, B., and R. Hentschke, Simulated phase behavior of model surfactant solutions, *Langmuir*, 16, 1626-1633, 2000.

Frenkel, D., and B. Smit, *Understanding Molecular Simulations: From Algorithms to Applications*, 2nd ed., Academic Press, San Diego, 2001.

Funari, S. S., M. C. Holmes, and G. J. T. Tiddy, Microscopy, X-Ray-Diffraction, and Nmr-Studies of Lyotropic Liquid-Crystal Phases in the C22eo6/Water System - a New Intermediate Phase, *Journal of Physical Chemistry*, 96, 11029-11038, 1992.

Funari, S. S., M. C. Holmes, and G. J. T. Tiddy, Intermediate Lyotropic Liquid-Crystal Phases in the C16eo6/Water System, *Journal of Physical Chemistry*, 98, 3015-3023, 1994.

Ganguly, R., V. K. Aswal, P. A. Hassan, I. K. Gopalakrishnan, and J. V. Yakhmi, Sodium Chloride and Ethanol Induced Sphere to Rod Transition of Triblock Copolymer Micelles, *J. Phys. Chem. B*, 109, 5653-5658, 2005.

Gelbart, W. M., and A. BenShaul, The "new" science of "complex fluids", *Journal of Physical Chemistry*, 100, 13169-13189, 1996.

Guerin, C. B. E., and I. Szleifer, Self-Assembly of Model Nonionic Amphiphilic Molecules, *Langmuir*, 15, 7901-7911, 1999.

Gutberlet, T., U. Dietrich, G. Klose, and G. Rapp, X-Ray Diffraction Study of the Lamellar-Hexagonal Phase Transition in Phosholipid/Surfactant Mixtures, *Journal of Colloid and Interface Science*, 203, 233-494, 1998.

Hamley, I. W., *Introduction to Soft Matter*, John Wiley & Sons, LTD, Chichester, 2000.

Hayashi, S., and S. Ikeda, Micelle Size and Shape of Sodium Dodecyl-Sulfate in Concentrated Nacl Solutions, *Journal of Physical Chemistry*, 84, 744-751, 1980.

Heerklotz, H., A. Tsamaloukas, K. Kita-Tokarczyk, P. Strunz, and T. Gutberlet, Structural, Volumetric, and Thermodynamic Characterization of a Micellar Sphere-to-Rod Transition, *J. Am. Chem. Soc.*, 126, 16544-16552, 2004.

Holmberg, K., B. Jonsson, B. Kronberg, and B. Lindman, Surfactants and Polymers in Aqueous Solution, *John Wiley & Sons, LTD*, 2002.

Imai, M., A. Kawaguchi, A. Saeki, K. Nakaya, T. Kato, K. Ito, and Y. Amemiya, Fluctuations of lamellar structure prior to a lamellar[over -->]gyroid transition in a nonionic surfactant system, *Physical Review E*, 62, 6865 LP - 6874, 2000.

Israelachvili, J., *Intermolecular & Surface Forces*, 2nd ed., 450 pp., Academic Press, London, 1991.

Israelachvili, J. N., *Intermolecular and Surface Forces*, 5th ed., London, 1995.

Israelachvili, J. N., D. J. Mitchell, and B. W. Ninham, Theory of Self-Assembly of Hydrocarbon Amphiphiles into Micelles and Bilayers, *Journal of the Chemical Society-Faraday Transactions Ii*, 72, 1525-1568, 1976.

Jonsson, B., O. Edholm, and O. Teleman, Molecular-Dynamics Simulations of a Sodium Octanoate Micelle in Aqueous-Solution, *Journal of Chemical Physics*, 85, 2259-2271, 1986.

Jury, S., P. Bladon, M. Cates, S. Krishna, M. Hagen, N. Ruddock, and P. Warren, Simulation of amphiphilic mesophases using dissipative particle dynamics, *Physical Chemistry Chemical Physics*, 1, 2051-2056, 1999.

Kim, H., and K. Lim, A model on the temperature dependence of critical micelle concentration, *Collids and Surfaces A: Physicochem. Eng. Aspects*, 235, 121-128, 2004.

Kim, S. Y., A. Z. Panagiotopoulos, and M. A. Floriano, Ternary oil-water-amphiphile systems: self-assembly and phase equilibria, *Molecular Physics*, 100, 2213-2220, 2002.

Kronberg, B., M. Costas, and R. Silveston, Thermodynamics of the hydrophobic effect in surfactant solutions-micellization and adsorption, *Pure & Appl. Chem.*, 67, 897-902, 1995.

Larson, R. G., Monte Carlo lattice simulation of amphiphilic systems in two and three dimensions, *The Journal of Chemical Physics*, 89, 1642-1650, 1988.

Larson, R. G., Monte-Carlo Simulation of Microstructural Transitions in Surfactant Systems, *Journal of Chemical Physics*, 96, 7904-7918, 1992.

Larson, R. G., Monte Carlo simulations of the phase behavior of surfactant solutions, *Journal De Physique Ii*, 6, 1441-1463, 1996.

Larson, R. G., L. E. Scriven, and H. T. Davis, Monte-Carlo Simulation of Model Amphiphilic Oil-Water Systems, *Journal of Chemical Physics*, 83, 2411-2420, 1985.

Leermakers, F. A. M., and J. M. H. M. Scheutjens, Statistical thermodynamics of association colloids. I. Lipid bilayer membranes, *The Journal of Chemical Physics*, 89, 3264-3274, 1988.

Lesemann, M., H. Nathan, T. P. DiNoia, C. F. Kirby, M. A. McHugh, J. H. vanZanten, and M. E. Paulaitis, Self-Assembly at High Pressures: SANS Study of the Effect of Pressure on Microstructure of $C_8E_5$ Micelles in Water, *Ind. Eng. Chem. Res.*, 42, 6425-6430, 2003.

Lin, C. E., Determination of critical micelle concentration of surfactants by capillary electrophoresis, *Journal of Chromatography A*, 1037, 467-478, 2004.

Lin, J. M., M. Nakagawa, K. Uchiyama, and T. Hobo, Determination of critical micelle concentration of SDS in formamide by capillary electrophoresis, *Chromatographia*, 50, 739-744, 1999.

Linse, P., Micellization of poly(ethylene oxide)-poly(propylene oxide) block copolymers in aqueous solution, *Macromolecules*, 26, 4437-4449, 1993a.

Linse, P., Phase behavior of poly(ethylene oxide)-poly(propylene oxide) block copolymers in aqueous solution, *J. Phys. Chem.*, 97, 13896-13902, 1993b.

Linse, P., Micellization of Poly(ethylene oxide)-Poly(propylene oxide) Block Copolymer in Aqueous Solution: Effect of Polymer Impurities, *Macromolecules*, 27, 2685-2693, 1994a.

Linse, P., Micellization of Poly(ethylene oxide)-Poly(propylene oxide) Block Copolymers in Aqueous Solution: Effect of Polymer Polydispersity, *Macromolecules*, 27, 6404-6417, 1994b.

Lisal, M., C. K. Hall, K. E. Gubbins, and A. Z. Panagiotopoulos, Self-assembly of surfactants in a supercritical solvent from lattice Monte Carlo simulations, *Journal of Chemical Physics*, 116, 1171-1184, 2002.

Liu, H., Y. Gao, and Z. Hu, Determination of critical micelle concentration values by capillary electrophoresis, *Journal of Analytical Chemistry*, 62, 176-178, 2007.

Lyons, C. J., B. P. Moore, C. D. Distel, E. Elbing, B. A. W. Coller, and I. R. Wilson, The Critical Micelle Concentration of Teric-X10 by the Small Rod-in-Free-Surface Method of Surface-Tension Measurement, *Colloids and Surfaces a-Physicochemical and Engineering Aspects*, 74, 287-292, 1993.

Mackie, A. D., K. Onur, and A. Z. Panagiotopoulos, Phase equilibria of a lattice model for an oil-water-amphiphile mixture, *Journal of Chemical Physics*, 104, 3718-3725, 1996.

Mackie, A. D., A. Z. Panagiotopoulos, and I. Szleifer, Aggregation behavior of a lattice model for amphiphiles, *Langmuir*, 13, 5022-5031, 1997.

Maillet, J. B., V. Lachet, and P. V. Coveney, Large scale molecular dynamics simulation of self-assembly processes in short and long chain cationic surfactants, *Physical Chemistry Chemical Physics*, 1, 5277-5290, 1999.

McMillan, W. G., and J. E. Mayer, The Statistical Thermodynamics of Multicomponent Systems, *The Journal of Chemical Physics*, 13, 276-305, 1945.

Mehta, S. K., K. K. Bhasin, R. Chauhan, and S. Dham, Effect of temperature on critical micelle concentration and thermodynamic behavior of dodecyldimethylethylammonium bromide and dodecyltrimethylammonium chloride in aqueous media, *Colloids and Surfaces a-Physicochemical and Engineering Aspects*, 255, 153-157, 2005.

Mitchell, D. J., G. J. T. Tiddy, L. Waring, T. Bostock, and M. P. McDonald, Phase-Behavior of Polyoxyethylene Surfactants with Water - Mesophase Structures and Partial Miscibility (Cloud Points), *Journal of the Chemical Society-Faraday Transactions I*, 79, 975-1000, 1983.

Mysels, E. K., and K. J. Mysels, Conductimetric Determination of Critical Micelle Concentration of Surfactants in Salt Solutions, *Journal of Colloid Science*, 20, 315-&, 1965.

Nagarajan, R., Molecular packing parameter and surfactant self-assembly: The neglected role of the surfactant tail, *Langmuir*, 18, 31-38, 2002.

Nagarajan, R., and E. Ruckenstein, Aggregation of amphiphiles as micelles or vesicles in aqueous media, *Journal of Collid and Interface Science*, 71, 580-604, 1979.

Nagarajan, R., and E. Ruckenstein, Theory of surfactant self-assembly: a predictive molecular thermodynamic approach, *Langmuir*, 7, 2934-2969, 1991.

Nakahara, Y., T. Kida, Y. Nakatsuji, and M. Akashi, New fluorescence method for the determination of the critical micelle concentration by photosensitive monoazacryptand derivatives, *Langmuir*, 21, 6688-6695, 2005.

Nakamura, H., A. Sano, and K. Matsuura, Determination of critical micelle concentration of anionic surfactants by capillary electrophoresis using 2-naphthalenemethanol as a marker for micelle formation, *Analytical Sciences*, 14, 379-382, 1998.

Nesmerak, K., and I. Nemcova, Determination of critical micelle concentration by electrochemical means, *Analytical Letters*, 39, 1023-1040, 2006.

Palmer, B. J., and J. Liu, Simulations of micelle self-assembly in surfactant solutions, *Langmuir*, 12, 746-753, 1996.

Panagiotopoulos, A. Z., M. A. Floriano, and S. K. Kumar, Micellization and phase separation of diblock and triblock model surfactants, *Langmuir*, 18, 2940-2948, 2002.

Paula, S., W. Sues, J. Tuchtenhagen, and A. Blume, Thermodynamics of Micelle Formation as a Function of Temperature: A High Sensitivity Titration Calorimetry Study, *J. Phys. Chem.*, 99, 11742-11751, 1995.

Pena, L. E., and J. Peters, L., Self-preserving conditioning shampoo formulation. in *United States Patent*, The Upjohn Company, Kalamazoo, Mich., USA, 1988.

Priev, A., S. Zalipsky, R. Cohen, and Y. Barenholz, Determination of critical micelle concentration of lipopolymers and other amphiphiles: Comparison of sound velocity and fluorescent measurements, *Langmuir*, 18, 612-617, 2002.

Puvvada, S., and D. Blankschtein, Molecular-thermodynamic approach to predict micellization, phase behavior and phase separation of micellar solutions. I. Application to nonionic surfactants, *The Journal of Chemical Physics*, 92, 3710-3724, 1990.

Rajagopalan, R., Simulations of self-assembling systems, *Current Opinion in Colloid & Interface Science*, 6, 357-365, 2001.

Rosen, M. J., *Surfactants and Interfacial Phenomena*, 3rd ed., Wiley, New York, 2004.

Sakya, P., J. M. Seddon, R. H. Templer, R. J. Mirkin, and G. J. T. Tiddy, Micellar cubic phases and their structural relationships: The nonionic surfactant system C12EO12/water, *Langmuir*, 13, 3706-3714, 1997.

Scheutjens, J. M. H. M., and G. J. Fleer, Statistical theory of the adsorption of interacting chain molecules. 1. Partition function, segment density distribution, and adsorption isotherms, *J. Phys. Chem.*, 83, 1619-1635, 1979.

Scheutjens, J. M. H. M., and G. J. Fleer, Statistical theory of the adsorption of interacting chain molecules. 2. Train, loop, and tail size distribution, *J. Phys. Chem.*, 84, 178-190, 1980.

Service, R. F., How far can we push chemical self-assembly, *Science*, 309, 95-95, 2005.

Shelley, J. C., and M. Y. Shelley, Computer simulation of surfactant solutions, *Current Opinion in Colloid & Interface Science*, 5, 101-110, 2000.

Siperstein, F. R., and A. D. Mackie, Phase behavior of a model surfactant-solvent system at intermediate and high densities, *Colloids and Surfaces a-Physicochemical and Engineering Aspects*, 270, 277-284, 2005.

Stirton, A. J., J. K. Weil, A. Stawitzke, and S. James, Synthetic detergents from animal fats. Disodium alpha-sulfopalmitate and sodium oleyl sulfate, *Journal of the American Oil Chemists' Society*, 29, 198, 1952.

Szleifer, I., Statistical Thermodynamics of Amphiphilic Aggregates, Ph.D. thesis, Hebrew University, Jerusalem, 1988.

Szleifer, I., Protein Adsorption on Surfaces with Grafted Polymers: A Theoretical Approach, *Biophysical Journal*, 72, 595-612, 1997.

Szleifer, I., A. Benshaul, and W. M. Gelbart, Chain Organization and Thermodynamics in Micelles and Bilayers .2. Model-Calculations, *Journal of Chemical Physics*, 83, 3612-3620, 1985.

Tanford, C., *The Hydrophobic Effect*, Wiley - Interscience, New York, 1980.

Widom, B., Some Topics in Theory of Fluids, *Journal of Chemical Physics*, 39, 2808-&, 1963.

Wijmans, C. M., E. Eiser, and D. Frenkel, Simulation study of intra- and intermicellar ordering in triblock-copolymer systems, *Journal of Chemical Physics*, 120, 5839-5848, 2004.

Wijmans, C. M., and P. Linse, Modeling of Nonionic Micelles, *Langmuir*, 11, 3748-3756, 1995.

Zhao, D. Y., J. L. Feng, Q. S. Huo, N. Melosh, G. H. Fredrickson, B. F. Chmelka, and G. D. Stucky, Triblock copolymer syntheses of mesoporous silica with periodic 50 to 300 angstrom pores, *Science*, 279, 548-552, 1998.

Zoeller, N., L. Lue, and D. Blankschtein, Statistical-Thermodynamic Framework to Model Nonionic Micellar Solutions, *Langmuir*, 13, 5258-5275, 1997.

# Chapter 3

## HYBRID AND MESOPOROUS

## MATERIALS

### Introduction

Porous and high-surface-area materials are of great interest and application in many fields of science and technology. Their pores can be of two types: open or closed. The closed pores are completely surrounded by the material framework, which isolates them from the outside and makes them useless for several important operations where the accessibility to a given fluid is required, such as catalysis, adsorption, and filtration. Materials with open pores can be of different natures; they can present a single connection to the surface, or they can be open on both sides. In some cases, open pores can also be interconnected. Figure 3.1 gives a schematic example of a cross section belonging to a generic porous particle.

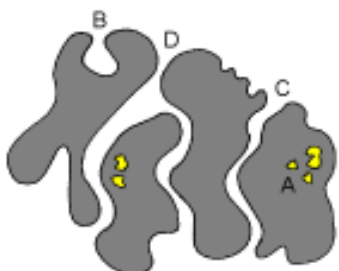

**Fig 3.1.** Cross section of a hypothetical porous particle presenting various examples of porosity: closed pores (A); blind pores (B); through pores (C); and interconnected pores (D).

There are different possible ways to classify porous materials. One is related to the definition of pore size furnished by the International Union of Pure and Applied Chemistry (IUPAC) [*Sing et al.*, 1985]. According to the IUPAC, porous materials can be divided into:

-   microporous materials, if their pore size is below 2 nm,
-   mesoporous materials, if their pore size is between 2 and 50 nm,
-   macroporous materials, if their pore size is bigger than 50 nm.

Generally, their porosity, being the ratio between the volume of the pore space and the total volume of the material, is between 0.2 and 0.95. Another classification takes into account their material constituents (organic or inorganic; ceramic or metal) or their properties (selectivity, adsorption capacity, thermal and chemical stability, etc.). Porous materials can also be classified according to their degree of order: fully crystalline (zeolites); ordered on a mesoscopic length scale, but amorphous on the atomic length scale (surfactant-templated materials); or fully disordered (silica gels) [*Schüth*, 2003]. Zeolites constitute the most important family of microporous solids because of their excellent properties, such as activity, selectivity, thermal stability, narrow pore-size distribution and high surface area which make them the most used materials in catalysis [*Corma*, 1997]. Activated carbons share with zeolites many of these features, and, thanks to their relative low cost, find large applications in many industrial processes, although the reactivity of carbon, in some cases, can limit their use. Despite these remarkable chemical and physical properties, they cannot be used when the size of the reactants is bigger than the dimension of the pores, being usually not more than 10 Å.

To overcome this limitation, there was an attempt to increase their pore-size up to some nanometers to obtain a mesoporous material, by dealumination of zeolites with $SiCl_4$ or steam. This solution has been found useful for only a few cases, as the percentage of mesoporosity generated is too low, and the resistance of the solid obtained is not high enough for most applications [*Corma*, 1997].

In the early 1990s, the researchers of Mobil Oil Corporation published an alternative way to synthesize mesoporous materials by taking benefit of the spontaneous tendency to organize of amphiphilic molecules. Hybrid systems of aqueous solutions composed by a surfactant and a silica precursor were able, under some given physical and chemical conditions, to self-assemble into very well organized mesophases. From such mesophases, by removing the organic template, ordered mesoporous materials were fabricated [*Beck et al.*, 1992]. Such an innovation opened a huge interest towards the supramolecular templating of functionalized porous materials, in many different industrial fields.

In this chapter, attention will be focused on the main characteristics of hybrid and mesoporous materials. Their synthesis, characterization, and main applications will be presented, with particular interest given to the choice of the organic template and inorganic (siliceous or non-siliceous) precursor. In section 3.1, we introduce the hybrid materials whose synthesis will be discussed in section 3.2. We will stress the importance of self-assembly as a necessary condition to observe the formation of ordered structures. Mesoporous materials, obtained by separating the organic template from the inorganic framework, are discussed in section 3.3, and the techniques to characterize them in section 3.4. Finally, the most important applications involving these ordered materials are reported in section 3.5. A brief overview of the methods to model hybrid and mesoporous materials by molecular simulation will be presented in section 3.6.

## 3.1 Hybrid Organic-Inorganic Materials.

Templating the synthesis of porous solids is commonly performed by following two different ways: one is soft-matter templating or *endotemplating*, and the other is hard-matter templating or *exotemplating* [*Hoffmann et al.*, 2006; *Schüth*, 2003]. Endotemplating creates the voids filled by the soft template around which the solid framework grows; when the template is removed, a porous structure is left. In the exotemplating, the template is a structure providing a scaffold with voids, which are filled by an inorganic precursor; when the scaffold is removed a powder material or a porous solid is formed, according to the three-dimensional connectivity of the template [*Schüth*, 2003].

The synthesis of hybrid materials considered in this context is an example of endotemplating, and in particular of supramolecular endotemplating. In fact, the templates used are not single molecules, like in the synthesis of zeolites [*Sastre*, 2006], but aggregates of molecules, and in particular of amphiphiles. Single molecules template the synthesis of microporous materials, but would not furnish a mesoporous architecture. The necessary condition to observe the formation of an ordered structure is given by the interactions established between the template and the inorganic precursor, which have to be very favorable to avoid a possible phase separation.

There are many definitions for hybrid organic-inorganic materials. In the most general description, hybrid materials mix organic and inorganic components at the molecular scale. They result from the combination of an organic template (an amphiphile) acting as a structure-directing agent with an inorganic precursor (generally silicate, organosilicate or aluminosilicate). The role of the template and the interactions established with the inorganic precursor are crucial to observe the formation of liquid crystals and the condensation of the precursor around them.

The scientific community has been focusing its attention on such ordered hybrid structures since the early 1990s. However, the interest towards the family these materials belong to, namely composite materials, started much earlier, in the 1940s. In those years, when the polymer industries were also quickly growing, it was realized that by introducing glass fibers in a light-weight, low-strength material, a stronger material can be obtained because the fibers act as barriers against the propagation of structural defects. The so-called glass fiber reinforced polymers (GFRP) were born, and, in approximately 60 years, they have become 90% of the composites market [*Bensaude-Vincent*, 2001].

A step ahead in the production of composites came from the space programs in the 1950s. The space race between the USA and URSS in those years was the main input for the development of the recently discovered carbon and boron nitride fibers. The latter had some relevance in military applications, but, mainly for their high cost, they did not find much space in other markets. Due to their better capabilities and lower cost, carbon fibers took the lead in the 1960s. The increasing use of a variety of fibers, led to a general definition of composites: materials combining two heterogeneous phases. Since the good mechanical properties of heterogeneous structures strictly rely on the quality of the interactions between the components, it was clear that studying the interfacial phenomena had to be the next step in materials research concerning composites [*Bensaude-Vincent*, 2001].

In the 1970s with the development of ceramic matrix composites (CMC) as an alternative to the metal-matrix composites (MMC), the scientific research in the field of composite materials changed its perspective. Instead of looking at composites as a combination of two different phases, the new point of view was to consider them as an association of a (metal or ceramic) matrix and reinforcing fibers. This classification was again modified in 1980s, when the synergy effects between the components constituting the composite materials became part of their standard definition.

In the 1990s, the attention was focused on smaller and smaller scales and the hybrid materials, constituting the latest generation of composite materials, were for the first time synthesized. Clearly, as the scales involved in the construction of these structures became smaller and smaller, the usual methods of making them (such as, lithography, etching, and micromoulding) reached physical limitations [*Edler*, 2004]. Therefore, instead of using a top-down approach where the accuracy of the mechanical tools to shape and cut materials was not sufficient at length scales of few nanometers, a bottom-up approach was developed. In a bottom-up approach to nanotechnology, simple building blocks arrange themselves into more complex structures, according to the nature of the components involved and the interactions established.

Hybrid organic-inorganic materials are obtained by a bottom-up self-assembly where an organic template and an inorganic precursor interact very favorably. In Nature, it is not difficult to find examples of hybrid materials. Bones closely associate inorganic and organic components: collagen protein fibers form the organic phase which is reinforced by small, rod-like crystals of hydroxyapatite $(Ca_{10}(HPO_4)_6(OH)_2)$, an inorganic calcium phosphate-based solid. MMC and CMC often present similar organizations as small particles of zirconia or ceramic SiC are used as a reinforcement in aluminum. As a matter of fact, in the design of hybrid materials, the aim is to mimic Nature in the assembling of stable structures. The natural behavior of amphiphiles to self-assemble into complex systems is crucial to understand the formation of these materials, whose importance has increased in the last fifteen years because they represent the key-step for the formation of ordered periodic mesoporous materials being of fundamental interest in catalysis [*Linssen et al.*, 2003], molecular separation [*Lee et al.*, 2001], adsorption [*Lei et al.*, 2004], and in many other industrial operations.

Sanchez *et al.* have classified hybrid materials into two classes, according to the nature of the interface and the interactions between the organic and the inorganic species [*Sanchez et al.*, 2001]. Class 1 includes hybrid systems where one of the components is entrapped within a network of the other component, such as organic dyes or biomolecules included in the porous solid matrixes [*Avnir et al.*, 1984; *Levy et* *al.*, 1989; *Reisfeld*, 1990]. In this case, the interactions between the hosting network and the entrapped species are very weak. In Class 2 are gathered the hybrid materials showing a stronger chemical bond between the inorganic and organic parts, such as a covalent or ion-covalent bond. The frontier between class 1 and class 2 is not often simple to locate, and hybrid systems with class 1 and class 2 characteristics can be obtained. In both cases, the degree of organization and the properties of the nanostructure obtained are strictly connected to the chemical nature of the constituting components, but also to the interactions between them at the interface.

## 3.2  Synthesis and Mechanism of Formation.

A procedure to synthesize hybrid and then mesoporous materials was first reported in a patent in 1969 [*Chiola et al.*, 1971]. Di Renzo *et al.* prepared a solid by following the instructions given in this patent 28 years earlier [*Di Renzo et al.*, 1997], and found that it led to a mesoporous material which Beck and coworkers had already patented in 1992 in the laboratories of Mobil Oil Corporation [*Beck et al.*, 1992]. It is worth saying that in the original patent only a few of the peculiar characteristics of these porous solids were identified. It was the group of Beck that really recognized the engineering value and the extraordinary features of this new family of mesoporous materials, which they called M41S. Large pores (from 2 to 10 nm), high surface areas (also above 1000 m$^2$/g) and hexagonal (MCM-41), cubic (MCM-48) or lamellar (MCM-50) structures constitute the *identity card* of this family, whose most representative member is the hexagonally ordered MCM-41[1].

The first ordered hybrid materials were synthesized from long-chain ionic surfactants (such as cetyltrimethyl ammonium bromide, CTAB) in a basic solution from a variety of silica sources[2] over a broad time/temperature regime (usually between 70°C and 150°C) [*Beck et al.*, 1992]. The resulting mesoporous materials, presenting amorphous silica walls, was obtained by calcination of the organic soft template at around 540 °C for one hour in flowing nitrogen and six hours in flowing air. Beck and coworkers suggested a liquid-crystal templating (LCT) mechanism for the formation of these materials. In the LCT mechanism, the condensation of the silica precursor is not the dominant factor for the formation of the mesoporous structure [*Zhao et al.*, 1996]. The amphiphilic molecules organize themselves in liquid crystal phases independently of the inorganic crystallization, and the polymerization of the silica precursor takes place around the self-assembled aggregate acting as a template [*Beck and Vartuli*, 1996]. The LCT is the dominant mechanism when the surfactant concentration is so high that the liquid crystals are already formed when the inorganic precursor starts organizing itself around them [*Attard et al.*, 1995].

---

[1] *MCM* states for Mobil Composition of Matter.
[2] Such as sodium silicate ($Na_2O \cdot SiO_2$), tetramethylammonium silicate (($CH_3$)$_4$N(OH)$\cdot$2SiO$_2$) and tetraethyl orthosilicate (($C_2H_5O$)$_4$Si).

Alternatively, it could be the introduction of the inorganic species that drives the configuration of the liquid crystal phases by a cooperative mechanism [*Monnier et al.*, 1993]. In this case, a silica source is added to the solution at low concentration of amphiphiles, and the ordered liquid crystal phase obtained is the result of a phase separation between a solvent-rich phase and an amphiphile-rich phase, where the concentration of the amphiphile is high enough to observe the formation of hexagonally ordered structures. In Figure 3.2 the two mechanisms are reported.

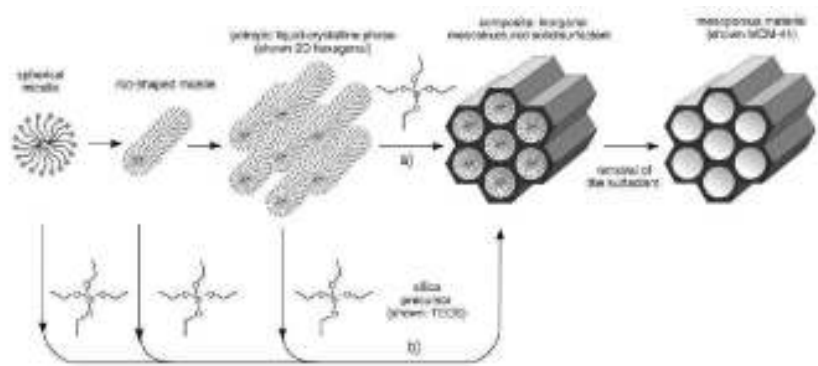

**Fig. 3.2**. Possible mechanistic pathways for the formation of a mesoporous hybrid material: (a) liquid crystal phase initiated and (b) silicate anion initiated [*Hoffmann et al.*, 2006].

Palmqvist [*Palmqvist*, 2003] analyzed the main factors affecting the synthesis of mesoporous materials, such as type of inorganic precursor, surfactant, concentration, *pH* and temperature, and concluded in agreement with Patarin *et al.* [*Patarin et al.*, 2002] that none of the mechanisms suggested so far can provide a general or exclusive reaction pathway as they depend too much on the experimental conditions of synthesis. One of these factors, the interactions at the interface between the surfactant and the inorganic precursor, as mentioned by Soler-Illia *et al.* [*Soler-illia et al.*, 2002], is key in understanding such discrepancies between different formation mechanisms.

The assembly of such mesostructures is very rapid, of the order of a minute or even faster [*Linden et al.*, 1999], but the shape of the walls is not defined until the condensation of silica in the framework is finished. After some minutes, however, the silica condensation has reduced the flexibility of the walls and any attempt to add, for instance, swelling agents to increase the pore size, would be completely useless [*Linden et al.*, 2000].

Since 1992, many synthesis techniques have been developed [*Corma*, 1997; *Sayari and Hamoudi*, 2001], involving different kinds of surfactants (cationic, anionic, nonionic, and block copolymers), silica sources and solutions (basic or acid). Neutral amine surfactants and neutral inorganic precursors were used to prepare mesoporous metal oxides with larger wall thickness being very useful to improve the thermal and hydrothermal stability of the final structure [*Tanev and Pinnavaia*, 1995]. However, these neutral amine surfactants as well as many ionic surfactants can be very expensive and, sometimes, even toxic, which can limit their applications [*Yu et al.*, 2004]. Therefore, shortly after the first synthesis of MCM-41, different research groups tried to find cheaper and safer structure directing agents.

Attard and coworkers used diblock copolymers of the type $C_{12}EO_8$ and $C_{16}EO_8$ as templates with a pure silica precursor (TMOS) under acidic conditions [*Attard et al.*, 1995]. The resulting materials, with a cubic or hexagonal arrangement, presented a pore size of around 3 nm; also lamellar mesophases have been obtained. Three years later the group of Stucky in Santa Barbara, California, obtained new ordered hybrid mesostructures with different triblock copolymers, such as $EO_{20}PO_{70}EO_{20}$ (P123), $EO_{106}PO_{70}EO_{106}$ (F107), or $EO_{132}PO_{50}EO_{132}$ (F108), as structure directing agents [*Zhao et al.*, 1998]. In both cases, that is in the synthesis performed by Attard and by Stucky, non-ionic polymeric surfactants were used, and today they seem to be the most popular templates for this kind of chemistry [*Fryxell*, 2006]. The use of block copolymers lowered the time of synthesis, typically performed under acidic conditions, to 6-12 hours from the seven days usually needed in the first synthesis in the 1990s [*Fryxell*, 2006]. The use of microwaves can even reduce the synthesis time to less than one hour [*Newalkar et al.*, 2004].

The length of the chain permits materials with bigger nanopores (up to 30 nm) or with thicker walls to be synthesized. These walls were amorphous as were the ones observed in the first materials synthesized by Beck, but in this case the solvophilic heads ($EO_x$) were long enough to interconnect the neighboring mesopores. After the removal of the template, micropores and even small mesopores with diameters between 2 and 3 nm could be observed between the ordered pores of the mesostructure [*Imperor-Clerc et al.*, 2000; *Kruk et al.*, 2000]. Stucky and coworkers called this family of materials SBA, whose most representative exponent is SBA-15, which shows a hexagonally ordered structure.

## 3.3 Mesoporous Materials.

Mesoporous materials are obtained from hybrid materials by removing the surfactant template. The inorganic framework obtained presents a porous structure with a cubic, hexagonal, or lamellar order. It is very interesting to note that prior to the removal of the template, the mesostructures are still conformationally functional [*Fryxell*, 2006]. In other words, a hexagonal phase can still be converted to a cubic phase, or a lamellar phase to a hexagonal phase, by properly tuning the reaction conditions towards the desired mesostructure. Like the microporous crystalline zeolites, the mesoporous materials also have a very large surface area, ordered pore systems, and very narrow pore size distributions [*Corma*, 1997]. Nevertheless, they differ from zeolites because of two main reasons. Their pores are bigger and their walls are not crystalline, although this last affirmation is not completely true for a particular class of mesoporous materials which will be analyzed later on.

The removal of the organic template is usually performed by calcination or solvent extraction. The first way is cheaper, but can lead to structural problems depending on the inorganic precursor used and on the eventual presence of functional groups. Silica is a very good precursor, being resistant and stable enough to maintain its amorphous structure at the temperatures of calcination (500 – 600°C). Many other precursors could crystallize at these temperatures, and this would create several defects in the material because of the strong curvature of the pores, being best compatible with amorphous wall structures [*Schüth*, 2003]. If functional organic groups are present, then the calcination becomes very risky to perform because the organic group could be destroyed by the high temperatures. In this case, a solvent extraction is preferred, unless the material is functionalized after the removal of the template, that is, by grafting the organic groups into the channels walls during a post-synthesis process.

A post-synthesis is an alternative to the co-condensation (direct synthesis) to obtain organically functionalized mesoporous materials. Figure 3.3 is a schematic representation of the channel walls of a hexagonally ordered mesoporous material when an organosilica precursor is grafted into the mesopores of a pure silica structure. In this case, the first synthesis is performed with a pure silica precursor, such as TEOS or TMOS, and, once the template has been removed by calcination, an organosilane of the type R-Si(EtO)$_3$ is grafted onto the walls of the pore during the second synthesis.

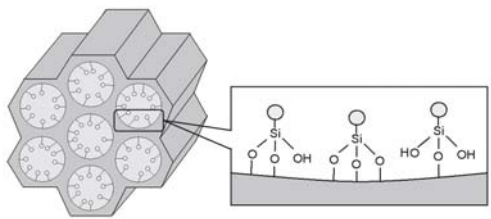

**Fig. 3.3**. Organization of the functionalizing organic groups grafted onto the channel walls of the mesopores. The light circles represent the functional organic group.

This procedure presents the advantage of preserving the functionalities from the high temperatures reached during the calcination, since they are added afterwards, but it also presents some problems related to the distribution of the organosilanes inside the pores. In particular, if such groups react preferentially at the pore opening at the beginning of the post-synthesis, then the diffusion of further molecules into the center of the pores can be affected and even give rise to the complete obstruction of the pores [*Hatton et al.*, 2005; *Hoffmann et al.*, 2006]. Therefore, this synthesis is more suitable when the mesopores are relatively large when compared to the organic functionalities to graft, namely when long block copolymers are used as templates.

However, as previously observed, the use of block copolymers usually leads to the formation of microporous channels connecting the mesopores [*Imperor-Clerc et al.*, 2000]. This microporosity is actually not always required and can even represent a problem for many applications in nanotechnology and for studies on the behavior of matter in confined spaces. This means that if we use a block copolymer as a template to obtain bigger pores in which organosilanes can be more homogenously grafted, then we have to think how to treat the inorganic framework to eliminate or reduce the degree of microporosity. The addition of salt, namely sodium chloride, during the synthesis reduces the microporosity [*Newalkar and Komarneni*, 2001].

An alternative way to overcome this problem is suggested in Figure 3.4, where the addition of the organic functionalities is performed thanks to a co-condensation or one-pot synthesis.

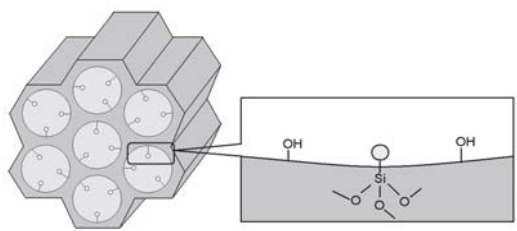

**Fig. 3.4**. Organization of the functionalizing organic groups in the channel walls of the mesopores as a result of a co-condensation with a terminal organosilica precursor. The light circles represent the functional organic group.

The template, the pure silica precursor, and the organosilane with a terminal organic group, are added simultaneously to the solution. The final structure presents the organic functionalities covalently bonded to the pore walls, and the pore blocking should be easily prevented by properly choosing the organic group. Moreover, the distribution of the organic groups into the pores should be much more homogenous than in the post-synthesis method. From this point of view, the co-condensation presents a very important advantage over the grafting process [*Hatton et al.*, 2005].

However, as we have already underlined above, the presence of the organic groups at this stage of the synthesis, almost completely prevents the possibility to perform a calcination, and a solvent extraction seems to be the only choice to remove the template. This option would not be too bad if the concentration of the precursor bringing the functionality was high enough to give rise to a high content of organic units attached to the pores. This would compensate the higher cost of a solvent extraction compared to the calcination. Unfortunately, this is not what happens, as the degree of order of the products decreases by increasing the concentration of the organosilicas, which can even lead to a totally disordered phase [*Hoffmann et al.*, 2006]. This is due to the fact that the organosilanes with terminal organic groups have to be co-assembled with TEOS to form a stable periodic mesoporous structure, limiting the organic content of the material to around 25% with respect to the silicon wall sites [*Hatton et al.*, 2005]. Therefore, if the concentration of the organosilica precursor is increased, the concentration of TEOS should increase as well, and a silica condensation into an amorphous phase becomes the most probable scenario.

Between 1999 and 2000, three different research groups published an innovative way to incorporate organic groups into the mesoporous structure [*Asefa et al.*, 2000; *Inagaki et al.*, 1999; *Melde et al.*, 1999]. This new class of mesoporous materials was called periodic mesoporous organosilicas (PMOs), and presented some interesting differences with respect to the functionalized mesoporous materials obtained with a direct synthesis with terminal organosilica precursors, or with a post-synthesis. First of all, the organic functionality is not anymore inside the channels of the material, but completely incorporated in the walls. The organosilica precursor used is of the type $(EtO)_3$-Si-R-Si$(EtO)_3$, where the organic functionality is covalently bonded to two silica centers and it is part of the pore walls, as schematically illustrated in Figure 3.5.

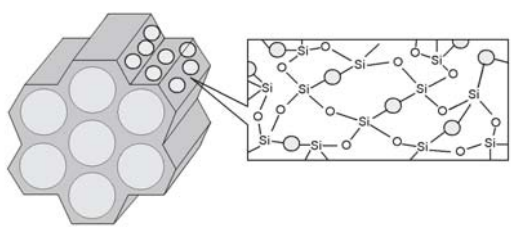

**Fig. 3.5**. Organization of the functionalizing organic groups in the channel walls of the mesopores as a result of a co-condensation with a bridging organosilica precursor. The light circles represent the functional organic group.

The use of such a precursor furnished several advantages: (1) since there is no need to co-assemble it with a second silica precursor, like TEOS, the distribution of the organic groups is more homogenous; (2) the pore blockage is strongly reduced or completely eliminated, although the load of organic groups is higher than with terminal organosilica precursors [*Liang et al.*, 2005]; (3) the pore walls can present a partial or complete crystalline order thanks to the higher flexibility given by the organic groups which permit the wall to follow the curvature of the cylindrical template [*Inagaki et al.*, 2002; *Yang et al.*, 1998]. The last feature, according to the functional organic group added to the framework, opens new potential applications as novel catalysts and advanced materials for sorption/separation processes [*Liang et al.*, 2005]. The accessibility of the organic functionalities has been proved by performing the bromination of the organic groups [*Asefa et al.*, 2000; *Melde et al.*, 1999].

### 3.3.1 Inorganic Functionalization and Non-siliceous Frameworks.

Organic functionalization is not the only modification that can be applied to mesoporous materials. The first inorganic functionalization was actually performed by the same group who patented these materials [*Beck et al.*, 1992; *Kresge et al.*, 1992]. They incorporated Al into the mesoporous framework by a metal doping approach to obtain aluminosilicates, which are still largely used for their catalysis properties: acid sites, in particular the Brønsted ones (from the tetrahedrally coordinated Al), are the active locus for most hydrocarbon reactions [*Zhao et al.*, 1996]. To maximize the presence of aluminium after the removal of the template, Khimyak and coworkers performed several solvent extractions in periodic mesoporous materials obtained by direct synthesis with silica precursors containing a bridging organic -$CH_2$-$CH_2$ group [*Hughes et al.*, 2005]. They proved that a much lower content of HCl in ethanol in comparison with the traditional extractions, was key to reach this aim.

Transition metals have also been incorporated directly into mesoporous silica, such as Nb [*Nowak et al.*, 2004]. The metal doping approach, although very easy to perform, leads to a quite random distribution of the metals throughout the wall structure, and therefore not all the metal catalyst may be found at the interface, where it should stay to be useful during catalysis. To overcome this problem, post-calcination metallization can be performed. In this case, the metals are added to the silica framework during a post-synthesis treatment, and their presence inside the walls can be completely prevented. However, the bond between the metal and the interface is not as strong as in the previous case, and leachability can become a serious problem [*Fryxell*, 2006]. Lanthanide elements, for instance, can be added by a post-synthesis treatment. These elements give the mesoporous material interesting photoluminescence properties [*Antochshuk et al.*, 2000].

Since the innovative fabrication of the first mesoporous materials, research groups have concentrated their efforts on the use of silica as inorganic precursor for the synthesis of ordered mesoporous materials. Some attempts were done to fabricate non-siliceous mesostructures in 1994, but the results were not completely satisfactory with the oxides selected (Pb) as the template could not be removed without structural collapse of the framework and then no mesoporous materials, but rather just mesostructures were able to be synthesized [*Huo et al.*, 1994]. This was mainly due to three factors: (1) the framework was not fully condensed, but rather a salt-like structure, and the removal of the template caused its collapse [*Stein et al.*, 1995]; (2) the non-siliceous materials were redox-unstable [*Schüth*, 2003]; or (3) the curvature of the pore walls imposed by the template was not compatible with the crystallization of the non-siliceous precursor, being more unstable than silica during calcination [*Schüth*, 2001].

Some advances permitted to overcome these problems [*Ciesla et al.*, 1994], especially by introducing a post-synthesis nanocasting approach by means of which a negative replica of the siliceous framework was created [*Ryoo et al.*, 1999]. This way was practicable only for those materials whose mesopores were interconnected, namely for three-dimensional structures, otherwise a simple bunch of independent *spaghetti* were obtained. Stable mesoporous structures were synthesized by using Nb oxides [*Antonelli and Ying*, 1996b], Ta oxides [*Antonelli and Ying*, 1996a], or Mn oxides [*Tian et al.*, 1997]. Such non-siliceous mesoporous materials have important features in many applications, such as in catalysis as ion-exchangers or supports for the hydrocarbon re-forming processes. Despite this fact, research studies in the field of the synthesis of non-siliceous mesostructures are still not as developed as for the synthesis of siliceous mesostructures. One reason is that many research groups working on mesostructures have a background in zeolite chemistry and therefore are very familiar with the chemistry of silicon and aluminum, but much less with the chemistry of other elements, as suggested by Schüth [*Schüth*, 2001]. It is also true that siliceous mesoporous materials are much more stable and cheaper than the non-siliceous ones, and moreover present larger surface areas and higher porosity. Therefore, at the moment, siliceous mesoporous materials are still considered more attractive. Nevertheless, taking into account how large the range of properties covered by non-siliceous materials is, it can be expected that many interesting applications will be soon accessible at large industrial scales.

## 3.4 Characterization

A satisfactory characterization of hexagonally ordered structures requires at least three independent experimental techniques [*Ciesla and Schüth*, 1999]: X-ray diffraction (XRD), transmission electron microscopy (TEM), and adsorption analysis. These techniques have also been combined with others, such as electron diffraction patterns [*Kresge et al.*, 1992], Fourier transform infrared spectroscopy [*Chen et al.*, 1993], and thermogravimetric analysis [*Zelenák et al.*, 2006].

In Figures 3.6 and 3.7, the results of a TGA and FTIR spectroscopy on a sample of SBA-15 made using pluronic P123 as structure directing agent are given. The two methods of characterization have been performed at the Department of Inorganic Chemistry, University of Košice, Slovak Republic, under the supervision of Professor Zeleňák.

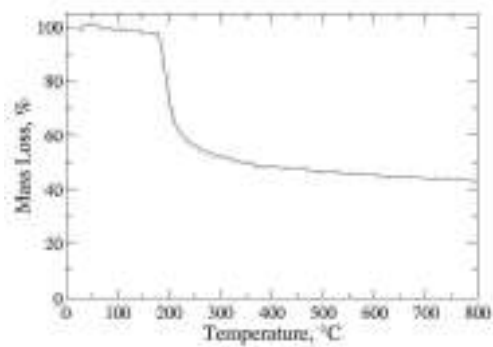

**Fig. 3.6**. Thermogravimetric analysis of an as-synthesized SBA-15 made using P123. The heating rate was 9°C/min. Courtesy of Vladimír Zeleňák, University of Košice, Slovak Republic.

During a TGA, the weight loss of a given material is measured as a function of the temperature. As the material is heated, the mass decreases because of the evaporation of the physisorbed water, of the water generated by the condensation of part of free silanols, and for the decomposition of the block-copolymer into $CO_2$ and CO. In our experiment, performed by using 10 mg of as-synthesized SBA-15 made using P123, approximately 50% of the weight loss is observed between 180°C and 440°C. At the end of the experiment, the sample presents less than half of its original weight.

The FTIR spectroscopy measurements performed on a as-synthesized and calcined sample of SBA-15 is shown in Figure 3.7.

At high wave numbers (above 3400 cm$^{-1}$), the bands referring to the O-H vibrations of the silanol groups dominate, whereas the vibrations of Si-O-Si are observed at lower wave numbers (below 1000 cm$^{-1}$). The bands around 3000 cm$^{-1}$ and 1300-1400 cm$^{-1}$ are due to the C-H stretching and bending vibrations, respectively. After calcination, such peaks are drastically reduced or eliminated, because the organic template has been removed.

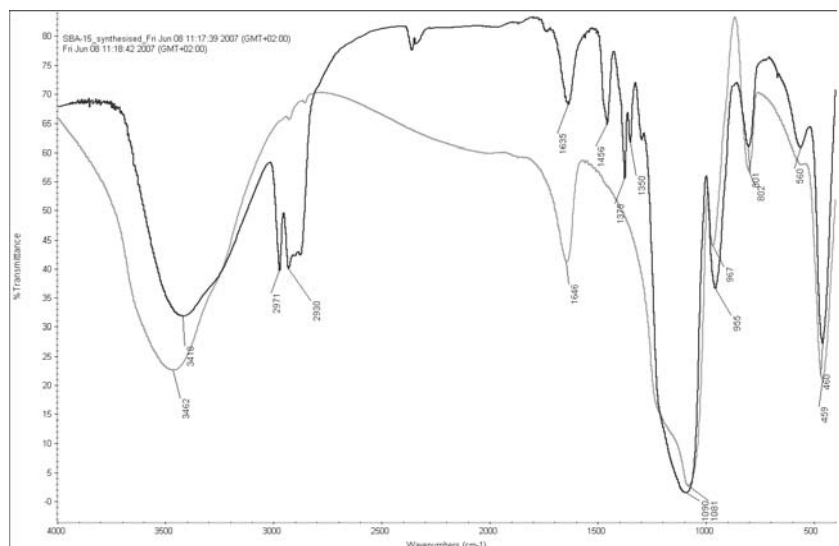

**Fig. 3.7**. Fourier transform infrared spectroscopy (FTIR) of an as-synthesized (dark curve) and a calcined (light curve) sample of SBA-15 made using P123. Courtesy of Vladimir Zeleňák, University of Košice, Slovak Republic.

The XRD pattern of MCM-41 reported in the work by Beck *et al.* shows 3-5 peaks between 2θ=2° and 6°, but samples presenting more peaks out of this range were also observed [*Edler and White*, 1997]. These peaks are produced because of the ordered hexagonal arrangement of the silica channels and are commonly named as (100), (110), (200), (210), and (300), being the directions on the x, y, and z axis, respectively. More often, only three or four well resolved peaks are detected, and sometimes even one peak is considered enough to prove the presence of a hexagonally array. The SBA-15 materials synthesized by the group of Stucky, show peaks at lower angles, namely between 2θ=1° and 3° [*Zhao et al.*, 1998]. XRD is not able to quantify the purity of the materials: even with many structural defects it is possible to obtain a pattern which could be easily associated with a hexagonal symmetry [*Schacht et al.*, 1998]. Therefore, to clarify the pore structure of these mesoporous materials, TEM is used, but additional techniques are needed to quantify the pore size and the wall thickness.

Adsorption of nitrogen, oxygen, or argon, has been usually performed to determine the surface area and to characterize the pore size distribution of solid catalysts [*Branton et al.*, 1993; *Schmidt et al.*, 1995]. The typical nitrogen isotherm at 77 K for MCM-41 with pore diameter around 4 nm, is a type IV isotherm, and shows two distinctive features: a sharp step at a relative pressure $0.41 < P/P_0 < 0.46$, and the absence of adsorption hysteresis [*Branton et al.*, 1993]. However, by modifying the temperature [*Rathousky et al.*, 1995], the adsorbate [*Branton et al.*, 1994], or the pore size [*Llewellyn et al.*, 1994; *Morishige et al.*, 1997], it is easy to observe hysteresis loops which are also theoretically predicted by the non-local density functional theory (NLDFT) applied to the study of nitrogen in confined media [*Ravikovitch et al.*, 1995]. The agreement between theory and experimental results support the physical scenario of capillary condensation in an open cylindrical capillary [*Ciesla and Schüth*, 1999].

For the calculation of the pore size distributions in cylindrical pores, different techniques can be used, and usually nitrogen and argon, adsorbed at low temperatures, are the most used probes for such characterizations. Kruk and coworkers performed nitrogen adsorption measurements at several relative pressures for different MCM-41 samples templated by surfactants of variable chain length. They showed that the pore size increases with the chain length of the template, compared different methods to calculate it in MCM-41, and also proposed a new procedure based on geometrical considerations [*Kruk et al.*, 1997]. These considerations led to the calculation of the diameter of the pore by considering the ratio of the pore volume to the pore wall volume in an infinite array of hexagonally ordered cylindrical aggregates. The value of the lattice spacing from XRD measurements is also needed.

A different approach is based on the NLDFT, which can be used for microporous and mesoporous materials. Ravikovitch and coworkers applied the NLDFT to study equilibrium phase transitions in cylindrical channels of MCM-41, and proved that the theory is able to properly describe the thermodynamics of nitrogen in confined mesopores, and to predict the thermodynamics limits for the adsorption-desorption hysteresis loop [*Ravikovitch et al.*, 1995]. The same researchers, some years later, tested the theory against Monte Carlo simulations for a system with Lennard-Jones fluids adsorbed in slit-shaped and cylindrical pores ranging from the microporous to the mesoporous region. They showed that the NLDFT, with a proper selection of the intermolecular interaction parameters, is able to quantitavely predict the structure of confined fluids at the solid surface and in pores, adsorption isotherms, the conditions for phase equilibrium and spinodal transitions [*Ravikovitch et al.*, 2001].

The most common procedure applied in the calculation of the pore-size distributions in mesoporous materials is the Barrett-Joyner-Halenda (BJH) method, which is based on the Kelvin equation [*Barrett et al.*, 1951]. Barrett and coworkers used nitrogen isotherms to determine the pore volume and the area distributions in porous solids. They showed that these distributions can be related to the desorption isotherms with respect to the pore radius if the equilibrium between the gas phase and the adsorbed phase is assumed to depend on two mechanisms: (1) physical adsorption on the pore walls, and (2) capillary condensation in the inner capillary volume. The BJH method overestimates the relative pressure at desorption, and underestimates the pore diameters by approximately 1 nm [*Ravikovitch et al.*, 1995].

The calculation of the wall thickness can be performed by determining the difference between the lattice parameter obtained by XRD and the pore size obtained by nitrogen adsorption measurements [*Ciesla and Schüth*, 1999]. Studies published by different researchers proved that the wall thickness remains constant at a value of around 1 nm, over a quite broad range of pore diameters, that is between 2.5 to 10 nm [*Beck et al.*, 1992; *Corma et al.*, 1997]. The pore walls are characterized by focusing on the structural properties and on the surface chemistry. The most established technique for this characterization is solid-state nuclear magnetic resonance (NMR) spectroscopy, which is able to determine the local atomic ordering of the amorphous walls [*Zhao et al.*, 2002b]. The solid surfaces are also analyzed by the adsorption of polar and unpolar molecules in order to establish their hydrophilic or hydrophobic properties. As a matter of fact, the relative hydrophobic nature of MCM-41 was proved by the adsorption of several substances, such as benzene [*Beck et al.*, 1992] and water [*Llewellyn et al.*, 1995].

The combination of all the results from different characterization methods furnished two structural models with amorphous walls for MCM-41 (Figure 3.8): one, obtained from a molecular dynamics simulation, presents a cylindrical pore structure [*Feuston and Higgins*, 1994]; the other shows a hexagonally shaped pore structure [*Behrens and Stucky*, 2003]. Interestingly, both of them have been observed through TEM [*Chenite et al.*, 1995; *Walker and Zasadzinski*, 1995].

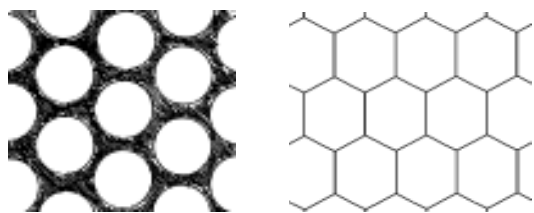

**Fig. 3.8**. Cylindrical (left) and hexagonal (right) pore structures, according to the results from the characterization techniques for MCM-41.

## 3.5 Applications.

The structural characteristics of mesoporous materials are key to their fields of application. They present very large surface areas, high porosity, and narrow pore size distribution. Generally, the functionalization of the channel walls is a necessary requisite for the material to be used in some given applications, such as catalysis or adsorption. The combination of the structural properties and the possibility to be functionalized, are key factors in adsorption, catalysis, separation, optics.

Mesoporous materials have been deeply analyzed as selective adsorbents of gases, liquids, and metals. The functionalized mesoporous silica materials are largely used as adsorbents to remove trace amounts of mercury and cadmium in various gas streams [*Luan et al.*, 2005], and with thiol- and amine-functionalizing groups they have been widely used in the removal of heavy metal ions from waste water [*Liu et al.*, 2000].

In many cases, these adsorbents can be regenerated by washing with alkaline or acid solution to recover both the adsorbents and the adsorbent dyes, like in the case of adsorbents prepared by grafting amino- and carboxylic-functional groups onto MCM-41 for the removal of blue dyes from wastewater [*Ho et al.*, 2003]. Mesoporous silica materials can also be used in the selective adsorption/desorption of biomolecules, such as proteins [*Yiu et al.*, 2001].

Pure silica mesoporous materials are electrically neutral and their catalytic activity is quite limited. For this reason, most of the interest has been focused on aluminum [*Corma et al.*, 1995], titanium [*Morey et al.*, 1996], zirconium [*Tuel et al.*, 1996], vanadium [*Shylesh and Singh*, 2005], cobalt [*Ghattas*, 2006], mesoporous silica materials. Trivalent atoms, such as $Al^{3+}$ or $Fe^{3+}$, can be incorporated onto the channel walls of MCM-41 instead of silicon atoms by leaving a negative charged framework [*Tuel and Gontier*, 1996]. Such a negative charge can be balanced by those protons providing the acid sites for the catalysis. Depending on the nature of the metal ions, the number of acid sites can vary. If the silicon atoms are substituted with tetravalent metal ions, such as $Ti^{4+}$, $V^{4+}$, or $Zr^{4+}$, the resulting material can be used in the oxidation of organic compounds [*Luan et al.*, 1996].

Mesoporous structures containing aluminum, are often used as supports for acidic catalytic reaction, however, for reactions requiring a very high acidity, then one of the other aforementioned structures with metal ions is more suitable [*Ciesla and Schüth*, 1999]. The weakness of the acid sites decreases the application of mesoporous materials in many petrochemical reactions, although their big pore size could help the diffusion of bulky substances often found in fine chemicals synthesis [*Taguchi and Schüth*, 2005].

A very hydrothermal stable MCM-41 material containing palladium has been used in the low-temperature catalytic combustion of CO for the treatment of the cold-start emissions in automobiles, showing a very good catalytic activity [*Liu et al.*, 2005]. A mesoporous silica material containing tungsten has been used as catalyst in the allylic oxidation of cyclohexene in the presence of hydrogen peroxide [*Ke and Liu*, 2007]. It was shown that this catalyst can be recovered and re-used after the reaction, preserving almost completely its original activity.

The large inner surface area of mesoporous materials and the possibility to functionalize their pore walls are of very interesting applications in the area of sorption. Liu and coworkers modified the surface of SBA-15 with triethanolamine (TEA) to obtain an adsorbent to separate $CO_2$ from its mixture with $CH_4$ and other gases [*Liu et al.*, 2007]. The modification with TEA does not change the structural order of the material, but enhances its selectivity for $CO_2$, by increasing of seven times the separation coefficient between $CO_2$ and $CH_4$. Moreover, the presence of TEA allows to regenerate the saturated material at ambient temperature, by keeping the stability and the good separation performance of the material.

Xu *et al.* studied the adsorption of $CO_2$ in an MCM-41 material modified with polyethylenimine (PEI) [*Xu et al.*, 2003]. They obtained different results according to the synthesis performed to functionalize the MCM-41 with PEI: they showed that a co-condensation gives better results, although the ones obtained with a post-synthesis are still very good as the adsorption capacity, in this case, is around thirty times higher than that of a non-functionalized MCM-41. They have also showed that by increasing the Si/Al ratio of MCM-41, the adsorption capacity of $CO_2$ increases as well, and it can increase even more if polyethylene glycol is added.

Several publications over the last years have treated mesoporous solids as ion conductors for batteries or fuel cells [*Mamak et al.*, 2002], electron conductors [*Emons et al.*, 2002], or as sensors [*Yamada et al.*, 2002]. In particular, Li and Nogami showed the importance of a cubic mesostructure for proton conductivity, which increases by increasing the water content in the pore. The conductivity in cylindrical pores was, on the contrary, very low and independent of the water content [*Li and Nogami*, 2002]. Zhou and coworkers published several papers where mesoporous materials were used in sensor applications [*Yamada et al.*, 2002].

Moreover, the ordered channels of mesoporous structures can provide an excellent confined space for the control of the size and shape of nanoscale materials. Huang *et al.* succeeded in synthesizing silver nanowires inside the channels of SBA-15 [*Huang et al.*, 2000]. After this, the group of Stucky employed the same ordered material to fabricate palladium and gold nanowires.

In high performance liquid chromatography (HPLC), SBA-15 with particularly large pores (templated by a $C_{18}$ surfactant) can be used to separate peptides and proteins [*Zhao et al.*, 2002a].

### 3.6  Modeling Hybrid and Mesoporous Materials.

The first attempt to simulate the formation of mesostructures by using surfactants as templates is due to Bhattacharya and Mahanti [*Bhattacharya and Mahanti*, 2001]. They carried out off-lattice MC simulations in two dimensions (2D) to study the self-assembly of ionic surfactants in the presence of neutral host particles added to pre-formed micelles or directly to the surfactant solution. The dependence of the final structures on the size, interaction strengths and concentration of the host particles was studied. They showed that host particles with purely repulsive interactions with the surfactant, decrease the effective available volume and change the effective concentration and size distribution of micelles, and explained the ordering of host particles by excluded volume effects. They further showed that when the host particle interacts with the surfactant, no ordered structures are formed. It should be pointed out that the solvent was not taken into account in these simulations, and may be the reason for the absence of the formation of ordered structures.

The study of Siperstein and Gubbins represented the first simulations in 3D of a water-like surfactant solution including an inorganic precursor [*Siperstein and Gubbins*, 2003]. In this work, lattice MC simulations were used to analyze the phase behavior and the structural order of a surfactant-inorganic oxide-solvent system in which the inorganic oxide and the solvent were modeled by one single site, and the surfactants by several connected sites. Siperstein and Gubbins showed that the formation of ordered liquid crystal phases was observed at high surfactant concentrations (above 40%), and their structures depended on the system composition and strength of the interactions. They reported phase diagrams for systems where partial and complete miscibility between the inorganic oxide and the solvent existed. In both cases, they showed the existence of an immiscibility gap in the ternary system, where hexagonal and lamellar structures were formed in a phase with high surfactant concentration. Reasonable agreement with experimental evidence was observed, as lamellar phases were obtained for low silica/surfactant ratios, and hexagonal phases for high silica/surfactant ratios. A similar approach has recently been used to model the formation of mesocellular foams [*Bhattacharya and Gubbins*, 2005] and silica nanoparticles [*Jorge et al.*, 2005].

Other attempts to model the formation of mesoporous materials using kinetic Monte Carlo (kMC) [*Schumacher et al.*, 2004] and molecular dynamics (MD) [*Coasne and Pellenq*, 2004; *Feuston and Higgins*, 1994; *Sonwane et al.*, 2005] have focused on the silica condensation or mainly on the surface relaxation, assuming the geometry of the mesostructure. kMC and MD studies have the advantage of providing a detailed description of the material, although they are not useful when the question that concerns us is the ability of the system to form ordered phases. Using atomistic simulations where the surfactants, solvent and silica sources are explicitly described, would be too computationally expensive to reach any meaningful result with current computing capabilities.

Schumacher and Seaton proposed a method to model hybrid organic-inorganic adsorbents based on pure-silica mesostructures generated by kinetic Monte Carlo (kMC) simulations of the synthesis [*Schumacher et al.*, 2006]. By using a simplified representation of a templating cylindrical micelle, they defined the interaction potentials for the formation of a silica layer around such aggregates by following the reaction path of the hydrothermal synthesis and calcination. The amorphous pore walls generated seem to be realistic and the isotherms obtained from the adsorption of ethane and carbon dioxide in samples of organic-modified MCM-41 show a very good agreement with experiments, although the degree of condensation of silica inside the pore wall is slightly lower than the experimental value. [*Schumacher and Seaton*, 2005].

The adsorption of some complex hydrocarbon molecules has been modeled by Fox and Bates, who studied the behavior of linear, branched, and cyclic hydrocarbons, and mixtures of them in model pores of MCM-41 by configurational bias Monte Carlo simulations. The material is modeled as a rigid framework of connected oxygen atoms interacting with the hydrocarbon chains through a Lennard-Jones potential, and the guest chains are modeled as sequences of $CH$, $CH_2$, or $CH_3$ pseudo-atoms. Their model pores present a hexagonal shape that can be easily modified in order to include different degrees of roughness [*Fox and Bates*, 2005]. They obtained good agreement with experiments for isotherms calculated for hexane, while branched and cyclic hydrocarbon experimental isotherms were not available for comparison. The effect of roughness was found to affect only the low-pressure adsorption, leaving unaffected the high-pressure adsorption, in agreement with previous works including much smaller chains [*He and Seaton*, 2003].

The role of surface roughness was also considered by other researchers [*Coasne et al.*, 2006]. Coasne and coworkers used two atomistic model pores of MCM-41. One presents a regular cylindrical section, whereas the other reproduces the morphology of a mesopore obtained from coarse-grained simulations which mimics the synthesis of a mesoporous material [*Siperstein and Gubbins*, 2003]. Both models are obtained by carving the pore out of a silica block, and the interface was then modeled by using an atomistic approach. Adsorption isotherms, small angle neutron scattering spectra, and isosteric heat curves are used to compare the results from the two pores. They found that the model pore obtained from the self-assembly of MCM-41 is too rough at a molecular scale, but it efficiently reproduces the surface order at a mesoscale length. The other model pore shows the opposite behavior: it is smooth at the molecular scale, but too regular at the mesoscale.

# References Chapter 3

Antochshuk, V., A. S. Araujo, and M. Jaroniec, Functionalized MCM-41 and CeMCM-41 materials synthesized via interfacial reactions, *Journal of Physical Chemistry B*, 104, 9713-9719, 2000.

Antonelli, D. M., and J. Y. Ying, Synthesis and characterization of hexagonally packed mesoporous tantalum oxide molecular sieves, *Chemistry of Materials*, 8, 874-881, 1996a.

Antonelli, D. M., and J. Y. Ying, Synthesis of a stable hexagonally packed mesoporous niobium oxide molecular sieve through a novel ligand-assisted templating mechanism, *Angewandte Chemie-International Edition in English*, 35, 426-430, 1996b.

Asefa, T., N. Coombs, O. Dag, C. Yoshina-Ishii, M. J. MacLachlan, and G. A. Ozin, Periodic mesoporous organosilicas (PMOs) with functional organic groups inside the channel walls., *Abstracts of Papers of the American Chemical Society*, 219, U883-U883, 2000.

Attard, G. S., J. C. Glyde, and C. G. Goltner, Liquid-Crystalline Phases as Templates for the Synthesis of Mesoporous Silica, *Nature*, 378, 366-368, 1995.

Avnir, D., D. Levy, and R. Reisfeld, The Nature of the Silica Cage as Reflected by Spectral Changes and Enhanced Photostability of Trapped Rhodamine-6g, *Journal of Physical Chemistry*, 88, 5956-5959, 1984.

Barrett, E. P., L. G. Joyner, and P. P. Halenda, The Determination of Pore Volume and Area Distributions in Porous Substances .1. Computations from Nitrogen Isotherms, *Journal of the American Chemical Society*, 73, 373-380, 1951.

Beck, J. S., and J. C. Vartuli, Recent advances in the synthesis, characterization and applications of mesoporous molecular sieves, *Current Opinion in Solid State & Materials Science*, 1, 76-87, 1996.

Beck, J. S., J. C. Vartuli, W. J. Roth, M. E. Leonowicz, C. T. Kresge, K. D. Schmitt, C. T. W. Chu, D. H. Olson, E. W. Sheppard, S. B. Mccullen, J. B. Higgins, and J. L. Schlenker, A New Family of Mesoporous Molecular-Sieves Prepared with Liquid-Crystal Templates, *Journal of the American Chemical Society*, 114, 10834-10843, 1992.

Behrens, P., and G. Stucky, Ordered Molecular Arrays as Templates: A New Approach to the Synthesis of Mesoporous Materials, *Angew Chem Int Ed Engl*, 32, 696-699, 2003.

Bensaude-Vincent, B., The Construction of a Discipline: Materials Science in the U.S.A, *Historical Studies in the Physical and Biological Sciences*, 31, part 2, 223-248, 2001.

Bhattacharya, A., and S. D. Mahanti, Self-assembly of ionic surfactants and formation of mesostructures, *Journal of Physics-Condensed Matter*, 13, 1413-1428, 2001.

Bhattacharya, S., and K. E. Gubbins, Modeling triblock surfactant-templated mesostructured cellular foams, *Journal of Chemical Physics*, 123, -, 2005.

Branton, P. J., P. G. Hall, and K. S. W. Sing, Physisorption of Nitrogen and Oxygen by Mcm-41, a Model Mesoporous Adsorbent, *Journal of the Chemical Society-Chemical Communications*, 1257-1258, 1993.

Branton, P. J., P. G. Hall, K. S. W. Sing, H. Reichert, F. Schüth, and K. K. Unger, Physisorption of Argon, Nitrogen and Oxygen by Mcm-41, a Model Mesoporous Adsorbent, *Journal of the Chemical Society-Faraday Transactions*, 90, 2965-2967, 1994.

Chen, C., H. Li, and M. Davis, Studies on mesoporous materials. I. Synthesis and characterization of MCM-41, *Microporous Materials*, 2, 17-26, 1993.

Chenite, A., Y. Le Page, and A. Sayari, Direct TEM Imaging of Tubules in Calcined MCM-41 Type Mesoporous Materials, *Chem. Mater.*, 7, 1015-1019, 1995.

Chiola, V., J. E. Ritsko, and C. D. Vanderpool, US Patent 3556725, 1971.

Ciesla, U., D. Demuth, R. Leon, P. Petroff, G. Stucky, K. Unger, and F. Schuth, Surfactant Controlled Preparation of Mesostructured Transition-Metal Oxide Compounds, *Journal of the Chemical Society-Chemical Communications*, 1387-1388, 1994.

Ciesla, U., and F. Schüth, Ordered mesoporous materials, *Microporous and Mesoporous Materials*, 27, 131-149, 1999.

Coasne, B., F. R. Hung, R. J. M. Pellenq, F. R. Siperstein, and K. E. Gubbins, Adsorption of sample gases in MCM-41 materials: The role of surface roughness, *Langmuir*, 22, 194-202, 2006.

Coasne, B., and R. J. M. Pellenq, Grand canonical Monte Carlo simulation of argon adsorption at the surface of silica nanopores: Effect of pore size, pore morphology, and surface roughness, *Journal of Chemical Physics*, 120, 2913-2922, 2004.

Corma, A., From microporous to mesoporous molecular sieve materials and their use in catalysis, *Chemical Reviews*, 97, 2373-2419, 1997.

Corma, A., Q. B. Kan, M. T. Navarro, J. PerezPariente, and F. Rey, Synthesis of MCM-41 with different pore diameters without addition of auxiliary organics, *Chemistry of Materials*, 9, 2123-2126, 1997.

Corma, A., A. Martinez, V. Martinezsoria, and J. B. Monton, Hydrocracking of Vacuum Gas-Oil on the Novel Mesoporous Mcm-41 Aluminosilicate Catalyst, *Journal of Catalysis*, 153, 25-31, 1995.

Di Renzo, F., H. Cambon, and R. Dutartre, A 28-year-old synthesis of micelle-templated mesoporous silica, *Microporous Materials*, 10, 283-286, 1997.

Edler, K. J., Soap and sand: construction tools for nanotechnology, *Phil. Trans. R. Soc. Lond. A*, 2635-2651, 2004.

Edler, K. J., and J. W. White, Further improvements in the long-range order of MCM-41 materials, *Chemistry of Materials*, 9, 1226-1233, 1997.

Emons, T. T., J. Q. Li, and L. F. Nazar, Synthesis and characterization of mesoporous indium tin oxide possessing an electronically conductive framework, *Journal of the American Chemical Society*, 124, 8516-8517, 2002.

Feuston, B. P., and J. B. Higgins, Model Structures for Mcm-41 Materials - a Molecular-Dynamics Simulation, *Journal of Physical Chemistry*, 98, 4459-4462, 1994.

Fox, J. P., and S. P. Bates, Adsorption and structure of hydrocarbons in MCM-41: a computational study, *Langmuir*, 21, 4746-4754, 2005.

Fryxell, G. E., The synthesis of functional mesoporous materials, *Inorganic Chemistry Communications*, 9, 1141-1150, 2006.

Ghattas, M. S., Cobalt-modified mesoporous FSM-16 silica: Characterization and catalytic study, *Microporous and Mesoporous Materials*, 97, 107-113, 2006.

Hatton, B., K. Landskron, W. Whitnall, D. Perovic, and G. A. Ozin, Past, present, and future of periodic mesoporous organosilicas - The PMOs, *Accounts of Chemical Research*, 38, 305-312, 2005.

He, Y. F., and N. A. Seaton, Experimental and computer simulation studies of the adsorption of ethane, carbon dioxide, and their binary mixtures in MCM-41, *Langmuir*, 19, 10132-10138, 2003.

Ho, K. Y., G. McKay, and K. L. Yeung, Selective adsorbents from ordered mesoporous silica, *Langmuir*, 19, 3019-3024, 2003.

Hoffmann, F., M. Cornelius, J. Morell, and M. Froba, Silica-based mesoporous organic-inorganic hybrid materials, *Angew Chem Int Ed Engl*, 45, 3216-3251, 2006.

Huang, M. H., A. Choudrey, and P. D. Yang, Ag nanowire formation within mesoporous silica, *Chemical Communications*, 1063-1064, 2000.

Hughes, B. J., J.-B. Guilbaud, M. Allix, and Y. Z. Khimyak, Synthesis of periodic mesoporous organosilicas with incorporated aluminium, *Journal of Materials Chemistry*, 15, 4728-4733, 2005.

Huo, Q. S., D. I. Margolese, U. Ciesla, P. Y. Feng, T. E. Gier, P. Sieger, R. Leon, P. M. Petroff, F. Schuth, and G. D. Stucky, Generalized Synthesis of Periodic Surfactant Inorganic Composite-Materials, *Nature*, 368, 317-321, 1994.

Imperor-Clerc, M., P. Davidson, and A. Davidson, Existence of a microporous corona around the mesopores of silica-based SBA-15 materials templated by triblock copolymers, *Journal of the American Chemical Society*, 122, 11925-11933, 2000.

Inagaki, S., S. Guan, Y. Fukushima, T. Ohsuna, and O. Terasaki, Novel mesoporous materials with a uniform distribution of organic groups and inorganic oxide in their frameworks, *Journal of the American Chemical Society*, 121, 9611-9614, 1999.

Inagaki, S., S. Guan, T. Ohsuna, and O. Terasaki, An ordered mesoporous organosilica hybrid material with a crystal-like wall structure, *Nature*, 416, 304-307, 2002.

Jorge, M., S. M. Auerbach, and P. A. Monson, Modeling spontaneous formation of precursor nanoparticles in clear-solution zeolite synthesis, *Journal of the American Chemical Society*, 127, 14388-14400, 2005.

Ke, I. S., and S. T. Liu, Synthesis and catalysis of tungsten oxide in hexagonal mesoporous silicas (W-HMS), *Applied Catalysis a-General*, 317, 91-96, 2007.

Kresge, C. T., M. E. Leonowicz, W. J. Roth, J. C. Vartuli, and J. S. Beck, Ordered Mesoporous Molecular-Sieves Synthesized by a Liquid-Crystal Template Mechanism, *Nature*, 359, 710-712, 1992.

Kruk, M., M. Jaroniec, C. H. Ko, and R. Ryoo, Characterization of the porous structure of SBA-15, *Chemistry of Materials*, 12, 1961-1968, 2000.

Kruk, M., M. Jaroniec, and A. Sayari, Adsorption study of surface and structural properties of MCM-41 materials of different pore sizes, *Journal of Physical Chemistry B*, 101, 583-589, 1997.

Lee, B., Y. Kim, H. Lee, and J. Yi, Synthesis of functionalized porous silicas via templating method as heavy metal ion adsorbents: the introduction of surface hydrophilicity onto the surface of adsorbents, *Microporous and Mesoporous Materials*, 50, 77-90, 2001.

Lei, J., J. Fan, C. Z. Yu, L. Y. Zhang, S. Y. Jiang, B. Tu, and D. Y. Zhao, Immobilization of enzymes in mesoporous materials: controlling the entrance to nanospace, *Microporous and Mesoporous Materials*, 73, 121-128, 2004.

Levy, D., S. Einhorn, and D. Avnir, Applications of the Sol-Gel Process for the Preparation of Photochromic Information-Recording Materials - Synthesis, Properties, Mechanisms, *Journal of Non-Crystalline Solids*, 113, 137-145, 1989.

Li, H. B., and M. Nogami, Pore-controlled proton conducting silica films, *Advanced Materials*, 14, 912-914, 2002.

Liang, Y., M. Hanzlik, and R. Anwander, Periodic mesoporous organosilicas: mesophases control *via* binary surfactant mixtures, *Journal of Materials Chemistry*, 16, 1238-1253, 2005.

Linden, M., P. Agren, S. Karlsson, P. Bussain, and H. Amenitsch, Solubilization of oil in silicate-surfactant mesostructures, *Langmuir*, 16, 5831-5836, 2000.

Linden, M., J. Blanchard, S. Schacht, S. A. Schunk, and F. Schüth, Phase behavior and wall formation in Zr(SO4)(2)/CTABr and TiOSO4/CTABr mesophases, *Chemistry of Materials*, 11, 3002-3008, 1999.

Linssen, T., K. Cassiers, P. Cool, and E. F. Vansant, Mesoporous templated silicates: an overview of their synthesis, catalytic activation and evaluation of the stability, *Advances in Colloid and Interface Science*, 103, 121-147, 2003.

Liu, A. M., K. Hidajat, S. Kawi, and D. Y. Zhao, A new class of hybrid mesoporous materials with functionalized organic monolayers for selective adsorption of heavy metal ions, *Chemical Communications*, 1145-1146, 2000.

Liu, S., L. D. Kong, X. W. Yan, Q. Z. Li, and A. He, Synthesis of hierarchically structured MCM-41 with high hydrothermal stability and its application in environmental Catalysis, *Nanoporous Materials Iv*, 156, 379-384, 2005.

Liu, X. W., L. Zhou, X. Fu, Y. Sun, W. Su, and Y. P. Zhou, Adsorption and regeneration study of the mesoporous adsorbent SBA-15 adapted to the capture/separation of CO2 and CH4, *Chemical Engineering Science*, 62, 1101-1110, 2007.

Llewellyn, P. L., Y. Grillet, F. Schüth, H. Reichert, and K. K. Unger, Effect of pore size on adsorbate condensation and hysteresis within a potential model adsorbent: M41S, *Microporous Materials*, 3, 345-349, 1994.

Llewellyn, P. L., F. Schüth, Y. Grillet, F. Rouquerol, J. Rouquerol, and K. K. Unger, Water Sorption on Mesoporous Aluminosilicate Mcm-41, *Langmuir*, 11, 574-577, 1995.

Luan, Z., J. A. Fournier, J. B. Wooten, D. E. Miser, and M. J. Chang, Functionalized mesoporous SBA-15 silica molecular sieves with mercaptopropyl groups: Preparation, characterization and application as adsorbents, *Nanoporous Materials Iv*, 156, 897-906, 2005.

Luan, Z. H., J. Xu, H. Y. He, J. Klinowski, and L. Kevan, Synthesis and spectroscopic characterization of vanadosilicate mesoporous MCM-41 molecular sieves, *Journal of Physical Chemistry*, 100, 19595-19602, 1996.

Mamak, M., N. Coombs, and G. A. Ozin, Practical solid oxide fuel cells with anodes derived from self-assembled mesoporous-NiO-YSZ, *Chemical Communications*, 2300-2301, 2002.

Melde, B. J., B. T. Holland, C. F. Blanford, and A. Stein, Mesoporous sieves with unified hybrid inorganic/organic frameworks, *Chemistry of Materials*, 11, 3302-3308, 1999.

Monnier, A., F. Schuth, Q. Huo, D. Kumar, D. Margolese, R. S. Maxwell, G. D. Stucky, M. Krishnamurty, P. Petroff, A. Firouzi, M. Janicke, and B. F. Chmelka, Cooperative Formation of Inorganic-Organic Interfaces in the Synthesis of Silicate Mesostructures, *Science*, 261, 1299-1303, 1993.

Morey, M., A. Davidson, and G. Stucky, A new step toward transition metal incorporation in cubic mesoporous materials: Preparation and characterization of Ti-MCM-48, *Microporous Materials*, 6, 99-104, 1996.

Morishige, K., H. Fujii, M. Uga, and D. Kinukawa, Capillary Critical Point of Argon, Nitrogen, Oxygen, Ethylene, and Carbon Dioxide in MCM-41, *Langmuir*, 13, 3494-3498, 1997.

Newalkar, B. L., H. Katsuki, and S. Komarneni, Microwave-hydrothermal synthesis and characterization of microporous-mesoporous disordered silica using mixed-micellar-templating approach, *Microporous and Mesoporous Materials*, 73, 161-170, 2004.

Newalkar, B. L., and S. Komarneni, Control over microporosity of ordered microporous-mesoporous silica SBA-15 framework under microwave-hydrothermal conditions: Effect of salt addition, *Chemistry of Materials*, 13, 4573-4579, 2001.

Nowak, I., M. Ziolek, and M. Jaroniec, Synthesis and characterization of polymer-templated mesoporous silicas containing niobium, *Journal of Physical Chemistry B*, 108, 3722-3727, 2004.

Palmqvist, A. E. C., Synthesis of ordered mesoporous materials using surfactant liquid crystals or micellar solutions, *Current Opinion in Colloid & Interface Science*, 8, 145-155, 2003.

Patarin, J., B. Lebeau, and R. Zana, Recent advances in the formation mechanisms of organized mesoporous materials, *Current Opinion in Colloid & Interface Science*, 7, 107-115, 2002.

Rathousky, J., A. Zukal, O. Franke, and G. Schulz-Ekloff, Adsorption on Mcm-41 Mesoporous Molecular-Sieves .2. Cyclopentane Isotherms and Their Temperature-Dependence, *Journal of the Chemical Society-Faraday Transactions*, 91, 937-940, 1995.

Ravikovitch, P. I., S. C. O'Domhnaill, A. V. Neimark, F. Schüth, and K. K. Unger, Capillary hysteresis in nanopores: Theoretical and experimental studies of nitrogen adsorption on MCM-41, *Langmuir*, 11, 4765-4772, 1995.

Ravikovitch, P. I., A. Vishnyakov, and A. V. Neimark, Density functional theories and molecular simulations of adsorption and phase transitions in nanopores, *Physical Review E*, 6401, -, 2001.

Reisfeld, R., Spectroscopy and Applications of Molecules in Glasses, *Journal of Non-Crystalline Solids*, 121, 254-266, 1990.

Ryoo, R., S. H. Joo, and S. Jun, Synthesis of highly ordered carbon molecular sieves via template-mediated structural transformation, *Journal of Physical Chemistry B*, 103, 7743-7746, 1999.

Sanchez, C., G. J. D. A. Soler-Illia, F. Ribot, T. Lalot, C. R. Mayer, and V. Cabuil, Designed hybrid organic-inorganic nanocomposites from functional nanobuilding blocks, *Chemistry of Materials*, 13, 3061-3083, 2001.

Sastre, G., A computational chemistry insight in the role of structure directing agents in the synthesis of zeolites, *Physical Chemistry Chemical Physics*, 9, 1052-1058, 2006.

Sayari, A., and S. Hamoudi, Periodic mesoporous silica-based organic - Inorganic nanocomposite materials, *Chemistry of Materials*, 13, 3151-3168, 2001.

Schacht, S., M. Janicke, and F. Schuth, Modeling X-ray patterns and TEM images of MCM-41, *Microporous and Mesoporous Materials*, 22, 485-493, 1998.

Schmidt, R., M. Stocker, E. Hansen, D. Akporiaye, and O. H. Ellestad, Mcm-41 - a Model System for Adsorption Studies on Mesoporous Materials, *Microporous Materials*, 3, 443-448, 1995.

Schumacher, C., J. Gonzalez, M. Perez-Mendoza, P. A. Wright, and N. A. Seaton, Modelling and experiment towards the design of mesoporous organic-inorganic hybrid adsorbents, *Recent Advances in the Science and Technology of Zeolites and Related Materials, Pts a - C*, 154, 386-393, 2004.

Schumacher, C., J. Gonzalez, M. Perez-Mendoza, P. A. Wright, and N. A. Seaton, Design of hybrid organic/inorganic adsorbents based on periodic mesoporous silica, *Industrial & Engineering Chemistry Research*, 45, 5586-5597, 2006.

Schumacher, C., and N. A. Seaton, Modeling of organically functionalized mesoporous silicas for the design of adsorbents, *Adsorption-Journal of the International Adsorption Society*, 11, 643-648, 2005.

Schüth, F., Non-siliceous mesostructured and mesoporous materials, *Chemistry of Materials*, 13, 3184-3195, 2001.

Schüth, F., Endo- and exotemplating to create high-surface-area inorganic materials, *Angewandte Chemie-International Edition*, 42, 3604-3622, 2003.

Shylesh, S., and A. P. Singh, Vanadium-containing ordered mesoporous silicates: Does the silica source really affect the catalytic activity, structural stability, and nature of vanadium sites in V-MCM-41?, *Journal of Catalysis*, 233, 359-371, 2005.

Sing, K. S. W., D. H. Everett, R. A. W. Haul, L. Moscou, R. A. Pierotti, J. Rouquerol, and T. Siemieniewska, Reporting Physisorption Data for Gas Solid Systems with Special Reference to the Determination of Surface-Area and Porosity (Recommendations 1984), *Pure and Applied Chemistry*, 57, 603-619, 1985.

Siperstein, F. R., and K. E. Gubbins, Phase separation and liquid crystal self-assembly in surfactant-inorganic-solvent systems, *Langmuir*, 19, 2049-2057, 2003.

Soler-illia, G. J. D., C. Sanchez, B. Lebeau, and J. Patarin, Chemical strategies to design textured materials: From microporous and mesoporous oxides to nanonetworks and hierarchical structures, *Chemical Reviews*, 102, 4093-4138, 2002.

Sonwane, C. G., C. W. Jones, and P. J. Ludovice, A model for the structure of MCM-41 incorporating surface roughness, *Journal of Physical Chemistry B*, 109, 23395-23404, 2005.

Stein, A., M. Fendorf, T. P. Jarvie, K. T. Mueller, A. J. Benesi, and T. E. Mallouk, Salt Gel Synthesis of Porous Transition-Metal Oxides, *Chemistry of Materials*, 7, 304-313, 1995.

Taguchi, A., and F. Schüth, Ordered mesoporous materials in catalysis, *Microporous and Mesoporous Materials*, 77, 1-45, 2005.

Tanev, P. T., and T. J. Pinnavaia, A Neutral Templating Route to Mesoporous Molecular-Sieves, *Science*, 267, 865-867, 1995.

Tian, Z. R., W. Tong, J. Y. Wang, N. G. Duan, V. V. Krishnan, and S. L. Suib, Manganese oxide mesoporous structures: Mixed-valent semiconducting catalysts, *Science*, 276, 926-930, 1997.

Tuel, A., and S. Gontier, Synthesis and characterization of trivalent metal containing mesoporous silicas obtained by a neutral templating route, *Chemistry of Materials*, 8, 114-122, 1996.

Tuel, A., S. Gontier, and R. Teissier, Zirconium containing mesoporous silicas: New catalysts for oxidation reactions in the liquid phase, *Chemical Communications*, 651-652, 1996.

Walker, S. A., and J. A. Zasadzinski, Self-Assembly of Silicate/Surfactant Mesoporous Materials, *Mater. Res. Symp. Proc.*, 371, 93-98, 1995.

Xu, X. C., C. S. Song, J. M. Andresen, B. G. Miller, and A. W. Scaroni, Preparation and characterization of novel CO2 "molecular basket" adsorbents based on polymer-modified mesoporous molecular sieve MCM-41, *Microporous and Mesoporous Materials*, 62, 29-45, 2003.

Yamada, T., H. S. Zhou, H. Uchida, M. Tomita, Y. Ueno, T. Ichino, I. Honma, K. Asai, and T. Katsube, Surface photovoltage NO gas sensor with properties dependent on the structure of the self-ordered mesoporous silicate film, *Advanced Materials*, 14, 812-815, 2002.

Yang, P. D., D. Y. Zhao, D. I. Margolese, B. F. Chmelka, and G. D. Stucky, Generalized syntheses of large-pore mesoporous metal oxides with semicrystalline frameworks, *Nature*, 396, 152-155, 1998.

Yiu, H. H. P., C. H. Botting, N. P. Botting, and P. A. Wright, Size selective protein adsorption on thiol-functionalised SBA-15 mesoporous molecular sieve, *Physical Chemistry Chemical Physics*, 3, 2983-2985, 2001.

Yu, C., B. Tian, X. Liu, J. Fan, H. Yang, and D. Y. Zhao, *Nanoporous Materials*, 14-46 pp., Imperial College Press, London, 2004.

Zelenák, V., V. Hornebecq, S. Mornet, O. Schaf, and P. Llewellyn, Mesoporous Silica Modified with Titania: Structure and Thermal Stability, *Chem. Mater.*, 18, 3184-3191, 2006.

Zhao, D. Y., J. L. Feng, Q. S. Huo, N. Melosh, G. H. Fredrickson, B. F. Chmelka, and G. D. Stucky, Triblock copolymer syntheses of mesoporous silica with periodic 50 to 300 angstrom pores, *Science*, 279, 548-552, 1998.

Zhao, J. W., F. Gao, Y. L. Fu, W. Jin, P. Y. Yang, and D. Y. Zhao, Biomolecule separation using large pore mesoporous SBA-15 as a substrate in high performance liquid chromatography, *Chemical Communications*, 752-753, 2002a.

Zhao, Q., W. H. Chen, S. J. Huang, Y. C. Wu, H. K. Lee, and S. B. Liu, Acidity characterization of MCM-41 materials using solid-state NMR spectroscopy, *Nanoporous Materials Iii*, 141, 453-458, 2002b.

Zhao, X. S., G. Q. M. Lu, and G. J. Millar, Advances in mesoporous molecular sieve MCM-41, *Industrial & Engineering Chemistry Research*, 35, 2075-2090, 1996.

*Probable impossibilities are to be preferred*
*to improbable possibilities.*
*Aristotelis*

# Chapter 4

## MODEL AND METHODOLOGY:

## MONTE CARLO SIMULATION AND

## QUASI CHEMICAL THEORY

### Introduction

In this chapter, the model and the methodology adopted in the study of self-assembling hybrid materials are presented. Both Monte Carlo (MC) simulations and quasi chemical theory (QCT) are developed on the basis of the same coarse-grained description, representing a ternary system where a surfactant and an inorganic precursor interact in a solvent to form periodical ordered structures. The model used in this work reflects only the essential physical and chemical properties of a real, and hence much more complex, amphiphilic solution. This choice is in agreement with the questions posed at the beginning of our research study. Which aggregation behavior do the self-assembling amphiphilic systems show? Which structures do they form at equilibrium? Why and how are their equilibrium phase behavior and the morphology of the resulting materials affected by different inorganic precursors and/or by different surfactants? Of course, by focusing our attention on the global structural organization, the price to pay is the loss of knowledge at an atomistic level because, in such a coarse-grained model, groups of atoms are arranged together in a simplified manner, as we will see later on in this chapter.

To make the system evolve from a completely random initial configuration to the final equilibrium configuration, MC simulations in the canonical ensemble (*NVT*) have been performed. MC methods find a wide application in the simulation of the behavior of physical and mathematical systems, and they are especially useful when the system studied is characterized by a very large number of interacting variables, such as self-assembling materials, liquids, biological structures. In molecular modeling, MC methods propose an alternative way in the analysis of physical phenomena with respect to molecular dynamics (MD). Instead of studying a given system according to the dynamic of atoms and molecules interacting for a given period of time under the laws of physics, the MC method generates time-independent configurations of the same system according to the Boltzmann probability distribution. The average of the properties calculated from each configuration, also called ensemble average, is equivalent to the average calculated over a given period of time (ergodic hypothesis).

In lattice simulations, continuous space is replaced with a discretized network in which each site is occupied by only one particle [*Larson et al.*, 1985], and the thermodynamic properties of the system are calculated by averaging the properties over a very large number of different time-independent configurations, according to the Metropolis algorithm [*Metropolis et al.*, 1953].

The results concerning the equilibrium phase behavior obtained by performing MC simulations have been compared with the QCT, a lattice-based mean field approximation. This theory was previously applied by other researchers in similar, although less complex, amphiphilic systems to compare the results of their simulations [*Kim et al.*, 2002; *Larson et al.*, 1985; *Mackie et al.*, 1996; *Mackie et al.*, 1995]. QCT presents a better treatment for the configurational energy than the one assumed by the regular solution theory, by allowing a non-random distribution of pairs of particles [*Tompa*, 1956]. In regular solution theory, despite the fact that the energy of mixing is not zero, the particles are considered to be completely independent from each other [*Tompa*, 1956]. Guggenheim separated the regular solutions from the strictly regular solutions, in which the energies of interaction between pairs of particles are no longer independent of their positions [*Guggenheim*, 1952], and the excess entropy of mixing, although not zero, is very small [*Tompa*, 1956]. The QCT is an approximation for the strictly regular solutions, assuming the independence of pairs of particles. It has been also defined as first approximation, in contrast to the zeroth approximation which postulates a complete random organization of the particles.

A complete random distribution of particles in a system where the configurational energy is not zero, can be physically reasonable to assume at very high temperatures, or when the components in the system show a similar behavior with the solvent. In our systems, the temperature is always below the critical point, and the surfactant and the hybrid particles have quite different properties. Therefore, the effect of non-randomness becomes an important factor to be taken into account.

## 4.1 Monte Carlo Simulation Method.

MC simulation methods are widely used to analyze the physical behavior and the equilibrium properties of many different kinds of macroscopic systems. They differ from other techniques of simulation, such as MD, by being stochastic, namely by being based on a non-deterministic approach to the study of a given phenomenon. In statistical mechanics, the MC method is applied to calculate the average properties of a system in a given *ensemble*.

A statistical ensemble (or just ensemble) can be defined as a theoretical set of a very large number of mutually non-interacting systems, each representing a possible configuration (or state) of the real system of interest. An ensemble where the number of particles, the volume, and the temperature are kept constant for all the systems is called the canonical ensemble, and it is usually referred to as the *NVT* ensemble. According to which properties are kept constant, other ensembles can be defined, such as the *NVE* (microcanonical) ensemble, and the *μVT* (grand-canonical) ensemble. In this work, the MC simulation method has been applied in the *NVT* ensemble, and we will deal with only this one.

Two postulates are of fundamental importance. One is the ergodic hypothesis, and the other is the postulate of equal *a priori* probability. The former introduces the equivalence between an ensemble-average and a time-average; the latter says that states of equal energy have the same probability to be observed. The importance of the ergodic hypothesis is in the fact that we can completely disregard the time-dependent interactions of the single atoms or molecules with each other, and still the macroscopic (statistical) behavior of the system can be evaluated. The postulate of equal *a priori* probability is essential to derive the expression of the Boltzmann distribution.

### 4.1.1 MC in the *NVT* ensemble.

A given many-body system is said to be in the *NVT* ensemble if it can share its energy with a large heat bath, by keeping constant the temperature. The distribution of the total energy is given by the partition function, representing the *bridge* between statistical mechanics and thermodynamics. As a matter of fact, all the thermodynamic properties of interest, are directly related to the partition function of a given ensemble.

The partition function in the *NVT* ensemble is:

$$Q(N,V,T) \equiv \frac{1}{N!h^{3N}} \int \exp\left(-\frac{H(\Theta)}{k_B T}\right) d\Theta \tag{4.1.1}$$

where $k_B$ is the Boltzmann constant, $T$ the absolute temperature, $h$ the Planck constant, and the function $H(\Theta)$ is the Hamiltonian of the system $\Theta$. In the most general case, $H(\Theta)$ is the sum of a kinetic energy and a potential energy contributions:

$$H(\mathbf{p}_N, \mathbf{r}_N) = \sum_{i=1}^{N} \frac{\mathbf{p}_i^2}{2m_i} + U(\mathbf{r}_N) \tag{4.1.2}$$

where $\mathbf{p}_N = (p_1, p_2, ..., p_N)$ and $\mathbf{r}_N = (r_1, r_2, ..., r_N)$ are the momenta and the coordinates of particles 1, 2, …, $N$ of the system $\Theta$, respectively, and $U(\mathbf{r}_N)$ is the potential energy.

In the following equation, a general property of a given system $\Theta$ is averaged in the *NVT* ensemble:

$$\langle \zeta \rangle_{NVT} = \frac{\int \zeta(\Theta) \exp\left(-\frac{H(\Theta)}{k_B T}\right) d\Theta}{Q(N,V,T)} \tag{4.1.3}$$

where $\zeta(\Theta)$ is the general property to be averaged. In our case $\zeta(\Theta)$ is just a function of the coordinates of the particles in the system, namely $\zeta(\Theta) = \zeta(\mathbf{r}_N)$.

The integration of equation (4.1.3) is the result of two integrations: one over the momenta and the other one over the positions of the particles. The first integration can be solved analytically (see Appendix A) and its contribution is equal to:

$$\int \exp\left(-\sum_{i=1}^{N} \frac{\mathbf{p}_i^2}{2k_B T m_i}\right) d\mathbf{p}_N = \int \exp\left(-\frac{\mathbf{p}_1^2}{2k_B T m_1}\right) d\mathbf{p}_1 \cdots$$

$$\cdots \int \exp\left(-\frac{\mathbf{p}_N^2}{2k_B T m_N}\right) d\mathbf{p}_N = \left(2\pi n k_B T\right)^{3N/2}$$

(4.1.4)

with $m_1 = \ldots = m_N = m$ .

Therefore, by substituting in equation (4.1.3) the Hamiltonian (4.1.2) and the result of the integral (4.1.4) , we obtain:

$$\langle \varsigma \rangle_{NVT} = \frac{\int \varsigma(\mathbf{r}_N) \exp\left(-\frac{U(\mathbf{r}_N)}{k_B T}\right) d\mathbf{r}_N}{\frac{1}{N! h^{3N}} \int \exp\left(-\frac{U(\mathbf{r}_N)}{k_B T}\right) d\mathbf{r}_N}$$

(4.1.5)

More synthetically:

$$\langle \varsigma \rangle_{NVT} = C \cdot \frac{\int \varsigma(\mathbf{r}_N) \exp\left(-\beta U(\mathbf{r}_N)\right) d\mathbf{r}_N}{Z(N,V,T)}$$

(4.1.6)

where $C$ is a constant, $\beta = \frac{1}{k_B T}$ , $U(\mathbf{r}_N)$ is the potential energy of the system, and

$Z(N,V,T) \equiv \int \exp\left(-\beta U(\mathbf{r}_N)\right) d\mathbf{r}_N$ is the configurational part of the partition function.

The MC method is then used to solve the ratio of the integrals in equation (4.1.6) in order to evaluate the generic ensemble (macroscopic) property $\langle \varsigma \rangle_{NVT}$ .

To clarify the utility of the MC method, let us consider the following integral:

$$I = \int_a^b F(x)\,dx \cong \sum_{j=1}^M F(x_j)\frac{(b-a)}{M} \tag{4.1.7}$$

The MC method approximates the real value of $I$ with the sum of the contributions $A_j \equiv F(x_j)(b-a)$ estimated at different values of $x \in [a,b]$ selected randomly a number $M$ of times. In other words, the real area below the curve defined by the function $F(x)$, with $x \in [a,b]$, is calculated by generating $M$ random areas $A_j$, each with a uniform probability $\frac{1}{M}$:

$$I \cong \sum_{j=1}^M A_j \frac{1}{M} = \frac{1}{M}\sum_{j=1}^M A_j \tag{4.1.8}$$

This approximation should furnish the exact value of $I$ for $M \to \infty$, but it is not generally effective when the argument of the integral is a function of the type reported in equation (4.1.6). In these cases, the choice of a uniform probability distribution function (*p.d.f.*) is no more satisfactory, because it would give the same weight to any point of the configuration space, regardless of its Boltzmann factor. It would be much more effective to use a *p.d.f.* allowing the sampling of more points of the configuration space whose Boltzmann factor is not too small.

Such a sampling is usually defined as importance sampling, and the *p.d.f.* chosen to sample the configuration space is exactly the Boltzmann distribution, representing the probability of a given conformation to appear in the sample:

$$P_B(\mathbf{r}_N) = \frac{\exp(-\beta U(\mathbf{r}_N))}{\int \exp(-\beta U(\mathbf{r}_N))\mathrm{d}\mathbf{r}_N} \tag{4.1.9}$$

Therefore, by introducing $P_B(\mathbf{r}_N)$ into the discrete form of equation (4.1.6), the ensemble average of the property $\langle \zeta \rangle_{NVT}$ becomes:

$$\langle \zeta \rangle_{NVT} = \frac{\int \zeta(\mathbf{r}_N)\exp(-\beta U(\mathbf{r}_N))\mathrm{d}\mathbf{r}_N}{\int \exp(-\beta U(\mathbf{r}_N))\mathrm{d}\mathbf{r}_N} \cong \frac{1}{M}\sum_{j=1}^M \zeta(\mathbf{r}_j) \tag{4.1.10}$$

The ensemble average of the generic property $\langle\varsigma\rangle_{NVT}$ is then obtained by randomly calculating such a property in a very large number of points of the configuration space, sampled with the Boltzmann weight $\exp(-\beta U(\mathbf{r}_N))$. The algorithm commonly applied for this purpose is the Metropolis algorithm [*Metropolis et al.*, 1953] which constructs a chain of successive configurations, the first one being randomly chosen and the following ones accepted or rejected according to their relative level of energy. In particular, if the energy of the new configuration is lower than the energy of the old one, then the new configuration is accepted and it becomes the starting configuration for the next ring of the chain. Otherwise, if the energy of the new configuration is higher than the energy of the old one, then the new configuration is accepted with probability $\exp(-\beta\Delta U(\mathbf{r}_N))$, that is, only if the ratio between the Boltzmann factors of the new and the old configurations is bigger than a given random number $R$, namely:

$$\frac{\exp(-\beta U(\mathbf{r}_N))_{new}}{\exp(-\beta U(\mathbf{r}_N))_{old}} = \exp(-\beta\Delta U(\mathbf{r}_N)) > R \tag{4.1.11}$$

If the new configuration is rejected, the contributions of the old configuration should be counted anyway, in order to evaluate properly the average properties of the system. The detailed balance between the following probabilities should be satisfied:

$$P(old \rightarrow new) = P(new \rightarrow old) \tag{4.1.12}$$

where $P(old \rightarrow new)$ is the probability to move from the old to the new configuration, and $P(new \rightarrow old)$ is the probability to move back from the new to the old configuration. The demonstration of equation (4.1.12) is reported in Appendix B.

## 4.2 Model

The lattice model used in this work represents the basic features of a system in which a surfactant, an inorganic precursor, and a solvent are present. The idea to model surfactant solutions in a lattice box is due to Larson [*Larson et al.*, 1985] who studied the aggregation behavior of surfactants in systems where water-like and oil-like solvents were present. In this model, space is organized into a cubic network of sites and the surfactant chains are described as sequences of connected solvophobic and solvophilic segments, each unit occupying one single site in the lattice box. A site on the amphiphilic chain is connected to any of its $z=26$ nearest or diagonally-nearest neighbors, while solvent and oil molecules occupied one single site each.

In our work, many characteristics of a real hybrid system have been disregarded, first because it would not be practical to include all the details in a simulation as it would be very expensive (in CPU time) to obtain meaningful results; and second because the governing factors alone give a very clear picture of the real phenomenon and its physical trend. Therefore, although simple, the selected model for this work is able to describe the phase behavior and the formation of ordered liquid crystal structures at equilibrium with dilute phases where free monomers or micelles are found.

Two different kinds of surfactants have been modeled: a linear surfactant, $H_4T_4$, made up of four tail segments $T$ and four head segments $H$; and a branched-head surfactant made up by a linear tail of four or five segments $T$, and a branched head of four segments $H$ disposed as illustrated in Figure 4.1. The surfactant with four tail segments was called $T_4HH_3$, whereas the surfactant with five tail segments $T_5HH_3$. The tail segments constitute the solvophobic part of the amphiphile, whereas the head segments constitute the solvophilic one. This double character is modeled by choosing appropriate interactions between the surfactant particles and the solvent. The linear surfactant can be considered a model structure directing agent for the synthesis of SBA-15-like mesoporous materials, where the mesopores are interconnected by micropores; whereas the branched-head surfactants are useful to model the self-assembling of MCM-41-like materials, where the mesopores are not interconnected.

To clarify the role of the head and tail segments for the linear surfactant, it should be noted that a head group represents one oxyethylene unit (-C-C-O-) and one tail group $T$ represents about two or three $CH_2$ groups. This means that our surfactant is approximately equivalent to the real amphiphilic diblock copolymer $CH_3$-$(CH_2)y$-$(O$-$CH_2$-$CH_2)_4$-OH, where $y$ is between 7 and 11 [*Fodi and Hentschke*, 2000]. As for the branched-head surfactant, the aim was to choose a chain that could more realistically represent the head size of an ionic surfactant, being bigger than any of the tail segments. Charges and counterions are included in the approximation of the model.

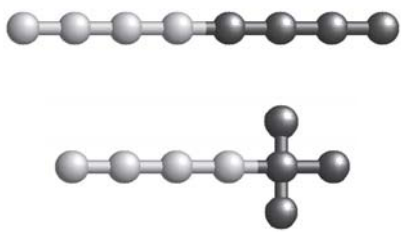

**Fig. 4.1.** Model surfactants $H_4T_4$ (top) and $T_4HH_3$ (down). Dark shading represents the surfactant tail, light shading represents the surfactant head.

The inorganic precursors that we have included in our study are: (1) pure silica precursors, such as tetra-alkoxysilanes of the type $Si(OR)_4$, and (2) organosilica precursors (OSPs), such as trialkoxyorganosilanes of the type $R'$-$Si(OR)_3$ or $(RO)_3Si$-$R'$-$Si(OR)_3$. The former are modeled according to their miscibility with the solvent; the latter are modeled also by considering the solvophobic/solvophilic nature of the organic group and its position (terminal or bridging) with respect to the silica source. Such changes in the structure of the precursors permit the analysis of how the equilibrium phase behavior and the morphology of the liquid crystal structures are affected.

A pure silica precursor is represented by $I_2$, whereas terminal OSPs were obtained by modifying $I_2$ with $IT$ or $IH$, where $T$ and $H$ are equivalent to a surfactant tail and head segment, respectively. $IT$ (and $IH$) precursors have a higher solvophobicity (solvophilicity) than the pure silica precursor, leading to quite remarkable changes in the phase diagrams and therefore in the final structures obtained. Bridging OSPs present an organic functional group in between two silica sources.

The inorganic precursors modeled in this work are listed in Table 4.1, according to all the possible modifications in the structure. In particular, the soluble and the insoluble silica sources are indicated with $I$ and $I'$ respectively.

**Table 4.1.** Inorganic precursors modeled.

|  | Soluble I | Insoluble I' |
|---|---|---|
| Pure Silica | $I_2$ | $I'_2$ |
| Terminal OSPs | $IT$ – $IH$ | $I'T$ – $I'H$ |
| Bridging OSPs | $ITI$ – $IHI$ | $I'TI'$ – $I'HI'$ |

In Figure 4.2, the inorganic precursors modeled in this work are reported.

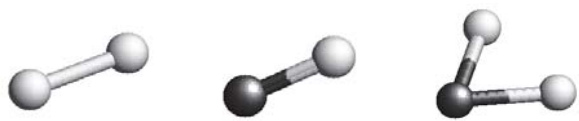

**Fig. 4.2**. Model inorganic precursors. Dark shading represents the organic group (*T* or *H*), light shading represents the silica source (*I* or *I′*).

The solvent beads occupying single empty sites in the lattice box are denoted by *S*. The solvent is not explicitly modeled and the surfactant heads are completely miscible with it. All segments lie on a cubic lattice, allowing bonds to have any of the 26 following orientations: (±1, 0, 0), (0, ±1, 0), (0, 0, ±1), (±1, ±1, 0), (±1, 0, ±1), (0, ±1, ±1), and (±1, ±1, ±1). This implies that bond-lengths were not constant in the simulation, but they were allow to fluctuate between 1 and $\sqrt{3}$ .

In this type of models, there is only one independent interaction for each pair of site types. For binary systems, the derivation can be found in most textbooks on polymer solutions [*Tompa*, 1956], which can be easily extended to multicomponent systems. The global interchange energy, $\omega_{ij}$ , between different types of sites is given by:

$$\omega_{ij} = \varepsilon_{ij} - \frac{1}{2}\left(\varepsilon_{ii} + \varepsilon_{jj}\right) \qquad (4.2.1)$$

with $i \neq j$ and $\varepsilon_{ij}$ being the individual interaction energies of different types of sites. The derivation of this equation is shown in Appendix C. The individual values of the interaction parameters and the global interchange energies are reported in Table 4.2 and Table 4.3, respectively, where the soluble inorganic and the insoluble organic components are indicated with *I* and *I′* respectively.

**Table 4.2.** Individual interaction parameters ($\varepsilon_{ij}$) between inorganic oxide and solvent.

| | Complete miscibility | | | | | Partial miscibility | | | |
|---|---|---|---|---|---|---|---|---|---|
| | **I** | **H** | **T** | **S** | | **I′** | **H** | **T** | **S** |
| **I** | 0 | | | | **I′** | -2 | | | |
| **H** | -2 | 0 | | | **H** | -3 | 0 | | |
| **T** | 0 | 0 | -2 | | **T** | -2 | 0 | -2 | |
| **S** | 0 | 0 | 0 | 0 | **S** | 0 | 0 | 0 | 0 |

**Table 4.3.** Global interchange energies ($\omega_{ij}$) between inorganic oxide and solvent.

| | Complete miscibility | | | | | Partial miscibility | | | |
|---|---|---|---|---|---|---|---|---|---|
| | **I** | **H** | **T** | **S** | | **I′** | **H** | **T** | **S** |
| **I** | 0 | | | | **I′** | 0 | | | |
| **H** | -2 | 0 | | | **H** | -2 | 0 | | |
| **T** | 1 | 1 | 0 | | **T** | 0 | 1 | 0 | |
| **S** | 0 | 0 | 1 | 0 | **S** | 1 | 0 | 1 | 0 |

The surfactant–solvent interactions have been proposed in previous works by other researchers [*Larson et al.*, 1985; *Mackie et al.*, 1997]. Two different cases are presented to simulate miscibility of the inorganic precursor in the solvent: in the first case, the inorganic precursor and the solvent are completely miscible ($\omega_{IS}$=0), in the second they are as immiscible as surfactant heads and tails ($\omega_{I'S}$=$\omega_{HT}$=1), as in previous work [*Siperstein and Gubbins*, 2003]. It is computationally convenient to define all individual interactions with the solvent, $\varepsilon_{is}$, as zero, and calculate the individual interaction parameters $\varepsilon_{ij}$ for the rest of the pairs using the selected global interchange energies.

In this work, the dimensionless temperature is defined using the head-tail interchange energy by $T^*=k_B T/\omega_{HT}$, where $k_B$ is the Boltzmann constant, $T$ is the absolute temperature and $\omega_{HT}$ is the surfactant head-tail interaction energy.

## 4.3 Simulation Method.

Lattice MC simulations in the *NVT* ensemble have been used to model the phase behavior and the formation of ordered structures for ternary systems. The initial configuration has been chosen randomly as it does not affect the final equilibrium of the system, and its total energy $U$ (normalized by $k_BT$) calculated. Periodic boundary conditions were applied to a fully occupied three-dimensional lattice in which the precursor and surfactant chains were moved by reptation and configurational bias moves (partial and complete regrowth). Since the solvent molecules are not explicitly modeled, they do not need to be moved: it is assumed that they occupy those sites left available by the chain-beads after a movement.

In a reptation move, a chain-end was chosen at random and displaced forward by one site to one of the *25* remaining neighboring sites. If this site was not free, the movement was rejected, otherwise the second bead of the chain was moved to the position previously occupied by the first bead, and so on until all the chain was moved. The move was accepted or rejected according to the following probability $p$:

$$p = \min\left[1, \exp(-\beta \Delta U)\right] \tag{4.3.1}$$

where $\Delta U$ represents the difference between the energy of the new configuration and that of the old one. If $\Delta U$ is negative, the movement was accepted, otherwise it was accepted with the probability $p$ evaluated above; i.e., a random number, $\alpha$, between *0* and *1* was compared with the probability $p$, and if $\alpha < p$, the particle was displaced to its new position, according to the Metropolis algorithm. This successful or unsuccessful movement constitutes a MC step and it was repeated till the simulations was considered finished. In Figure 4.3 a schematic representation of a reptation move is given.

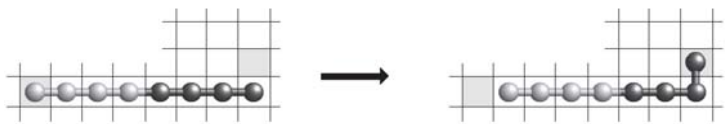

**Fig. 4.3**. Schematic representation of a reptation move performed by one surfactant chain.

Chain regrowth using the configurational bias method, first introduced by Rosenbluth in 1955 [*Frenkel and Smit*, 2002], has been found very useful to speed up the simulations, although it has a high computational cost. Performing a bias move involves displacing a chain (or part of it) from one place in the lattice box to another, as illustrated in Figure 4.4.

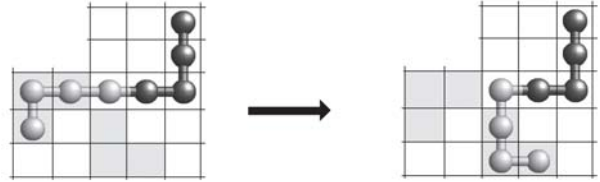

**Fig. 4.4.** Schematic representation of a partial regrowth performed by three of the four tail segments of a surfactant chain.

This unreal movement is justified by the ergodic hypothesis stating the equivalence between time average and average over the statistical ensemble.

In practice, we perform this movement according to the following procedure [*Frenkel and Smit*]:

1. A chain is randomly selected and deleted.

2. A random site in the lattice box is chosen to insert the first particle of the deleted chain.

3. If this site is occupied, the movement is rejected, otherwise the first particle is allocated.

4. The energy $e_1(n)$ of the first particle in the new configuration and its Rosenbluth weight $w_1(n) = k \cdot \exp[-\beta \cdot e_1(n)]$, where $k$ is the number of trial directions[1], are calculated.

5. The second particle is inserted in one of the $k$ possible random directions. We choose one of them, say, $n$, with a probability:

$$p_2(n) = \frac{\exp[-\beta \cdot e_2(n)]}{w_2(n)} \qquad (4.3.2)$$

where $w_2(n)$ is defined as:

$$w_2(n) = \sum_{j=1}^{k} \exp[-\beta \cdot e_2(j)] \qquad (4.3.3)$$

---

[1] During the configurational bias moves, only *10* different directions, out of the *26* possible directions given by the selected lattice coordination number, were used to grow the chain. These *10* directions were selected randomly for each segment to be grown.

6.  The last step is repeated till all the chain has been built, and the Rosenbluth factor for the new configuration can be calculated.

7.  The Rosenbluth factor is calculated as:

$$W_{new} = \prod_{i=1}^{l} w_i(n) \qquad (4.3.4)$$

where $l$ is the length of the chain.

8.  To calculate the Rosenbluth factor for the old configuration, the old chain, that is the one previously deleted, is re-built. For each particle $i$ the Rosenbluth factor is given by

$$w_i(o) = \sum_{j=1}^{k} \exp[-\beta \cdot e_i(j)] \,. \qquad (4.3.5)$$

9.  When all the chain has been retraced, the Rosenbluth factor for the old configuration is calculated as:

$$W_{old} = \prod_{i=1}^{l} w_i(o) \,. \qquad (4.3.6)$$

10. The movement from the old to the new configuration is accepted with a probability given by:

$$p = \min[1, W_{new}/W_{old}]. \qquad (4.3.7)$$

The derivation of the above probability is reported in Appendix D.

The simulation was considered finished if no significant changes were observed in the total energy of the system and in the densities of each phase. Along with these verifications, a visual inspection of the phases formed was useful to determine if the last configuration obtained for a given system could be considered the most probable configuration sampled. In systems which may be difficult to sample, especially when micelles were found to be at equilibrium with liquid crystals, we verified the validity of our results by modifying the simulation box, the global concentration of the mixture, the temperature, and we also followed the random walk of a bunch of chains in the box during the simulation.

Typical simulations use a total number of steps of the order of *$10^9$ – $10^{10}$*. A typical mix of the MC moves used is 80% reptation, 10% partial regrowth, and 10% complete regrowth. This combination, being the best compromise between two factors, the CPU time required for each step and the percentage of accepted movements, was chosen by analyzing the equilibration for several possible combinations of MC moves, and by comparing them on the basis of the above factors.

In Figure 4.5, this comparison is reported for a binary system in which only surfactant chains are present. $10^9$ MC steps have been performed for each combination at *$T^*$*=8.0. Single complete and partial regrowth moves take around *10* times and *4* times more than a single reptation step, respectively. However, the degree of equilibration obtained by using a configurational bias move, over the same number of MC steps, is significantly better. In other words, a higher number of MC steps would be necessary to obtain the same equilibration state by using a simple reptation. Therefore, we decided to combine reptation with partial and complete regrowth in order to increase simulation speed (in terms of CPU time) and sampling efficiency.

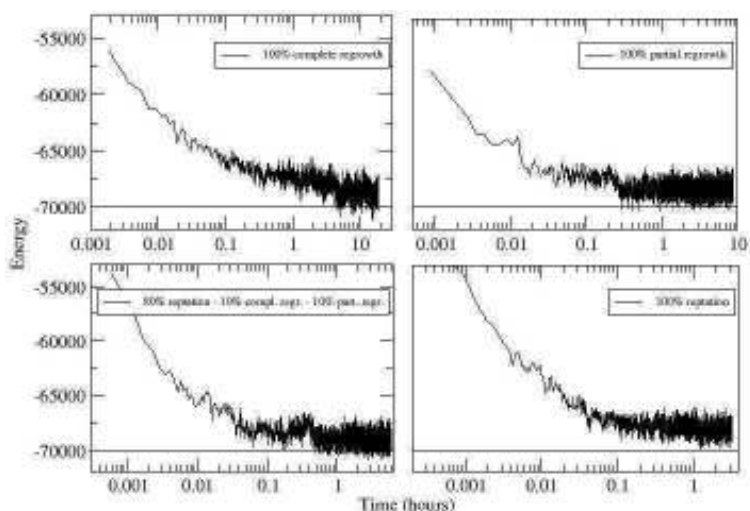

**Fig. 4.5**. Energy equilibration for an $H_4T_4$/solvent system at *$T^*$*=8.0, for different combinations of movements. A logarithmic scale is used for the horizontal axis.

Direct interfacial simulations were performed at the reduced temperature $T^*$=8.0 in a box of size 24×24×100 for at least 3×10⁹ MC steps. In some cases, simulations were carried out in larger boxes of size 24×24×200 to better observe the formation of two phases at equilibrium, and to reduce the effect of the interface on the bulk of one or both of them. Single-phase simulations were performed in a 40³ box to study in more detail the structures of the liquid crystal phases obtained, and to calculate the radial distribution functions. In such cases, because of the low solvent volume fraction (1-5%), and large number of surfactant chains, the number of MC steps was at least 70×10⁹. Cubic boxes of sizes ranging between 20³ and 48³ were also used to study the properties of the self-assembled structures. In some cases, different box sizes were analyzed to determine if observable phenomena were related with a particular box size.

To generate the initial configuration for the direct interfacial simulations, inorganic oxide and surfactant chains were randomly distributed in a concentrated region of the box where their total concentration was around 60%; in the single phase simulations the chains were allocated filling all the spaces in the lattice and the system was allowed to evolve at a very high temperature ($T^*$=10⁴) for 2×10⁶ MC steps: this created a completely random redistribution of the chains and the initial configuration for the simulations. The initial and the final configuration for a direct interfacial simulation are reported in Figure 4.6.

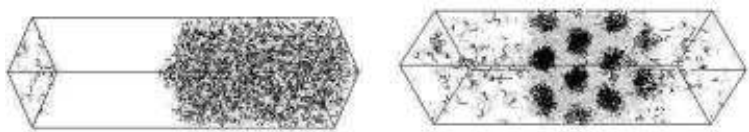

**Fig. 4.6.** Initial (left) and final (right) configuration for a $H_4T_4/I_2$/solvent system, simulated for 6×10⁹ MC steps at $T^*$=8.0. The surfactant tails are represented by dark segments, whereas the surfactant heads and the inorganic precursor by light segments. Global concentrations: $H_4T_4$ 25%, $I_2$ 5%.

After the system reached equilibrium, the ternary phase diagrams for the surfactant–inorganic–solvent system were obtained by simulating the system at different global concentrations and averaging the composition of each phase along the $z$ direction, obtained from density profiles far from the interfaces.

## 4.4 Quasi Chemical Theory.

The regular solution theory postulates the random distribution of beads even though the enthalpy of mixing is not equal to zero. Only in the limiting condition of very high temperatures, the random distribution of atoms can be considered a real scenario for such solutions. This approximation, usually called the *zero order approximation*, is improved by the quasichemical theory assuming the independence of pairs of beads. The improvement is in the fact that we are not considering anymore a random distribution of beads, but rather a random distribution of pairs of beads. An even better approximation would assume a random distribution of triplets, quadruplets or bigger clusters of beads since the probability of a bead to belong to several pairs (or clusters) would be reduced. In the following, we review the QCT for a general mixture to apply it to our specific problem.

The quasi-chemical approximation for a lattice model postulates that the pairs are connected by an equation reminiscent of the law of chemical equilibrium:

$$\frac{\left(\frac{1}{2}\left(zX_{ij}\right)\right)^2}{\left(zX_{ii}\right)\left(zX_{jj}\right)} = \exp\left[\frac{-\omega_{ij}}{k_BT}\right] \tag{4.4.1}$$

where:

$z$ is the lattice coordination number,

$zX_{ij}$ is the number of the nearest-neighbor *i-j* pairs,

$\omega_{ij}$ is the global interchange energy between *i* and *j* sites,

$k_B$ is the Boltzmann constant,

$T$ is the absolute temperature.

The number of nearest-neighbor pairs is also related by the following equations of site-balance:

$$zX_{ii} = \frac{zN_i - zX_{ij}}{2} \tag{4.4.2}$$

stating that the number of the nearest-neighbor *i-i* pairs is given by subtracting from all the nearest neighbors ($zN_i$) the *i-j* pairs ($zX_{ij}$). $N_i$ is the total number of *i*-sites. Factor ½ avoids double counting.

For an ideal solution:

$$zX_{ii}^{id} = \frac{1}{2}N_i\left(x_i z\right) \tag{4.4.3}$$

The amount of *i-i* nearest-neighbor pairs is given by the number $N_i$ of *i* sites multiplied by the probability of finding any other nearest-neighbor site of type *i*, given by the fraction of the coordination number ($x_i z$). The factor ½ avoids double counting. The above probability does not consider how the particles would prefer to be associated according to the interactions they form.

Equation (4.4.3) can be re-written in the following way:

$$X_{ii}^{id} = \frac{1}{2}x_i^2 N \tag{4.4.4}$$

where $N$ represents the total number of sites in the lattice box.

A similar equation is derived for $X_{ij}^{id}$ where the factor ½ is not present as we are counting over different sites:

$$X_{ij}^{id} = x_i x_j N \tag{4.4.5}$$

Combining equation (4.4.4) and (4.4.5), we obtain:

$$X_{ij}^{id} = x_i x_j N = \sqrt{\frac{2X_{ii}^{id}}{N}\frac{2X_{jj}^{id}}{N}}N = 2\left(X_{ii}^{id}X_{jj}^{id}\right)^{\frac{1}{2}} \tag{4.4.6}$$

Therefore, for an ideal solution:

$$\frac{\left(\frac{1}{2}X_{ij}^{id}\right)^2}{X_{ii}^{id}X_{jj}^{id}} = 1 \tag{4.4.7}$$

Equation (4.4.7) can be obtained from equation (4.4.1) by setting $\omega_{ij}$ equal to zero, namely when the solution is ideal or athermal ($\Delta H_{mix}$=0). However, even for regular solutions ($\Delta H_{mix} \neq 0$) equation (4.4.1) can be used as a first approximation (usually referred to as the *zeroth approximation*) to calculate the thermodynamic properties of the mixture.

Instead of a random mixing where all the sites are independent of each other, the so called *first order approximation* treats the numbers of pairs as completely independent. In this case, the number of nearest-neighbor sites are modified by introducing a quantity $k$ defined as:

$$X_{ij} = kX_{ij}^{id} \tag{4.4.8}$$

By definition, $k$ cannot be negative. Then, from equations (4.4.2) and (4.4.3) we obtain:

$$X_{ii} = \frac{1}{2}x_i(1-kx_j)N, \quad X_{jj} = \frac{1}{2}x_j(1-kx_i)N \tag{4.4.9}$$

If $k$=1, we go back to the zeroth approximation. If $k$ is equal to zero, there would not be any *i-j* pair, that is the component (*i* or *j*) would be pure. It is also worth noting that $(1-kx_i) \geq 0$ and $(1-kx_j) \geq 0$. Therefore $0 \leq k \leq 2$.

In our ternary system, we are dealing with four different kinds of sites: *H*, *T*, *I* (or *I'*) and *S*. Therefore there are ten possible nearest-neighbor pairs to take into account [*Larson et al.*, 1985]. They are summarized in Table 4.4.

Table 4.4. Nearest-neighbor pairs for our lattice model.

| | **H** | **T** | **I** | **S** |
|---|---|---|---|---|
| **H** | $X_{HH}$ | $X_{HT}$ | $X_{HI}$ | $X_{HS}$ |
| **T** | | $X_{TT}$ | $X_{TI}$ | $X_{TS}$ |
| **I** | | | $X_{II}$ | $X_{IS}$ |
| **S** | | | | $X_{SS}$ |

It is also useful to calculate the total number of contacts, $q_{ai}$, formed by a given type of site *a* (*H*, *T*, *I* or *S*) belonging to a given component *i* (surfactant (1), inorganic precursor (2) or solvent (3)) with its nearest neighbors which do not belong to the same chain. In Table 4.5 we report these values for the systems containing a soluble silica source, being the same for the analogous systems with an insoluble silica source.

**Table 4.5**. Number of nearest neighbors for the three components in the systems studied.

| | $H_4T_4/I_2/S$ | $H_4T_4/IH/S$ | $H_4T_4/IT/S$ | $H_4T_4/IHI/S$ | $H_4T_4/ITI/S$ |
|---|---|---|---|---|---|
| $H_4T_4$ | $zq_{h,1} = 4z - 7$ | $zq_{h,1} = 4z - 7$ | $zq_{h,1} = 4z - 7$ | $zq_{h,1} = 4z - 7$ | $zq_{h,1} = 4z - 7$ |
| | $zq_{t,1} = 4z - 7$ | $zq_{t,1} = 4z - 7$ | $zq_{t,1} = 4z - 7$ | $zq_{t,1} = 4z - 7$ | $zq_{t,1} = 4z - 7$ |
| | $zq_{i,1} = 0$ | $zq_{i,1} = 0$ | $zq_{i,1} = 0$ | $zq_{i,1} = 0$ | $zq_{i,1} = 0$ |
| | $zq_{s,1} = 0$ | $zq_{s,1} = 0$ | $zq_{s,1} = 0$ | $zq_{s,1} = 0$ | $zq_{s,1} = 0$ |
| $Ix$ | $zq_{h,2} = 0$ | $zq_{h,2} = z - 1$ | $zq_{h,2} = 0$ | $zq_{h,2} = z - 2$ | $zq_{h,2} = 0$ |
| | $zq_{t,2} = 0$ | $zq_{t,2} = 0$ | $zq_{t,2} = z - 1$ | $zq_{t,2} = 0$ | $zq_{t,2} = z - 2$ |
| | $zq_{i,2} = 2z - 2$ | $zq_{i,2} = z - 1$ | $zq_{i,2} = z - 1$ | $zq_{i,2} = 2z - 2$ | $zq_{i,2} = 2z - 2$ |
| | $zq_{s,2} = 0$ | $zq_{s,2} = 0$ | $zq_{s,2} = 0$ | $zq_{s,2} = 0$ | $zq_{s,2} = 0$ |
| $S$ | $zq_{h,3} = 0$ | $zq_{h,3} = 0$ | $zq_{h,3} = 0$ | $zq_{h,3} = 0$ | $zq_{h,3} = 0$ |
| | $zq_{t,3} = 0$ | $zq_{t,3} = 0$ | $zq_{t,3} = 0$ | $zq_{t,3} = 0$ | $zq_{t,3} = 0$ |
| | $zq_{i,3} = 0$ | $zq_{i,3} = 0$ | $zq_{i,3} = 0$ | $zq_{i,3} = 0$ | $zq_{i,3} = 0$ |
| | $zq_{s,3} = z$ | $zq_{s,3} = z$ | $zq_{s,3} = z$ | $zq_{s,3} = z$ | $zq_{s,3} = z$ |

From now on, we will refer to the $H_4T_4/I_2/S$ system unless otherwise specified.

The total number of nearest neighbors for each chain is defined as:

$$\sum_\alpha zq_{\alpha i} = zq_i \qquad (4.4.10)$$

The following relation, equivalent to equation (4.4.2), between nearest-neighbor pairs of sites can be written:

$$zX_{\alpha\alpha} = \frac{\sum_{i=1}^{3} zq_{\alpha i} N_i - \sum_{\beta \neq \alpha} zX_{\alpha\beta}}{2} \qquad (4.4.11)$$

Equation (4.4.11) gives the total number of nearest-neighbor pairs in the lattice box and can be written for each $\alpha\alpha$ pair by using the data in Table 4.5. Developing equation (4.4.11) for the system where a pure silica precursor is present, the $X_{\alpha\alpha}$ values are obtained and reported as follows.

$$2X_{HH} + X_{HT} + X_{HI} + X_{HS} = q_{H,1}N_1$$
$$2X_{TT} + X_{HT} + X_{TI} + X_{TS} = q_{T,1}N_1$$
$$2X_{II} + X_{HI} + X_{TI} + X_{IS} = q_{I,2}N_2 \qquad (4.4.12)$$
$$2X_{SS} + X_{HS} + X_{TS} + X_{IS} = q_{S,3}N_3$$

Equations (4.4.12) represent the balance of the number of sites in the lattice box. To solve the above system, we add six more equations from the quasi-chemical equilibrium:

$$4X_{\alpha\alpha}X_{\beta\beta} = \eta_{\alpha\beta}^2 X_{\alpha\beta}^2 \qquad (4.4.13)$$

where

$$\eta_{\alpha\beta} = \exp\frac{\omega_{\alpha\beta}}{k_B T} = \exp\frac{\omega_{\alpha\beta}}{T^*} \qquad (4.4.14)$$

The difference in the chemical potential for each component will be calculated by differentiating the free energy of mixing obtained from the Gibbs-Helmholtz equation.

The enthalpy of mixing can be written as [*Tompa*, 1956]:

$$\Delta H\big|_{mix} = zk_B T \sum_{\alpha,\beta}\left(X_{\alpha\beta} - \sum_{i=1}^{3} X_{\alpha\beta}^{i,0}\right)\ln\eta_{\alpha\beta} \qquad (4.4.15)$$

where $X_{\alpha\beta}^{i,0}$ refers to the number of $\alpha$–$\beta$ contacts of the pure component *i*.

Integrating the enthalpy of mixing, we obtain the free energy of mixing:

$$\frac{\Delta G}{T}\bigg|_{mix} = \int \Delta H\big|_{mix}\, d\left(\frac{1}{T}\right) \qquad (4.4.16)$$

The integration of equation (4.4.16) is reported in Appendix E.

Therefore:

$$\left.\frac{\Delta G}{RT}\right|_{mix} = \left.\frac{\Delta G^*}{RT}\right|_{mix} + \frac{1}{2}z\sum_\alpha\left\{\left(\sum_{i=1}^3 q_{\alpha,i}n_i\right)\ln\frac{X_{\alpha\alpha}}{X^*_{\alpha\alpha}} - \sum_{i=1}^3\left(q_{\alpha,i}n_i\right)\ln\frac{X^{i,0}_{\alpha\alpha}}{X^{i,0^*}_{\alpha\alpha}}\right\} \tag{4.4.17}$$

where the superscript $^*$ represents the conditions in the athermal solution. In particular, the free energy of mixing for an athermal solution, $\left.\dfrac{\Delta G^*}{RT}\right|_{mix}$, is given by [*Tompa*, 1956]:

$$\left.\frac{\Delta G^*}{RT}\right|_{mix} = \sum_{i=1}^3 n_i\ln\varphi_i + \frac{1}{2}z\sum_{i=1}^3 q_i n_i\ln\frac{\xi_i}{\varphi_i} \tag{4.4.18}$$

with: $\varphi_i = \dfrac{r_i N_i}{\displaystyle\sum_{j=1}^3 r_j N_j}$  and  $\xi_i = \dfrac{q_i N_i}{\displaystyle\sum_{j=1}^3 q_j N_j}$ .

In our case, $r_1 = 8$, $r_2 = 2$ and $r_3 = 1$, for surfactant, inorganic precursors and solvent respectively.

The change in the chemical potential for the three components is the partial molar Gibbs free energy:

$$\frac{\Delta\mu_i}{RT} = \left(\frac{\partial\,\Delta G/RT}{\partial n_i}\right)_{T,n_{j\neq i}} = \frac{\Delta\mu_i^*}{RT} + \frac{1}{2}z\sum_\alpha q_{\alpha,i}\ln\frac{X_{\alpha\alpha}X^{i,0^*}_{\alpha\alpha}}{X^{i,0}_{\alpha\alpha}X^*_{\alpha\alpha}} \tag{4.4.19}$$

The change in the chemical potential for athermal solutions, $\dfrac{\Delta\mu_i^*}{RT}$, is given by [*Tompa*, 1956]:

$$\frac{\Delta\mu_i^*}{RT} = \ln\varphi_i + \frac{1}{2}zq_i\ln\frac{\xi_i}{\varphi_i} \tag{4.4.20}$$

The number of contacts for the general solution can be obtained from equations (4.4.11) and (4.4.13), by substituting $X_{\alpha\beta}$ :

$$X_{\alpha\alpha}^{\frac{1}{2}} \sum_{\beta} \frac{X_{\beta\beta}^{\frac{1}{2}}}{\eta_{\alpha\beta}} = \frac{1}{2} \sum_{i=1}^{3} q_{\alpha,i} N_i \tag{4.4.21}$$

Equations (4.4.21) must be solved numerically. However, they can be further simplified when all $\eta_{\alpha\beta} = 1$, namely when the solution is athermal. In this case, we obtain the following set of equations for our ternary system:

$$
\begin{aligned}
\left(X_{HH}^{*}\right)^{\frac{1}{2}}\left\{\left(X_{HH}^{*}\right)^{\frac{1}{2}}+\left(X_{TT}^{*}\right)^{\frac{1}{2}}+\left(X_{ll}^{*}\right)^{\frac{1}{2}}+\left(X_{SS}^{*}\right)^{\frac{1}{2}}\right\} &= \frac{1}{2}\left\{q_{H,1}N_1+q_{H,2}N_2+q_{H,2}N_2\right\} \\
\left(X_{TT}^{*}\right)^{\frac{1}{2}}\left\{\left(X_{HH}^{*}\right)^{\frac{1}{2}}+X_{TT}^{*\frac{1}{2}}+\left(X_{ll}^{*}\right)^{\frac{1}{2}}+\left(X_{SS}^{*}\right)^{\frac{1}{2}}\right\} &= \frac{1}{2}\left\{q_{T,1}N_1+q_{T,2}N_2+q_{T,2}N_2\right\} \\
\left(X_{ll}^{*}\right)^{\frac{1}{2}}\left\{\left(X_{HH}^{*}\right)^{\frac{1}{2}}+\left(X_{TT}^{*}\right)^{\frac{1}{2}}+\left(X_{ll}^{*}\right)^{\frac{1}{2}}+\left(X_{SS}^{*}\right)^{\frac{1}{2}}\right\} &= \frac{1}{2}\left\{q_{I,1}N_1+q_{I,2}N_2+q_{I,2}N_2\right\} \\
\left(X_{SS}^{*}\right)^{\frac{1}{2}}\left\{\left(X_{HH}^{*}\right)^{\frac{1}{2}}+\left(X_{TT}^{*}\right)^{\frac{1}{2}}+\left(X_{ll}^{*}\right)^{\frac{1}{2}}+\left(X_{SS}^{*}\right)^{\frac{1}{2}}\right\} &= \frac{1}{2}\left\{q_{S,1}N_1+q_{S,2}N_2+q_{S,2}N_2\right\}
\end{aligned}
\tag{4.4.22}
$$

Summing all the above equations with $\sum_{\alpha} q_{\alpha,i} = q_i$, we obtain:

$$\left\{\left(X_{HH}^{*}\right)^{\frac{1}{2}}+\left(X_{TT}^{*}\right)^{\frac{1}{2}}+\left(X_{II}^{*}\right)^{\frac{1}{2}}+\left(X_{SS}^{*}\right)^{\frac{1}{2}}\right\}^2 = \frac{1}{2} \sum_{i=1}^{3} q_i N_i \tag{4.4.23}$$

Therefore:

$$\sum_{\beta} \left(X_{\beta\beta}^{*}\right)^{1/2} = \frac{1}{2} \sum_{i=1}^{3} q_i N_i \tag{4.4.24}$$

And equation (4.4.21) for athermal solutions becomes:

$$X_{\alpha\alpha}^{*} = \frac{\dfrac{1}{2}\left\{\displaystyle\sum_{i=1}^{3} q_{\alpha,i} N_i\right\}^2}{\displaystyle\sum_{i=1}^{3} q_i N_i} \tag{4.4.25}$$

If the component $i$ is pure, then equation (4.4.25) takes the form:

$$X_{\alpha\alpha}^{i,0^*} = \frac{\frac{1}{2}\left(q_{\alpha,i}N_i\right)^2}{q_i N_i} \tag{4.4.26}$$

Equation (4.4.21) can be simplified when the component $i$ is pure and the chain is symmetric. In this case, the number of nearest neighbors, $zq_{\alpha i}$, is the same for both the beads of a given chain, and we can write:

$$\left(X_{\alpha\alpha}^{i,0}\right)^{\frac{1}{2}}\left\{\left(X_{\alpha\alpha}^{i,0}\right)^{\frac{1}{2}} + \frac{\left(X_{\beta\beta}^{i,0}\right)^{\frac{1}{2}}}{\eta_{\alpha\beta}}\right\} = \left(X_{\beta\beta}^{i,0}\right)^{\frac{1}{2}}\left\{\left(X_{\beta\beta}^{i,0}\right)^{\frac{1}{2}} + \frac{\left(X_{\alpha\alpha}^{i,0}\right)^{\frac{1}{2}}}{\eta_{\alpha\beta}}\right\} \tag{4.4.27}$$

Therefore, the number of contacts formed by a pure component is the same for each type of site when its chain is symmetric. In particular:

$$X_{HH}^{1,0} = X_{TT}^{1,0} \qquad X_{HH}^{2,0} = X_{II}^{2,0} \qquad X_{TT}^{2,0} = X_{II}^{2,0} \tag{4.4.28}$$

Regarding the pure bridging precursors, whose chains are not symmetric, the number of nearest neighbors of the inorganic source, $I$, is twice as the number of nearest neighbors of the organic source $H$ or $T$. Therefore, equations (4.4.28) are not valid anymore, and the relation between the same kinds of contacts for pure components can be calculated from equation (4.4.21).

In particular:

$$2\frac{X_{HH}^{2,0}}{X_{II}^{2,0}} + \frac{1}{\eta_{HI}}\left(\frac{X_{HH}^{2,0}}{X_{II}^{2,0}}\right)^{\frac{1}{2}} = \frac{q_{H,2}}{q_{I,2}} \simeq \frac{1}{2}$$

$$\tag{4.4.29}$$

$$2\frac{X_{TT}^{2,0}}{X_{II}^{2,0}} + \frac{1}{\eta_{TI}}\left(\frac{X_{TT}^{2,0}}{X_{II}^{2,0}}\right)^{\frac{1}{2}} = \frac{q_{T,2}}{q_{I,2}} \simeq \frac{1}{2}$$

Each one of the equations (4.4.29) gives two solutions depending on the value of the interactions established between the inorganic and organic beads:

$$\frac{X_{HH}^{2,0}}{X_{II}^{2,0}} = \frac{1 + 4\eta_{HI}^2 \pm \sqrt{1 + 8\eta_{HI}^2}}{8\eta_{HI}^2}$$

(4.4.30)

$$\frac{X_{TT}^{2,0}}{X_{II}^{2,0}} = \frac{1 + 4\eta_{TI}^2 \pm \sqrt{1 + 8\eta_{TI}^2}}{8\eta_{TI}^2}$$

and both can be meaningful. However, given that the chains of the bridging precursors contain two *I* beads and only one organic bead, the number of contacts *I-I* should not be lower than the one between the organic beads. Considering the values assumed by $\eta_{\alpha\beta}$, the values of the ratios in (4.4.30) are the following:

$$\left.\frac{X_{HH}^{2,0}}{X_{II}^{2,0}}\right|_{IHI} = \left.\frac{X_{HH}^{2,0}}{X_{II}^{2,0}}\right|_{I'HI'} = 0.19 \qquad \left.\frac{X_{TT}^{2,0}}{X_{II}^{2,0}}\right|_{ITI} = 0.25 \qquad \left.\frac{X_{TT}^{2,0}}{X_{II}^{2,0}}\right|_{I'TI'} = 0.24$$

(4.4.31)

Therefore, in the most general case equation (4.4.21) takes the following form:

$$\left(X_{\alpha\alpha}^{i,0}\right)^{\frac{1}{2}} \left\{ \left(X_{\alpha\alpha}^{i,0}\right)^{\frac{1}{2}} + \frac{\left(X_{\beta\beta}^{i,0}\right)^{\frac{1}{2}}}{\eta_{\alpha\beta}} \right\} = X_{\alpha\alpha}^{i,0} \left(\frac{\eta_{\alpha\beta} + c}{\eta_{\alpha\beta}}\right) = \frac{1}{2} q_{\alpha,i} N_i$$

(4.4.32)

where *c* is a constant whose value is always *1*, except for bridging precursors. In this case, *c* is the square root of the ratios in (4.4.31).

Therefore:

$$X_{\alpha\alpha}^{i,0} = \frac{1}{2} q_{\alpha,i} N_i \left(\frac{\eta_{\alpha\beta}}{\eta_{\alpha\beta} + c}\right)$$

(4.4.33)

The values of $X_{\alpha\alpha}^*$, $X_{\alpha\alpha}^{i,0^*}$ and $X_{\alpha\alpha}^{i,0}$, and the solution of the system (4.4.21) can be substituted in equation (4.4.19) to calculate the chemical potential.

$$\frac{\Delta\mu_i}{RT} = \frac{\Delta\mu_i^*}{RT} + \frac{1}{2}z\sum_\alpha q_{\alpha,i}\ln\left(\frac{2(\eta_{\alpha\beta}+c)X_{\alpha\alpha}}{\eta_{\alpha\beta}}\frac{q_{\alpha,i}N_i}{\left(\sum\limits_{i=1}^{3}q_{\alpha,i}N_i\right)^2}\frac{\sum\limits_{i=1}^{3}q_iN_i}{q_iN_i}\right) \qquad (4.4.34)$$

At equilibrium, the chemical potential of the concentrated phase is equal to the chemical potential of the dilute phase. The concentrations at equilibrium can be calculated by solving a non-linear system of three equations:

$$\begin{aligned}\Delta\mu_1^{(conc)} &= \Delta\mu_1^{(dil)}\\ \Delta\mu_2^{(conc)} &= \Delta\mu_2^{(dil)}\\ \Delta\mu_3^{(conc)} &= \Delta\mu_3^{(dil)}\end{aligned} \qquad (4.4.35)$$

where the subscripts *1*, *2*, and *3* indicate the surfactant, the precursor, and the solvent, respectively, in the concentrated (*conc*) or dilute (*dil*) phase.

The number of unknowns is given by the number of concentrations of each component in both phases, that is *6*. However, this number is reduced to *4* by the two physical constraints $\sum\limits_{i=1}^{3}\varphi_i^{(dil,conc)}=1$. Therefore, we need to fix only one concentration (usually the surfactant concentration in the rich phase) to solve the system (4.4.35), whose solutions have been found by applying the Levenberg-Marquardt (LM) algorithm for least-squares estimation of nonlinear parameters [*Marquardt*, 1963]. The LM method optimizes the solution of a given set of (non-linear) functions by minimizing the sum of their squares. Like other optimization techniques, the LM method needs an initial guess of the unknowns involved. Its better robustness, compared to the Gauss-Newton algorithm (GNA), permits a minimum to be found even when this estimate is not so close to the final solution. The improvement with respect to the GNA is in the application of a gradient-descent technique that approaches the minimum by calculating the negative of the gradient of the functions at a given point.

Three functions are defined, $M_i = \frac{\Delta\mu_i^{(conc)} - \Delta\mu_i^{(dil)}}{\Delta\mu_i^{(conc)}}$ and three unknown concentrations. It is more convenient to use vector notation:

$$\underline{M}^T = \left( M_1, M_2, M_3 \right)$$
$$\underline{\phi}^T = \left( \phi_1^{(dil)}, \phi_2^{(conc)}, \phi_2^{(dil)} \right) \tag{4.4.36}$$

The method consists in finding the concentration vector $\underline{\phi}$ for which the objective function $S(\underline{\phi}) = \underline{M}^T \underline{M} = \sum_{i=1}^{3} \left( M_i(\underline{\phi}) \right)^2$ obtains its minimal value.

# Appendix A

Integration of the kinetic component of the Hamiltonian

For a given single particle, with momentum $p_1$ and mass $m_1$, the kinetic contribute to the Hamiltonian can be integrated as follows:

$$I = \int \exp\left(-\frac{p_1^2}{2kTm_1}\right)dp_1 = \int e^{\left(-Cp_1^2\right)}dp_1 \qquad (A.1)$$

where $C = \dfrac{1}{2kTm_1}$

With a variable substitution $x = \sqrt{C}\,p_1$, we can write:

$$I^2 = \frac{1}{C}\left\{\int e^{\left(-x^2\right)}dx\right\}^2 = \frac{1}{C}\left\{\int e^{\left(-x^2\right)}dx \cdot \int e^{\left(-y^2\right)}dy\right\} \qquad (A.2)$$

The integral that we want to solve now is:

$$\iint e^{-\left(x^2+y^2\right)}dxdy \qquad (A.3)$$

Let's change to polar coordinates:

$$x = r\cos\phi$$
$$y = r\sin\varphi \tag{A.4}$$

and the Jacobian associated is:

$$J = \begin{Vmatrix} \dfrac{\partial x}{\partial r} & \dfrac{\partial x}{\partial \phi} \\[2mm] \dfrac{\partial y}{\partial r} & \dfrac{\partial y}{\partial \phi} \end{Vmatrix} = \begin{Vmatrix} \cos\phi & -r\sin\varphi \\ \sin\varphi & r\cos\varphi \end{Vmatrix} = r \tag{A.5}$$

so the incremental area element is $dA = dxdy = rdrd\varphi$ and the double integral becomes:

$$\iint e^{-\left(x^2 + y^2\right)} dxdy = \int\limits_0^{2\pi}\int\limits_0^{\infty} e^{-r^2} rdrd\varphi = 2\pi\int\limits_0^{\infty} e^{-r^2} rdr \tag{A.6}$$

Therefore:

$$2\pi\int\limits_0^{\infty} e^{-r^2} rdr = 2\pi\left(-\frac{1}{2}\right)\int\limits_0^{-\infty} e^{u} du = -\pi\left[e^{u}\right]_0^{-\infty} = -\pi(0-1) = \pi \tag{A.7}$$

The value of the integral $I$ is:

$$I^2 = \frac{1}{C}\pi \tag{A.8}$$

$$I = \sqrt{\frac{\pi}{C}} = \sqrt{2kTm_1\pi} \tag{A.9}$$

For $N$ particles in a three-dimensional space with $m_1 = m_2 = ... = m$, the integration gives:

$$I' = \left(\sqrt{2kTm\pi}\right)^{3N} \tag{A.10}$$

# Appendix B

Detailed balance for the Metropolis algorithm

The Metropolis algorithm is a sufficient condition to satisfy the following detailed balance:

$$P(old \rightarrow new) = P(new \rightarrow old) \tag{B.1}$$

In particular:

$$P(old \rightarrow new) = P(old) \cdot P_{gen}(old \rightarrow new) \cdot P_{acc}(old \rightarrow new) \tag{B.2}$$

and

$$P(new \rightarrow old) = P(new) \cdot P_{gen}(new \rightarrow old) \cdot P_{acc}(new \rightarrow old) \tag{B.3}$$

where $P(old)$ is the probability to be in the old configuration , $P_{gen}(old \rightarrow new)$ is the probability of generating the new configuration from the old configuration, and $P_{acc}(old \rightarrow new)$ is the probability of accepting the new configuration.

If $U(\mathbf{r}_N)_{new} < U(\mathbf{r}_N)_{old}$, the probability to accept the new configuration is 1 and the probability to move from the old to the new configuration is given by:

$$P(old \rightarrow new) = P(old) \cdot P_{gen}(old \rightarrow new) \cdot P_{acc}(old \rightarrow new) =$$

$$(\text{B.4})$$

$$= \frac{\exp(-\beta U(\mathbf{r}_N))_{old}}{\sum_{i=1}^{N} \exp(-\beta U(\mathbf{r}_i))} \cdot P_{gen}(old \rightarrow new) \cdot 1$$

The probability to move from the new configuration to the old one is:

$$P(new \rightarrow old) = P(new) \cdot P_{gen}(new \rightarrow old) \cdot P_{acc}(new \rightarrow old) =$$

$$= \frac{\exp(-\beta U(\mathbf{r}_N))_{new}}{\sum_{i=1}^{N} \exp(-\beta U(\mathbf{r}_i))} \cdot P_{gen}(new \rightarrow old) \cdot \exp[-\beta(-\Delta U(\mathbf{r}_N))] = \qquad (\text{B.5})$$

$$= P_{gen}(new \rightarrow old) \cdot \frac{\exp(-\beta U(\mathbf{r}_N))_{old}}{\sum_{i=1}^{N} \exp(-\beta U(\mathbf{r}_i))}$$

Therefore:

$$\frac{P(old \rightarrow new)}{P(new \rightarrow old)} = \frac{\dfrac{\exp(-\beta U(\mathbf{r}_N))_{old}}{\sum_{i=1}^{N} \exp(-\beta U(\mathbf{r}_i))} \cdot P_{gen}(old \rightarrow new)}{P_{gen}(new \rightarrow old) \cdot \dfrac{\exp(-\beta U(\mathbf{r}_N))_{old}}{\sum_{i=1}^{N} \exp(-\beta U(\mathbf{r}_i))}} = \frac{P_{gen}(old \rightarrow new)}{P_{gen}(new \rightarrow old)} \qquad (\text{B.6})$$

Since in these MC moves the generation of a given new configuration is as random as the regeneration of the old configuration, then $P_{gen}(old \rightarrow new) = P_{gen}(new \rightarrow old)$ and the detailed balance in equation (B.1) results satisfied.

If $U(\mathbf{r}_N)_{new} > U(\mathbf{r}_N)_{old}$, it is easy to prove the same result.

# Appendix C

Energy of a multicomponent system in a cubic lattice

The energy of a multicomponent system, $E$, where all components are restricted to a lattice with coordination $z$ with nearest neighbor interactions, is given by the sum of the interactions of all the pairs in the lattice,

$$E = \sum_{i=1}^{k} N_{ii}\varepsilon_{ii} + \frac{1}{2}\sum_{i=1}^{k}\sum_{j=1}^{k} N_{ij}\varepsilon_{ij}\left(1-\delta_{ij}\right) \tag{C.1}$$

where $N_{ij}$ is the total number of contacts between species $i$ and $j$ on a lattice for some given configuration and $\varepsilon_{ij}$ is the interaction energy for each contact between species $i$ and $j$. The delta function needs to be included to avoid double counting the pure component that is considered in the first summation.

The total number of contacts of each species, $i$, is given by the total number of sites occupied by that specie, $n_i$, and the lattice coordination number,

$$zn_i = N_{ii} + \sum_{j=1}^{k} N_{ij} = 2N_{ii} + \sum_{j=1}^{k} N_{ij}\left(1-\delta_{ij}\right) \tag{C.2}$$

Notice that $N_{ii}$ must be counted twice. Combining equations (C.1) and (C.2) one obtains,

$$E = \frac{1}{2} \sum_{i=1}^{k} \left( zn_i - \sum_{j=1}^{k} N_{ij} \left( 1 - \delta_{ij} \right) \right) \varepsilon_{ii} + \frac{1}{2} \sum_{i=1}^{k} \sum_{j=1}^{k} N_{ij} \varepsilon_{ij} \left( 1 - \delta_{ij} \right) \tag{C.3}$$

which can be simplified as follows,

$$E = \frac{1}{2} \sum_{i=1}^{k} zn_i \varepsilon_{ii} + \frac{1}{2} \sum_{i=1}^{k} \sum_{j=1}^{k} N_{ij} \left( 1 - \delta_{ij} \right) \left( \varepsilon_{ij} - \varepsilon_{ii} \right) \tag{C.4}$$

considering that $\varepsilon_{ij} = \varepsilon_{ji}$ then,

$$E = \frac{1}{2} \sum_{i=1}^{k} zn_i \varepsilon_{ii} + \frac{1}{2} \sum_{i=1}^{k-1} \sum_{j=i+1}^{k} N_{ij} \left( 2\varepsilon_{ij} - \varepsilon_{ii} - \varepsilon_{jj} \right) \tag{C.5}$$

If we define an interchange energy, $\omega_{ij} = \varepsilon_{ij} - \frac{1}{2} \left( \varepsilon_{ii} + \varepsilon_{jj} \right)$, then equation (C.5) becomes,

$$E = \frac{1}{2} \sum_{i=1}^{k} zn_i \varepsilon_{ii} + \sum_{i=1}^{k-1} \sum_{j=i+1}^{k} N_{ij} \omega_{ij} \tag{C.6}$$

Since the number of each species in the lattice is constant the first term is also constant, which represents a thermodynamically irrelevant shift in the zero of the energy. The second term is a summation over all the pairs of unlike contacts of the number of such contacts multiplied by the corresponding interchange energy. Hence, for this $N$ component system, there will be $N(N-1)/2$ independent energy parameters.

# Appendix D

Acceptance probability in configurational bias Monte Carlo

In configurational bias MC simulations, a movement is accepted according to the following probability:

$$P(old \rightarrow new) = P(old) \cdot P_{gen}(old \rightarrow new) \cdot P_{acc}(old \rightarrow new) \tag{D.1}$$

where *P(old)* is the probability to be in the configuration *(old)*, $P_{gen}(old \rightarrow new)$ is the probability of generating the configuration *(new)* from the configuration *(old)*, and $P_{acc}(old \rightarrow new)$ is the probability of accepting the configuration *(new)*. To satisfy the detailed balance, the probability in equation (D.1) has to be equal to the probability of moving back to the configuration *(old)* from the configuration *(new)*:

$$P(new \rightarrow old) = P(new) \cdot P_{gen}(new \rightarrow old) \cdot P_{acc}(new \rightarrow old) = P(old \rightarrow new) \tag{D.2}$$

Therefore:

$$P(old) \cdot P_{gen}(old \rightarrow new) \cdot P_{acc}(old \rightarrow new) =$$

$$= P(new) \cdot P_{gen}(new \rightarrow old) \cdot P_{acc}(new \rightarrow old) \tag{D.3}$$

The ratio between the probabilities of accepting a movement is given by:

$$\frac{P_{acc}(old \rightarrow new)}{P_{acc}(new \rightarrow old)} = \frac{P(new)}{P(old)} \cdot \frac{P_{gen}(new \rightarrow old)}{P_{gen}(old \rightarrow new)}$$ (D.4)

The probability of finding a macroscopic system in the particular state *(new)* or *(old)* is given by:

$$P(new) = \frac{\exp(-\beta U(new))}{\sum_i \exp(-\beta U_i)}$$ (D.5)

$$P(old) = \frac{\exp(-\beta U(old))}{\sum_i \exp(-\beta U_i)}$$ (D.6)

where $\beta = \frac{1}{kT}$ and $U$ is the energy of the system in a given configuration.

Therefore:

$$\frac{P(new)}{P(old)} = \frac{\exp(-\beta U(new))}{\exp(-\beta U(old))} = \exp(-\beta \Delta U)$$ (D.7)

The probability of generating a new state from the old one is given by the product of the probabilities of the single beads constituting the chain of length $l$ to me moved in the new configuration:

$$P_{gen}(old \rightarrow new) = \prod_{i=1}^{l} p_i(old \rightarrow new) =$$
$$= \prod_{i=1}^{l} \frac{\exp(-\beta e_i(new))}{\sum_{j=1}^{k} \exp(-\beta e_i(j))} = \frac{\exp(-\beta U(new))}{W(new)}$$ (D.8)

where $k$ represents the trial directions for the bead $i$ to be built, and $e_i(new)$ and $e_i(j)$ its energy in the configuration *(new)* and *(j)*, respectively. $W(new)$ is the Rosenbluth weight of the new configuration, and it is defined as the product of the contributions of each single bead in the chain:

$$W(new) = \prod_{i=1}^{l} w_i(new)$$ (D.9)

Since the first bead is inserted at random, its contribution to the Rosenbluth weight is:

$$w_1(new) = k \exp\left(-\beta e_1(new)\right)$$ (D.10)

For all the other beads, there are $k$ trial directions to be built, and their contribution is:

$$w_i(new) = \sum_{j=1}^{k} \exp\left(-\beta e_i(j)\right)$$ (D.11)

Similarly, the probability to regenerate the configuration *(old)* from the configuration *(new)*, is given by:

$$P_{gen}(new \to old) = \prod_{i=1}^{l} p_i(new \to old) =$$
$$= \prod_{i=1}^{l} \frac{\exp\left(-\beta e_i(old)\right)}{\exp\left(-\beta e_i(old)\right) + \sum_{j=2}^{k} \exp\left(-\beta e_i(j)\right)}$$ (D.12)

The denominator in equation (D.12) simply states that of $k$ directions, one is the position of bead $i$ in the configuration *(old)*, and the other *k-1* directions represent neighboring positions it could have occupied.
Therefore:

$$P_{gen}(new \to old) = \frac{\exp\left(-\beta U(old)\right)}{\prod_{i=1}^{l} w_i(old)}$$ (D.13)

where:

$$w_i(old) = \exp\left(-\beta e_i(old)\right) + \sum_{j=2}^{k} \exp\left(-\beta e_i(j)\right) \tag{D.14}$$

and $W(old) \equiv \prod_{i=1}^{l} w_i(old)$ is the Rosenbluth weight of the configuration *(old)*.

Then, from equation (D.4) we obtain:

$$\frac{P_{acc}(old \rightarrow new)}{P_{acc}(new \rightarrow old)} = \frac{\exp(-\beta U(new))}{\exp(-\beta U(old))} \cdot \frac{\exp(-\beta U(old))}{W(old)} \cdot \frac{W(new)}{\exp(-\beta U(new))} = \tag{D.15}$$

$$= \frac{W(new)}{W(old)}$$

Therefore the trial move from *(old)* to *(new)* is accepted with a probability equal to:

$$P_{acc}(old \rightarrow new) = \min\left[1, \frac{W(new)}{W(old)}\right] \tag{D.16}$$

When we perform a reptation move, the first bead of the chain is randomly allocated on one of the possible *z-1* free neighbors, and the other beads follow the first one. It means that they are not randomly allocated, but they have only one possible direction to move to. It means that the probability of generating a new configuration, $P_{gen}(old \rightarrow new)$, is equal to the probability to regenerating the old one, $P_{gen}(new \rightarrow old)$, and:

$$\frac{P_{acc}(old \rightarrow new)}{P_{acc}(new \rightarrow old)} = \exp\left(-\beta \Delta U\right) \tag{D.17}$$

The move is accepted with the following probability:

$$P_{acc}(old \rightarrow new) = \min\left[1, \exp\left(-\beta \Delta U\right)\right] \tag{D.18}$$

# Appendix E

Free energy of mixing for chains containing different kinds of contact points

To compute the free energy of mixing, we integrate the Gibbs-Helmholtz equation:

$$\left.\frac{\Delta G}{k_B T}\right|_{mix} = \int \Delta H\big|_{mix}\, d\!\left(\frac{1}{k_B T}\right) \tag{E.1}$$

where the enthalpy of mixing is given by [*Tompa*, 1956]:

$$\Delta H\big|_{mix} = z k_B T \sum_{\alpha,\beta} \left( X_{\alpha\beta} - \sum_{i=1}^{3} X_{\alpha\beta}^{i,0} \right) \ln \eta_{\alpha\beta} \tag{E.2}$$

with $\eta_{\alpha\beta} = \exp\dfrac{\omega_{\alpha\beta}}{k_B T}$.

The derivative of $\eta_{\alpha\beta}$ with respect to $\dfrac{1}{k_B T}$ is equal to:

$$\frac{d\,\eta_{\alpha\beta}}{d\!\left(\dfrac{1}{k_B T}\right)} = \omega_{\alpha\beta} \exp\frac{\omega_{\alpha\beta}}{k_B T} = k_B T \ln\!\left(\eta_{\alpha\beta}\right) \eta_{\alpha\beta} \tag{E.3}$$

From the simultaneous application of equations (E.1), (E.2), and (E.3), we obtain:

$$\left. \frac{\Delta G}{k_B T} \right|_{mix} = z \int \sum_{\alpha,\beta} \left( X_{\alpha\beta} - \sum_{i=1}^{3} X_{\alpha\beta}^{i,0} \right) \frac{d\eta_{\alpha\beta}}{\eta_{\alpha\beta}} \tag{E.4}$$

We can rewrite equation (E.4) as follows:

$$\left. \frac{\Delta G}{k_B T} \right|_{mix} = z \int \sum_{\alpha,\beta} X_{\alpha\beta} \frac{d\eta_{\alpha\beta}}{\eta_{\alpha\beta}} - z \int \left( \sum_{i=1}^{3} X_{\alpha\beta}^{i,0} \right) \frac{d\eta_{\alpha\beta}}{\eta_{\alpha\beta}} \tag{E.5}$$

Since the integral of a sum of functions is equal to the sum of the integrals of each function, we rewrite equation (E.5) in a more convenient way:

$$\left. \frac{\Delta G}{k_B T} \right|_{mix} = z \int \sum_{\alpha,\beta} X_{\alpha\beta} \frac{d\eta_{\alpha\beta}}{\eta_{\alpha\beta}} - z \sum_{i=1}^{3} \int X_{\alpha\beta}^{i,0} \frac{d\eta_{\alpha\beta}}{\eta_{\alpha\beta}} \tag{E.6}$$

Therefore, we need to solve only one of the two integrals of (E.6), as they are mathematically identical. We choose to solve $\int \sum_{\alpha,\beta} X_{\alpha\beta} \dfrac{d\eta_{\alpha\beta}}{\eta_{\alpha\beta}}$ .

For this purpose, we first calculate the total derivative of $X_{\alpha\alpha}$ from the equation of quasi-chemical equilibrium (4.4.1).

The partial derivatives of $X_{\alpha\alpha}$ are calculated with respect to $\eta_{\alpha\beta}$, $X_{\alpha\beta}$, and $X_{\beta\beta}$:

$$\frac{\partial X_{\alpha\alpha}}{\partial \eta_{\alpha\beta}} = 2 \frac{X_{\alpha\alpha}}{\eta_{\alpha\beta}} \; ; \qquad \frac{\partial X_{\alpha\alpha}}{\partial X_{\alpha\beta}} = 2 \frac{X_{\alpha\alpha}}{X_{\alpha\beta}} \; ; \qquad \frac{\partial X_{\alpha\alpha}}{\partial X_{\beta\beta}} = -\frac{X_{\alpha\alpha}}{X_{\beta\beta}} \tag{E.7}$$

Then, the total derivative of $X_{\alpha\alpha}$ is:

$$dX_{\alpha\alpha} = 2 \frac{X_{\alpha\alpha}}{\eta_{\alpha\beta}} d\eta_{\alpha\beta} + 2 \frac{X_{\alpha\alpha}}{X_{\alpha\beta}} dX_{\alpha\beta} - \frac{X_{\alpha\alpha}}{X_{\beta\beta}} dX_{\beta\beta} \tag{E.8}$$

By dividing all by $X_{\alpha\alpha}$, we obtain:

$$\frac{dX_{\alpha\alpha}}{X_{\alpha\alpha}} + \frac{dX_{\beta\beta}}{X_{\beta\beta}} = 2 \frac{d\eta_{\alpha\beta}}{\eta_{\alpha\beta}} + 2 \frac{dX_{\alpha\beta}}{X_{\alpha\beta}} \tag{E.9}$$

In particular:

$$\frac{d\eta_{\alpha\beta}}{\eta_{\alpha\beta}} = \frac{1}{2}\frac{dX_{\alpha\alpha}}{X_{\alpha\alpha}} + \frac{1}{2}\frac{dX_{\beta\beta}}{X_{\beta\beta}} - \frac{dX_{\alpha\beta}}{X_{\alpha\beta}} \tag{E.10}$$

By substituting $\dfrac{d\eta_{\alpha\beta}}{\eta_{\alpha\beta}}$ in the integral $\displaystyle\iint\sum_{\alpha,\beta}X_{\alpha\beta}\frac{d\eta_{\alpha\beta}}{\eta_{\alpha\beta}}$ of the equation (E.6), we obtain:

$$\iint\sum_{\alpha,\beta}X_{\alpha\beta}\frac{d\eta_{\alpha\beta}}{\eta_{\alpha\beta}} = \frac{1}{2}\iint\sum_{\alpha,\beta}X_{\alpha\beta}\frac{dX_{\alpha\alpha}}{X_{\alpha\alpha}} + \frac{1}{2}\iint\sum_{\alpha,\beta}X_{\alpha\beta}\frac{dX_{\beta\beta}}{X_{\beta\beta}} - \iint\sum_{\alpha,\beta}X_{\alpha\beta}\frac{dX_{\alpha\beta}}{X_{\alpha\beta}} \tag{E.11}$$

From equation (4.4.11), we know:

$$\sum_{\beta\neq\alpha}X_{\alpha\beta} = \sum_{i=1}^{3}q_{\alpha i}N_i - 2X_{\alpha\alpha} \tag{E.12}$$

Then, equation (E.11) becomes:

$$
\begin{aligned}
&\frac{1}{2}\iint\sum_{\alpha,\beta}X_{\alpha\beta}\frac{dX_{\alpha\alpha}}{X_{\alpha\alpha}} + \frac{1}{2}\iint\sum_{\alpha,\beta}X_{\alpha\beta}\frac{dX_{\beta\beta}}{X_{\beta\beta}} - \iint\sum_{\alpha,\beta}X_{\alpha\beta}\frac{dX_{\alpha\beta}}{X_{\alpha\beta}} = \\[2mm]
&\frac{1}{2}\iint\left(\sum_{i=1}^{3}(q_{\alpha i}N_i) - 2X_{\alpha\alpha}\right)\frac{dX_{\alpha\alpha}}{X_{\alpha\alpha}} + \frac{1}{2}\iint\left(\sum_{i=1}^{3}(q_{\beta i}N_i) - 2X_{\beta\beta}\right)\frac{dX_{\beta\beta}}{X_{\beta\beta}} - \sum_{\alpha,\beta}X_{\alpha\beta} + C_1 = \\[2mm]
&\left\{\frac{1}{2}\iint\sum_{i=1}^{3}(q_{\alpha i}N_i)\frac{dX_{\alpha\alpha}}{X_{\alpha\alpha}} + \frac{1}{2}\iint\sum_{i=1}^{3}(q_{\beta i}N_i)\frac{dX_{\beta\beta}}{X_{\beta\beta}}\right\} - \left\{X_{\alpha\alpha} + X_{\beta\beta}\right\} - \sum_{\alpha,\beta}X_{\alpha\beta} + C_2 = \\[2mm]
&\frac{1}{2}\sum_{\alpha}\iint\sum_{i=1}^{3}(q_{\alpha i}N_i)\frac{dX_{\alpha\alpha}}{X_{\alpha\alpha}} - \sum_{\alpha}X_{\alpha\alpha} - \sum_{\alpha,\beta}X_{\alpha\beta} + C_3
\end{aligned}
\tag{E.13}
$$

where $C_1$, $C_2$, and $C_3$ are constants of integration.
Therefore, the free energy of mixing is equal to:

$$\left.\frac{\Delta G}{k_B T}\right|_{mix} = \left\{\frac{1}{2}z\sum_\alpha\sum\int\sum_{i=1}^3(q_{\alpha i}N_i)\frac{dX_{\alpha\alpha}}{X_{\alpha\alpha}} - z\sum_\alpha X_{\alpha\alpha} - z\sum_{\alpha,\beta}X_{\alpha\beta}\right\} +$$

$$-\left\{\frac{1}{2}z\sum_\alpha\sum\int\sum_{i=1}^3(q_{\alpha i}N_i)\frac{dX_{\alpha\alpha}}{X_{\alpha\alpha}} - z\sum_\alpha X_{\alpha\alpha} - z\sum_{\alpha,\beta}X_{\alpha\beta}\right\}^{i,0} + C_4 \qquad \text{(E.14)}$$

From equation (E.12) we know that:

$$\left(\sum_\alpha X_{\alpha\alpha} + \sum_{\alpha,\beta}X_{\alpha\beta}\right) = \sum_{i=1}^3 q_i N_i = \left(\sum_\alpha X_{\alpha\alpha} + \sum_{\alpha,\beta}X_{\alpha\beta}\right)^{i,0} \qquad \text{(E.15)}$$

Then (E.14) can be simplified to the form:

$$\left.\frac{\Delta G}{k_B T}\right|_{mix} = \left\{\frac{1}{2}z\sum_\alpha\sum\int\sum_{i=1}^3(q_{\alpha i}N_i)\frac{dX_{\alpha\alpha}}{X_{\alpha\alpha}}\right\} - \left\{\frac{1}{2}z\sum_\alpha\sum\int\sum_{i=1}^3(q_{\alpha i}N_i)\frac{dX_{\alpha\alpha}}{X_{\alpha\alpha}}\right\}^{i,0} + C_4 \qquad \text{(E.16)}$$

So, we obtain:

$$\left.\frac{\Delta G}{k_B T}\right|_{mix} = \frac{1}{2}z\sum_\alpha\left(\sum_{i=1}^3(q_{\alpha i}N_i)\ln\frac{X_{\alpha\alpha}}{X_{\alpha\alpha}^{i,0}}\right) + C \qquad \text{(E.17)}$$

The constant of integration $C$ can be determined by considering that it does not change when the free energy is calculated for an athermal solution:

$$\left.\frac{\Delta G^*}{k_B T}\right|_{mix} = \frac{1}{2}z\sum_\alpha\left(\sum_{i=1}^3(q_{\alpha i}N_i)\ln\frac{X_{\alpha\alpha}^*}{X_{\alpha\alpha}^{i,0*}}\right) + C \qquad \text{(E.18)}$$

By eliminating the constant $C$ between equations (E.17) and (E.18), we obtain:

$$\left.\frac{\Delta G}{RT}\right|_{mix} = \left.\frac{\Delta G^*}{RT}\right|_{mix} + \frac{1}{2}z\sum_\alpha\left(\sum_{i=1}^3(q_{\alpha i}n_i)\ln\frac{X_{\alpha\alpha}}{X_{\alpha\alpha}^{i,0}}\right) - \frac{1}{2}z\sum_\alpha\left(\sum_{i=1}^3(q_{\alpha i}n_i)\ln\frac{X_{\alpha\alpha}^*}{X_{\alpha\alpha}^{i,0^*}}\right) =$$

$$\left.\frac{\Delta G^*}{RT}\right|_{mix} + \frac{1}{2}z\sum_\alpha\left(\sum_{i=1}^3(q_{\alpha i}n_i)\ln\frac{X_{\alpha\alpha}X_{\alpha\alpha}^{i,0^*}}{X_{\alpha\alpha}^{i,0}X_{\alpha\alpha}^*}\right)$$

(E.19)

In terms of number of moles:

$$\left.\frac{\Delta G}{RT}\right|_{mix} = \left.\frac{\Delta G^*}{RT}\right|_{mix} + \frac{1}{2}z\sum_\alpha\left(\sum_{i=1}^3(q_{\alpha i}n_i)\ln\frac{X_{\alpha\alpha}X_{\alpha\alpha}^{i,0^*}}{X_{\alpha\alpha}^{i,0}X_{\alpha\alpha}^*}\right)$$

(E.20)

The chemical potential of the generic component $i$ is given by the derivative of equation (E.20) with respect to the number of moles of $i$, by keeping constant the number of the moles of the other components:

$$\frac{\Delta\mu_i}{RT} = \left(\frac{\partial\,\Delta G/RT}{\partial n_i}\right)_{T,n_{j\neq i}} = \frac{\Delta\mu_i^*}{RT} + \frac{1}{2}z\sum_\alpha\left(q_{\alpha,i}\ln\frac{X_{\alpha\alpha}X_{\alpha\alpha}^{i,0^*}}{X_{\alpha\alpha}^{i,0}X_{\alpha\alpha}^*}\right)$$

(E.21)

# References Chapter 4

Fodi, B., and R. Hentschke, Simulated phase behavior of model surfactant solutions, *Langmuir*, 16, 1626-1633, 2000.

Frenkel, D., and B. Smit, *Understanding Molecular Simulation*, 2nd ed., San Diego, 2002.

Guggenheim, E. A., *Mixtures*, Clarendon Press, Oxford, 1952.

Kim, S. Y., A. Z. Panagiotopoulos, and M. A. Floriano, Ternary oil-water-amphiphile systems: self-assembly and phase equilibria, *Molecular Physics*, 100, 2213-2220, 2002.

Larson, R. G., L. E. Scriven, and H. T. Davis, Monte-Carlo Simulation of Model Amphiphilic Oil-Water Systems, *Journal of Chemical Physics*, 83, 2411-2420, 1985.

Mackie, A. D., K. Onur, and A. Z. Panagiotopoulos, Phase equilibria of a lattice model for an oil-water-amphiphile mixture, *Journal of Chemical Physics*, 104, 3718-3725, 1996.

Mackie, A. D., A. Z. Panagiotopoulos, and S. K. Kumar, Monte-Carlo Simulations of Phase-Equilibria for a Lattice Homopolymer Model, *Journal of Chemical Physics*, 102, 1014-1023, 1995.

Mackie, A. D., A. Z. Panagiotopoulos, and I. Szleifer, Aggregation behavior of a lattice model for amphiphiles, *Langmuir*, 13, 5022-5031, 1997.

Marquardt, D. W., An Algorithm for Least-Squares Estimation of Nonlinear Parameters, *Journal of the Society for Industrial and Applied Mathematics*, 11, 431-441, 1963.

Metropolis, N., A. W. Rosenbluth, M. N. Rosenbluth, A. H. Teller, and E. Teller, Equation of State Calculations by Fast Computing Machines, *Journal of Chemical Physics*, 21, 1087-1092, 1953.

Siperstein, F. R., and K. E. Gubbins, Phase separation and liquid crystal self-assembly in surfactant-inorganic-solvent systems, *Langmuir*, 19, 2049-2057, 2003.

Tompa, H., *Polymer Solutions*, Butterworths Scientific, London, 1956.

*Things should be made as simple as possible,*
*but not any simpler.*
*Albert Einstein*

# Chapter 5

## MACROSCOPIC PHASE SEPARATION IN

## AMPHIPHILIC SYSTEMS

### Introduction

Above the critical micelle concentration, surfactants are able to form micelles of different sizes and shapes according to some given thermodynamic constraints, and inter- and intra-aggregates interactions [*Israelachvili*, 1995]. In Chapter 2, we observed that when the *cmc* is reached, the concentration of free monomers no longer changes and the concentration of the aggregates starts increasing. Therefore, it is reasonable and spontaneous to address the following questions: what is going to happen if we keep on increasing the surfactant concentration? To which extent can we increase such a concentration before some macroscopic changes become evident? Can micellization and macroscopic phase separation coexist?

In this chapter, we try to answer to the above posed questions by presenting the results of MC simulations we have performed in binary and ternary amphiphilic systems, with particular attention to the equilibrium phase diagrams obtained by using different model precursors. After a general introduction on the phase behavior of the amphiphilic systems and their ability to form liquid crystal phases, in section 5.2 we report the phase diagrams calculated by performing MC simulations and by applying QCT.

According to the interactions established between the precursors with the solvent and the amphiphile, different ternary phase diagrams have been obtained. A two-phase region is generally observed where a concentrated phase at high content of surfactant is at equilibrium with a dilute solvent-rich phase [*Patti et al.*, 2006]. Some of the analyzed ternary systems are able to form ordered liquid crystal phases being key in the formation of periodic mesoporous silica or organosilica materials, which will be treated in more detail in Chapters 6 and 7. In section 5.3, the phase diagrams obtained from MC simulations are compared to the ones calculated with the quasi-chemical theory [*Guggenheim*, 1952], by focusing our attention on the difference in the size of the miscibility gaps, that is, on the driving force for the phase separation.

In section 5.4, we analyze the possibility of observing the simultaneous presence of a macroscopic phase separation and a micellization in a given ternary amphiphilic system. Our results indicate that if in the solvent-rich phase the surfactant concentration is high enough, micelles can be observed at equilibrium with a dense phase presenting liquid crystals. In some cases, liquid crystals are obtained in both phases.

## 5.1 Phase Behavior of Amphiphilic Systems.

At the critical micelle concentration, surfactants are able to form micelles. Typical aggregation numbers are in the order of tens of amphiphilic molecules, with diameters of a few nanometers [*Israelachvili*, 1995; *Rosen*, 2004]. However, such a behavior is not of general validity as long-tail surfactants, for instance, do not show any micellization because of their low solubility in the solvent. In this case, even at low concentrations, what we observe is the formation of a solvent-rich phase at equilibrium with a surfactant-rich phase, without the formation of micelles. In Figure 5.1, a snapshot of a model binary surfactant-solvent system showing a crude phase separation is reported. $H_2T_{16}$, the model diblock surfactant forming a separate bulk phase, presents a very long solvophobic tail and it does not form micelles as already reported in another work [*Panagiotopoulos et al.*, 2002].

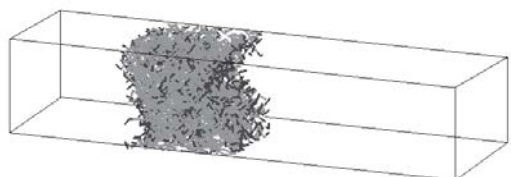

**Fig. 5.1**. Phase separation in a binary surfactant-solvent system. Surfactant: $H_2T_{16}$. Surfactant concentration: 20% (vol). Temperature: $T^*$=8.0. Simulation box: 24×24×100. Dark shading: head segments. Light shading: tail segments. The solvent is not shown.

When the surfactant solubility is high enough, then the formation of micelles can be observed at concentrations above the *cmc*, whose value depends on several chemical and physical properties, such as:

a) Amphiphile chemical structure. Namely, the hydrocarbon chain length, the polarity of the solvophilic group, the valence of counterions in ionic surfactants.

b) Temperature. To a first approximation, *cmc* can be assumed independent of temperature [*Holmberg et al.*, 2002], but theoretical models have calculated this dependence for ionic, non-ionic, and zwitterionic surfactants, and proved the existence of some characteristic differences [*Kim and Lim*, 2004].

c) Co-solutes. The addition of salt lowers the *cmc*, especially in long-chain ionic surfactants [*Rosen*, 2004].

By replacing the long-tail surfactant $H_2T_{16}$ with a surfactant presenting a shorter tail, that is more solvophilic, such as $H_4T_4$, then micellization can occur, as shown in Figure 5.2, where two snapshots obtained by simulations at different temperatures, but at the same surfactant concentration, are reported. At first glance, a difference can be observed in the concentration of free monomers in the two simulation boxes, and hence a dependence on temperature of the aggregates distribution in terms of size and shape can be argued. In particular, the structure and the concentration of surfactant as well as the temperature of the system deeply affect the possibility to observe micelles in a given amphiphilic solution.

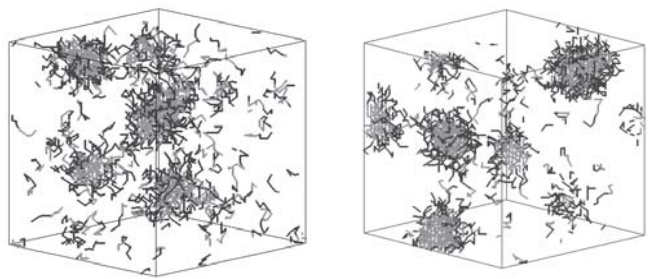

**Fig. 5.2**. Micellization in a binary surfactant-solvent system. Surfactant: $H_4T_4$. Surfactant concentration: 5% (vol). Simulation box: $40^3$. Reduced temperature: $T^*=7.5$ (left), $T^*=6.5$ (right). Dark shading: head segments. Light shading: tail segments. The solvent is not shown.

Surfactants can also show different behaviors when they form micelles, according to the considerations exposed in Chapter 2. The symmetric surfactant $H_4T_4$ gives rise to spherical micelles (as can be inferred from Figure 5.2, although we will verify this affirmation in Chapter 6) and these micelles practically do not change their shape at low to medium surfactant concentrations. On the other hand, the asymmetric surfactant $H_3T_6$ starts by forming spherical micelles at low surfactant concentrations, and then forms rod-like micelles at higher surfactant concentrations [*Al-Anber et al.*, 2003]. The transition from a spherical micellar structure to a rod-like structure is usually referred to as a second critical micelle concentration [*Porte et al.*, 1984]. In Figure 5.3, a rod-like configuration for the amphiphilic system with $H_3T_6$ is reported.

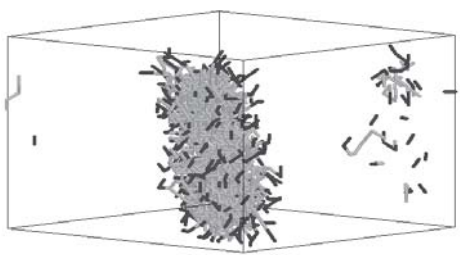

**Fig. 5.3**. Micellization in a binary surfactant-solvent system. Surfactant: $H_3T_6$. Surfactant concentration: 3% (vol). Simulation box: $40^3$. Reduced temperature: $T^*=8.0$. Light shading: tail segments. Dark shading: head segments. The solvent is not shown.

It is clear that it is not possible to generalize completely the micellization behavior of surfactants, but at least one important aspect is evident from the above considerations: surfactants (whose tail group is not excessively long or solvophobic) at their limit of solubility as free monomers, that is the *cmc*, do not form a bulk phase coexisting with the solvent, but organize into aggregates of variable shape and size [*Gelbart and BenShaul*, 1996]. Such a behavior satisfies the hydrophobic effect, that is the necessity for the hydrocarbon tails to be screened from the contact with the solvent, and also permits the solvation of the head groups. Both phenomena are also satisfied below the *cmc*, when the free monomers tend to adsorb at the liquid surface by leaving only the head groups in contact with the liquid, and well above the *cmc*, when the high surfactant concentration modifies the macroscopic properties of the system.

Therefore, the effect of micellization is not so different compared to the effect of a phase separation, as both aim to reduce the free energy of the system by finding a compromise between the hydrophobic effect and the solvation of the heads. However, from a thermodynamic point of view, the micellization is not a true phase separation because micelles are finite-size aggregates: there is phase separation on a microscopic length scale, but not on a macroscopic one [*Mackie et al.*, 1996]. In a macroscopic phase separation, the distinctive aspect is a sudden modification in the physical properties. The physical properties of a dilute solution of surfactants start changing at the *cmc* for the formation of the first aggregates, and this change is continuous and gradual with the increasing total surfactant concentration. When such a concentration is high enough, the micelles start packing together in macroscopic ordered arrangements, according to their original shape and size. These packing arrangements are usually referred to as *liquid crystals* [*Rosen*, 2004].

### 5.1.1 Liquid Crystals.

Liquid crystals (LCs) are intermediate phases (mesophases) between solids and liquids [*Khoo and Simoni*, 1998]. They can flow like liquids, but preserve a molecular order comparable to that of a solid crystalline. The molecules constituting a LC phase are commonly called *mesogens*. LCs can be divided into thermotropic LCs and lyotropic LCs. Thermotropic LCs exhibit phase transition from the liquid or solid phase into the LC phase as temperature is modified; whereas in lyotropic LCs the phase transition into the LC phase depends also on the concentration of the mesogens and the properties of the solvent. In lyotropic LCs, the role of the solvent is of fundamental importance as it provokes the aggregation of the mesogens into micellar aggregates, and then their packing into a well-defined macroscopic arrangement, being the LC phase.

Therefore, a lyotropic LC phase can be obtained from an aqueous solution of amphiphilic molecules (the aforementioned mesogens) by increasing the surfactant concentration to values above the *cmc*. In particular, spherical micelles pack together into cubic LC phases, cylindrical micelles into hexagonal LC phases, and bilayers into lamellar LC phases [*Holmberg et al.*, 2002]. However, it is also possible to observe a transition from a given LC phase to another one by further increment of the surfactant concentration, as observed in Figure 5.4, where a lamellar phase is obtained from a hexagonal phase by increasing the surfactant concentration from 50% up to 75% by volume.

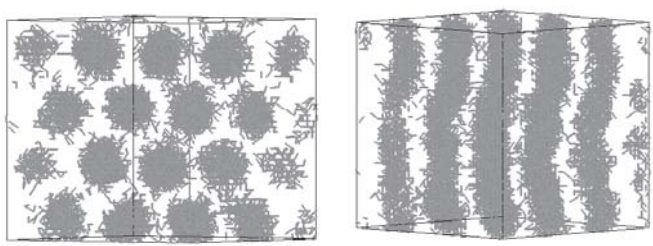

**Fig. 5.4**. Liquid crystal phases obtained from $H_4T_4$ at 50% (hexagonal, left) and 75% (lamellar, right) by volume. Simulation box: $40^3$. Reduced temperature: $T^*$=8.0. For clarity, only tail segments are shown.

The conditions at which the above mentioned LC phases are obtained are usually reported in phase diagrams showing the range of concentrations and temperatures needed for a given phase to be formed.

In Figure 5.5, a typical phase diagram for a binary amphiphilic system showing phase transitions from a simple isotropic micellar structure to more complex lyotropic LC phases is reported [*Balmbra et al.*, 1969].

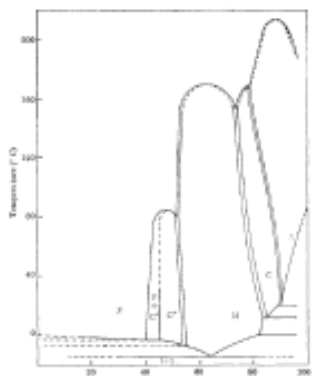

**Fig. 5.5**. Phase diagram for dodecyltrimethylammonium chloride and water. *F*: micelles; *C'*: cubic phases; *M*: hexagonal phases; *C*: gyroid phases; *N*: lamellar phases; *I*: ice; *S*: solid. Concentrations in weight%. [*Balmbra et al.*, 1969]. Solid lines : experimental boundaries. Dashed lines: interpolated boundaries.

In Figure 5.6, we report a schematic phase diagram presenting the LC phases obtained in a binary system containing the model surfactant $H_4T_4$. The simulations have been performed in a lattice box of size $40^3$ during $2\times10^{10}$ MC steps at different reduced temperatures between $T^*$=6.5 and $T^*$=8.5. A visual inspection confirmed the presence of hexagonally ordered cylinders or lamellae. Cubic and gyroid phases have not been detected. These structures depend on the box dimensions, and our aim in this research work was not to establish the (narrow) range of concentrations in which they appear.

The results presented in Figure 5.6 are in very good agreement with the results of the simulations run by other researchers [*Larson*, 1996; *Siperstein and Gubbins*, 2003], who detected the gyroid phase at a surfactant volume fraction of around 75%.

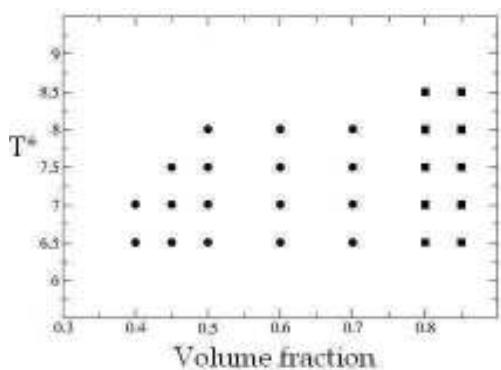

**Fig. 5.6.** Schematic $H_4T_4$/solvent phase diagram showing the range of formation of hexagonal and lamellar LC structures. Solid circles and squares represent hexagonal and lamellar LC phases, respectively.

We observe the formation of hexagonal phases for surfactant concentrations above 40% by volume, and lamellar phases for much higher concentrations (above 75%). The range of formation of ordered phases becomes smaller as the temperature is increased. In particular, no order is observed when the temperature is raised above $T^*$=9.0. In this case, the system is composed of a disordered solution containing free monomers and behaves like a liquid.

If the system is composed of three components, then an isothermal ternary phase diagram is commonly used to visualize the phases at equilibrium. The temperature is kept constant to make the analysis of the diagram easier, otherwise a three-dimensional representation would be needed. A ternary phase diagram is often presented as an equilateral triangle, with each vertex representing one of the three pure components, and each side representing the mutual solubility of pairs of components. Inside the triangle, all the points indicate thermodynamic equilibrium between the components distributed in one or more phases. In Figure 5.7, a schematic representation of an isothermal ternary phase diagram for an amphiphilic water solution is reported.

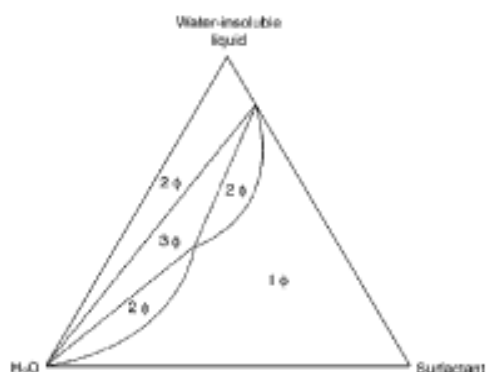

**Fig. 5.7.** Schematic ternary phase diagram for an amphiphilic water solution. The number of phases ($\phi$) is reported [*Rosen*, 2004].

The addition of a third component brings important modifications in the phase and aggregation behavior of a given amphiphilic aqueous solution. According to the nature of this component, and in particular to the interactions established with the surfactant and the solvent, the mechanism and the range of concentrations to synthesize LC phases are also modified. An appropriate choice of the third component can lead to the formation of ordered LC phases even when the total surfactant concentration in the system is well below the above mentioned values needed to observe hexagonal phases in binary systems. This fact is the consequence of the phase separation into a solvent-rich phase and a surfactant-rich phase, and is the topic of the following section.

## 5.2 Phase Separation in Amphiphilic Ternary Systems with Different Hybrid Precursors.

One of the aims in this research work, is to understand to which extent a given ordered LC phase can be obtained by modifying the characteristics of an amphiphilic system, and in particular the conditions needed to observe hexagonally ordered cylindrical aggregates, being key in the synthesis of hybrid mesoporous materials.

In this section, we analyze the phase behavior of a model amphiphilic solution constituted by a solvent, a surfactant, and a third component that we call hybrid precursor for containing an organic and an inorganic group. The morphological analysis of such structures is reported in Chapter 6 and Chapter 7. Here, we study how the nature of the hybrid precursor can affect the phase behavior of amphiphilic solutions, by focusing on the interactions formed with the solvent and with the surfactant.

Different hybrid precursors have been modeled according to their solubility with the solvent, their organic group, and the position of this organic group, as already described in Chapter 4. The peculiarity of the system is that the inorganic part of the precursor has favorable interactions with the solvophilic segments of the surfactant, similar to what is expected to be found in systems containing surfactants and inorganic oxide precursor. In the following, we first report the phase diagrams obtained for systems with pure inorganic precursors or terminal hybrid precursors, and then those with bridging hybrid precursors.

It should be noted that the nomenclature used here to identify the organosilica precursors has been chosen for uniformity with the literature [*Hatton et al.*, 2005], and it does not aim to stress on the position of the organic group in the precursor as a leading factor for the phase separation. The position of the organic group can affect the organization of the organic and inorganic beads in the corona of the aggregate, as we will see in Chapter 7, but it is not a dominating factor to determine the phase behavior of systems containing terminal or bridging precursors.

All the binary phase diagrams precursor/solvent showing a phase separation are reported in Appendix F. These diagrams have been obtained by simulating the binary systems at different reduced temperatures for $10^9$ MC steps in lattice boxes of size 24×24×100. The phase diagram of the system containing the precursor $I'T$ is exactly equal to the one containing the pure silica precursor $I'_2$, because the interactions established in the two systems are identical ($\omega_{I'T} = 0, \omega_{I'S} = \omega_{TS} = 1$). Therefore it has not been included.

Those precursors with an insoluble inorganic bead, $I'$, or a solvophilic organic group, $H$, show partial miscibility with the solvent. The repulsion between $I'$ and the solvent ($S$) or the strong attraction between the inorganic segment ($I'$) and the solvophilic one ($H$), are the driving force for the phase separation in these binary systems. On the other hand, the binary systems containing $I_2$ or the hybrid precursors $IT$ and $ITI$ do not show phase separation even at the very low reduced temperature $T^*$=2.0, because $I$ is completely miscible with the solvent and $\omega_{IT}$ is sufficiently repulsive to prevent a phase separation originated by an eventual $I$-$T$ association. The presence of a solvophobic group is not sufficient to give rise to a phase separation, unless the inorganic bead is insoluble in the solvent too, because the interactions between $I$ and $T$ are quite unfavorable ($\omega_{IT} = 1$) for the formation of a phase mainly composed by hybrid precursors.

Such a result confirms that the phase separation in the systems here presented originates from a purely energetic effect. In some other cases, despite the absence of any explicit interaction between the components, an entropic phase transition can be observed, as reported for colloidal, polymeric, and also amphiphilic systems, where the presence of endcaps or junctions between the chains lead to the formation of networks and can give rise to a phase separation between a junction-poor and a junction-rich phase [*Zilman et al.*, 2003].

Considering the total number of different precursors modeled, ten different phase diagrams have been calculated. We estimated the critical points in those systems presenting equilibrium data in the neighborhood of the critical point, by extrapolating the following equation [*Widom*, 1967]:

$$\rho_{surf} - \rho_{surf}^{C} = A \left| \rho_{prec} - \rho_{prec}^{C} \right|^{\beta^*} \tag{5.1}$$

where $A$ is a constant, $\rho_{surf}$ and $\rho_{prec}$ are the surfactant and the precursor concentrations in a given phase, respectively. The concentrations presenting the suffix $C$ are the concentrations at the critical point. The renormalized critical exponent, $\beta^*$, is approximately 1/3 for the model used, which belongs to the Ising universality class [*Fisher*, 1968; *Widom*, 1967]. Therefore, the binodal curve associated with this equation is practically cubic in the neighborhood of the critical point, as already observed [*Widom*, 1967]. We chose to use this value to calculate the concentrations at the critical point, as also suggested in other research works applying the same lattice model [*Mackie et al.*, 1995]. To estimate the coordinates of the critical points we extrapolated the coexisting densities from the two closest tie lines with our assumed value of $\beta^*=1/3$.

### 5.2.1    Pure Inorganic and Terminal Hybrid Precursors. Phase Diagrams.

In this section we present the phase diagrams of six different systems: $H_4T_4/I_2/S$, $H_4T_4/I'_2/S$, $H_4T_4/IT/S$, $H_4T_4/I'T/S$, $H_4T_4/IH/S$, and $H_4T_4/I'H/S$ at $T^*$=8.0. For each system, we report the phase diagrams obtained by performing lattice MC simulations, and by applying the quasi chemical theory (QCT). Generally, according to the global concentration in the system, and the nature of the precursor, a biphasic equilibrium can be observed between a surfactant-rich phase and a dilute solvent-rich phase.

The equilibrium points in the phase diagrams are obtained by analyzing the concentration profiles of the last configurations, as indicated in Figure 5.8 for a system with phase separation, and in Figure 5.9 for a system without phase separation, but with formation of micelles.

To guarantee a good estimate of the equilibrium compositions in systems presenting phase separation, the average densities are calculated in the bulk of the phases, that is, far from the interfaces. When it is not possible to clearly identify the bulk concentrations, then the length of the simulation box is increased or several composition profiles calculated from different configurations are averaged. The concentrations at equilibrium are then reported in the isothermal ternary phase diagrams and are connected by tie-lines.

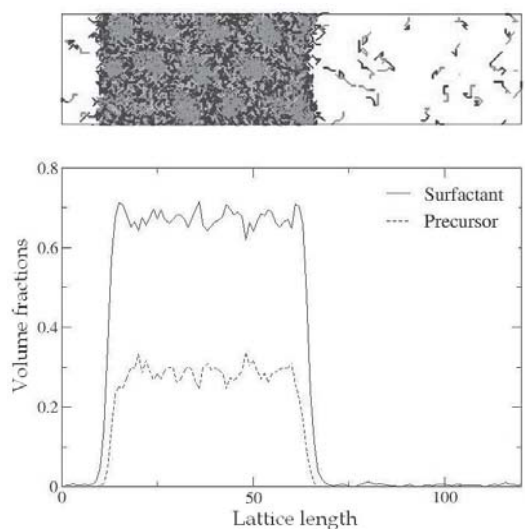

**Fig. 5.8.** Snapshot and concentration profiles of the system $H_4T_4/I_2/S$ at $T^*=8.0$. A phase separation is observed. Global concentrations: 30% $H_4T_4$, 12% $I_2$. Simulation box: 30×30×120. Dark shading: surfactant heads. Light shading: surfactant tails or precursor. The solvent is not shown.

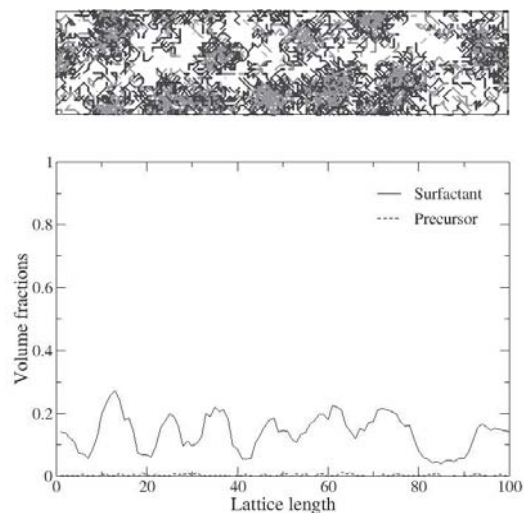

**Fig. 5.9**. Snapshot and concentration profiles of the system $H_4T_4/I_2/S$ at $T$*=8.0. No phase separation, but micellization is observed. Global concentrations: 14% $H_4T_4$, 0.35% $I_2$. Simulation box: 24×24×100. Dark shading: surfactant heads. Light shading: surfactant tails or precursor. The solvent is not shown.

Figures 5.10 and 5.11 show the phase diagrams for the systems presenting a precursor with a complete or a partial solubility with the solvent, respectively. In both cases, the phase separation predicted by MC simulations has been confirmed by QCT. In particular, the agreement between simulations and QCT is very good when the precursor is only partially miscible with the solvent, especially when the volume concentration of the surfactant in the surfactant-rich phase is lower than 55%, as no long-range order is observed. At higher concentrations, the agreement is still good even though the system gives rise to ordered LC phases which are not taken into account by QCT.

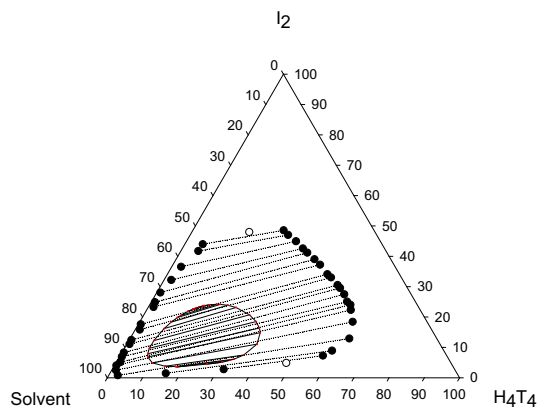

**Fig. 5.10.** Phase diagram of the system $H_4T_4/I_2/S$ at $T^*$=8.0. Solid circles and dotted lines: MC simulations. Solid lines: QCT. The empty circles represent the estimated location of the critical points, based on MC data.

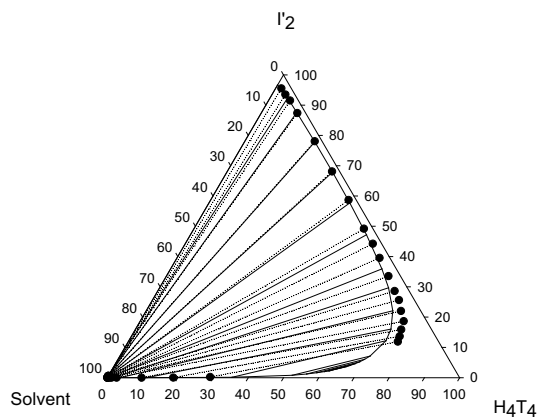

**Fig. 5.11**. Phase diagram of the system $H_4T_4/I'_2/S$ at $T^*$=8.0. Solid circles and dotted lines: MC simulations. Solid lines: QCT.

For the completely miscible precursor, the driving force for the phase separation is the strong interaction between its inorganic source, *I*, and the solvophilic heads of the surfactant, *H*. Such a phase separation is usually referred to as associative phase separation, whereas a segregative phase separation is the result of a very strong repulsion between two different components, and/or when the two components interact with the solvent in a different way [*Piculell and Lindman*, 1992]. The phase separation observed in the $H_4T_4/I'_2/S$ system is a typical example of segregative phase separation, and is the result of two factors. In the first place, there is a very strong repulsion between the precursor $I'_2$ and the solvent $S$. In the second place, there are also strong interactions between the inorganic segment and the surfactant heads. These driving forces lead to a lower content of solvent in the concentrated phase in comparison to the system with $I_2$.

Figure 5.12 shows the phase separation observed in a simulation box of size 24×24×100 for some mixtures of both completely and partially soluble pure inorganic precursors.

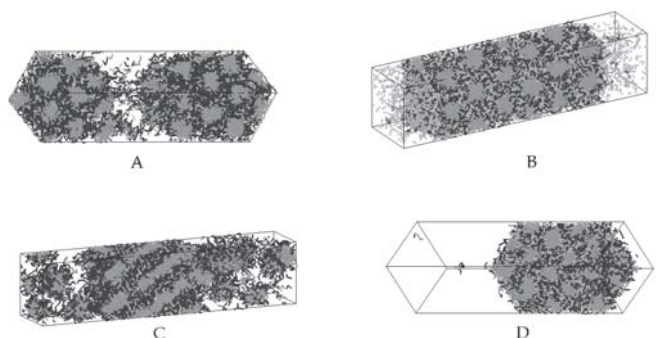

**Fig. 5.12**. Phase separation observed at $T^*$=8.0 for pure hybrid precursors in a lattice box of size 24×24×100. Global concentrations: 40% $H_4T_4$ - 5% $I_2$ (A), 38% $H_4T_4$ - 22% $I_2$ (B), 40% $H_4T_4$ - 5% $I'_2$ (C), 30% $H_4T_4$ - 12% $I'_2$ (D). Light shading represents the surfactant tails or the inorganic precursor, dark shading represents the surfactant heads. The solvent is not shown.

If one of the two beads in the hybrid precursor is substituted with a solvophilic group, $H$, the size of the miscibility gap becomes smaller regardless of the solubility of the precursor in the solvent, as observed in Figures 5.13 and 5.14.

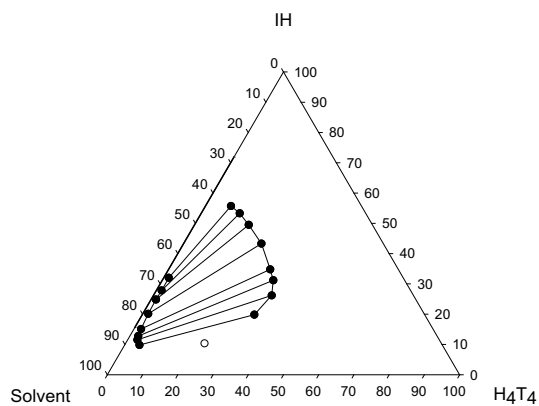

**Fig. 5.13**. Phase diagram of the system $H_4T_4/IH/S$ at $T^*$=8.0. Solid circles and solid lines: MC simulations. No miscibility gap has been found with QCT at this temperature. The empty circle represents the estimated location of the critical point, based on MC data.

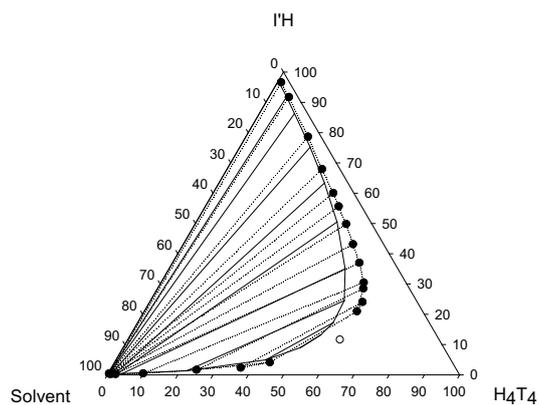

**Fig. 5.14**. Phase diagram of the system $H_4T_4/I'H/S$ at $T^*$=8.0. Solid circles and dotted lines: MC simulations. Solid lines: QCT. The empty circle represents the estimated location of the critical point, based on MC data.

Effectively, the driving force for the associative phase separation decreases, because the interactions formed between the surfactant heads and $IH$ are not as strong as the interactions between the surfactant heads and $I_2$, as reported in Table 4.2. The same considerations are still valid if we change $I'_2$ with $I'H$. In this case, the decrease of the driving force for the phase separation is also due to the increase of solubility of the precursor in the solvent, due to the presence of the solvophilic group $H$.

By observing the ternary phase diagram of the system $H_4T_4/IH/S$ in Figure 5.13, it seems that the precursor presents a complete solubility in the solvent. As a matter of fact, the critical temperature of the binary system $IH/S$ is approximately $T^*$=8.0, as can be inferred from Figure F.2 in Appendix F. The proximity of this critical point makes it difficult to observe this phase separation in the ternary phase diagram of the system $H_4T_4/IH/S$ in the region close to the *IH-S* axis (Figure 5.13). In the case of the QCT, an immiscibility gap is found however it is restricted to a very small area in which the highest surfactant concentration is close to 0.5% (Figure 5.15).

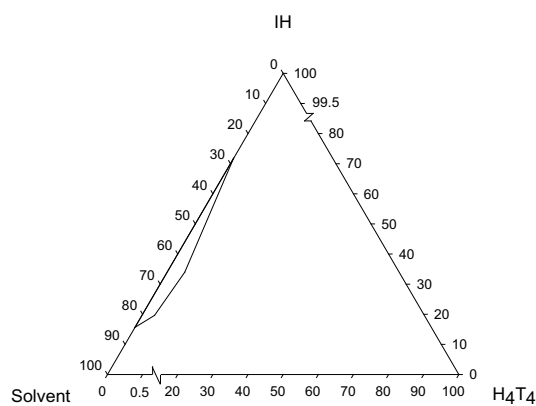

**Fig. 5.15**. Phase diagram of the system $H_4T_4/IH/S$ obtained with QCT at $T^*$=8.0.

The agreement between MC simulations and QCT is very good for the system $H_4T_4/I'H/S$, as observed in Figure 5.14. The main difference with the purely inorganic precursor $I'_2$, is that $I'H$ forms ordered aggregates in the dense phases at lower surfactant concentrations. Therefore deviations of the QCT from the equilibrium data of the simulations are observed at lower surfactant concentrations than with $I'_2$. Ordered liquid crystal phases are observed at surfactant concentrations higher than 50% by volume. Nevertheless the quantitative agreement between theory and simulations is still very satisfactory.

Figure 5.16 shows the phase separation observed in a simulation box of size 24×24×100 for some mixtures of both complete and partial soluble terminal solvophilic precursors.

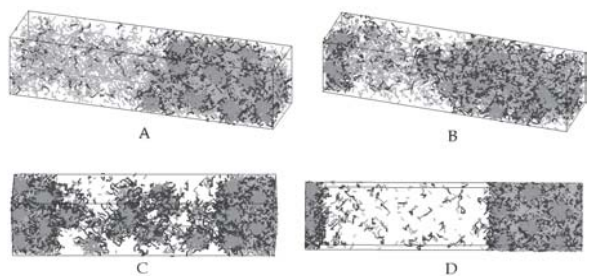

**Fig. 5.16.** Phase separation observed at $T^*=8.0$ for solvophilic terminal precursors in a $24\times24\times100$ lattice box. Global concentrations: 15% $H_4T_4$ - 20% $IH$ (A), 20% $H_4T_4$ - 15% $IH$ (B), 30% $H_4T_4$ - 10% $I'H$ (C), 20% $H_4T_4$ - 15% $I'H$ (D). Light shading represents the surfactant tails or the inorganic precursor, dark shading represents the surfactant heads. The solvent is not shown.

The phase diagrams of the systems with a solvophobic terminal group, $T$, are reported in Figures 5.17 and 5.18.

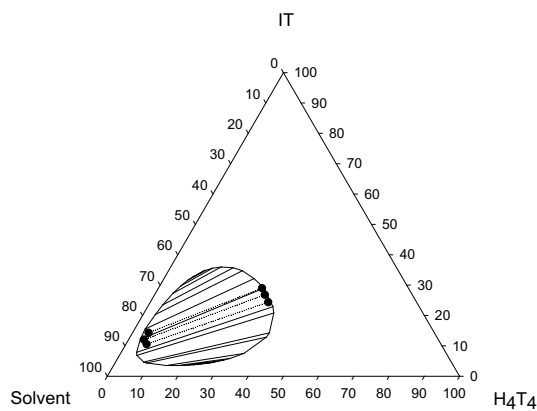

**Fig. 5.17.** Phase diagram of the system $H_4T_4/IT/S$ at $T^*=8.0$. Solid circles and dot lines: MC simulations. Solid lines: QCT.

The quantitative agreement between theory and simulations is remarkably good for both systems $H_4T_4/IT/S$ and $H_4T_4/I'T/S$, as no ordered liquid crystal phases are observed at any global concentration along the coexistence line. Despite the high surfactant concentration in the concentrated phase of the system containing $I'T$, we do not observe the formation of any ordered structure. Such a result is not associated with the interactions established between the beads $I'$ and $T$, that are different to the interactions between $I$ and $T$ (simulations with $\omega_{I'T} = 1$ did not give ordered structures either), but more probably with the strong insolubility of the precursor $I'T$ in the solvent.

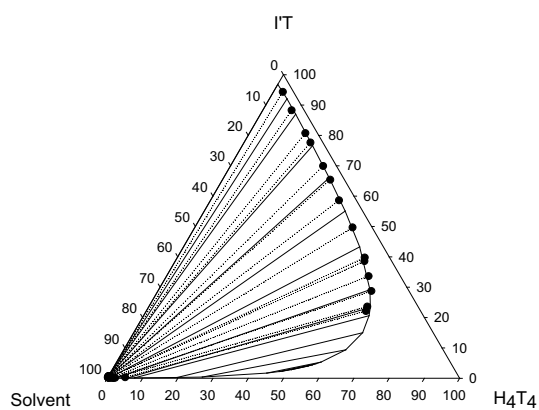

**Fig. 5.18**. Phase diagram of the system $H_4T_4/I'T/S$ at $T^*=8.0$. Solid circles and dotted lines: MC simulations. Solid lines: QCT.

Figure 5.19 shows the phase separation observed in a simulation box of size 24×24×100 for two solutions of both complete and partial soluble terminal solvophobic precursors.

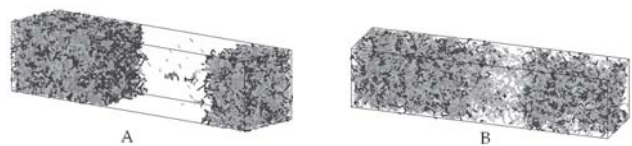

**Fig. 5.19**. Phase separation observed at $T^*=8.0$ for solvophobic terminal precursors in a lattice box of size 24×24×100. Global concentrations: 30% $H_4T_4$ - 20% $I'T$ (A), 20% $H_4T_4$ - 20% $IT$ (B). Light shading represents the surfactant tails or the inorganic precursor, dark shading represents the surfactant heads. The solvent is not shown.

### 5.2.2    Bridging Hybrid Precursors. Phase Diagrams.

We have modeled the bridging hybrid precursors with chains formed by one solvophilic ($H$) or solvophobic ($T$) segment in between two segments of type $I$ or $I'$. The phase diagrams of four different systems ($H_4T_4/IHI/S$, $H_4T_4/I'HI'/S$, $H_4T_4/ITI/S$, and $H_4T_4/I'TI'/S$) obtained by performing lattice MC simulations, and by applying the QCT are calculated at $T^*=8.0$.

As a general trend, we observe that a higher concentration of surfactant is achieved in the concentrated phase in comparison with the corresponding terminal precursors, and the miscibility gaps result to be bigger. The change in the shape of the miscibility gap can be explained by considering the nature of the driving force for the phase separation, that is the strong interaction between the $I$ (or $I'$) segments and the surfactant head segments. If we increase the number of $I$ ($I'$) beads in a given precursor, the phase separation will be enhanced as the surfactant (heads) will experiment a stronger attraction towards the hybrid precursor. On the other hand, if we substitute a bridging precursor with a terminal one, the attraction between surfactant and precursor will be weaker and the amount of surfactant in the concentrated phase will be less.

The phase diagram of the system containing the bridging organic precursor $I'HI'$ is very similar to the phase diagram of the system containing $I'_2$, because the decrease in the driving force for the segregative phase separation (higher solubility for $I'HI'$ than for $I'_2$) is compensated by an increase in the driving force for the associative phase separation, due to the presence of an extra $H$ group in the precursor. When we change $I'_2$ with $I'TI'$, the solubility of the precursor in the solvent does not change, but its attraction towards the surfactant heads decreases due to the presence of the solvophobic group $T$, and the segregative driving force becomes weaker.

The phase diagrams obtained with solvophilic bridging precursors are reported in Figures 5.20 and 5.21.

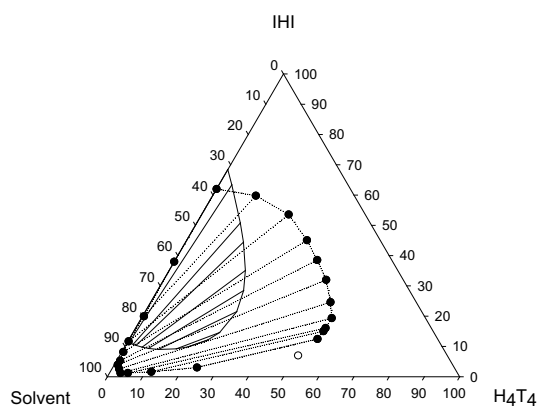

**Fig. 5.20**. Phase diagram of the system $H_4T_4/IHI/S$ at $T^*=8.0$. Solid circles and dot lines: MC simulations. Solid lines: QCT. The empty circle represents the estimated location of the critical point, based on MC data.

When the solvophilic precursor *IHI* is used, the biphasic region is smaller than in the $H_4T_4/I'HI'/S$ system, but it is bigger than in the analogous terminal precursor, *IH* (Figure 5.13). At the same reduced temperature, $T^*=8.0$, both precursors are able to phase separate, but only by using *IHI* is it possible to observe ordered LC phases in the concentrated phase, as the surfactant concentration at equilibrium with the solvent-rich phase is higher than the one achieved in the concentrated phase of the system containing *IH*. This point clarifies why a phase separation between a solvent-rich phase and a surfactant-rich phase is of fundamental importance for such ternary amphiphilic systems. In fact, thanks to a macroscopic phase separation, and according to the nature of the components in the mixture, a LC can form even when the global surfactant concentration in the system is very low. The agreement between MC simulations and QCT is quite poor as ordered LC phases have been found over all the range of surfactant-rich mixtures.

A good agreement between MC simulations and QCT predictions can be observed when the inorganic segment is only partially soluble in the solvent, and in particular for those surfactant concentrations in the concentrated phases lower than 50% by volume. At higher concentrations, the miscibility gap predicted by QCT becomes slightly smaller than the one calculated by MC simulations, as ordered LC phases form.

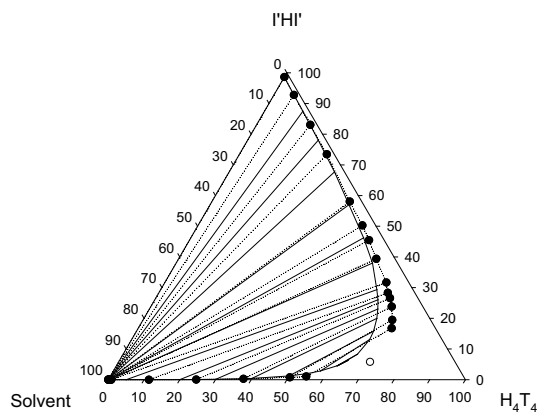

**Fig. 5.21**. Phase diagram of the system $H_4T_4/I'HI'/S$ at $T^*$=8.0. Solid circles and dotted lines: MC simulations. Solid lines: QCT. The empty circle represents the estimated location of the critical point, based on MC data.

When the surfactant concentration in the dense phase is above 50%, a different tendency in the slope of the tie-lines is detected. Compared to the tie-lines at lower surfactant concentrations, their slope seems to favor a higher amount of hybrid precursor in the dense phase, and a second point of intersection, called pole [*Mertslin et al.*, 1961], between the tie-lines is detected. According to Campbell *et al.* [*Campbell et al.*, 1963], the isothermal tie-lines of a given ternary phase diagram, if adequately extended, should meet in one pole, usually outside the triangular diagram. This supposition has found support in some experimental ternary phase diagrams, such as the ethanol/water/$n$-hexane diagram [*Mertslin et al.*, 1961], and also in the empirical rule developed by Tarassenkov [*Tarassenkov*, 1946], who also affirmed that the straight line connecting the critical point to the pole is the tangent of the binodal curve at the critical point. Tarassenkov tested his method with six different ternary systems, showing a reasonable agreement with the experimental results. Although these works declared that deviations from this behavior should be considered as experimental errors, other experimental phase diagrams, such as the ethanol/chloroform/water diagram [*Zollweg*, 1971], show a quite different trend in which at least two poles can be seen. A theoretical approach to this issue was presented for ternary liquid mixtures by Widom [*Widom*, 1967], who calculated the coordinates of the tie-lines intersection under some given approximations.

In our ternary amphiphilic systems, although we did not enter into the details concerning the orientation of the equilibrium tie-lines, we observed one or two poles of intersection between the tie-lines in a given phase diagram. This is possibly connected to the presence of ordered structures which are not observed at lower surfactant concentrations. However, the fact that such a behavior is also noticed in the tie-lines calculated with the QCT, which does not predict the formation of ordered phases, is probably due to a second cause not involving the self-assembly, but mere energetic considerations based on the origin of the phase separation in these systems. At low surfactant concentrations, the driving force for the phase separation is mainly due to the immiscibility of the inorganic source $I'$ in the solvent. When the surfactant concentration increases, then the strong interactions established between the surfactant heads and the inorganic precursor acquire more weight and can become dominant at very high surfactant concentrations. As a matter of fact, the slope of the tie-lines in this region leads to a higher content of hybrid precursor in the surfactant-rich phase, compared to the case in which the pole would not have changed its position. Although less evident, a similar change in the slope of the tie-lines can be also observed in the phase diagrams of those systems with a hybrid precursor presenting an insoluble inorganic source and ordered aggregates in the dilute phases. In these cases, two poles are also generally observed.

Figure 5.22 shows the phase separation observed in simulation boxes of size 24×24×100 for some solutions of both complete and partially soluble bridging solvophilic precursors.

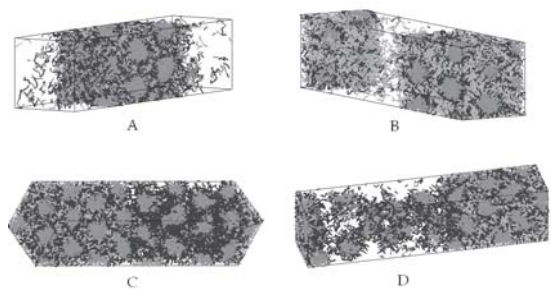

**Fig. 5.22.** Phase separation observed at $T^*$=8.0 for solvophilic bridging precursors in 24×24×100 lattice boxes. Global concentrations: 30% $H_4T_4$ - 10% $IHI$ (A), 25% $H_4T_4$ - 20% $IHI$ (B), 60% $H_4T_4$ - 10% $I'HI'$ (C), 40% $H_4T_4$ - 10% $I'HI'$ (D). Light shading represents the surfactant tails or the inorganic precursor, dark shading represents the surfactant heads. The solvent is not shown.

In Figures 5.23 and 5.24, the phase diagrams obtained for systems presenting solvophobic bridging precursors are reported.

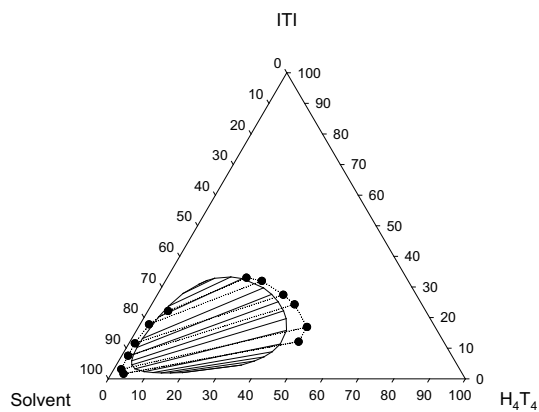

**Fig. 5.23**. Phase diagram of the system $H_4T_4/ITI/S$ at $T^*=8.0$. Solid circles and dotted lines: MC simulations. Solid lines: QCT.

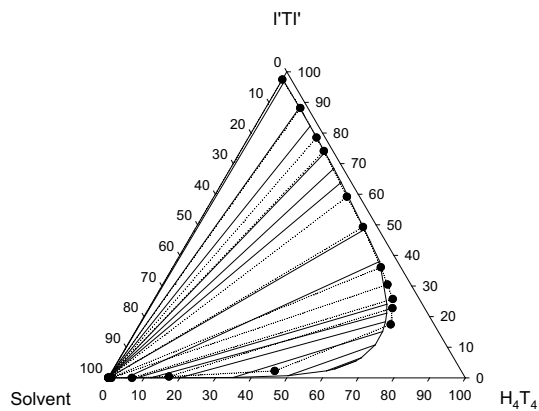

**Fig. 5.24**. Phase diagram of the system $H_4T_4/I'TI'/S$ at $T^*=8.0$. Solid circles and dotted lines: MC simulations. Solid lines: QCT.

As mentioned previously, the system with *ITI* presents a bigger miscibility gap than that observed in the system with the analogous terminal precursor, *IT*. However, this is not the only remarkable difference between the two systems. Spherical aggregates are formed in the concentrated phase of the system with *ITI* which are not observed in the system with *IT*. This can be also argued from the quantitative discrepancy between the miscibility gap calculated with the simulations and the one obtained with the QCT. On the other hand, the phase diagram obtained with *I'TI'* shows a very similar trend to the one obtained with $I'T$, and its agreement with the phase diagram calculated with the QCT is very good. Figure 5.25 shows the phase separation observed in 24×24×100 simulation boxes for two solutions of both completely and partially soluble bridging solvophobic precursors.

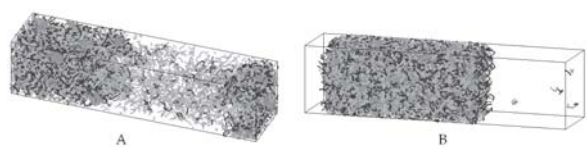

**Fig. 5.25**. Phase separation observed at $T^*$=8.0 for solvophobic bridging precursors in 24×24×100 lattice boxes. Global concentrations: 20% $H_4T_4$ - 20% $ITI$ (A), 25% $H_4T_4$ - 25% $I'TI'$ (B). Light shading represents the surfactant tails or the inorganic precursor, dark shading represents the surfactant heads. The solvent is not shown.

## 5.3 QCT and MC Simulations.

In the previous sections, we observed that the phase diagrams calculated with the QCT (QCT-diagrams) are in very good agreement with the ones obtained with lattice MC simulations (MC-diagrams), especially when no micelles or ordered LC phases are formed. In this case, the scenario predicted by the simulations does not differ so much from the one predicted by the theory, as a phase separation without formation of ordered structures is observed.

Considering that MC simulations and QCT use the same model and the same parameters , and that the thermodynamic properties of the system are a result of the interactions between neighboring sites, in this section the QCT-diagrams and the MC-diagrams are compared by calculating the number of contacts formed at equilibrium between $H$, $T$, $I$ (or $I'$), and $S$ beads. As expected, the closer the number of the contacts calculated with both methods, the better the agreement between simulation and theory. It should be noted that we are counting the sum of the total number of contacts in both phases in equilibrium at the same global composition, rather than the contacts in each separate phase.

The presence of an interface can represent a problem when comparing the number of contacts if its volume is not negligible with respect to the total volume of the simulation box. Therefore, in the following comparison, we are considering those systems whose phases at equilibrium share an interface occupying only a small percentage of the total box volume, in between 5% and 10%, and being as far as possible from the critical region, ensuring that the interface is as sharp as possible.

In Table 5.1, we report the number of contacts formed in MC simulations divided by the number of contacts calculated with QCT, when ordered phases are observed in the system $H_4T_4/I'_2/S$. In this case, the number of contacts between beads of the same type, that is $X_{HH}$, $X_{TT}$, $X_{I'I'}$ and $X_{SS}$, is higher when calculated from MC simulations than from QCT. Regarding the other kinds of contacts, namely those between beads of different types, a different trend is observed: only the number of contacts between $H$ and $I'$ sites, $X_{HI'}$, is higher in MC than in QCT; all the other contacts show an opposite tendency in order to satisfy the balance of sites (see equation 4.4.12).

**Table 5.1**. Ratio between the number of contacts between different kinds of beads obtained from MC simulations and QCT, for the systems with pure insoluble hybrid precursor. Global concentrations are considered.

| System | $\dfrac{HH_{MC}}{HH_{QCT}}$ | $\dfrac{HT_{MC}}{HT_{QCT}}$ | $\dfrac{HI'_{MC}}{HI'_{QCT}}$ | $\dfrac{HS_{MC}}{HS_{QCT}}$ | $\dfrac{TT_{MC}}{TT_{QCT}}$ | $\dfrac{TI'_{MC}}{TI'_{QCT}}$ | $\dfrac{TS_{MC}}{TS_{QCT}}$ | $\dfrac{I'I'_{MC}}{I'I'_{QCT}}$ | $\dfrac{I'S_{MC}}{I'S_{QCT}}$ | $\dfrac{SS_{MC}}{SS_{QCT}}$ |
|---|---|---|---|---|---|---|---|---|---|---|
| 4% $I'_2$ - 30% $H_4T_4$ | 1.46 | 0.63 | 1.45 | 0.82 | 2.20 | 0.51 | 0.49 | 1.63 | 0.50 | 1.08 |
| 6% $I'_2$ - 30% $H_4T_4$ | 1.50 | 0.57 | 1.38 | 0.68 | 2.06 | 0.48 | 0.45 | 1.34 | 0.72 | 1.06 |
| 8% $I'_2$ - 30% $H_4T_4$ | 1.48 | 0.56 | 1.30 | 0.64 | 2.04 | 0.52 | 0.42 | 1.23 | 0.92 | 1.06 |
| 10% $I'_2$ - 30% $H_4T_4$ | 1.48 | 0.54 | 1.29 | 0.57 | 2.08 | 0.55 | 0.39 | 1.18 | 0.90 | 1.06 |
| 15% $I'_2$ - 30% $H_4T_4$ | 1.33 | 0.52 | 1.23 | 0.65 | 1.94 | 0.61 | 0.56 | 1.22 | 0.67 | 1.03 |
| 20% $I'_2$ - 30% $H_4T_4$ | 1.22 | 0.49 | 1.17 | 0.91 | 1.83 | 0.65 | 0.96 | 1.21 | 0.59 | 1.02 |
| 5% $I'_2$ - 40% $H_4T_4$ | 1.53 | 0.64 | 1.48 | 0.78 | 2.27 | 0.49 | 0.43 | 1.49 | 0.53 | 1.17 |

This result is coherent with the formation of ordered aggregates in MC simulations, where tail-rich cores are surrounded by a corona of head segments in contact with the beads of the hybrid precursor. Since QCT does not predict the formation of microphase separated regions, the density of contacts between equal beads cannot be as high as in the simulations.

Moreover, the number of contacts between the solvent beads and the other three types of beads, that is, $X_{TS}$, $X_{HS}$ and $X_{I'S}$, is smaller when calculated with the simulations than with the theory. Since these types of contacts are mostly present in the surfactant-rich phase, as can be observed in the phase diagrams of Figure 5.11, there is more solvent admitted in this phase when applying QCT. This also explains the restriction of the miscibility gap in the QCT-diagrams with respect to the MC-diagrams.

On the other hand, when no structural order is observed, the results of MC simulations and QCT are quite similar, as shown in Table 5.2 for the system $H_4T_4/I'T/S$ which does not show the formation of any ordered LC phase.

**Table 5.2**. Ratio between the number of contacts between different kinds of beads obtained from simulations and QCT, for the system $H_4T_4/I'T/S$. Global concentrations are considered.

| System | $\frac{HH_{MC}}{HH_{QCT}}$ | $\frac{HT_{MC}}{HT_{QCT}}$ | $\frac{HI'_{MC}}{HI'_{QCT}}$ | $\frac{HS_{MC}}{HS_{QCT}}$ | $\frac{TT_{MC}}{TT_{QCT}}$ | $\frac{TI'_{MC}}{TI'_{QCT}}$ | $\frac{TS_{MC}}{TS_{QCT}}$ | $\frac{I'I'_{MC}}{I'I'_{QCT}}$ | $\frac{I'S_{MC}}{I'S_{QCT}}$ | $\frac{SS_{MC}}{SS_{QCT}}$ |
|---|---|---|---|---|---|---|---|---|---|---|
| 10% $I'T$ - 10% $H_4T_4$ | 1.15 | 0.86 | 0.94 | 1.25 | 1.10 | 0.94 | 0.93 | 0.99 | 1.64 | 1.00 |
| 10% $I'T$ - 15% $H_4T_4$ | 1.16 | 0.81 | 0.97 | 1.24 | 1.17 | 0.95 | 0.78 | 1.06 | 1.34 | 1.00 |
| 15% $I'T$ - 15% $H_4T_4$ | 1.23 | 0.89 | 0.98 | 1.03 | 1.13 | 0.97 | 0.74 | 1.04 | 1.17 | 1.01 |

The miscibility gaps of both QCT and MC simulations becomes practically identical, as well as $X_{HH}$, $X_{I'T'}$, $X_{SS}$ and $X_{HI'}$. The number of contacts between the surfactant tails, $X_{TT}$, is still higher when calculated with MC simulations because, although there is not any long range order in the concentrated phase, small aggregates can form and peaks of tails concentration are reported. All the other contacts fluctuate since the total number of contacts is conserved on a lattice.

Both QCT and MC simulations have been previously applied to study oil-water-amphiphile systems with symmetric or asymmetric short surfactants [*Kim et al.*, 2002; *Larson et al.*, 1985; *Mackie et al.*, 1996]. In these studies, only one independent interaction parameter was used, as water was modeled by a single $H$ segment and oil by one or more $T$ segments. The comparison between theory and simulations shows an opposite trend in the relative size of the miscibility gaps compared to our results. In particular, the surfactant concentration in the concentrated phase was observed to be higher when calculated with QCT than with MC simulations [*Kim et al.*, 2002]. The system studied in our research work requires six different independent interaction parameters ($\omega_{HT}$, $\omega_{HI}$, $\omega_{HS}$, $\omega_{TI}$, $\omega_{TS}$, and $\omega_{IS}$), and hence displays a much richer behavior from the aforementioned model. The presence of the inorganic units, $I$ or $I'$, being strongly attracted by the surfactant heads, introduces new and more complex features with respect to a system where only two kinds of beads ($H$ and $T$) were considered.

The apparent discordance between the results of the two kinds of systems finds a possible explication in the equations of site balance for the ternary $H_4T_4/T/H$ system, given below:

$$2X_{HH} + X_{HT} = q_{H,1}N_1 + q_{H,3}N_3$$
$$2X_{TT} + X_{HT} = q_{T,1}N_1 + q_{T,2}N_2$$

(5.2)

where the subscripts *1*, *2*, and *3*, refer to the surfactant ($H_4T_4$), oil ($T$), and water ($H$), respectively. The number of contacts $X_{TT}$ and $X_{HH}$ result to be higher from MC simulations because ordered aggregates are formed, and tail segments tend to stay together as well as head segments. Since $X_{TT}$ and $X_{HH}$ result to be smaller when QCT is applied, then $X_{HT}$ has to be higher to satisfy the sites balance (5.2). Therefore, considering that the contacts between *H* and *T* sites are mainly formed in the surfactant-rich phase, the surfactant concentration is higher when calculated with QCT than with MC simulations.

## 5.4 Micellization and Phase Separation.

The ternary systems analyzed so far, generally show a miscibility gap in the phase diagrams, being the evidence of a phase separation between a (dilute) solvent-rich phase and a (concentrated) surfactant-rich phase. According to the model hybrid precursor, the concentrated phase can present ordered liquid crystals, and, in some of these cases, the corresponding dilute phase shows a surfactant concentration high enough for the formation of micelles or more complex aggregates, as observed in sections 5.2.1 and 5.2.2. As a matter of fact, a visual inspection confirmed to us the formation of ordered aggregates, especially spherical micelles, in the dilute phase.

Such a result is not of general validity for the amphiphilic systems, as reported by other researchers [*Panagiotopoulos et al.*, 2002], whose work focused on binary surfactant/solvent solutions. They showed that the same model for surfactants of varying head and tail lengths either forms micelles or phase separates, but never both, and suggest that the experiments showing the opposite are a consequence of the particular solvation properties of water.

Our system is more complex as a third component, that is the hybrid precursor, is added. Moreover, the dilute phase, being almost totally composed by surfactant and solvent as the precursor concentration is lower than 2%, is typical of a binary $H_4T_4/S$ system which forms micelles [*Al-Anber et al.*, 2005; *Floriano et al.*, 1999]. Therefore, the simultaneous presence of micellization and phase separation is reasonable. Other researchers studied very similar ternary amphiphilic systems with particular attention to their phase behavior [*Bhattacharya and Gubbins*, 2005; *Siperstein and Gubbins*, 2003], and, although in their work it was not explicitly mentioned, their ternary phase diagrams are such that micellization and phase separation can simultaneously occur.

Nevertheless, we preferred to verify in more detail the validity of our results by (*a*) changing the size of the lattice box; (*b*) simulating the systems at different global concentrations belonging to the same tie-line; (*c*) decreasing gradually the temperature from $T^*$=10.0 to $T^*$=8.0; and (*d*) following the displacement of some randomly chosen chains in the lattice box during the simulation.

Such verifications are additional tests to confirm that these systems are properly sampled since the simultaneous presence of micelles and liquid crystals in two separate phases could be considered as a local minimum of energy in which the system is trapped. We could have performed the same verifications for all the systems studied, including those showing either micellization or phase separation, as it would have been a more careful procedure to ensure the achievement of the equilibrium than the examination of the total energy of the system and a visual inspection of its final configuration. Nevertheless, we think it is reasonable to leave such a detailed analysis for the most *critical* systems, and to consider the others at equilibrium if the total energy does not undergo any significant change.

In the following, we report the results of the above mentioned points (*a*) to (*d*) for the systems $H_4T_4/I'HI'/S$ (60% $H_4T_4$, 10% $I'HI'$) and $H_4T_4/I'H/S$ (35%$H_4T_4$, 10% $I'H$), showing equilibrium between two LC phases, and a LC phase with spherical micelles, respectively. In Figures 5.26 and 5.27, the final configurations of both systems in lattice boxes of different size are compared. For the system $H_4T_4/I'H/S$, an enlargement of both phases is also presented in order to better appreciate the presence of micelles in the dilute phase, and of the hexagonally ordered liquid crystals in the concentrated phase.

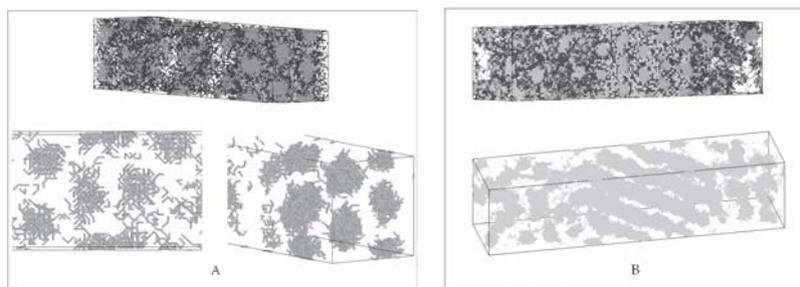

**Fig. 5.26.** Final configurations for the system $H_4T_4/I'H/S$ (35%$H_4T_4$, 10% $I'H$) at $T^*$=8.0. Lattice boxes: 24×24×100 (A) and 35×35×140 (B). Light shading represents the surfactant tails or the inorganic precursor, dark shading represents the surfactant heads. In Fig. A, an enlargement of the surfactant tails in the dilute and concentrated phases is given. Fig. B has been rotated to better observe the presence of spherical aggregates micelles. The solvent is not shown.

In Figure 5.27, the system $H_4T_4/I'HI'/S$ shows the same hexagonal arrangement in the dilute and in the concentrated phases. The presence of the precursor $I'HI'$, which accumulates only in the solvent-rich phases because of its immiscibility, helps to distinguish the two phases.

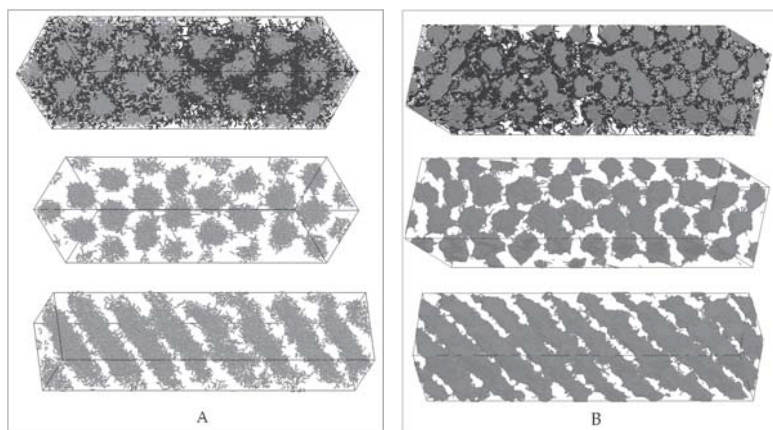

**Fig. 5.27.** Final configurations for the system $H_4T_4/I'HI'/S$ (60%$H_4T_4$, 10% $I'HI'$) at $T^*$=8.0. Lattice boxes: 24×24×100 (A) and 30×30×125 (B). Light shading represents the surfactant tails or the inorganic precursor, dark shading represents the surfactant heads. Only the surfactant tails are shown to better appreciate the order in the two phases. The solvent is not shown.

In Figure 5.28, the final configurations of the same systems at different global concentrations are reported. The global concentrations have been selected from the same equilibrium tie-line connecting the dilute and concentrated phases of Figures 5.26 and 5.27. Again, two LC phases are observed for the system with the bridging precursor, and spherical micelles at equilibrium with a LC phase for the system with the terminal precursor. In Figure 5.28 (B), the lattice box has been rotated to better appreciate the micelles in the dilute phase.

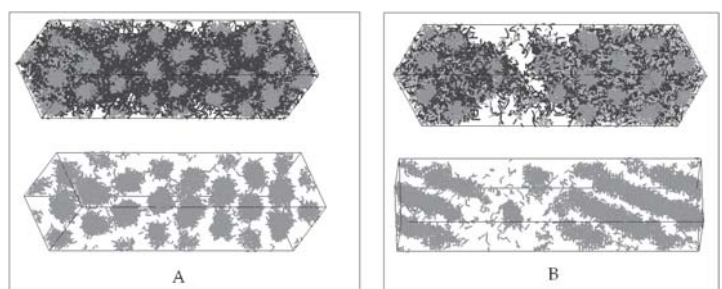

**Fig. 5.28**. Final configurations for the systems (*A*) $H_4T_4/I'HI'/S$ (55%$H_4T_4$, 5% $I'HI'$), and (*B*) $H_4T_4/I'H/S$ (42%$H_4T_4$, 16% $I'H$) at $T^*$=8.0. Lattice box: 24×24×100. Light shading represents the surfactant tails or the inorganic precursor, dark shading represents the surfactant heads. Only the surfactant tails are shown to better appreciate the order in the two phases. The solvent is not shown.

The total energy of both the analyzed systems has been calculated for decreasing values of the reduced temperature, as observed in Fig. 5.29. Starting from $T^*$=10.0, the temperature has been gradually lowered by a $\Delta T^*$=0.2, till $T^*$=8.0. At each reduced temperature, 2×10$^{10}$ MC steps have been performed to allow the systems to equilibrate, and the total number of MC steps was 2.2×10$^{11}$. The snapshots of the final configurations are observed in Figure 5.30.

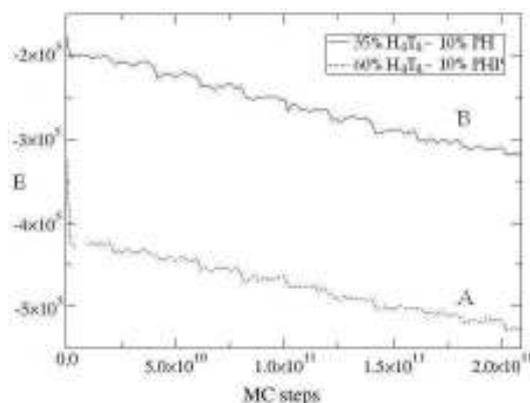

**Fig. 5.29**. Variation of the energy, $E$, of the systems (*A*) $H_4T_4/I'HI'/S$ (60%$H_4T_4$, 10% $I'HI'$), and (*B*) $H_4T_4/I'H/S$ (35%$H_4T_4$, 10% $I'H$), at different reduced temperatures. Lattice box: 24×24×100.

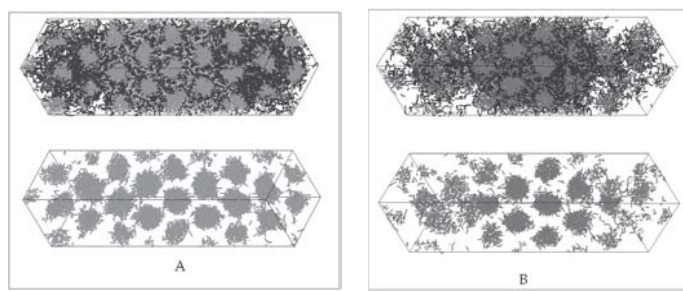

**Fig. 5.30**. Final configurations for the systems (*A*) $H_4T_4/I'HI'/S$ (60%$H_4T_4$, 10% $I'HI'$), and (*B*) $H_4T_4/I'H/S$ (35%$H_4T_4$, 10% $I'H$). The final reduced temperature ($T^*$=8.0) has been achieved starting from $T^*$=10.0. Lattice box: 24×24×100. Light shading represents the surfactant tails or the inorganic precursor, dark shading represents the surfactant heads. Only the surfactant tails are shown to better appreciate the order in the two phases. The solvent is not shown.

The last verification performed is based on the following consideration. When the chemical potentials of the components in different phases are not the same, the system has not yet reached the thermodynamic equilibrium, and there is a net mass transfer from the phase at higher chemical potential to the phase at lower chemical potential. When the system reaches equilibrium, the mass transfer does not stop, but becomes bi-directional, namely the molecules can move indifferently from one phase to the other. On average, the mass transfer from the dilute phase to the concentrated phase is equal to the mass transfer in the opposite direction. This is usually defined as dynamic equilibrium, because the observable properties of the system are not changing, but there is a constant movement of molecules between the phases.

Therefore, we followed the displacement of a bunch of randomly selected surfactant chains in the lattice box once the system reached equilibrium (or what was supposed to be at equilibrium). To be precise, we did not follow the displacement of the whole chain, but just of one of its segments, being approximately located in the middle, in order to speed up the calculations. A simulation of $10^{10}$ MC steps was performed and the position of the chain was registered every $10^7$ MC steps, namely a total of $10^3$ times. Since the phases can also displace during the calculation, we have compared the displacement of the selected chains with regard to the concentration profiles of all the chains in the box. The results are shown in Figure 5.31 for both the systems $H_4T_4/I'HI'/S$ (60% $H_4T_4$, 10% $I'HI'$), and $H_4T_4/I'H/S$ (35% $H_4T_4$, 10% $I'H$).

In both cases, the selected chains appear in the two phases. However, to be completely sure that the chains were really moving from one phase to the other and viceversa, we also calculated their position during the simulation. Figure 5.32 illustrates that effectively the chains are moving from the dilute to the concentrated phase and viceversa, and not just from a phase at higher chemical potential to a phase at lower chemical potential, that would correspond to a non-equilibrium situation.

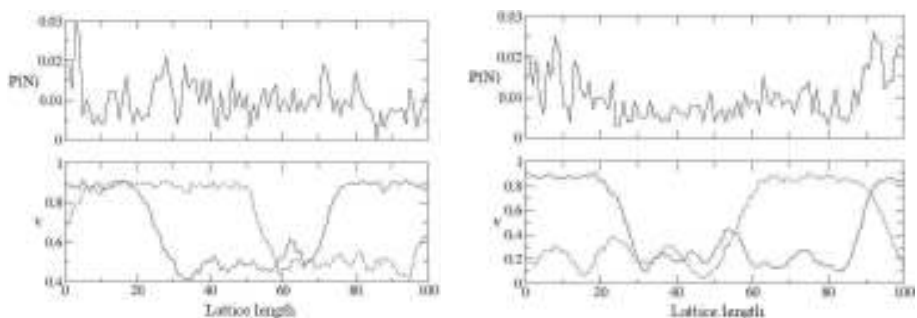

**Fig. 5.31**. Chain displacement in the systems (left) $H_4T_4/I'HI'/S$ (60% $H_4T_4$, 10% $I'HI'$), and (right) $H_4T_4/I'H/S$ (35% $H_4T_4$, 10% $I'H$), and relative chain concentration profiles. *P(N)* is the probability for a chain to be at a given lattice length, and $v$ is the volume fraction of all the chains in the box. The graphs at the top show the number of times a given chain happened to be at a given lattice length, and the graphs at the bottom represent the concentration profile of all the chains in the box at the beginning (solid line) and at the end (dashed line) of the simulation.

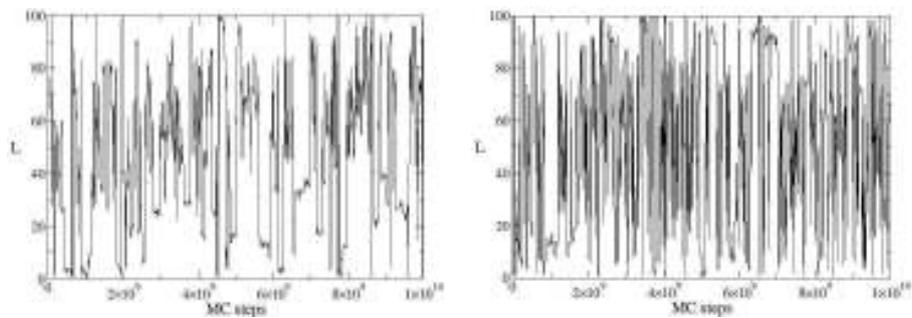

**Fig. 5.32**. Position of a selected chain of the systems (right) $H_4T_4/I'HI'/S$ (60% $H_4T_4$, 10% $I'HI'$) and (left) $H_4T_4/I'H/S$ (35% $H_4T_4$, 10% $I'H$) versus the simulation steps. $L$ is the lattice length.

# Appendix F

Binary Phase Diagrams Hybrid Precursor/Solvent

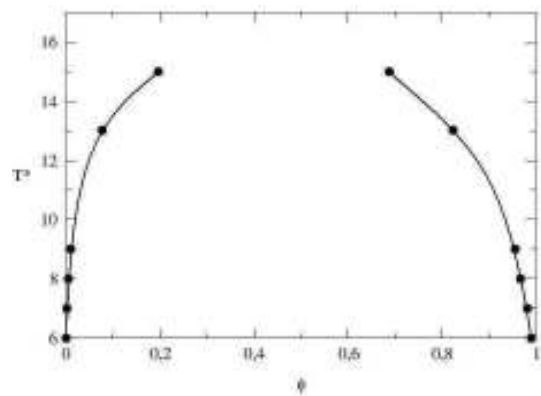

**Fig. F.1**. Phase diagram of the binary system $I'_2/S$. $\phi$ is the precursor volume fraction. The solid lines are guides for the eyes.

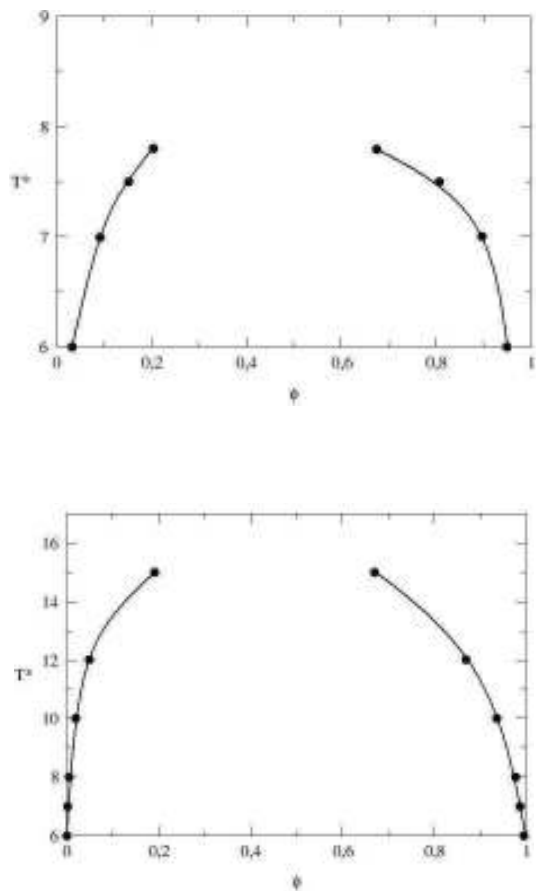

**Fig. F.2**. Phase diagram of the binary systems $IH/S$ (top) and $I'H/S$ (down). $\phi$ is the precursor volume fraction. The solid lines are guides for the eyes.

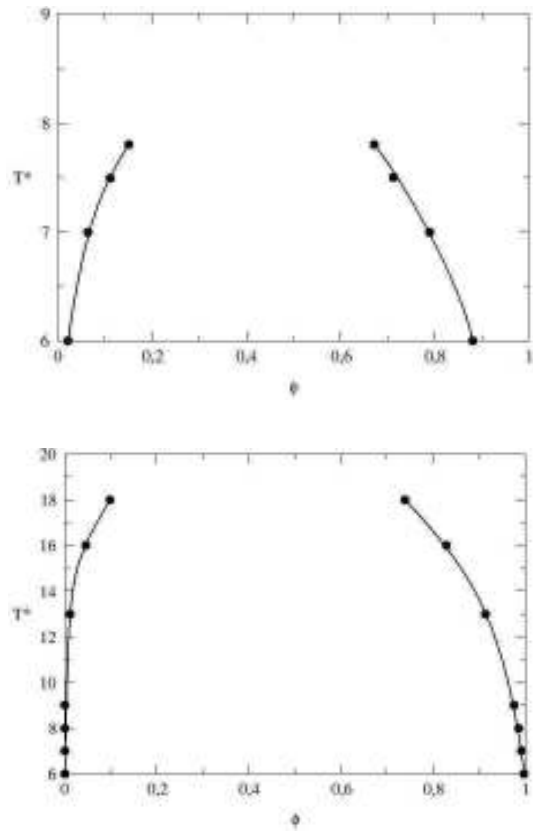

**Fig. F.3**. Phase diagram of the binary systems *IHI*/*S* (top) and *I′HI′*/*S* (down). $\phi$ is the precursor volume fraction. The solid lines are guides for the eyes.

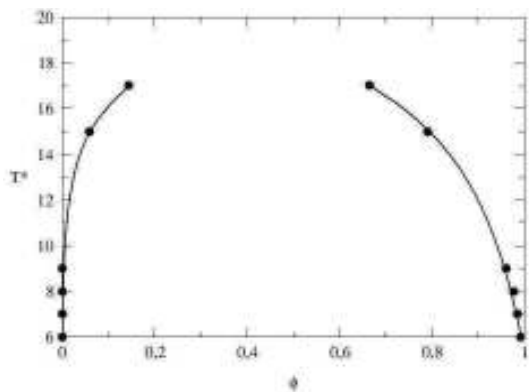

**Fig. F.4.** Phase diagram of the binary system $l'Tl'/S$. $\phi$ is the precursor volume fraction. The solid lines are guides for the eyes.

# Appendix G

**Equilibrium Data**

$$H_4T_4/I_2/S$$

| MC | | | | QCT | | | |
|---|---|---|---|---|---|---|---|
| Concentrated Phase | | Dilute Phase | | Concentrated Phase | | Dilute Phase | |
| Surfactant | Precursor | Surfactant | Precursor | Surfactant | Precursor | Surfactant | Precursor |
| 57.5788 | 7.2650 | 31.6692 | 2.7481 | 30.0000 | 5.0874 | 26.1085 | 4.4684 |
| 59.3416 | 8.9076 | 15.8829 | 1.4335 | 32.0000 | 5.7138 | 18.1208 | 3.5368 |
| 62.2272 | 12.9030 | 2.5825 | 0.7921 | 33.0000 | 6.1080 | 17.6670 | 3.6220 |
| 60.4307 | 18.4216 | 1.2153 | 2.4566 | 35.0000 | 7.4503 | 14.5702 | 3.7729 |
| 57.9644 | 22.3090 | 0.5570 | 4.0148 | 37.0000 | 11.0333 | 10.3530 | 4.8782 |
| 55.5170 | 25.0627 | 0.6790 | 6.8789 | 36.0000 | 14.8218 | 8.0540 | 6.8231 |
| 53.5032 | 27.4678 | 0.5382 | 8.2812 | 35.0000 | 16.3567 | 7.5391 | 7.8834 |
| 51.2915 | 29.5266 | 0.7791 | 11.0645 | 33.0000 | 18.4621 | 7.1215 | 9.6799 |
| 50.1085 | 30.5302 | 0.6230 | 12.4353 | 30.0000 | 20.7071 | 7.1207 | 12.1877 |
| 47.0000 | 33.0000 | 1.3021 | 15.8257 | 25.0000 | 23.1062 | 8.0525 | 16.1679 |
| 45.4282 | 34.0972 | 0.7063 | 17.5821 | 20.0000 | 24.1306 | 10.0043 | 19.8108 |
| 41.7807 | 37.2173 | 1.4782 | 23.2788 | 19.0000 | 24.1716 | 10.5811 | 20.5077 |
| 39.3092 | 39.0168 | 1.0931 | 24.8553 | 13.0000 | 22.5548 | 15.6082 | 23.7067 |
| 36.1545 | 41.1838 | 1.0341 | 28.0269 | 12.0000 | 21.8537 | 16.8581 | 23.9912 |
| 34.2355 | 42.5603 | 2.0718 | 32.1798 | 9.0000 | 18.2374 | 22.2613 | 23.8519 |
| 31.1074 | 44.9281 | 2.6762 | 36.5346 | | | | |
| 27.7927 | 47.0189 | 4.9613 | 41.7268 | | | | |
| 25.7837 | 48.5498 | 5.1117 | 43.9794 | | | | |

$H_4T_4/I'_2/S$

| MC | | | | QCT | | | |
|---|---|---|---|---|---|---|---|
| Concentrated Phase | | Dilute Phase | | Concentrated Phase | | Dilute Phase | |
| Surfactant | Precursor | Surfactant | Precursor | Surfactant | Precursor | Surfactant | Precursor |
| 77.6281 | 11.7158 | 47.1855 | 0.6309 | 66.0000 | 3.6396 | 61.8061 | 2.2541 |
| 77.0739 | 13.9489 | 35.9649 | 0.3350 | 67.0000 | 4.0742 | 61.8075 | 2.2599 |
| 76.7804 | 15.0361 | 19.9425 | 0.1263 | 68.0000 | 4.6048 | 60.6774 | 2.0080 |
| 75.8165 | 16.5085 | 4.1824 | 0.0063 | 69.0000 | 5.2552 | 58.6700 | 1.6258 |
| 74.8000 | 18.6591 | 2.6284 | 0.0000 | 73.0000 | 12.2112 | 35.3707 | 0.1464 |
| 73.6690 | 20.2257 | 1.3485 | 0.0000 | 71.0000 | 7.1484 | 52.0849 | 0.8161 |
| 72.3524 | 22.0269 | 1.1640 | 0.0079 | 70.0000 | 6.0691 | 55.9752 | 1.2269 |
| 69.9752 | 25.6349 | 0.6120 | 0.0087 | 72.5000 | 16.0428 | 11.2734 | 0.0077 |
| 67.2365 | 28.6236 | 0.1157 | 0.0108 | 70.0000 | 22.0991 | 3.7329 | 0.0050 |
| 63.0253 | 33.5559 | 0.0434 | 0.0289 | 65.0000 | 29.6726 | 1.6030 | 0.0000 |
| 57.4955 | 39.5645 | 0.0402 | 0.0310 | 60.0000 | 35.9259 | 0.1439 | 0.0145 |
| 53.2436 | 44.2454 | 0.0356 | 0.0445 | 50.0000 | 47.1732 | 0.0050 | 0.0922 |
| 48.2292 | 49.2130 | 0.0000 | 0.0715 | 40.0000 | 57.7023 | 0.0000 | 0.1229 |
| 39.1392 | 58.6806 | 0.0000 | 0.1350 | 30.0000 | 67.8867 | 0.0000 | 0.2115 |
| 29.7280 | 68.1019 | 0.0000 | 0.2251 | 20.0000 | 77.8409 | 0.0000 | 0.3626 |
| 19.8418 | 78.0748 | 0.0000 | 0.3522 | 10.0000 | 87.5827 | 0.0000 | 0.4715 |
| 10.2280 | 87.3943 | 0.0000 | 0.5787 | 5.0000 | 92.3617 | 0.0000 | 0.5051 |
| 6.0882 | 91.5443 | 0.0000 | 0.5620 | 0.0000 | 96.6525 | 0.0000 | 0.5168 |
| 3.7825 | 93.5210 | 0.0000 | 0.6166 | | | | |
| 1.6358 | 95.5247 | 0.0000 | 0.4419 | | | | |
| 0.0000 | 96.8355 | 0.0000 | 0.5350 | | | | |

$H_4T_4/IH/S$ - MC

| Concentrated Phase | | Dilute Phase | |
|---|---|---|---|
| Surfactant | Precursor | Surfactant | Precursor |
| 33.5622 | 26.2074 | 2.7830 | 11.5267 |
| 31.8094 | 19.8495 | 4.2245 | 9.8187 |
| 31.5297 | 31.1921 | 2.4079 | 12.7717 |
| 28.8532 | 34.7946 | 2.0340 | 15.1070 |
| 22.0872 | 43.3102 | 1.5556 | 20.1319 |
| 15.4095 | 49.5000 | 1.3825 | 24.8392 |
| 10.9375 | 53.2986 | 1.4873 | 27.8414 |
| 7.2489 | 55.7339 | 1.4313 | 32.0191 |

$H_4T_4/I'H/S$

| MC | | | | QCT | | | |
|---|---|---|---|---|---|---|---|
| Concentrated Phase | | Dilute Phase | | Concentrated Phase | | Dilute Phase | |
| Surfactant | Precursor | Surfactant | Precursor | Surfactant | Precursor | Surfactant | Precursor |
| 60.3723 | 20.9922 | 44.0529 | 4.1353 | 54.0000 | 13.2192 | 50.3400 | 9.0247 |
| 60.3909 | 24.0709 | 36.6597 | 2.3751 | 55.5000 | 17.2506 | 42.5325 | 4.9562 |
| 58.3934 | 28.5089 | 24.4815 | 1.6777 | 55.0000 | 24.5134 | 21.8524 | 1.3245 |
| 57.5347 | 30.5069 | 9.9338 | 0.5036 | 50.0000 | 35.0480 | 4.0067 | 0.4269 |
| 53.0331 | 37.0302 | 2.2330 | 0.2634 | 40.0000 | 50.4819 | 2.6780 | 0.4482 |
| 48.1564 | 43.1713 | 1.2902 | 0.3788 | 30.0000 | 63.3498 | 2.1340 | 0.4678 |
| 42.8767 | 49.8358 | 0.7487 | 0.3979 | 20.0000 | 75.1846 | 0.9850 | 0.4320 |
| 37.8577 | 55.6502 | 0.6690 | 0.3388 | 10.0000 | 86.3022 | 0.4786 | 0.4174 |
| 34.1146 | 60.0576 | 0.5208 | 0.5164 | 5.0000 | 91.8003 | 0.2771 | 0.4231 |
| 26.9660 | 67.9570 | 0.4630 | 0.4745 | 0.0000 | 97.0890 | 0.0100 | 0.5000 |
| 17.6183 | 78.6780 | 0.4429 | 0.4145 | | | | |
| 5.6961 | 91.7163 | 0.2894 | 0.4372 | | | | |
| 1.0349 | 96.6163 | 0.3142 | 0.4630 | | | | |
| 0.0000 | 97.8117 | 0.0000 | 0.5710 | | | | |

$H_4T_4/IT/S$

| MC | | | | QCT | | | |
|---|---|---|---|---|---|---|---|
| Concentrated Phase | | Dilute Phase | | Concentrated Phase | | Dilute Phase | |
| Surfactant | Precursor | Surfactant | Precursor | Surfactant | Precursor | Surfactant | Precursor |
| 33.4924 | 24.3643 | 5.8569 | 10.5641 | 27.0000 | 4.4087 | 24.4100 | 3.9935 |
| 31.2984 | 26.8476 | 4.3899 | 12.0453 | 28.0000 | 4.6269 | 22.6129 | 3.7630 |
| 29.4404 | 29.0064 | 4.5756 | 14.2978 | 29.0000 | 4.8940 | 21.1275 | 3.6162 |
| | | | | 30.4002 | 5.2885 | 20.0000 | 3.5511 |
| | | | | 32.0000 | 5.8619 | 18.3672 | 3.4838 |
| | | | | 35.0611 | 7.4678 | 15.0000 | 3.4980 |
| | | | | 38.7451 | 12.8306 | 9.0000 | 4.3253 |
| | | | | 38.8593 | 14.2758 | 8.0000 | 4.6673 |
| | | | | 36.8702 | 20.7951 | 5.0000 | 6.8690 |
| | | | | 35.5687 | 22.8736 | 4.4000 | 7.8592 |
| | | | | 30.0000 | 29.0228 | 3.3250 | 12.1233 |
| | | | | 25.5858 | 32.1365 | 3.0000 | 15.5319 |
| | | | | 20.9971 | 34.7189 | 3.5000 | 20.2742 |
| | | | | 17.9056 | 35.6737 | 4.0000 | 23.4582 |
| | | | | 14.3491 | 35.9668 | 5.0000 | 27.2872 |
| | | | | 12.0896 | 35.5550 | 6.0000 | 29.7575 |
| | | | | 10.4646 | 34.8590 | 7.0000 | 31.5210 |
| | | | | 9.9781 | 34.6111 | 8.0000 | 32.7276 |
| | | | | 9.8310 | 34.4009 | 9.0000 | 33.6243 |

### $H_4T_4/I'T'/S$

| MC | | | | QCT | | | |
|---|---|---|---|---|---|---|---|
| Concentrated Phase | | Dilute Phase | | Concentrated Phase | | Dilute Phase | |
| Surfactant | Precursor | Surfactant | Precursor | Surfactant | Precursor | Surfactant | Precursor |
| 62.3415 | 22.1467 | 4.9210 | 0.3173 | 55.0000 | 4.0251 | 50.8492 | 2.8231 |
| 61.9996 | 23.6264 | 2.1441 | 0.1562 | 56.0000 | 4.4063 | 48.4741 | 2.3153 |
| 60.5985 | 28.7368 | 1.6612 | 0.0681 | 57.0000 | 4.8252 | 47.7208 | 2.1808 |
| 57.3423 | 33.6444 | 0.9668 | 0.0958 | 58.0000 | 5.3145 | 44.4602 | 1.6668 |
| 53.7103 | 38.5863 | 0.4364 | 0.1689 | 63.0000 | 9.3518 | 26.7671 | 0.3965 |
| 53.2005 | 39.7570 | 0.2807 | 0.1625 | 65.0000 | 14.9791 | 19.7247 | 0.2576 |
| 44.7724 | 49.6785 | 0.1812 | 0.1963 | 64.0000 | 20.6028 | 7.6836 | 0.1058 |
| 36.4844 | 58.6111 | 0.1157 | 0.3059 | 63.0000 | 23.3089 | 5.8147 | 0.1000 |
| 30.5773 | 65.3863 | 0.0331 | 0.3472 | 60.0000 | 29.1277 | 2.6314 | 0.0959 |
| 26.3083 | 69.9219 | 0.0031 | 0.2914 | 50.0000 | 43.2479 | 0.4981 | 0.1491 |
| 18.8386 | 77.6538 | 0.0141 | 0.4692 | 40.0000 | 55.0487 | 0.0623 | 0.2279 |
| 15.8457 | 80.7556 | 0.0000 | 0.5100 | 30.0000 | 66.0092 | 0.0072 | 0.3211 |
| 8.1808 | 88.2734 | 0.0000 | 0.6614 | 20.0000 | 76.5731 | 0.0004 | 0.4070 |
| 2.6612 | 94.3353 | 0.0000 | 0.7668 | 10.0000 | 86.9331 | 0.0000 | 0.4904 |
| 0.0000 | 97.1253 | 0.0000 | 0.4050 | 5.0000 | 92.0131 | 0.0000 | 0.5111 |
| | | | | 0.0000 | 96.7860 | 0.0000 | 0.5553 |

### $H_4T_4/IHI/S$

| MC | | | | QCT | | | |
|---|---|---|---|---|---|---|---|
| Concentrated Phase | | Dilute Phase | | Concentrated Phase | | Dilute Phase | |
| Surfactant | Precursor | Surfactant | Precursor | Surfactant | Precursor | Surfactant | Precursor |
| 53.4322 | 12.4688 | 23.9240 | 3.0633 | 22.4556 | 12.1341 | 18.0000 | 9.8639 |
| 53.8133 | 15.0568 | 11.6667 | 1.6580 | 24.6067 | 14.6054 | 15.0000 | 9.2126 |
| 53.9063 | 16.0962 | 5.2083 | 1.2247 | 25.9902 | 21.5114 | 10.0000 | 9.0168 |
| 54.0327 | 19.3651 | 3.1046 | 1.2357 | 24.4519 | 28.4520 | 7.0000 | 9.3524 |
| 51.0499 | 24.6362 | 2.0414 | 2.6042 | 21.6403 | 35.0964 | 5.0000 | 9.7732 |
| 46.1133 | 32.0000 | 0.9772 | 3.9484 | 16.4167 | 44.4639 | 3.0000 | 10.3628 |
| 40.3125 | 38.5547 | 1.0000 | 5.3105 | 12.3874 | 50.7736 | 2.0000 | 10.7146 |
| 34.1248 | 45.1031 | 0.4480 | 8.1709 | 3.6706 | 63.3572 | 0.5000 | 11.2632 |
| 24.6745 | 53.5952 | 0.1929 | 11.6995 | 0.0076 | 68.4289 | 0.0010 | 11.4274 |
| 12.2639 | 59.7847 | 0.4216 | 19.9554 | 0.0000 | 67.9061 | 0.0000 | 9.7133 |
| 0.0000 | 62.0000 | 0.0000 | 38.0000 | | | | |

$H_4T_4/I'HI'/S$

| MC | | | | QCT | | | |
|---|---|---|---|---|---|---|---|
| Concentrated Phase | | Dilute Phase | | Concentrated Phase | | Dilute Phase | |
| Surfactant | Precursor | Surfactant | Precursor | Surfactant | Precursor | Surfactant | Precursor |
| 70.9696 | 16.7678 | 54.9054 | 1.0650 | 64.0000 | 6.2918 | 61.8358 | 4.5239 |
| 69.7871 | 19.4673 | 50.4519 | 0.7639 | 65.6144 | 8.2903 | 62.0000 | 4.6216 |
| 67.4747 | 23.8068 | 37.7152 | 0.2938 | 66.6702 | 13.1224 | 58.0000 | 2.8473 |
| 65.5990 | 26.5668 | 24.6042 | 0.0363 | 66.3830 | 15.8112 | 50.0000 | 1.1997 |
| 64.1222 | 28.3095 | 11.3636 | 0.0298 | 65.2163 | 19.9753 | 40.0000 | 0.4147 |
| 62.0341 | 31.6812 | 0.6127 | 0.0198 | 63.9622 | 23.0447 | 30.0000 | 0.1365 |
| 55.3566 | 39.3953 | 0.5634 | 0.0106 | 62.5016 | 26.0298 | 20.0000 | 0.0427 |
| 50.2880 | 45.4664 | 0.4268 | 0.0217 | 60.5342 | 29.3523 | 5.0000 | 0.0063 |
| 46.0281 | 50.2922 | 0.2315 | 0.0158 | 55.0000 | 38.0000 | 0.1300 | 0.0043 |
| 38.6000 | 58.0579 | 0.0356 | 0.0000 | 48.3699 | 46.2269 | 0.1400 | 0.0060 |
| 24.4889 | 73.4446 | 0.0000 | 0.0000 | 38.9997 | 57.2002 | 0.1400 | 0.0095 |
| 15.0475 | 83.0357 | 0.0000 | 0.0000 | 29.4935 | 67.6857 | 0.0001 | 0.0129 |
| 5.5961 | 92.7778 | 0.0000 | 0.0306 | 19.8089 | 78.0134 | 0.0020 | 0.0145 |
| 0.0000 | 98.6359 | 0.0000 | 0.0372 | 10.7656 | 87.4640 | 0.0300 | 0.0174 |
| | | | | 0.0000 | 98.561 | 0.0000 | 0.0004 |

$H_4T_4/ITI/S$

| MC | | | | QCT | | | |
|---|---|---|---|---|---|---|---|
| Concentrated Phase | | Dilute Phase | | Concentrated Phase | | Dilute Phase | |
| Surfactant | Precursor | Surfactant | Precursor | Surfactant | Precursor | Surfactant | Precursor |
| 47.2447 | 12.1293 | 3.4488 | 1.6423 | 35.0000 | 4.4296 | 19.7883 | 1.9892 |
| 47.2000 | 16.9001 | 2.1050 | 3.1033 | 38.0000 | 5.7550 | 16.4371 | 1.8883 |
| 40.0321 | 24.3107 | 1.7886 | 7.5299 | 40.0000 | 7.1322 | 13.8897 | 1.9006 |
| 35.2847 | 27.4545 | 1.6570 | 11.5707 | 41.5000 | 8.8152 | 11.5179 | 2.0058 |
| 27.0392 | 31.9661 | 2.5287 | 17.7838 | 42.5000 | 11.3182 | 8.9114 | 2.2971 |
| 22.1786 | 33.0404 | 5.6852 | 22.1513 | 42.0000 | 15.8000 | 5.8720 | 3.2205 |
| | | | | 40.0000 | 19.6194 | 4.3023 | 4.5460 |
| | | | | 38.0000 | 22.1764 | 3.6146 | 5.8356 |
| | | | | 35.0000 | 25.1550 | 3.1131 | 7.9260 |
| | | | | 32.0000 | 27.5306 | 2.9392 | 10.2455 |
| | | | | 28.6312 | 29.6781 | 3.0000 | 13.1135 |
| | | | | 21.1879 | 32.7151 | 4.0000 | 20.1792 |
| | | | | 17.6902 | 33.2442 | 5.0000 | 23.6590 |
| | | | | 13.3910 | 32.7956 | 7.0000 | 27.8345 |
| | | | | 11.5907 | 32.0718 | 10.0000 | 30.8721 |

$$H_4T_4/I'TI'/S$$

| MC | | | | QCT | | | |
|---|---|---|---|---|---|---|---|
| Concentrated Phase | | Dilute Phase | | Concentrated Phase | | Dilute Phase | |
| Surfactant | Precursor | Surfactant | Precursor | Surfactant | Precursor | Surfactant | Precursor |
| 70.3809 | 17.4421 | 45.4980 | 2.2516 | 67.5922 | 6.5373 | 60.0000 | 2.2642 |
| 68.2000 | 22.7742 | 16.7837 | 0.4095 | 66.5598 | 5.4682 | 62.0000 | 2.8945 |
| 66.8118 | 25.7559 | 6.6362 | 0.0000 | 69.3916 | 10.7582 | 50.0000 | 0.6978 |
| 62.9496 | 30.5066 | 0.8589 | 0.0000 | 69.2591 | 15.1128 | 35.0000 | 0.1085 |
| 58.2562 | 36.1265 | 0.5704 | 0.0000 | 68.4681 | 18.3214 | 20.0000 | 0.0139 |
| 46.6849 | 49.2932 | 0.1587 | 0.0000 | 67.4183 | 21.1089 | 12.0000 | 0.0048 |
| 37.2435 | 59.1817 | 0.0000 | 0.0000 | 66.0123 | 24.0229 | 8.0000 | 0.0046 |
| 23.3342 | 74.0947 | 0.0000 | 0.0000 | 57.0000 | 38.0123 | 5.0000 | 0.0040 |
| 19.0582 | 78.5156 | 0.0000 | 0.0109 | 47.5345 | 48.4411 | 0.1000 | 0.0076 |
| 9.6832 | 88.1554 | 0.0000 | 0.1013 | 33.2885 | 63.8578 | 0.0100 | 0.0229 |
| 0.0000 | 97.4806 | 0.0000 | 0.0434 | 28.9755 | 68.3592 | 0.0010 | 0.0286 |
| | | | | 23.8894 | 73.6050 | 0.0002 | 0.0361 |
| | | | | 15.4282 | 82.2143 | 0.0000 | 0.0501 |
| | | | | 9.4121 | 88.2623 | 0.0000 | 0.0576 |
| | | | | 0.8848 | 96.7430 | 0.0000 | 0.0636 |

# References Chapter 5

Al-Anber, Z. A., J. B. Avalos, and A. D. Mackie, Prediction of the critical micelle concentration in a lattice model for amphiphiles using a single-chain mean-field theory, *Journal of Chemical Physics*, 122, -, 2005.

Al-Anber, Z. A., J. B. I. Avalos, M. A. Floriano, and A. D. Mackie, Sphere-to-rod transitions of micelles in model nonionic surfactant solutions, *Journal of Chemical Physics*, 118, 3816-3826, 2003.

Balmbra, R. R., J. S. Clunie, and J. F. Goodman, Cubic Mesomorphic Phases, *Nature*, 222, 1159-&, 1969.

Bhattacharya, S., and K. E. Gubbins, Modeling triblock surfactant-templated mesostructured cellular foams, *Journal of Chemical Physics*, 123, 2005.

Campbell, A. N., E. M. Kartzmark, and J. M. T. Gieskes, Vapor-Liquid Equilibria, Densities, and Refractivities in System Acetic Acid - Chloroform - Water at 25 Degrees C, *Canadian Journal of Chemistry-Revue Canadienne De Chimie*, 41, 407, 1963.

Fisher, M. E., Renormalization of Critical Exponents by Hidden Variables, *Physical Review*, 176, 257-&, 1968.

Floriano, M. A., E. Caponetti, and A. Z. Panagiotopoulos, Micellization in model surfactant systems, *Langmuir*, 15, 3143-3151, 1999.

Gelbart, W. M., and A. BenShaul, The "new" science of "complex fluids", *Journal of Physical Chemistry*, 100, 13169-13189, 1996.

Guggenheim, E. A., *Mixtures*, Clarendon Press, Oxford, 1952.

Hatton, B., K. Landskron, W. Whitnall, D. Perovic, and G. A. Ozin, Past, present, and future of periodic mesoporous organosilicas - The PMOs, *Accounts of Chemical Research*, 38, 305-312, 2005.

Holmberg, K., B. Jonsson, B. Kronberg, and B. Lindman, *Surfactants and Polymers in Aqueous Solution*, 2nd ed., John Wiley & Sons, LTD, Chichester, 2002.

Israelachvili, J. N., *Intermolecular and Surface Forces*, 2nd ed., Academic Press, London, 1995.

Khoo, I., and F. Simoni, *Physics of Liquid Crystalline Materials*, Philadelphia, 1998.

Kim, H. U., and K. H. Lim, A model on the temperature dependence of critical micelle concentration, *Colloids and Surfaces a-Physicochemical and Engineering Aspects*, 235, 121-128, 2004.

Kim, S. Y., A. Z. Panagiotopoulos, and M. A. Floriano, Ternary oil-water-amphiphile systems: self-assembly and phase equilibria, *Molecular Physics*, 100, 2213-2220, 2002.

Larson, R. G., Monte Carlo simulations of the phase behavior of surfactant solutions, *Journal De Physique Ii*, 6, 1441-1463, 1996.

Larson, R. G., L. E. Scriven, and H. T. Davis, Monte-Carlo Simulation of Model Amphiphilic Oil-Water Systems, *Journal of Chemical Physics*, 83, 2411-2420, 1985.

Mackie, A. D., K. Onur, and A. Z. Panagiotopoulos, Phase equilibria of a lattice model for an oil-water-amphiphile mixture, *Journal of Chemical Physics*, 104, 3718-3725, 1996.

Mackie, A. D., A. Z. Panagiotopoulos, and S. K. Kumar, Monte-Carlo Simulations of Phase-Equilibria for a Lattice Homopolymer Model, *Journal of Chemical Physics*, 102, 1014-1023, 1995.

Mertslin, R. V., N. I. Nikurashina, and V. A. Petrov, On the Properties of the Field of Separation of Ternary Systems Containing a Predominating Binary System, *Zhurnal Fizicheskoi Khimii*, 35, 2770-2774, 1961.

Panagiotopoulos, A. Z., M. A. Floriano, and S. K. Kumar, Micellization and phase separation of diblock and triblock model surfactants, *Langmuir*, 18, 2940-2948, 2002.

Patti, A., A. D. Mackie, and F. R. Siperstein, Molecular simulation study on the structure of templated porous materials obtained from different inorganic precursors, *Studies in Surface Science and Catalysis*, 160, 495, 2006.

Piculell, L., and B. Lindman, Association and Segregation in Aqueous Polymer/Polymer, Polymer/Surfactant, and Surfactant/Surfactant Mixtures - Similarities and Differences, *Advances in Colloid and Interface Science*, 41, 149-178, 1992.

Porte, G., Y. Poggi, J. Appell, and G. Maret, Large micelles in concentrated solutions. The second critical micellar concentration, *J. Phys. Chem.*, 88, 5713-5720, 1984.

Rosen, M. J., Surfactants and Interfacial Phenomena, *Wiley-Interscience*, 2004.

Siperstein, F. R., and K. E. Gubbins, Phase separation and liquid crystal self-assembly in surfactant-inorganic-solvent systems, *Langmuir*, 19, 2049-2057, 2003.

Tarassenkov, D. N., On the Distribution of the Co-Dissolving Constituent in the Ternary Cleaving Systems, *Zhurnal Obshchei Khimii*, 16, 1583-1588, 1946.

Widom, B., Plait Points in 2- and 3-Component Liquid Mixtures, *Journal of Chemical Physics*, 46, 3324-3332, 1967.

Zilman, A., T. Tlusty, and S. A. Safran, Entropic networks in colloidal, polymeric and amphiphilic systems, *Journal of Physics-Condensed Matter*, 15, S57-S64, 2003.

Zollweg, J. A., Shape of Coexistence Curve near a Plait Point in a 3-Component System, *Journal of Chemical Physics*, 55, 1430-&, 1971.

# Chapter 6

## MICROSCOPIC SEGREGATION IN HYBRID

## AMPHIPHILIC SYSTEMS

### Introduction

In the previous chapter, the phase separation observed in hybrid amphiphilic systems led to the following important conclusion: it is possible to synthesize ordered liquid crystal phases even at low surfactant concentrations, if an appropriate precursor is chosen that is able to drive the system towards a phase separation. In some cases, micelles and liquid crystals can coexist at equilibrium. Following the conclusions of Chapter 5, we are now interested in studying those hybrid structures formed as a result of a macroscopic phase separation between a solvent-rich and a surfactant-rich phase. According to some given conditions, these structures can be characterized by a microscopic segregation affecting their local order. For instance, when the surfactant concentration in such phases is high enough, spherical or cylindrical aggregates, or lamellar structures can be observed.

In Chapter 5, we observed how the choice of a terminal or bridging hybrid precursor affects the phase separation of an amphiphilic solution. The solubility of such a precursor in the solvent, and the presence and the position of a more or less solvophobic organic group, being key factors affecting the phase separation of the system, are also of fundamental importance to eventually observe the formation of ordered structures. Several cases are modeled where the composition corresponds to high surfactant concentration phases similar to that obtained from the synthesis of hybrid materials resulting from a phase separation.

When using terminal or bridging hybrid inorganic precursors, comparable to organosilicas of the type R-Si(OEt)$_3$ or (EtO)$_3$-Si-R-Si-(OEt)$_3$, respectively, we observe that the organic segment is generally well mixed with the inorganic precursor and surfactant heads, and no preferential location of the organic groups is observed. However, when the inorganic source is not soluble in the solvent, it is possible to detect an interesting penetration of the precursor into the corona of the spherical or cylindrical aggregates. This is especially evident for those bridging precursors including two insoluble inorganic sources, whose repulsive interactions with the solvent lead them closer to the solvophobic core.

When pure inorganic or hybrid precursors are used, we show that the behavior of the systems are analogous to those where cosurfactants or cosolvents are used, and that the lack of ordering in some cases can be explained by the change in solvent quality when using such precursors. A comparison of structural characterization of the different phases using several tools, such as aggregate size distribution, density profiles, and pair radial distribution function is presented.

In the following sections, we first introduce the tools used to analyze the results obtained from the study of the structures formed in hybrid amphiphilic systems with terminal or bridging precursors. Systems presenting pure inorganic or hybrid terminal precursors are analyzed in section 6.2, whereas those presenting bridging precursors are analyzed in section 6.3. The structural analysis of the hexagonally ordered cylinders, being of fundamental importance in the design and tailoring of mesoporous materials, will be treated in more detail in Chapter 7.

## 6.1 Structural Analysis.

In this section, a brief description of the methods used to analyze the structures obtained is given. A simple visual inspection provides information of the structure of the ordered phases. However, in order to identify the phases and their structure using quantitative information obtained from the simulations, we have determined several properties such as cluster size distribution, density profiles, and radial distribution functions.

### 6.1.1 Cluster Size Distribution.

When the system to analyze consists of small clusters, it is useful to calculate the cluster size distribution to determine the preferential size of aggregates and their dispersion. Aggregates or clusters were defined by collections of chains sharing at least one tail segment as a neighbor. This criterion has been used by several researchers to identify aggregates in surfactant solutions [*Panagiotopoulos et al.*, 2002]. The clusters were labeled using a simple algorithm where a linked list identifying the molecules that belong to the cluster is generated [*Allen and Tildesley*, 1987].

The cluster size distribution $P(N)$, represents the average fraction of clusters of size $N$ during the simulation. Using this definition, the average volume fraction occupied by clusters that contain $N$ surfactant chains of length $m$, $<\phi_N>$, is defined as:

$$<\phi_N> = NmP(N)/V \qquad (6.1)$$

where $V$ is the total volume of the simulation box. As strong correlations may exist in these systems, the cluster size distribution was calculated at intervals of $10^5$ steps. The averaged distributions were obtained for different simulation lengths until no significant evolution of the distribution was observed.

### 6.1.2 Composition Profiles.

The composition profiles, $\rho_i(r)$, of a given type of site, $i$, describe the composition in the $r$ direction for a given ordered structure, using as the origin the center of that structure. To calculate a composition profile, it is necessary to specify a given structure of the aggregate. For spherical aggregates, the composition profile can be calculated for concentric spherical shells around the center of the aggregate, for infinite cylindrical aggregates it is necessary to locate the line that defines the center of the aggregate, and calculate the composition in concentric cylindrical shells perpendicular to the central line.

The center of spherical aggregates was calculated as the center of mass of the cluster, giving equal weight to tails and head segments. To calculate the composition profile, the center of the cluster was approximated to the nearest site in the lattice, and the composition profile was obtained by counting the number of sites of a given nature in concentric shells of radius $r$, taking as the origin the cluster center, and dividing by the total number of sites in that concentric shell, considered as the volume of the shell. The volume of each shell, $V(r)$ was calculated at the beginning of the simulation, and is defined as the number of sites at a distance $d$, such that $r \leq d < r+1$. For spherical aggregates, density profiles need to be averaged for each aggregate size. For cylindrical aggregates, differences in the number of surfactants per unit length of the cylinder were not taken into account when calculating the density profiles.

For systems containing infinite cylinders in a hexagonal arrangement, the line defining the center of the cylindrical aggregate corresponds to the line connecting the centers of quasi-circular cross-sections. A cylinder cross-section was determined using the same criteria as for a spherical aggregate, but restricting the calculation to a single plane. A complete cylinder was composed of cross-sections that shared at least one tail segment as neighbors. The composition profiles were calculated by counting the number of sites of a given nature in concentric shells of radius $r$, that were perpendicular to the line defining the center of the aggregate. The profiles were determined at different points along the aggregate and averaged out, assuming that the composition profile is independent of the axial position of the cylinder.

When comparing systems with different global composition, it is sometimes useful to compare the composition distributions in a normalized way. We define the normalized density of sites of type $i$, $\rho_{i,N}(r)$, as the ratio of the composition profile $\rho_i(r)$ over the global composition, $\rho_i$. By definition, $\rho_{i,N}(r)$ should converge to unity at large values of $r$.

### 6.1.3    Radial Distribution Functions.

In many cases, it is not possible to define a structure for an aggregate that will allow us to determine the composition profile, nevertheless, it is possible to observe tail-rich domains and head-rich domains. To overcome this problem, and to characterize the systems where no ordered phases were observed, we calculated the radial distribution function. The radial distribution function, $g_{ij}(r)$, compares the local composition around a site occupied by a bead of type $i$, with the overall composition of sites occupied by beads of type $j$:

$$g_{ij}(r) = \frac{1}{\rho_j} \frac{\left\langle N_j(r) \right\rangle_i}{V(r)} \tag{6.2}$$

where $\left\langle N_j(r) \right\rangle_i$ is the average number of sites of type $j$ at a distance $r$ from a site of type $i$, and $V(r)$ is the volume of the shell of radius $r$. The average density of sites of type $j$ in the simulation box is $\rho_j$.

The calculation of $g_{ij}(r)$ does not imply the assumption of any particular structure, and it can provide information regarding the microphase separation as well as the composition of the continuous and discontinuous phases. Nevertheless, information of $g_{ij}(r)$ is sometimes difficult to interpret because the information about the local composition is averaged, regardless of whether the site is in the interface or in the center of a microstructure. In general, $g_{ij}(r)$ was calculated every $10^5$ configuration, and averaged over at least *100* configurations, for systems in cubic simulation boxes, after the system reached equilibrium.

### 6.1.4    Principal Radii of Gyration.

The principal radii of gyration are the square roots of the eigenvalues of the tensor of gyration defined as follows [*Rudnick and Gaspari*, 1987]:

$$T = \begin{pmatrix} T_{11} & T_{12} & T_{13} \\ T_{21} & T_{22} & T_{23} \\ T_{31} & T_{32} & T_{33} \end{pmatrix} \tag{6.3}$$

where:

$$T_{mn} = T_{nm} = \frac{1}{N} \sum_{p=1}^{N} \left( x_{p,m} - \left\langle x_m \right\rangle \right) \left( x_{p,n} - \left\langle x_n \right\rangle \right) \tag{6.4}$$

$N$ is the number of particles, whose coordinates in the three-dimensional space are represented by the vector $\vec{x}_p$. The quantities $x_{p,m}$ and $\left\langle x_m \right\rangle$ are the $m^{th}$ components of the vector $\vec{x}_p$ and of the center of mass of the cluster, respectively. The matrix defined in (6.3) can be diagonalized and its eigenvalues give the squares of the principal radii of gyration $R_1^2$, $R_2^2$, and $R_3^2$. In principle, if a given cluster had a perfect spherical shape, then the three radii would be identical.

A useful one-parameter to measure the deviation of the aggregates from sphericity is the so-called asphericity factor, defined as follows [*Rudnick and Gaspari*, 1987]:

$$A_d = \frac{\displaystyle\sum_{i>j}^{d}\left\langle\left(R_i^2 - R_j^2\right)^2\right\rangle}{(d-1)\left\langle\left(\displaystyle\sum_{i=1}^{d}R_i^2\right)^2\right\rangle} \tag{6.5}$$

where $d$=3 in our case. If the aggregate shows a perfect spherical shape, then $A_d$ is equal to zero; otherwise $A_d$ has values between *0* and *1*.

## 6.2  Structural Analysis in Ternary Amphiphilic Systems.

Table 6.1 summarizes the range of conditions at which the different structures were observed in the systems studied, and Figure 6.1 shows a collection of structures obtained in different systems. In the following, unless otherwise specified, the reduced temperature is set to $T^*$=8.0.

In general, it was observed that some of the properties calculated were sensitive to the system size for cubic simulation boxes smaller than $30^3$. Therefore, all the structural analysis given in this work corresponds to results obtained in cubic simulation boxes containing $40^3$ sites, unless otherwise specified.

The intervals indicated in Table 6.1 correspond exclusively to the systems containing a high surfactant and high inorganic concentration in equilibrium with a dilute phase. The type of structures formed were determined by a combination of visual inspection of the structures formed, and information obtained from the radial distribution functions, cluster size distribution, and radii of gyration.

The presence of micelles is characterized by a minimum in the cluster size distribution, and a maximum that does not depend on the simulation box size, and spherical micelles are identified as those having three equal (or very similar) radii of gyration. We do not identify regions with elongated micelles due to the problems encountered in distinguishing between adjacent spherical micelles touching each other and an elongated micelle.

**Table 6.1**. Summary of structures observed in the surfactant rich phase, in equilibrium with a dilute phase.

| System | Component | Volume fraction | | | |
|---|---|---|---|---|---|
| | | Disordered | Spherical | Hexagonal | Lamellar |
| $H_4T_4$-$I_2$-S | $H_4T_4$ | - | < 0.48 | 0.48-0.63 | - |
| | $I_2$ | - | > 0.30 | < 0.30 | - |
| $H_4T_4$-$IH$-S | $H_4T_4$ | - | Full | - | - |
| | $IH$ | - | Full | - | - |
| $H_4T_4$-$IT$-S | $H_4T_4$ | Full range | - | - | - |
| | $IT$ | Full range | - | - | - |
| $H_4T_4$-$I'_2$-S | $H_4T_4$ | <0.55 | - | 0.55-0.70 | >0.70 |
| | $I'_2$ | >0.40 | - | 0.25-0.40 | 0.12-0.25 |
| $H_4T_4$-$I'H$-S | $H_4T_4$ | <0.05 | 0.05-0.45 | >0.45 | - |
| | $I'H$ | >0.92 | 0.45-0.92 | <0.45 | - |
| $H_4T_4$-$I'T$-S | $H_4T_4$ | Full range | - | - | - |
| | $I'T$ | Full range | - | - | - |
| $H_4T_4$-$IHI$-S | $H_4T_4$ | - | < 0.41 | > 0.41 | - |
| | $IHI$ | - | > 0.40 | < 0.40 | - |
| $H_4T_4$-$ITI$-S | $H_4T_4$ | < 35 | > 0.35 | - | - |
| | $ITI$ | > 30 | < 0.30 | - | - |
| $H_4T_4$-$I'HI'$-S | $H_4T_4$ | < 0.45 | 0.45-0.55 | 0.55-0.70 | > 0.70 |
| | $I'HI'$ | > 0.50 | 0.40-0.50 | 0.17-0.40 | < 0.17 |
| $H_4T_4$-$I'TI'$-S | $H_4T_4$ | Full range | - | - | - |
| | $I'TI'$ | Full range | - | - | - |

Hexagonal and lamellar phases were identified by visual inspection of the configurations obtained during the simulation. We denote disordered phases as those where the first maximum in tail-tail radial distribution function, $g_{TT}$, is located at $r<11$, and that contain either aggregates with a broad distribution of sizes dependant on the box size, where none of the aggregate size can be clearly identified as dominant, and systems where practically all surfactants in the simulation box (more than *2500* chains) form a single aggregate, and no ordered phase is identified.

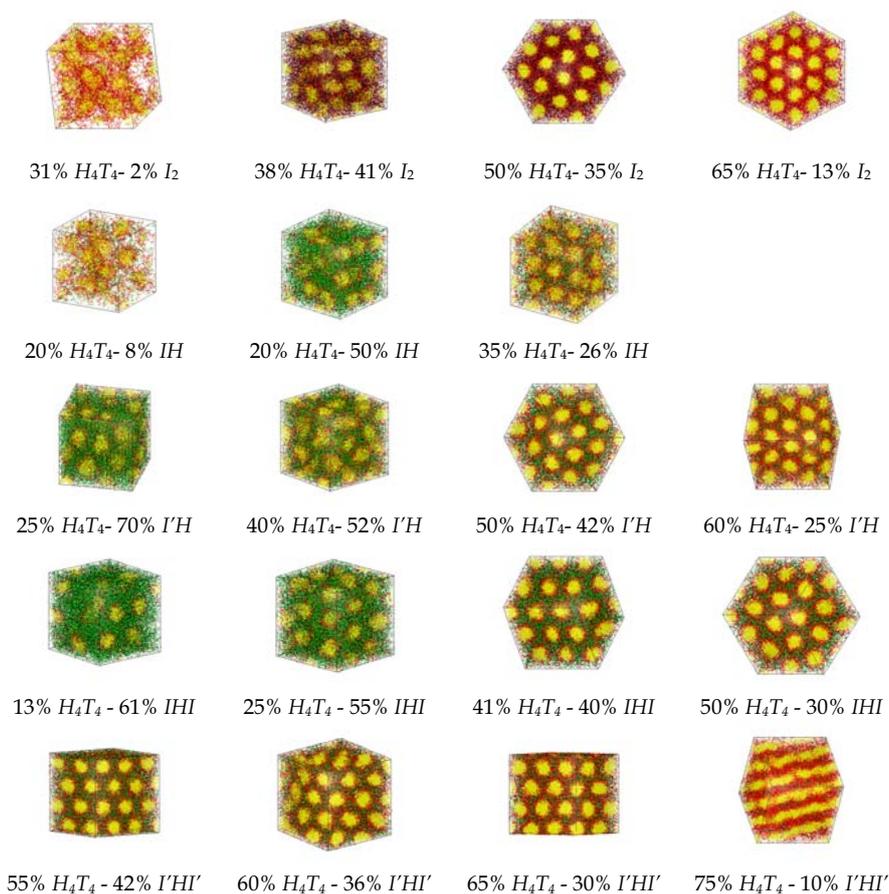

31% $H_4T_4$- 2% $I_2$     38% $H_4T_4$- 41% $I_2$     50% $H_4T_4$- 35% $I_2$     65% $H_4T_4$- 13% $I_2$

20% $H_4T_4$- 8% $IH$     20% $H_4T_4$- 50% $IH$     35% $H_4T_4$- 26% $IH$

25% $H_4T_4$- 70% $I'H$     40% $H_4T_4$- 52% $I'H$     50% $H_4T_4$- 42% $I'H$     60% $H_4T_4$- 25% $I'H$

13% $H_4T_4$ - 61% $IHI$     25% $H_4T_4$ - 55% $IHI$     41% $H_4T_4$ - 40% $IHI$     50% $H_4T_4$ - 30% $IHI$

55% $H_4T_4$ - 42% $I'HI'$     60% $H_4T_4$ - 36% $I'HI'$     65% $H_4T_4$ - 30% $I'HI'$     75% $H_4T_4$ - 10% $I'HI'$

**Fig. 6.1.** Representative structures obtained during the simulations of ternary systems with pure silica or hybrid precursors. $H$ segments belonging to a surfactant are shown in red, $T$ segments in yellow, $I$ (or $I'$) segments in black and $H$ segments belonging to a hybrid precursor are shown in green. At low surfactant concentration, spherical micelles are formed, and as surfactant concentration increases (greater than 40%), infinite cylinders in hexagonal arrangements can be observed. $T^*$=8.0.

## 6.2.1 Amphiphilic Systems with Pure Inorganic or Hybrid Terminal Precursors.

When a pure inorganic precursor is used ($I_2$ or $I'_2$), the ability of surfactants to self-assemble and form spherical aggregates is mainly influenced by the solubility of the precursor in the solvent. For the system where the precursor is not soluble ($H_4T_4$-$I'_2$-

*S*), self-assembled ordered phases are only observed at high surfactant concentration, and no ordered phases are formed at low surfactant concentration, as opposed to systems where the precursor is completely miscible in the solvent ($H_4T_4$-$I_2$-$S$), or binary systems containing only $H_4T_4$ –$S$.

Considering that the solvent volume fraction in the $H_4T_4$-$I'_2$-$S$ systems studied is between 1-2%, the results can be interpreted as if they were a binary system composed only of $H_4T_4$-$I'_2$. For the binary $H_4T_4$-$S$, the interchange energies are $\omega_{HS}$=0, and $\omega_{TS}$=1, and for the binary $H_4T_4$-$I'_2$ the interchange energies are $\omega_{HI}$=-2, and $\omega_{TI}$=1. Therefore, it is possible to say that as far as energetic considerations are concerned, $I'_2$ is a "better solvent" for $H_4T_4$ than *S*, and solubilizes the surfactant thus preventing the formation of spherical aggregates at low surfactant concentration, even when at the temperature studied it is possible to observe micelles in the $H_4T_4$-*S* system. Evidently, the better solvent quality of $I'_2$ over *S* is energetically based, and overcomes the unfavorable entropic contribution for having two connected beads as solvent, as opposed to one (for the solubility of surfactants in solvents with different chain lengths see the work of Kim *et al.* [*Kim et al.*, 2002]). At high surfactant concentration, the *H*-*T* repulsion dominates, and it is possible to observe ordered phases.

For hybrid precursors, the solubility of the inorganic segment in the solvent has less influence on the formation of ordered phases than the type of organic group attached to the inorganic segment. When the organic group is solvophobic, no ordered structures were formed, regardless of the surfactant concentration, but when the organic group is solvophilic, it is possible to observe the formation of ordered phases, like the ones shown in Figure 6.1 where the concentration of surfactant appears to be the determinant factor to decide which phase will be formed.

In the remainder of this section we present the structural analysis of the different phases formed using the tools described previously. At low surfactant concentrations, where finite size aggregates are formed, their aggregation number can be influenced by the type and concentration of precursor, and is generally between *60* and *90* surfactant chains (Figure 6.2).

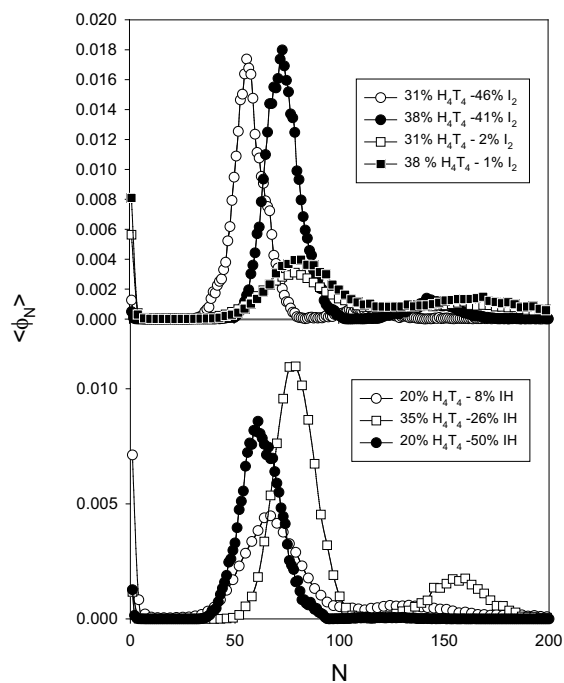

**Fig. 6.2**. Cluster size distributions for systems containing mainly spherical aggregates. The top figure is for the $H_4T_4$-$I_2$-$S$ system and the lower for $H_4T_4$-$IH$-$S$, at different concentrations along the coexistence line of the two phase region.

When $I_2$ or $IH$ is used as a precursor, the average aggregate size is practically independent of composition over a wide range of conditions, from systems containing a very small amount of precursor (1-2%), to systems containing up to 41% of $I_2$. Only systems containing a high amount of $I_2$ (46%) show a lower aggregation number. Therefore, in these systems, one can assume that the overall *solvent* properties only change when the precursor concentration is high enough. The sharper profiles observed at high $I_2$ concentration are consistent with the better solvent interpretation using the following argument. In systems composed mainly of $H_4T_4/S$, at 30% $H_4T_4/S$ the second critical micellar concentration (*cmc*) has been reached (where elongated micelles start to appear), which is evident from the cluster size distribution and the appearance of a broad distribution of clusters that contain more than *100* chains, as observed in Figure 6.3 for three different surfactant volume fractions.

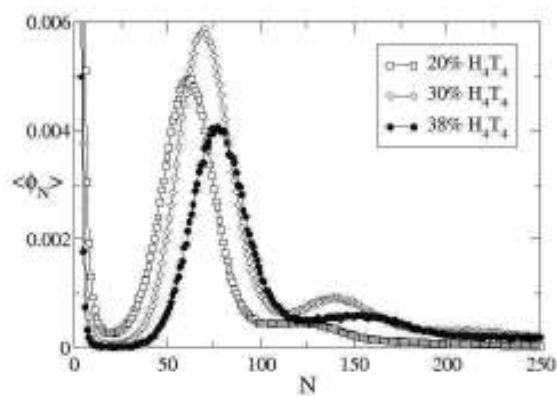

**Fig. 6.3**. Cluster size distribution of the binary system $H_4T_4/S$ at different surfactant concentrations.

On the other hand, in systems containing mainly $H_4T_4/I_2$, the second *cmc* has not been reached, and a sharper peak appears denoting principally the presence of spherical micelles. The second peak that appears in the cluster size distribution of Figure 6.2 containing 38% $H_4T_4$ and 41% $I_2$ does not indicate the presence of elongated micelles but rather the inability of the algorithm used for the calculation cluster size distributions to distinguish between two spherical micelles in contact at a single point, which is likely to be observed at such high surfactant concentrations, as reported in Figure 6.4.

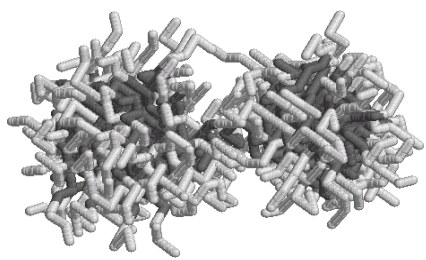

**Fig. 6.4**. Two spherical aggregates touching at a given point in the system with 38% $H_4T_4$ and 41% $I_2$.

A similar behavior is observed in systems containing *IH* as precursor, although increasing the precursor concentration while maintaining the surfactant concentration constant has little influence on the average aggregate size, but increasing the surfactant concentration, increases the average aggregate size. Probably in this case the difference in solvent quality is not significant enough to show a change in aggregate size, and the larger aggregates observed at high surfactant concentration are a result of the formation of cubic structures. It should be noted that two clear peaks are observed at these conditions, similar to the system containing 38% $H_4T_4$ and 41% $I_2$.

The changes in the aggregate size depending on surfactant concentration, are not observed in binary systems where increasing the surfactant concentration increases the number of aggregates but not significantly their size [*Mackie et al.*, 1997]. As a matter of fact, if we increase the surfactant concentration from 20% to 38%, the aggregation number increases approximately from *60* to *80* (Figure 6.3), but the first minimum in the tail density profiles, giving an idea of the size of the spherical aggregates, does not significantly change (Figure 6.5).

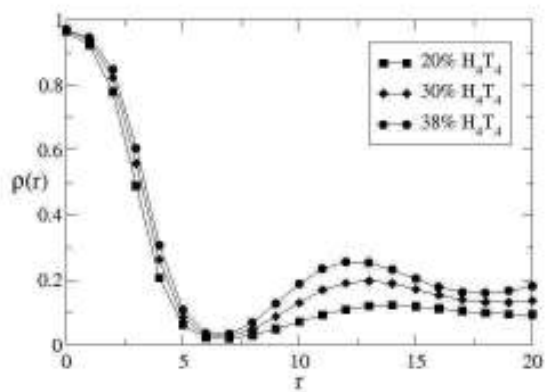

**Fig. 6.5.** Surfactant tail density distributions of binary systems obtained at different surfactant concentrations.

We consider that it is not possible to observe the same behavior in our ternary system because the changes in surfactant concentration are accompanied by changes in the precursor concentration, that changes also the *overall solvent quality*. The observed behavior is closer to what is observed in systems with cosurfactants and cosolvents, where the number of surfactant chains in an aggregate is reduced as the cosurfactant concentration increases, while the number of surfactant chains increases slightly by the presence of the cosolvent [*Chennamsetty et al.*, 2005]. Therefore, it can be said that $I_2$ is acting as a cosurfactant, while *IH* as a cosolvent, although no studies were done at low surfactant concentration to determine whether the precursors formed part of the micelle or are mainly located in the solvent.

In the cases where surfactant concentration is less than 30% by volume and the aggregates are far from each other, only one peak in the cluster size distribution is observed, but when the concentration of aggregates is high, as in the case of 35% $H_4T_4$ with 26% *IH*, several peaks are observed at integer multiples of the first peak. At these surfactant concentrations, by inspection we observe spherical aggregates periodically ordered, as shown in Figure 6.6.

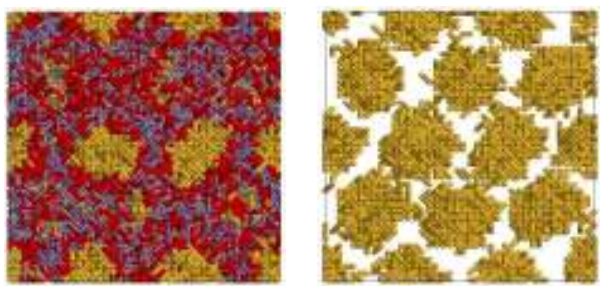

**Fig. 6.6.** Structures observed for the system $H_4T_4$ (35%)/ *IH* (26%)/S, in a $40^3$ simulation box at $T^*=8.0$, where it is possible to observe the ordered spherical aggregate structure, (left) showing the surfactant tails in yellow, surfactant heads in red and hybrid precursors in blue and grey, where the organic part of the hybrid precursor is shown in blue, and (right) only showing the surfactant tails.

The second peak mainly indicates that it is possible to have two spherical micelles touching each other, rather than the formation of elongated micelles as is found for certain head/tail ratios [*Panagiotopoulos et al.*, 2002; *Salaniwal et al.*, 2003]. This is also confirmed by visual inspection of isolated aggregates of different sizes obtained during the simulations, and when concentration profiles of aggregates of different sizes are analyzed. The presence of several peaks could be attributed to averaging the system properties over relatively short simulations [*Nelson et al.*, 1999], nevertheless, in our case, due to the high surfactant concentration where these peaks are observed, we consider them to represent the probability of observing two, or more, adjacent spherical aggregates, sharing at least one tail segment as neighbors.

Considering that the properties of the ordered materials will strongly depend on the structure of the self-assembled phases, it is important to analyze the density profile of spherical and cylindrical aggregates. The observed profiles are practically insensitive to the number of chains in the spherical aggregate, as long as the aggregate has the assumed shape. As the number of surfactant chains increases, the core radius increases slightly, consistently with the increase in volume of tail segments, in order to accommodate them, as shown in Figure 6.7, but the head segments distribution, inorganic segments distribution (not shown in the figure), and the distance to the *nearest* aggregate is practically independent of the aggregate size.

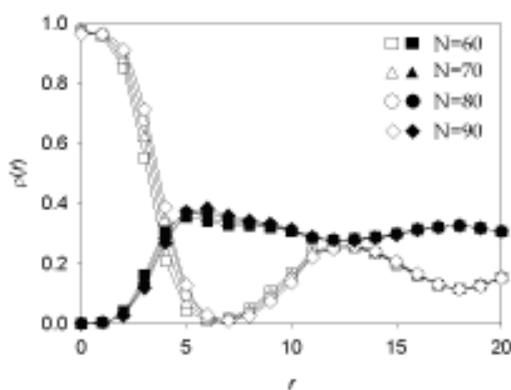

**Fig. 6.7.** Density distribution profiles of aggregates with different number of surfactant chains in a system containing 35% $H_4T_4$ and 26% *IH* at $T^*=8.0$. Open symbols correspond to surfactant tail density distributions, and closed symbols to precursor solvophilic group density distributions.

The distance to the nearest aggregate is assumed to be the position of the first peak in the *T*-segments density distributions and its position depends on the overall surfactant concentration, as shown in Figure 6.8, where we compared the surfactant tail density profiles for spherical aggregates containing *70* surfactant chains in systems with different precursors. The core structure is independent of the precursor and concentration, and the separation between micelles is approximately *13* lattice units, and decreases as the surfactant concentration increases.

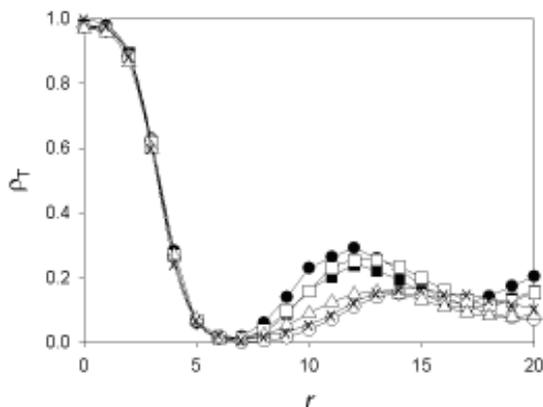

**Fig. 6.8.** Surfactant tail density profiles for spherical aggregates made up of *70* surfactant chains at $T^*$=8.0: (●) 38% $H_4T_4$ - 41% $I_2$, (■) 31% $H_4T_4$ - 46% $I_2$, (○) 20% $H_4T_4$ - 50% *IH*, (□) 35% $H_4T_4$ - 26% *IH*, (△) 20% $H_4T_4$ - 8% *IH*, and (×) 26% $H_4T_4$ - 70% *I'H*.

We found it convenient to compare the normalized density profiles of inorganic and solvophilic segments rather than the density profiles in absolute values, because the concentrations of these segments in a given simulation can be very different. In general, we found that the normalized profiles of solvophilic segments (*H* segments of the surfactant and precursor chains) and inorganic segments are similar. At low surfactant concentration, for phases containing a high precursor concentration, the maximum in *H* segments is closer to the core of the micelle than the peak of the inorganic segment distribution, and as the surfactant concentration increases, the two curves tend to merge (Figures 6.9a and 6.9b).

The only case where the *I* segments are closer to the core of the aggregate is when *I'H* precursors are used at low concentration, because the precursor is positioned at the interface between the core and the shell of the micelle, shielded from the solvent by the *H* segments, as can be inferred by analyzing Figure 6.9c. In this figure, we observe that the inorganic segment, *I'*, is able to penetrate the corona of the aggregate towards its solvophobic core. This fact is the result of two effects: the immiscibility of *I'* with the solvent, and the repulsion between solvophilic segments and tails. The *H* segments belonging to the precursor and to the surfactant create a barrier to protect *I'* by the contact with the solvent, and themselves by the contact with the surfactant tails.

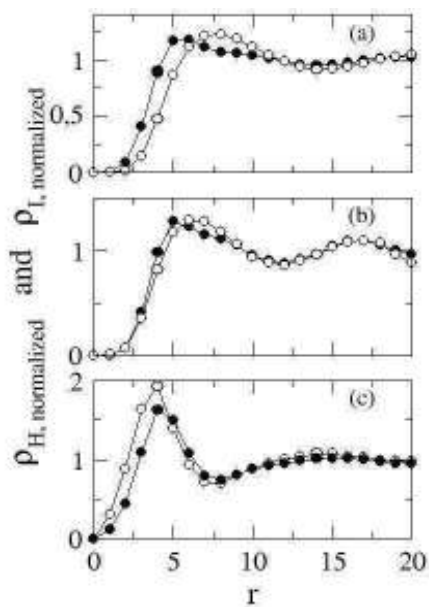

**Fig. 6.9.** Normalized density profiles of solvophilic (●) and inorganic (○) segments in spherical aggregates made up of *70* surfactant chains in systems containing different precursors at *T\*=8.0*: (a) 20% *$H_4T_4$* - 50% *IH*, (b) 38% *$H_4T_4$* - 41% *$I_2$*, and (c) 28% *$H_4T_4$* - 1% *I'H*.

This phenomenon is not observed with the other immiscible precursors, because no aggregates were observed. In all other cases, penetration of *I*-segments into the corona region is not favored. A similar behavior to the one observed in spherical aggregates is observed in cylindrical aggregates ordered in a hexagonal arrangement. Surfactant tail density distribution profiles in hexagonal phases indicate that as concentration increases, the cylinders become closer to each other (see Figure 6.10a).

These profiles seem to be more dependent on the surfactant concentration than on the nature and concentration of the precursor. The first peak in the density distribution of surfactant heads (Figure 6.10b) is slightly broader than the first peak in the density distribution of inorganic segments (Figure 6.10c), being consistent with the low penetration of the *I* segments into the corona of the micelles.

When the surfactant concentration is low enough (approximately 50%), it is possible to observe a split in the first peak of the head density profile, but not in the inorganic density profile. This split corresponds to the minimum in the surfactant tail density profile, indicating that the surfactant heads are surrounding the tail-rich cylindrical cores, and that inorganic segments are located farther from the center of the cylinders, similar to what is observed in most systems with spherical aggregates. As the surfactant concentration increases, the separation between cylinders decreases and the split in the surfactant head distributions disappears.

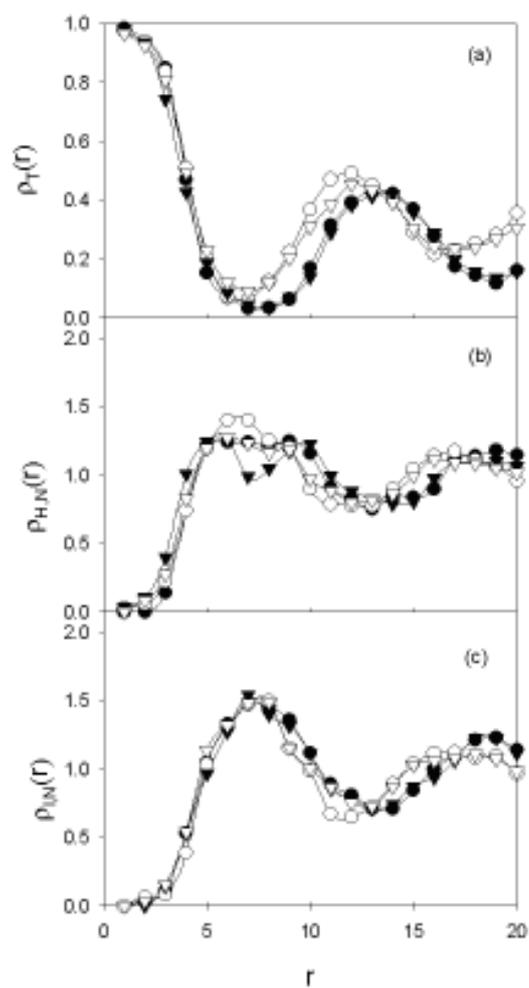

**Fig. 6.10.** Density distribution profiles of surfactant tails *(a)*, and normalized density distribution profiles of surfactant heads *(b)* and inorganic segment *(c)* in hexagonal phases observed in different systems at $T^* = 8.0$: (●) 52% $H_4T_4$ - 32% $I_2$; (○) 65% $H_4T_4$ - 13% $I_2$; (▼) 50% $H_4T_4$ - 42% $I'H$; and (∇) 60% $H_4T_4$ - 25% $I'H$.

The split in the density distribution of the head segments can be magnified if the concentration of precursor is increased, at a constant surfactant concentration, as seen in simulations at concentrations far from the two-phase coexistence line.

The density profiles provide direct information on the local composition of the aggregates, if the shape of the aggregate is correctly assumed, otherwise it can be misleading. Equivalent information to the density profiles can be obtained from radial distribution functions, without the need of specifying a particular structure for the aggregates. The calculation of the radial distribution function during the simulations is easier to implement than the calculation of the density profiles, as there is no need to identify specific clusters. The information lost when calculating the radial distribution function does not prevent us from reaching the same conclusions as from the density distributions with respect to the relative location of each type of segment.

The surfactant tail-tail distribution can be used to indicate the average distance between ordered structures, similar to the surfactant tail density profiles; the surfactant tail-head, and the surfactant tail-inorganic radial distribution functions can be used to determine the relative position of the head and inorganic segments relative to a given tail segment. When self-assembled structures are formed, tail segments concentrate in the core of the aggregate and the tail-head and tail-inorganic radial distribution function can be used to describe the location of head and inorganic segments relative to the center of the ordered structures. Figure 6.11 shows a comparison of these radial distribution functions for the system $H_4T_4/I_2/S$ at different compositions.

Although care should be taken in the interpretation of these figures, it is easy to see from the tail-tail radial distribution function that in all cases, tail segments are likely to be surrounded by other tail segments, as $g_{TT}$ is larger than unity for small distances, indicating the formation of tail-rich core structures. A similar behavior to the density distributions curves is observed, although it is difficult to see from Figure 6.11a that the location of the first peak becomes closer to the origin as the concentration of surfactant increases.

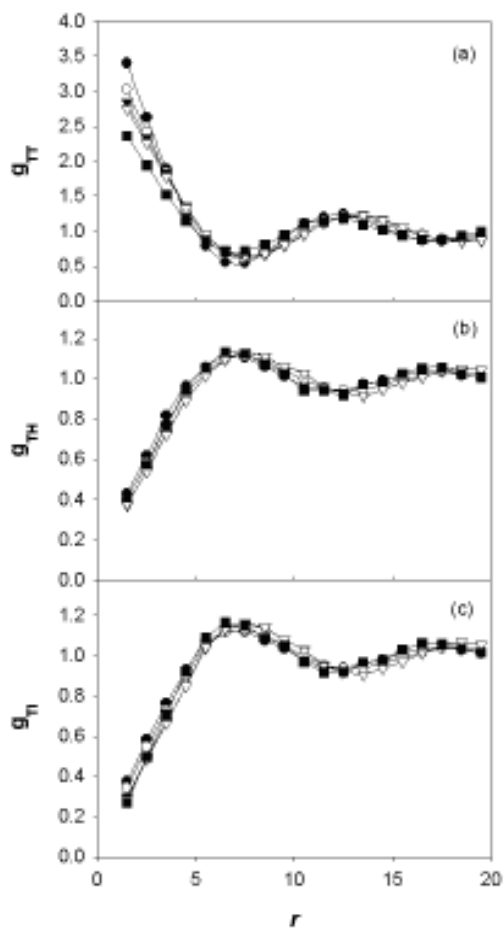

**Fig. 6.11.** Radial distribution functions, (a) $g_{TT}$, (b) $g_{TH}$, and (c) $g_{TI}$, for $H_4T_4/I_2/S$ at $T^*=8.0$ and different compositions: 31% $H_4T_4$ - 2% $I_2$; (●) 38% $H_4T_4$ - 41% $I_2$; (○) 45% $H_4T_4$ - 38% $I_2$; (▼) 50% $H_4T_4$ - 35% $I_2$; (▽) 52% $H_4T_4$ - 32% $I_2$; and (■) 60% $H_4T_4$ - 22% $I_2$.

The location of the maximum in $g_{TT}$ is shown in Figure 6.12, where one can clearly see that the first peak moves to smaller values of $r$ as surfactant concentration increases in the presence of ordered structures. The reduction in the values of $g_{TT}$ with concentration at short distances is probably due to the higher interfacial area between tails and other types of segment, rather than to a reduction in tail concentration in the core of the aggregates.

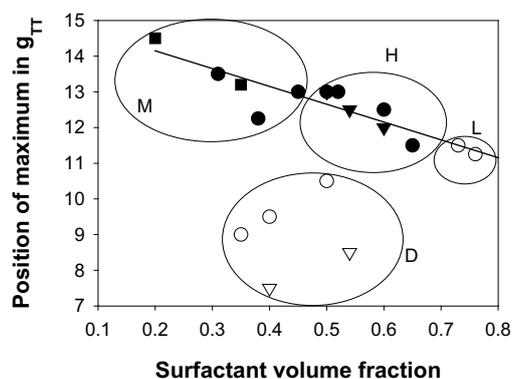

**Fig. 6.12.** Position of the first maximum in the surfactant tail-tail radial distribution function for systems with different precursors at $T^*=8.0$: (●) $I_5$; (○) $I'_5$; (▼) $I'H$; (▽) $I'T$; and (■) $IH$. The volume fraction of precursor and solvent are taken such that the overall composition in the simulation corresponds to points on the equilibrium lines of the corresponding phase diagrams. $M$ is where spherical or elongated aggregates are found, $H$ denotes hexagonal phases, $L$ denotes lamellar phases, and $D$ is for disordered phases.

A comparison of the position of the first maximum in $g_{TT}$ for different systems shown in Figure 6.12, indicates that the periodicity of all systems where ordered structures are formed is between a distance of *11* and *14* lattice segments. Disordered phases are characterized by a location of the first maximum in $g_{TT}$ at *r<11*. In ordered systems, a general trend is observed where the position of the first peak in $g_{TT}$ decreases with surfactant concentration, indicating that the structures are becoming closer to each other. The position of the first maximum in $g_{TT}$ when no ordered structures are observed depends strongly on the system, whereas for the cases where ordered structures are present, it is practically independent of the type of precursor. For systems containing *IT*, $g_{TT}$ decays monotonically, until it converges to unity, and no maximum is found.

The position of the peaks seems to vary continuously with increasing surfactant concentration, and no clear indication of the presence of one microphase or another can be obtained. The largest deviation from the general trend is observed for a surfactant volume fraction of 38%, where the maximum in $g_{TT}$ is located at a shorter distance than what would be expected from the general trend. When the system contains approximately 35% surfactant, spherical aggregates ordered in a cubic phase are expected for systems where only surfactant and solvent are present [*Larson*, 1989]. If cubic phases are present, then the properties of the system should be very sensitive to the simulation box size. Nevertheless, simulations in cubic boxes with sizes ranging from $20^3$ to $48^3$ show practically no difference in the position of the first peak. The other two radial distribution functions, $g_{TH}$ and $g_{TI}$ have similar shapes (Figures 6.11b and 6.11c), where at short distances they have values considerably smaller than unity, indicating that the inorganic or head segments are not likely to be mixed with the tail segments, and the first maximum in $g_{TH}$ and $g_{TI}$ corresponds to the first minimum in $g_{TT}$, and the location of the maximums in the $g_{TH}$ and $g_{TI}$ curves is in good agreement with the density distributions. In addition, when no ordered structures are formed, as in the case of $H_4T_4$-$I'T$ and $H_4T_4$-$I'_2$, it is possible to obtain some information about the local structure using radial distribution functions. In such cases, the shape of $g_{TT}$ is significantly different, the first peak is located at smaller values of $r$ and the function converges to unity faster than when ordered structures are observed. Therefore, a good indication of the formation of ordered structures is the location of the first peak in $g_{TT}$.

In general, when hybrid precursors are used, the radial distribution functions between the surfactant tail segments, and any of the two segments of the precursor are practically indistinguishable. Differences are only observed at short distances for systems containing $I'T$ as precursor, where the values of $g_{T(\text{surfactant})I'(\text{precursor})}$ are larger than those for $g_{T(\text{surfactant})I(\text{precursor})}$. Considering that no ordered phases were found in this case, it cannot be stated that the organic segment will be more accessible to adsorbed molecules. On the other hand, solvophilic and inorganic segments in hybrid precursors ($IH$ and $I'H$) are equally distributed with respect to the surfactant tails, suggesting that part of the solvophilic organic segments may not be accessible to adsorbed molecules in the pore network.

In general, the same information can be obtained from density distributions where a particular aggregate shape needs to be assumed, and radial distribution functions, where no assumption is made, but simple spherical averages are taken. This information can be used to scan in a simple way the compositions where ordered phases are formed and where self-assembled structures can yield templated materials, as well as in the cases where the distribution of particles in the corona of surfactant self-assembled systems is of interest.

### 6.2.2    Hybrid Amphiphilic Systems with Bridging Precursors.

At low to moderate surfactant concentrations, spherical aggregates are formed and the correspondent cluster size distribution ($CSD$) for the system containing the solvophilic precursor $IHI$ is shown in Figure 6.13.

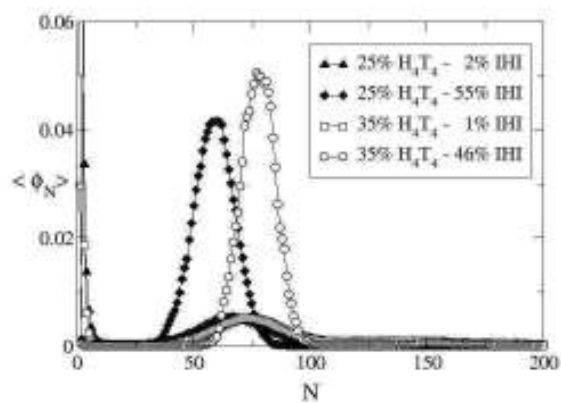

**Fig. 6.13**. Cluster size distributions for the system $H_4T_4$-*IHI-S* containing spherical aggregates, at different concentrations along the coexistence line of the two-phase region (see Fig. 4.20). $T^*$=8.0

The average cluster size can be affected by the concentration of the precursor and the surfactant, and is roughly in between *50* and *80* surfactant chains. If we compare the systems at the same surfactant concentration, as reported in Figure 6.13, a high content of *IHI* gives rise to sharper *CSD* profiles than those observed at low *IHI* concentrations. In other words, an increase in the *IHI* concentration leads to more defined spherical aggregates whose size, however, is not significantly modified, even at high precursor concentrations. On the other hand, if we increase the surfactant concentration from 25% to 35%, the average aggregation number increases as well, as observed, less significantly, in binary surfactant/solvent systems. The bridging solvophilic precursor is then acting as a co-surfactant which supports the aggregation of the primary surfactant [*Chennamsetty et al.*, 2005].

The *CSDs* of the system containing *ITI* are reported in Figure 6.14. In this case, we observe a different trend: by increasing the precursor concentration in systems presenting the same concentration of surfactant, the aggregation number decreases significantly, and the shape of the distribution profiles undergoes two interesting changes. The first minimum of the function increases, and the tail density profile goes faster to zero. Both factors suggest that the solvophobic bridging precursor, *ITI*, is deeply modifying the properties of the solution in such a way that the surfactant can be more easily dissolved in the solvent. In other words, the solvophobic effect leading to the formation of aggregates is weakened by the presence of *ITI*, and the necessity for the surfactants to self-assemble to shield their tails results reduced. For these reasons, we can say that the precursor *ITI* is acting as a co-solvent.

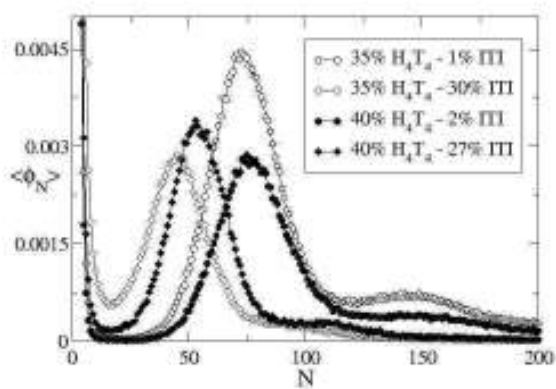

**Fig. 6.14**. Cluster size distributions for system *$H_4T_4$-ITI-S* containing spherical aggregates, at different concentrations along the coexistence line of the two phase region (see Fig. 5.20). *T*\*=8.0

In Figure 6.15, we report the tail and head density profiles of a system containing 35% *$H_4T_4$* and 46% *IHI* and forming spherical aggregates at different aggregation numbers. When the number of surfactant chains in a given cluster increases from *65* to *90*, the size of the aggregates is practically unaffected, as can be easily inferred by observing the minimum of the three tail distributions analyzed. Moreover, the distance to the nearest aggregates, being represented by the first peak in the tail density distributions, is not affected either. A slight difference between the three profiles, is observed inside the core, where the tail density of the aggregates with more surfactant chains is higher than that of the aggregates at lower aggregation number. Therefore, the inner structure of the core undergoes a modification to accommodate more or less surfactant tails, as the volume of a sphere changes with the radius cubed, but the size of the aggregate does not change.

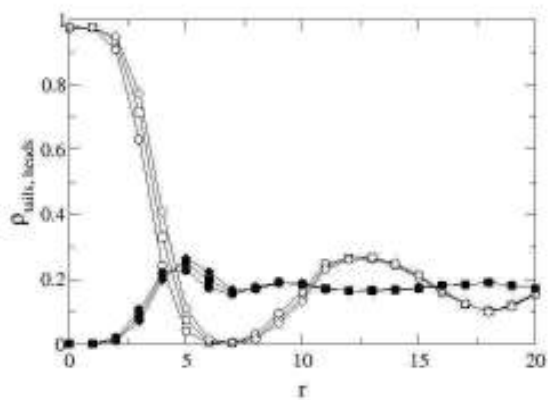

**Fig. 6.15**. Density distribution profiles of aggregates with different number of surfactant chains at $T^*$=8.0 in a system containing 35% $H_4T_4$ and 46% $IHI$: circles represent aggregates with *65*, squares aggregates with *78*, and diamonds aggregates with *90* surfactant chains. Open symbols correspond to surfactant tail density distributions, and closed symbols to surfactant head density distributions.

On the other hand, if we compare the size of aggregates at different concentrations, but at the same aggregation number, then the core structure is not affected at all, as observed in Figure 6.16 which has been obtained by changing surfactant and precursor concentrations and keeping constant the aggregation number of chains in the aggregates (*70* surfactant chains). In this case, the tail density profiles inside the core are identical for each of the systems considered, and then independent from the precursor and surfactant concentrations. On the other hand, the distance between neighboring aggregates increases as the surfactant concentration decreases, and is found to be from *12* to *15* lattice units.

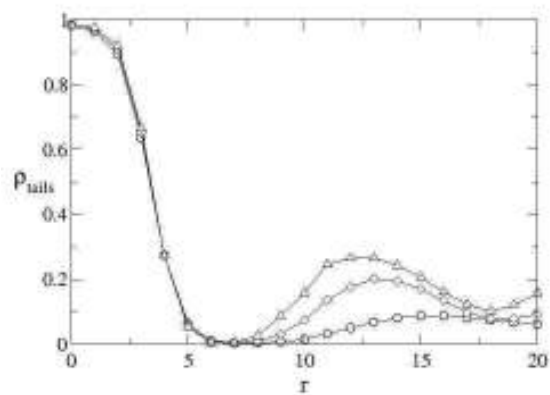

**Fig. 6.16.** Surfactant tail density profiles for spherical aggregates made up of *70* surfactant chains at *T\*=8.0*: (○) 13% $H_4T_4$ - 61% *IHI*, (△) 25% $H_4T_4$ - 55% *IHI*, and (◇) 35% $H_4T_4$ - 46% *IHI*.

Similar considerations are also valid for the solvophobic bridging precursor *ITI*. In Figure 6.17, the density distribution profiles of spherical aggregates with different aggregation numbers are reported for the system with 50% $H_4T_4$ and 15% *ITI*.

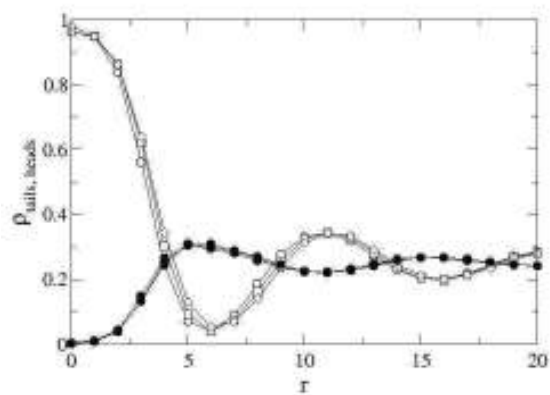

**Fig. 6.17.** Density distribution profiles of aggregates with different number of surfactant chains at *T\*=8.0* in a system containing 50% $H_4T_4$ and 15% *ITI*: circles represent aggregates with *65*, squares aggregates with *75*, and diamonds aggregates with *85* surfactant chains. Open symbols correspond to surfactant tail density distributions and closed symbols to surfactant head density distributions.

By increasing the number of surfactant chains from *65* to *85*, the core radius slightly increases in order to accommodate more chains, and the distance between the aggregates remains constant, that is approximately *11* lattice units. The tail distribution profiles at different surfactant and precursor concentrations for spherical aggregates made up of *70* chains, are given in Figure 6.18.

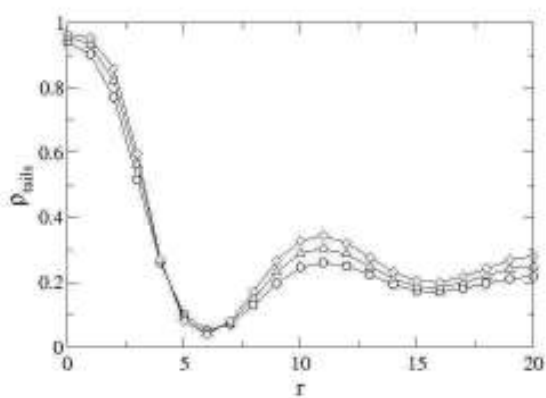

**Fig. 6.18.** Surfactant tail density profiles for spherical aggregates made up of *70* surfactant chains at $T^*$=8.0: ($\circ$) 40% $H_4T_4$ - 27% *ITI*, ($\triangle$) 45% $H_4T_4$ - 22% *ITI*, and ($\diamond$) 50% $H_4T_4$ - 15% *ITI*.

It should be noted that the first minimum in the tail density distribution profiles is not zero, as observed with the precursor *IHI* in Figure 6.16, but it reaches a given positive value. This result is more likely connected to the assumption of a spherical shape to calculate the density profiles of the aggregates, than to the presence of surfactant tails around the micelle corona. By comparing the asphericity factors for the systems considered in Figures 6.16 and 6.18, we observe that those tail distribution profiles presenting a first minimum higher than zero, are typical of systems containing slightly elongated aggregates.

In Figure 6.19, the asphericity factors of some of these systems are reported as function of the number of chains per aggregate. The shape of the aggregates obtained in the systems with *IHI* is closer to a sphere than that of the aggregates observed in the systems with *ITI*. This result is especially evident at high aggregation numbers, like the one considered in Figures 6.16 and 6.18, where elongated aggregates are observed.

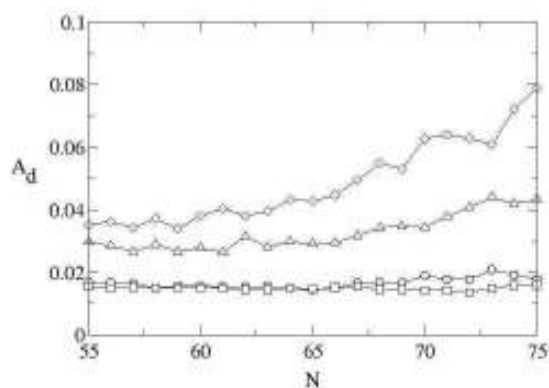

**Fig. 6.19.** Asphericity factors of the systems ($\circ$) 13% $H_4T_4$ - 61% *IHI*, ($\square$) 25% $H_4T_4$ - 55% *IHI*, ($\triangle$) 45% $H_4T_4$ - 22% *ITI*, and ($\diamond$) 40% $H_4T_4$ - 27% *ITI*, at different aggregation numbers *N*. $T^*$=8.0.

It is also interesting to note that close to the center of a given aggregate, the average tail density slightly decreases with increasing the *ITI* concentration. Such a dependence is not observed in systems containing *IHI* (Figure 6.16), which do not show any remarkable change in the aggregate core when the precursor concentration is modified. Therefore, the change of the density profiles in the inner aggregate core observed in Figure 6.18, must be due to the presence of the solvophobic group *T* in the precursor, as is explained by considering Figure 6.20.

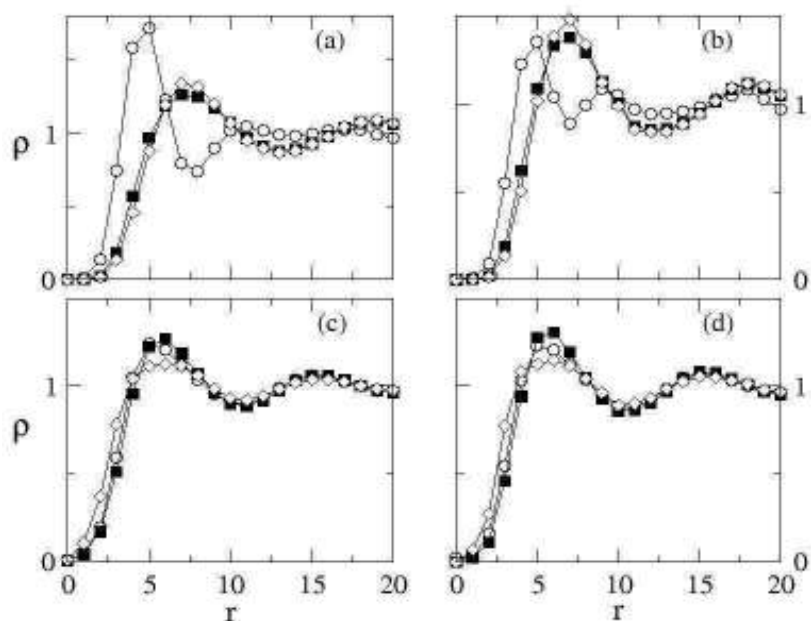

**Fig. 6.20.** Normalized density profiles of surfactant head (○), inorganic segment (■), and precursor organic group (◇), for spherical aggregates made up of *70* surfactant chains in systems containing different precursors at $T^*=8.0$: (a) 25% $H_4T_4$ - 55% *IHI*, (b) 35% $H_4T_4$ - 46% *IHI*, (c) 45% $H_4T_4$ - 22% *ITI*, and (d) 50% $H_4T_4$ - 15%*ITI*.

In this figure, we report the normalized density profiles of the beads belonging to the hybrid solvophilic and solvophobic precursors, whose average distributions are compared to the surfactant head profiles. Generally, there is no preferential location of one kind of bead with respect to the others around the micelle core: the precursors *IHI* and *ITI* are practically mixed with the surfactant heads. However, it is interesting to note that the solvophobic group of *ITI* is closer to the micelle core than the solvophilic group of *IHI*, which is completely surrounded by surfactant head segments. This fact is coherent with the interactions established between the solvophobic micelle core and the organic group of the precursor, being repulsive when such a group is solvophilic ($\omega_{HT}=1$). Since there is no repulsion between the surfactant tails and the solvophobic organic group of *ITI*, such a group can penetrate the micelle core more effectively.

By observing the density distributions of the beads belonging to *IHI*, they are well distinguished from the profile of the surfactant heads, and quite similar because of the strong attraction between the precursor *H* and *I* beads ($\omega_{HI}=-2$). The maximum of their profiles is located at a larger distance from the center of the micelle core than the maximum of the surfactant heads profile. This is due to the fact that *IHI* forms very favorable interactions with the micelle corona, but does not like the solvophobic core from which the precursor beads are separated by the surfactant heads. Systems with terminal solvophilic precursors show a similar trend, as observed in Figure 6.9, unless the inorganic source is insoluble in the solvent. In this case, the interactions established by the inorganic source, *I'*, with the solvophobic micelle core on one side and with the solvent on the other, lead to a deeper penetration of the precursor into the corona region. Precursors with a soluble inorganic source, for similar considerations, will stay closer to the interface between micelle and solvent.

In solvophobic bridging precursors, the role played in *I'H* by the inorganic source, is played by the solvophobic segment *T*, which prefers to stay closer to the micelle core than to the micelle-solvent interface. This explains why in correspondence with the micelle core, the average surfactant tail density decreases when the precursor concentration increases, as observed in Figure 6.18. In particular, the density distributions profiles in Figures 6.20 (c) and (d) show two important differences with the profiles in (a) and (b): the organic source *T* of *ITI* is closer to the micelle core than the solvophilic group *H* of *IHI*; and the density profiles of the beads belonging to *ITI* are not as similar as the ones belonging to *IHI*. Therefore, a system containing a solvophobic bridging precursor gives rise to an inorganic layer that is closer to the micelle core and broader than the one formed by using a solvophilic bridging precursor.

The hexagonally ordered rod-like aggregates observed in systems containing solvophilic precursors of the type *IHI* and *I'HI'*, present a behavior which can be considered very similar to the one observed in spherical aggregates. In particular, by increasing the surfactant concentration, the cylindrical aggregates become closer to each other, as reported in Figure 6.21 where the distance between two neighboring cylinders is between *12* and *14* lattice units according to the surfactant concentration. As observed with systems containing solvophilic terminal precursors (Figure 6.10), the tail density profiles are not significantly affected by the concentration and nature of the precursor.

In Figure 6.21, the density profiles of the surfactant tails and heads, and the normalized density profiles of the inorganic segment are reported. At first glance, we observe that the first peak in the density distributions of surfactant heads and inorganic beads is located where the tail density reaches its minimum. As a matter of fact, the core of the cylinders is surrounded by the surfactant heads and the precursor. However, if we focus on the heads profiles, it is interesting to note that the first peak is much broader than the one reported for the inorganic bead, and it is the combination of two consecutive peaks, which are separated by a relative minimum being located at approximately *7* lattice units from the center of the cylinder.

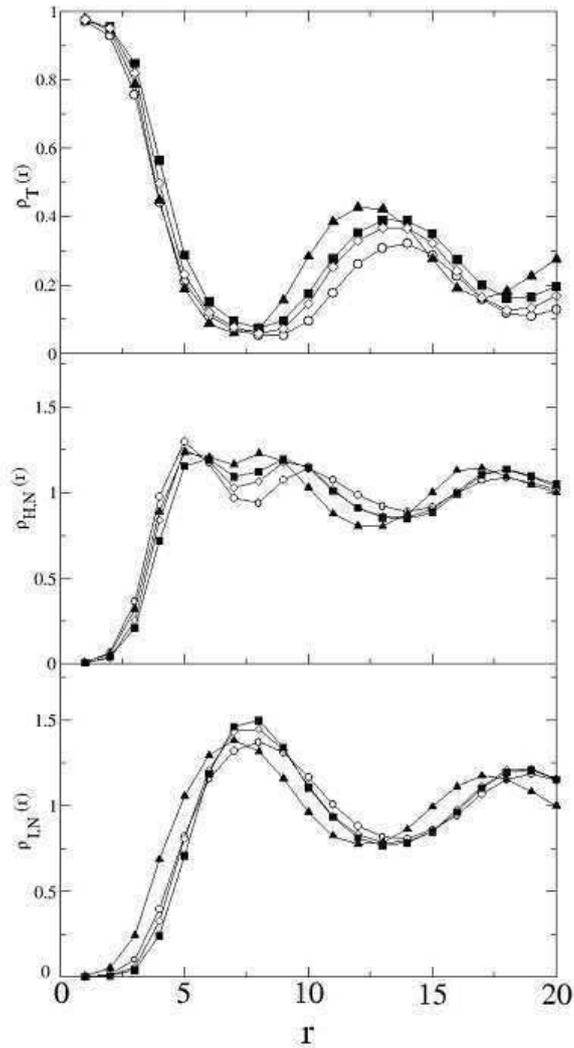

**Fig. 6.21.** Tail density profile and normalized head and inorganic segment density profiles in hexagonal phases observed in different systems at *T\**=8.0: (○) 41% $H_4T_4$ - 40% *IHI*; (◇) 48% $H_4T_4$ - 32% *IHI*; (■) 53% $H_4T_4$ - 25% *IHI*; and (▲) 55% $H_4T_4$ - 42% *I'HI'*.

At a distance of *5-6* lattice units from such a relative minimum, the tail density profiles show a maximum, giving the distance between the cores of two neighboring cylinders. Therefore, we can conclude that the inorganic precursor is well mixed with the surfactant heads, but preferentially far from the center of the cylinders, where the surfactant heads, belonging to neighboring cylinders, touch each other.

When the inorganic source is insoluble in the solvent, it can penetrate the corona a bit more, although this effect is balanced by the presence of a solvophilic group which prefers to stay closer to the surfactant heads than to the solvophobic core. In this case, the maximum in the tail distribution profile is closer to the center of the cylinder than when a soluble precursor is used. This result seems to be a compromise between the surfactant and precursor concentration, but the solubility of the inorganic unit, which can give rise to more compact structures by penetrating more deeply into the corona of the cylinders, seems to play an important role as well.

This point will be more extensively discussed in Chapter 7.

The tail-tail, tail-head, and tail-inorganic radial distribution functions of systems containing solvophilic precursors and presenting hexagonal ordered cylinders are reported in Figure 6.22, where the aforementioned conclusions regarding the effect on the structure of the solubility of the inorganic source, are confirmed. It seems that the concentration of surfactants does not affect the shape of the distributions as much as the solubility of the inorganic source. As a matter of fact, the systems containing *I'HI'* show a first peak in the $g_{TT}$ at a distance from the origin of approximately *12–13* lattice units, whereas the systems containing *IHI* at around *14* lattice units.

This means that the liquid crystal phases formed by hexagonally ordered cylinders result to be more compact when *I'HI'* is used as precursor, because of the presence of a insoluble inorganic source being located closer to the solvophobic core than the soluble inorganic source. The analysis of $g_{TH}$ and $g_{TI}$ furnishes the same conclusions and, as observed for terminal precursors, suggests that surfactant tails are not likely to be mixed with surfactant heads or inorganic beads in the cylinder core.

When no ordered structures are observed, the maximum in $g_{TT}$ is closer to the core of the aggregate and the curve converges to unity faster than for ordered structures (not shown).

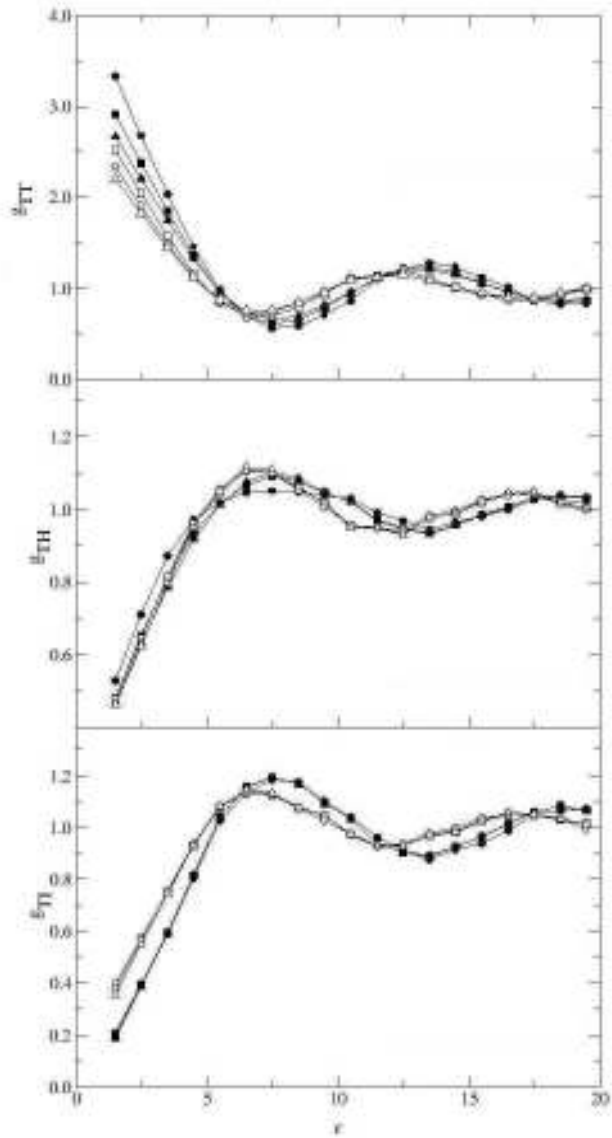

**Fig. 6.22.** Radial distribution functions, (a) $g_{TT}$, (b) $g_{TH}$, and (c) $g_{TI}$, at $T^*$=8.0 and different compositions: (●) 41% $H_4T_4$ - 40% $IHI$; (■) 48% $H_4T_4$ - 32% $IHI$; (▲) 53% $H_4T_4$ - 25% $IHI$; (□) 55% $H_4T_4$ - 42% $I'HI'$; (○) 60% $H_4T_4$ - 36% $I'HI'$; and (△) 65% $H_4T_4$ - 30% $I'HI'$.

# References Chapter 6

Allen, M. P., and D. Tildesley, *Computer Simulation of Liquids*, Clarendon Press, Oxford, 1987.

Chennamsetty, N., H. Bock, L. F. Scanu, F. R. Siperstein, and K. E. Gubbins, Cosurfactant and cosolvent effects on surfactant self-assembly in supercritical carbon dioxide, *Journal of Chemical Physics*, 9, 122, 2005.

Kim, S. Y., A. Z. Panagiotopoulos, and M. A. Floriano, Ternary oil-water-amphiphile systems: self-assembly and phase equilibria, *Molecular Physics*, 100, 2213-2220, 2002.

Larson, R. G., Self-Assembly of Surfactant Liquid-Crystalline Phases by Monte-Carlo Simulation, *Journal of Chemical Physics*, 91, 2479-2488, 1989.

Mackie, A. D., A. Z. Panagiotopoulos, and I. Szleifer, Aggregation behavior of a lattice model for amphiphiles, *Langmuir*, 13, 5022-5031, 1997.

Nelson, P. H., T. A. Hatton, and G. C. Rutledge, Asymmetric growth in micelles containing oil, *Journal of Chemical Physics*, 110, 9673-9680, 1999.

Panagiotopoulos, A. Z., M. A. Floriano, and S. K. Kumar, Micellization and phase separation of diblock and triblock model surfactants, *Langmuir*, 18, 2940-2948, 2002.

Rudnick, J., and G. Gaspari, The Shapes of Random-Walks, *Science*, 237, 384-389, 1987.

Salaniwal, S., S. K. Kumar, and A. Z. Panagiotopoulos, Competing ranges of attractive and repulsive interactions in the micellization of model surfactants, *Langmuir*, 19, 5164-5168, 2003.

*Io stimo più il trovar un vero, benché di cosa leggiera,*
*ché il disputar lungamente delle massime questioni,*
*senza conseguir verità nissuna.*
*Galileo Galilei.*

# Chapter 7

## TAILOR-DESIGNED ORDERED

## MESOPOROUS MATERIALS

### Introduction

The design of ordered mesoporous materials is a complex problem as many parameters can be varied independently. It is not the purpose of this chapter address general strategies for materials design, but to analyze the systems presented in the previous chapters and provide some guidelines for the design of materials based on those surfactants and precursors. In this chapter, we focus on those tunable parameters which permit an appropriate design of these materials, with particular attention to the choice of the silica precursors and to the structure directing agent. We have already observed that some important changes in the nature of the precursor can deeply affect the phase behavior of a given ternary amphiphilic system, and in some cases these changes can completely inhibit the formation of any self-assembled structure. As a matter of fact, the strong attraction between the template and the silica precursor is key to observe the organization of the inorganic framework around the liquid crystals, and if we change the properties of such a precursor, no ordered phases might be obtained. In particular, when the precursor is sufficiently solvophobic, the quality of the solvent is modified in such a way that the amphiphilic molecules do not feel anymore the necessity to shield their tails. On the other hand, if the solubility of the inorganic source decreases, then interesting structural modifications, such as in the distribution of the inorganic precursor around the pore, or in the pore diameter itself, can be detected.

The use of different silica precursors in the experiments is related to the possibility to functionalize the material by a direct synthesis or by a post-synthesis. In Chapter 3, we have underlined the pros and cons of each procedure by focusing our attention on the organization of the organic groups that are totally incorporated into the network of the silica matrix or available in the mesopore [*Hoffmann et al.*, 2006]. Clearly, our coarse-grained model is not able to enter into such details quantitavely, but we can still observe remarkable qualitative tendencies when terminal or bridging organosilica precursors are used. Such differences will be analyzed in this chapter to weigh the role of the precursor in tailoring the design of ordered mesoporous materials.

In addition, the role played by the templating agent is crucial for the synthesis of ordered mesoporous materials. In this chapter, we analyze the differences between those structures obtained by using a surfactant modeled as a linear chain of connected beads or a branched-head surfactant whose head (see Figure 4.1) makes our model surfactant closer to an ionic surfactant, by using the same precursor. The interest behind this modification is particularly related to the experimental evidence that block copolymers give rise to ordered materials whose mesopores are interconnected by micropores [*Imperor-Clerc et al.*, 2000]. On the other hand, ionic surfactants are able to template the synthesis of materials where micropores are not observed. We want to understand the role of the surfactant head structure in defining the wall thickness/pore size ratio in the porous material, or corona/core ratio in the micellar system.

In the first part of this chapter, we analyze the morphology of the hexagonally-ordered mesostructures with particular attention to (1) the distance between neighboring cylindrical aggregates, that is, between the pores of the resulting mesoporous materials; (2) the thickness of the pore walls as a function of the precursor concentration; (3) the distribution of the organic functional groups of the precursor chains; and (4) the pore size distribution.

In the second part of this chapter, we focus on the structural differences observed when a branched-head surfactant is used as opposed to a linear surfactant.

Finally, the possibility to combine in a many-scale approach the results obtained in the coarse grained simulation with fully atomistic simulations for the silica condensation reaction is presented. A cylindrical aggregate is isolated from the hexagonally ordered structure, and used as a templating aggregate for the condensation of a silica layer, and the consequent formation of a mesopore.

## 7.1 Differences between Terminal and Bridging Hybrid Precursors

In Chapter 3, we have underlined the benefits of functionalizing mesoporous materials, and some interesting applications have been reported. In this section, we want to analyze the structural differences observed in our simulations when a hexagonally-ordered mesoporous material is formed in the presence of terminal or bridging organosilica precursors.

Terminal organosilica precursors are used in a direct synthesis (*co-condensation*) or in a post-synthesis functionalization (*grafting*). In both cases, the organic group is likely to be available in the mesopores and is connected to the channel walls directly (one-pot synthesis) or through a silicon atom (post-synthesis). On the other hand, bridging organosilica precursors, included in the amphiphilic solution along with the surfactant, are completely incorporated into the silica network, and the organic functionality is covalently bonded with two silica atoms (see Figures 3.3, 3.4, and 3.5). When we compare the structures of the mesoporous materials obtained by using different precursors, we should also keep in mind that the results are strictly related to the one-pot synthesis path.

### 7.1.1 Separation between Cylindrical Aggregates

As observed in Chapter 6, the tail density profiles give us a measure of the distance between neighboring cylindrical aggregates in a hybrid material, or, equivalently, between neighboring pores in a mesoporous material. In general, if this distance decreases, the pores get closer and, if the pore diameter does not undergo any significant change (as usually observed for the same surfactant), then the thickness of the inorganic framework separating them can become thinner. However, we will see that the precursor concentration plays an important role in determining the thickness of the pore wall, and the separation between cylindrical aggregates might not significantly change even when the surfactant concentration increases.

In Figure 7.1, we report the surfactant tail density profiles calculated for two systems containing the pure silica precursor $I_2$. By increasing the surfactant concentration from 52% to 65%, the first peak of the tail distribution profiles, representing the radial distance between a given aggregate and the neighboring ones, is observed at a shorter lattice distance from the origin.

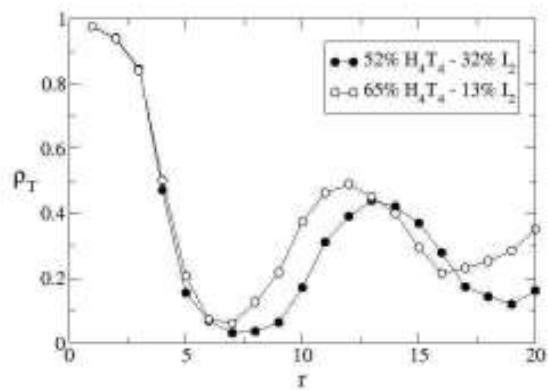

**Fig. 7.1.** Surfactant tail density profiles observed in hexagonally-ordered mesostructures containing a pure silica precursor.

A similar behavior is observed in Figure 7.2, where the tail density profiles of the terminal precursor $T'H$ are given.

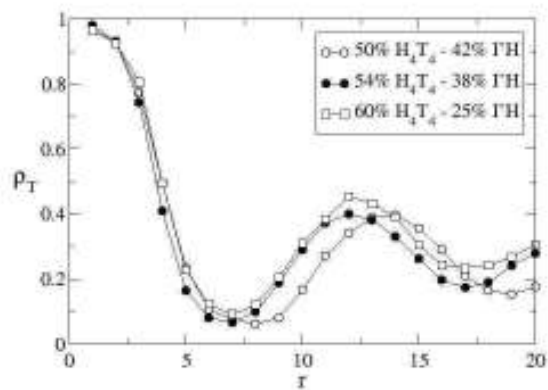

**Fig. 7.2.** Surfactant tail density profiles observed in hexagonally-ordered mesostructures containing a terminal organosilica precursor.

We observe that by increasing the surfactant concentration from 50% to 54%, the first peak of the curves gets closer to the origin, but if we further increase this concentration up to 60, then the distance remains constant to approximately *12* lattice units (LU).

The same situation is observed in Figure 7.3, with the bridging precursor *IHI*.

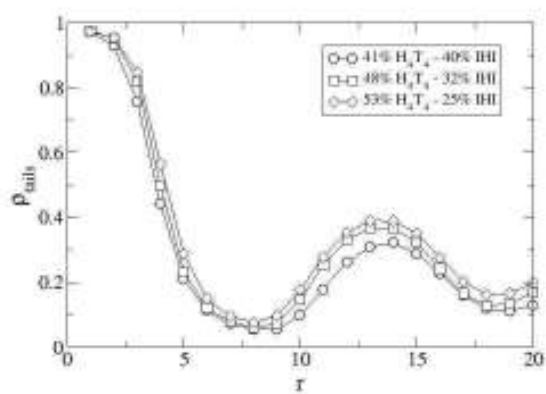

**Fig. 7.3.** Surfactant tail density profiles observed in hexagonally-ordered mesostructures formed with a solvophilic bridging precursor.

By increasing the surfactant concentration from 41% to 53%, the distance between the cylindrical aggregates remains practically constant, being closer to *14* LU for the system containing less surfactant, and to *13.5* LU for the others.

When we keep the inorganic precursor constant, it is interesting to observe that the surfactant concentration affects the distance between cylindrical aggregates, regardless the type of inorganic precursor, as shown in Figure 7.4.

Generally, an increase in the surfactant concentration has a direct consequence on the distance between cylindrical aggregates, that is, the higher this concentration, the closer the aggregates. However, such a dependence is limited by structural considerations imposed by the content of inorganic precursor. In particular, when the precursor concentration is sufficiently high, any increase of the surfactant concentration has a weaker effect on the distance between the cylindrical aggregates.

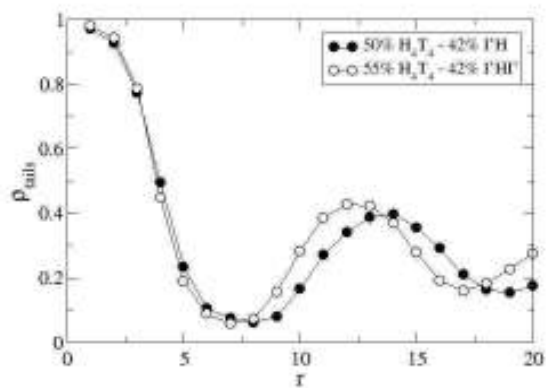

**Fig. 7.4.** Surfactant tail density profiles observed in hexagonally-ordered mesostructures formed with a terminal and a bridging precursor.

To summarize these results, in Figure 7.5, we report the position of the peak of the tail density distribution profiles ($P_{tail}$) as a function of the surfactant concentration for systems containing different precursor. As general trend, by increasing the surfactant concentration, the peak is located at shorter distances from the center of the aggregates and the final structure of the material results to be more compact.

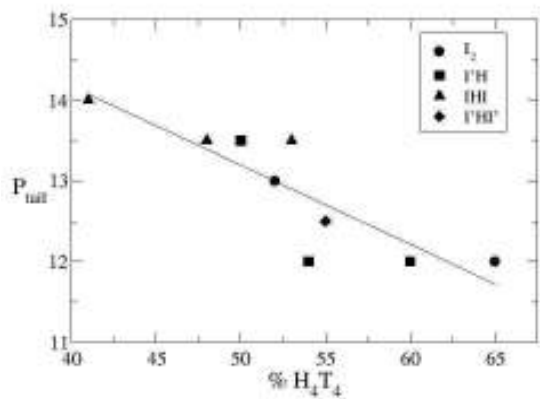

**Fig. 7.5.** Position of the peaks in the tail distribution profiles ($P_{tail}$) as a function of the concentration of the surfactant.

### 7.1.2    Inorganic Wall Thickness

We define the inorganic wall thickness as the distance between the first two points in the normalized density distribution profile of a given precursor, at which the normalized density is *1*. Such a definition, being completely arbitrary, consider as wall thickness that part of the material surrounding the pore where the most of the inorganic precursor concentrates.

The conclusions of the previous section are useful in the analysis of the dependence of the wall thickness on the surfactant and precursor concentration. In general, when the surfactant concentration increases, the pores get closer, and, if there are no significant changes in their diameter, the wall becomes thinner.

In Figure 7.6, we show the normalized density profiles of the precursor chains $I_2$ and *IHI*, in systems at the same precursor concentration (32%), but different surfactant concentration.

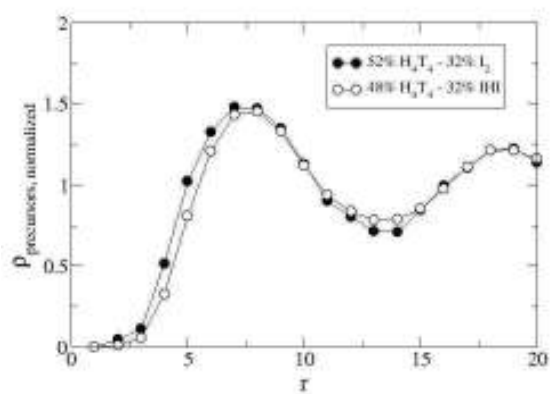

**Fig. 7.6.** Normalized density profiles of the precursors $I_2$ and *IHI* in the systems indicated in the legend box.

The distance between the cylindrical aggregates in the system with 52% of surfactant is smaller than that in the system with 48% of surfactant (see Figure 7.5), and the pore walls result to be slightly thicker. Figure 7.6 shows that the first peak in the precursor density distribution is slightly broader for the system containing a high surfactant concentration, which would suggest a thicker wall, independently that the cylinders are closer than at low surfactant concentration.

In Figure 7.7, some distribution profiles belonging to different precursors are given.

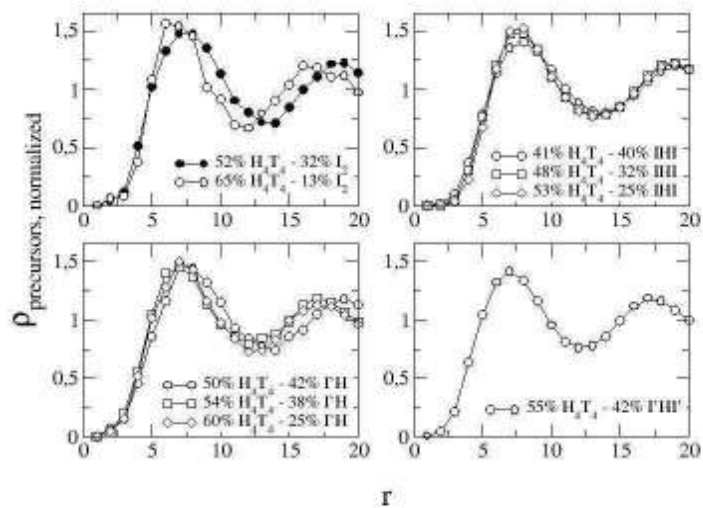

**Fig. 7.7.** Normalized density profiles of some precursors with or without functional organic groups.

By analyzing the density distribution profiles of the pure silica precursor, $I_2$, we observe a significant shrinkage of the wall thickness from *5.5* to *4.2* lattice units, when the surfactant concentration is increased from 52% to 65%. This effect is the consequence of the increase in the surfactant concentration which leads to closer pores, as observed in Figure 7.1, and is permitted by a much lower content of precursor in the system (19% less).

The systems containing terminal or bridging inorganic precursors experience a similar trend: when the surfactant concentration decreases and the precursor concentration increases, the pore walls become thicker, as shown in Figure 7.8, where the wall thickness is evaluated as a function of the precursor concentration for different types of precursors.

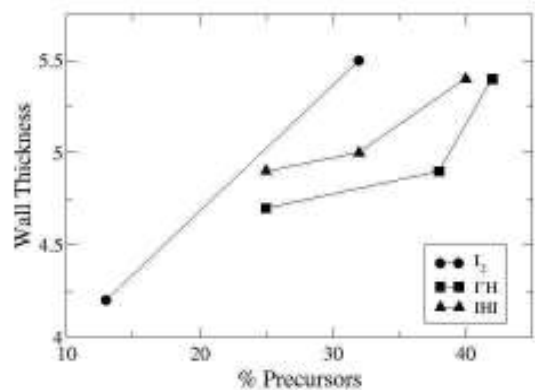

**Fig. 7.8.** Wall thickness of silica and organosilica precursors as a function of their concentration in the system.

The difference between the thickness of the pore walls of systems with *I'H* and *IHI* is mainly due to a different content of surfactant.

As general trend, by increasing the surfactant concentration and decreasing the precursor concentration, the wall thickness is reduced, as summarized in Table 7.1.

**Table 7.1**. Wall Thickness (WT) of the systems given in Figure 7.5

| System | WT | System | WT | System | WT | System | WT |
|---|---|---|---|---|---|---|---|
| 52% H₄T₄ | 5.5 | 41% H₄T₄ | 5.4 | 50% H₄T₄ | 5.4 | 55% H₄T₄ | 5.1 |
| 32% I₂ | | 40% IHI | | 42% I'H | | 42% I'HI' | |
| 65% H₄T₄ | 4.2 | 48% H₄T₄ | 5.0 | 54% H₄T₄ | 4.9 | | |
| 13% I₂ | | 32% IHI | | 38% I'H | | | |
| | | 53% H₄T₄ | 4.9 | 60% H₄T₄ | 4.7 | | |
| | | 25% IHI | | 25% I'H | | | |

In Figure 7.9, the dependence of the wall thickness of the systems listed in Table 1 on the concentration of the silica units (*I* or *I'*) is given.

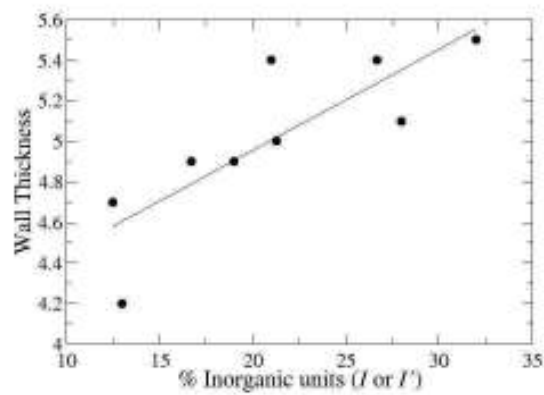

**Fig. 7.9.** Wall thickness of silica and organosilica precursors as a function of the concentration of the inorganic units in the related chains.

Increasing the concentration of the inorganic beads leads to thicker pore walls, and this dependence is almost linear. The largest deviations from the straight regression line are due to a very high surfactant concentration (65%) when the wall is 4.2 LU thick, or to a very high precursor concentration (42%) when the wall is 5.4 LU thick.

### 7.1.3    Distribution of the Organic Functional Groups

In this section, we outline the main factors affecting the distribution of the organic functional groups in the solvophilic corona of the cylindrical aggregates, by focusing on their position with respect to the silica units ($I$ or $I'$).

To evaluate the distribution of the inorganic precursor chains into the solvophilic corona of the aggregate, we compare their density distribution profiles with those of the surfactant head segments, as shown in Figure 7.10.

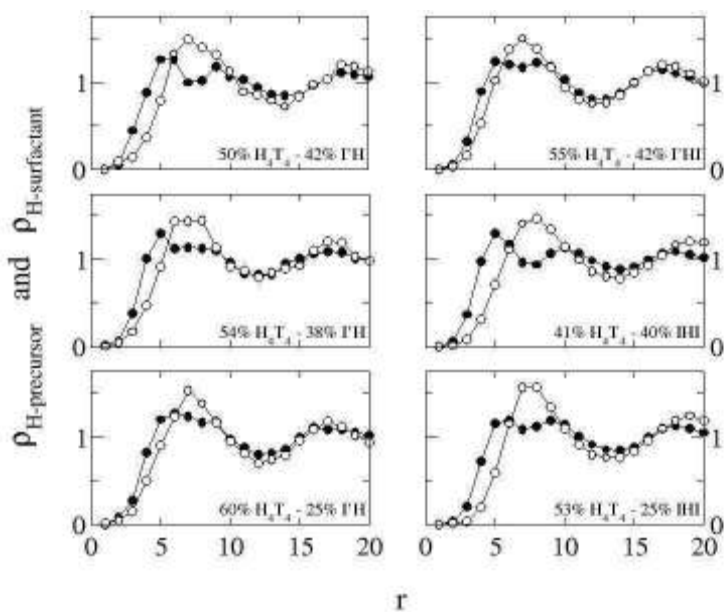

**Fig. 7.10.** Normalized density profiles of the solvophilic group of the inorganic precursors (blank circles) and surfactant heads (solid circles).

The first peak in the density profiles of the surfactant heads can show a split if the neighboring coronas are not particularly close, but its extension is practically constant. In correspondence of such a peak, the density distribution of the solvophilic group of the inorganic precursors *I′H*, *IHI*, and *I′HI′*, present a maximum. This means that the organic groups are mostly concentrated in the space available between the coronas of neighboring aggregates.

In order to know where and how the functional organic groups (always solvophilic, because the precursors with a solvophobic group did not show ordered liquid crystal phases) distribute themselves around the corona, we consider Figure 7.11, where the normalized density profiles of the organic and inorganic units constituting the precursor chains are shown.

Interestingly, the profiles of the organic functional groups belonging to the bridging precursors *I′HI′* or *IHI* do not present any significant difference from those of the silica units belonging to the same precursor chains. On the other hand, the density profiles of the organic groups belonging to the terminal precursor *I′H* (plots in the left column of Figure 7.11) present a slight difference with the corresponding profiles of the silica units. In particular, close to the solvophobic core of the cylindrical aggregates, the density of the inorganic unit $I'$ is slightly higher than that of the solvophilic group $H$. It seems that the precursor chains of the terminal precursors are distributed in such a way that their inorganic beads are oriented towards the solvophobic core of the aggregates to shield the solvophilic functional group from the proximity of the solvophobic core.

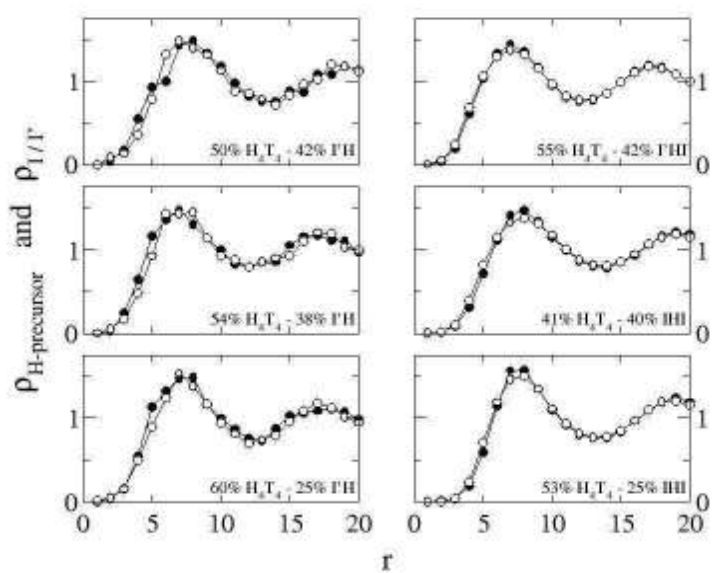

**Fig. 7.11.** Normalized density profiles of the solvophilic functional group of the precursor (blank circles), and of the silica unit (solid circles), belonging to terminal or bridging precursors.

The repulsive interactions formed between the solvophilic group $H$ and the solvophobic core of the cylinders, justify this evidence. This conclusion is extremely important for the design of functional mesoporous materials by one-pot synthesis, because the organic group, in this case, could not be available in the pore wall for some given applications, such as adsorption or catalysis.

The organic group of the bridging precursors would show a very similar distribution if it was not constrained between two silica beads, that is, if it was able to organize itself as the $H$ group of $I'H$.

It is worth to saying that the difference between the density profiles is not very significant because of the shortness of the chain modeled; longer chains might reveal a more interesting gap of density.

### 7.1.4    Pore Size Distribution

The main factor determining the pore size of materials obtained with a given surfactant is the size of the core, which depends on the length of the solvophobic segments. The first minimum of the tail distribution profiles gives information on the size of the solvophobic core of a given system, and it could be used to calculate the pore size of the resulting material. However, such a minimum is located approximately where the density distribution profiles of the inorganic precursor present a peak, that is, in the space occupied by the inorganic pore wall. This means that the pore radius has to be smaller than the distance of this peak from the center of the core, and that the minimum of the tail distribution profiles only gives a qualitative information of the pore size.

In this section, we calculated the pore diameter of the systems analyzed so far by considering the distribution of the inorganic precursor after the removal the template from the lattice box. To calculate the pore size distribution (PSD), for each site not occupied by the precursor beads, we calculated the biggest pseudo-spherical volume including this site which can be fitted inside a pore. Then, we averaged the results of *100* configurations sampled every $10^6$ MC steps.

It should be noted that the pore size distribution strictly depends on the more or less restrictive definition of the pseudo-spherical element. If we considered as the biggest volume that formed when all the concentric layers are completely free of precursor beads, then the PSD would present a peak at lower values of the pore diameter. On the other hand, if we accept that some sites in the outer shells can be occupied, then the PSD would present a peak at higher, and perhaps more realistic, values (see Figure 7.12). However, given that all the PSDs have been calculated by using the same definition of pseudo-spherical volume, a comparison between them is not affected by a more or less restrictive choice.

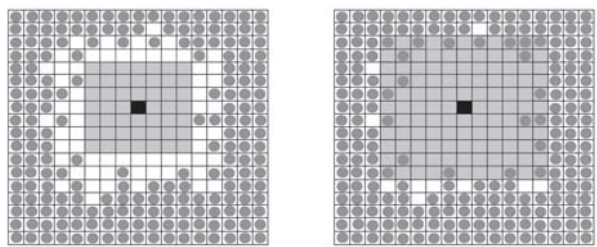

**Fig. 7.12.** Schematic representation in a two-dimensional lattice of the pseudo-spherical volume (light shaded squares) used in the calculation of the PSD. Left: all the concentric layers of the volume are free of precursor beads (solid circles); right: the outer shells of the volume can contain some precursor beads. The black square represents a site around which the concentric shells are considered.

In the next three figures, we compare a non-functionalized mesoporous material with a material containing three different organosilica precursors. In general, the pore size does not change significantly, but some interesting tendencies, potentially useful for the design and the pore tuning of these materials, can be observed.

In Figure 7.13, the pore size of the material assembled with *IHI* (*~9* lattice units) is slightly bigger than that of the material assembled with *I₂* (*~8.5* lattice units). This result is in agreement with the tail distribution profiles of the same systems, given in Figure 7.14, and the difference between their wall thickness (Table 7.1).

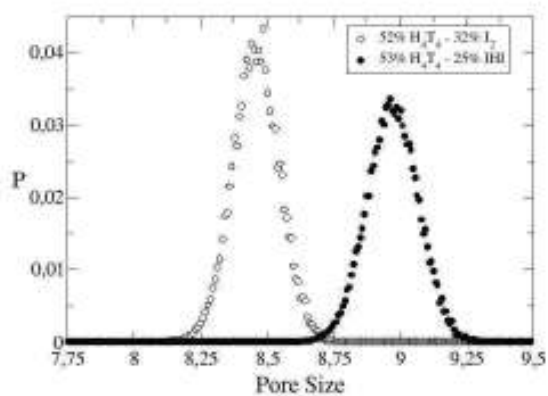

**Fig. 7.13.** Pore size distributions of a mesoporous material with a pure silica precursor, and functionalized with a solvophilic bridging precursor. The pore size is in lattice units.

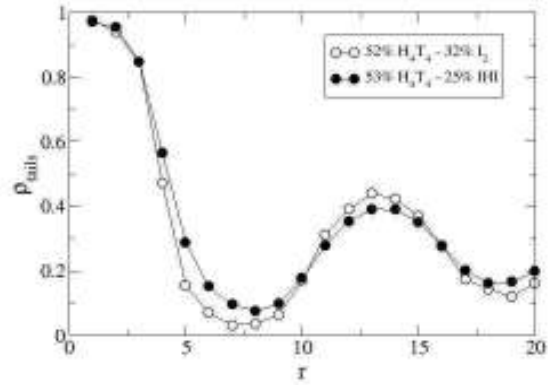

**Fig. 7.14.** Surfactant tail density profiles observed in hexagonally-ordered mesostructures assembled with a pure silica precursor or with a solvophilic bridging precursor.

The effect of the surfactant and precursor concentration on the pore size distribution is determinant, as observed in Figure 7.15, where the PSDs of systems containing *IHI* are compared.

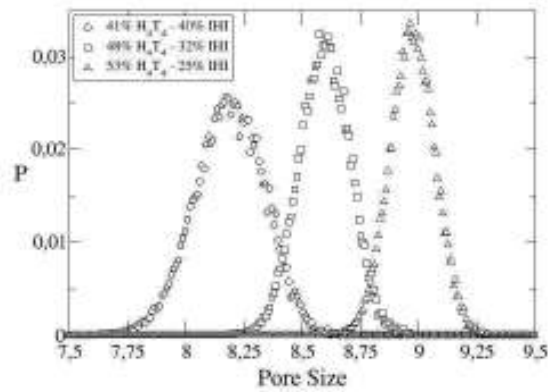

**Fig. 7.15.** Pore size distributions of systems containing a solvophilic bridging precursor. The pore size is in lattice units.

The system containing more surfactant and less precursor shows a bigger pore size. By increasing the precursor concentration, the wall thickness increases (see Table 7.1) and, since the distance between the cylindrical aggregates does not change (see Figure 7.3), the pore size decreases.

When we use the bridging precursor *I'HI'*, then the pore size decreases, as illustrated in Figure 7.16.

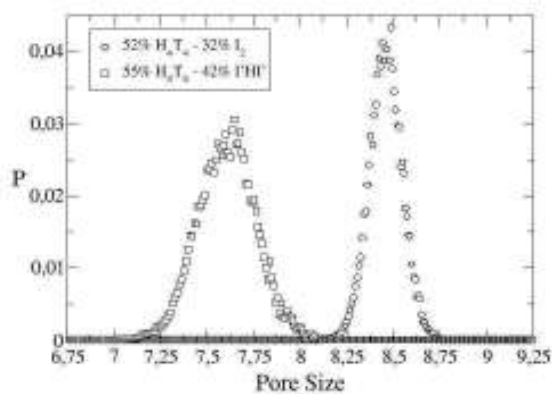

**Fig. 7.16.** Pore size distributions of a mesoporous material with a pure silica precursor, and functionalized with an insoluble solvophilic bridging precursor. The pore size is in lattice units.

In this case, the system containing $I_2$ has a thicker pore wall than that of the system with *I'HI'* (see Table 7.1), and, although its cylindrical aggregates are located at a slightly larger distance from each other, its pore size results to be smaller. Therefore, the difference between the precursor concentration in these two systems has a more important weight than the difference in the surfactant concentration.

By comparing the pore size distribution of a system containing *I'H* with that of the system containing $I_2$ (Figure 7.17), we observe that the material with the pure silica precursor has a larger pore diameter. The distance between the cylindrical aggregates is slightly shorter for the system containing *I'H*, and the pore walls are thicker in the system with $I_2$ (see Table 7.1). The balance between these two factors gives rise to such a small difference in the pore size distribution.

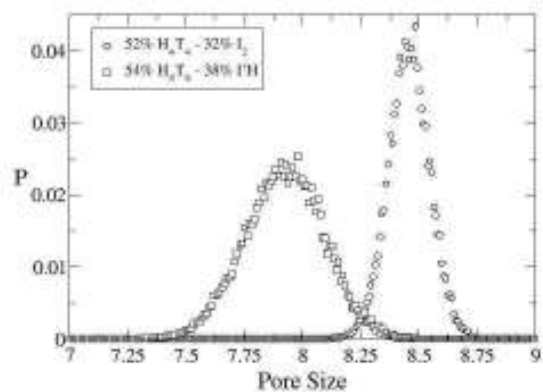

**Fig. 7.17.** Pore size distributions of a mesoporous material with a pure silica precursor, and functionalized with an insoluble solvophilic terminal precursor. The pore size is in lattice units.

The above results are confirmed when comparing the pore size distributions of systems containing *I'H* and *I'HI,'* at the same surfactant concentration (Figure 7.186). We observe that the pore size of the system containing a lower concentration of precursor is bigger, being consistent with the difference between the thickness of the related pore walls observed in Table 7.1.

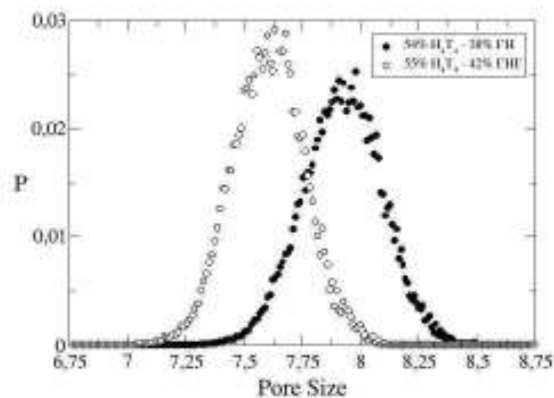

**Fig. 7.18.** Pore size distributions of systems containing a solvophilic bridging or terminal precursor. The pore size is in lattice units.

Finally, in Figure 7.19, we compare the PSDs of systems containing a terminal insoluble precursor, $I'H$.

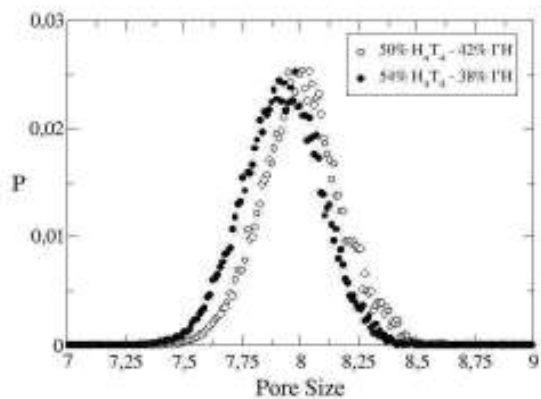

**Fig. 7.19.** Pore size distributions of systems containing a solvophilic terminal precursor. The pore size is in lattice units.

Practically, the pore size is identical, although a very slight reduction in the pore size can be detected when the surfactant concentration increases. This is due to the compromise between two opposing effects: the increase of surfactant concentration leading to a shorter distance between the cylindrical aggregates (see Figure 7.2), and the decrease of the precursor concentration leading to a thinner pore wall (Table 7.1). The effect of increasing the surfactant concentration seems to be more important in determining the final pore size distribution.

## 7.2 Branched-head Surfactants and Linear Surfactants

The surfactant architecture is of fundamental importance for the synthesis of mesoporous materials. In Chapter 3, we observed that depending on the structure directing agent it is possible to synthesize different types of mesoporous materials, such as MCM-41 or SBA-15. The former is usually synthesized by using ionic surfactants, whereas the latter by using block copolymers. An ionic surfactant presents a charged head whose geometry is quite different from that of the head belonging to a block copolymer.

In this section, we simplified the geometry of an ionic surfactant by modeling it as a branched-head diblock chain, and compare the structures obtained by two different types of surfactant: one being modeled by a linear chain, and the other by a branched-head chain with a linear solvophobic tail. The branched-head model is a simple way of increasing the size of the solvophilic part of the surfactant without increasing its length. In a continuum model this can be easily accomplished by changing the size of the head group, which would be very complicated to do in a lattice model.

The branched-head surfactants considered in this section and given in Figure 7.20, are made up of three head segments connected to a branched-head segment, and of four (Figure 7.19$a$) or five (Figure 7.19$b$) tail segments.

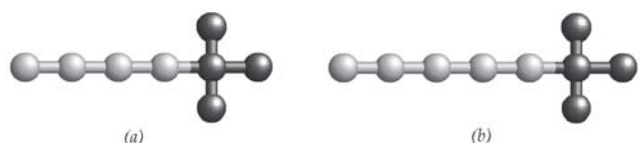

*(a)*          *(b)*

**Fig. 7.20**. Model branched-head surfactants: *(a)* $T_4HH_3$ and *(b)* $T_5HH_3$. Dark shading: heads; light shading: tails.

The aggregation behavior of a binary system containing $T_4HH_3$ with the same model solvent used in the simulations with $H_4T_4$, show the formation of spherical aggregates, but no hexagonally ordered phases. The difficulty in forming rod-like aggregates was probably due to the packing geometry of $T_4HH_3$: the branched-head accommodates its own beads in the corona differently than the linear surfactant $H_4T_4$. In particular, the head group area is bigger than that of the linear surfactant and the value of the packing parameter ($p = v/a_0 l$) is probably not high enough to make the formation of cylindrical aggregates possible [*Israelachvili*, 1991].

The surfactant $T_5HH_3$ in binary systems shows the formation of hexagonal phases at concentrations between 50% and 70% by volume, as reported in Figure 7.21.

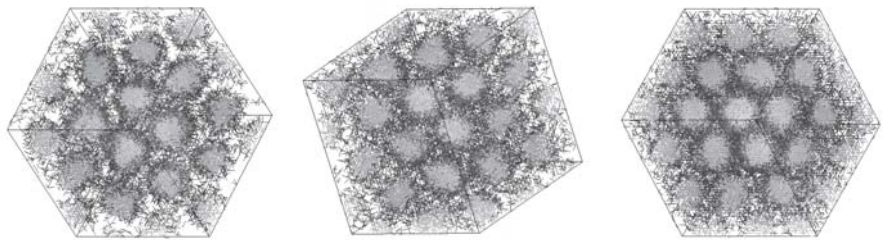

**Fig. 7.21**. Hexagonally ordered cylinders obtained at $T^*$=8.0 with $T_5HH_3$. Volume fractions: 50% (left), 58% (centre), 70% (right). Light shading: tail segments. Dark shading: head segments. The solvent is not shown.

In Figure 7.22, the ternary phase diagram of the system $T_5HH_3/I_2/S$, obtained at the reduced temperature $T^*$=8.0, is reported.

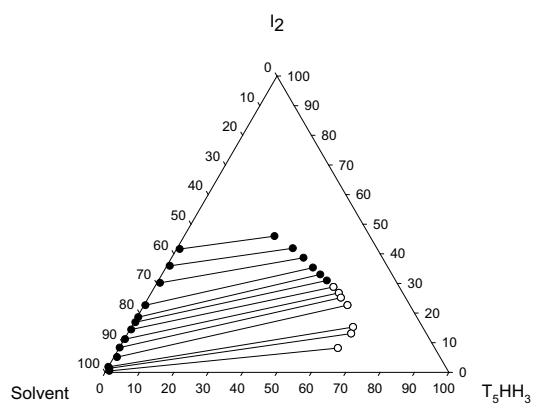

**Fig. 7.22**. Ternary phase diagram of the system $T_5HH_3/I_2/S$ obtained by MC simulations at $T^*$=8.0. The blank circles indicate the presence of hexagonally ordered cylindrical aggregates.

It is very similar to the phase diagram of the system $H_4T_4/I_2/S$ (Figure 5.10), although in this case the immiscibility gap is slightly bigger because the driving force for the phase separation is increased by the lower solubility of $T_5HH_3$ in the solvent-rich phase. The difference in the solubility is a consequence of a longer solvophobic tail than in $H_4T_4$.

Hexagonally ordered liquid crystal phases have been obtained as a result of the phase separation between a solvent-rich phase and a surfactant-rich phase, which contains approximately between 50% and 65% of surfactant and between 10% and 30% of pure silica precursor. Figure 7.23 shows a configuration, obtained after $80 \times 10^9$ MC steps, of a system with 50% of surfactant and 10% of inorganic precursor. The final surfactant-rich phase contains approximately 65% of surfactant, being more than enough to observe the formation of a hexagonal phase. To better appreciate the formation of the ordered structure, a snapshot without the template is also given.

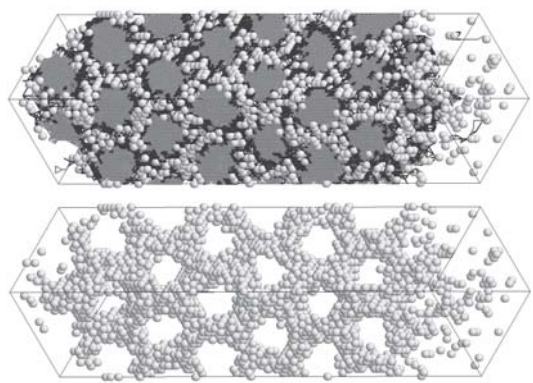

**Fig. 7.23**. Phase separation observed in the system $T_5HH_3/I_2/S$ at $T^*$=8.0 in a lattice box of size 24×24×100. Global concentrations: 50% $T_5HH_3$ - 10% $I_2$. Light shading represents the surfactant tails, dark shading represents the surfactant heads; the white spheres represent the inorganic precursor. The solvent is not shown. In the lower figure, the surfactant is not shown.

Two important factors have been analyzed by using single phase simulations: (1) the effect of the tail length on the pore diameter; and (2) the effect of the head architecture on the pore walls. In Figure 7.24, we report the PSD of two systems with the same volumetric concentration of surfactant (52%) and inorganic precursor (around 30%): one refers to a system containing the linear surfactant $H_4T_4$, and the other to a system containing the branched-head surfactant $T_5HH_3$.

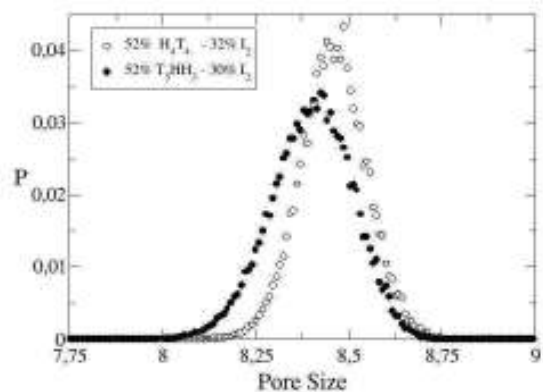

**Fig. 7.24.** Pore size distributions of systems containing a linear or a branched-head surfactant, in the presence of a pure silica precursor. The pore size is in lattice units.

The PSDs present a very similar trend, and the pore size seems to be practically independent of the choice of the surfactant and, in particular, on the increase of the number of tail beads. However, there is an opposite tendency to the one we would have expected: an increase of the tail length should give rise to a bigger pore, and we observe a slight decrease. Why?

The tail density distributions, given in Figure 7.25, indicate that the diameter of the solvophobic core is practically the same in both cases, but that the distance between cylinders is smaller for the branched-head surfactant. The smaller space between cylinders results in thinner walls, which is confirmed by looking at the inorganic density distributions (Figure 7.26). Therefore, the wall thickness and pore size can also be tuned when changing the head architecture.

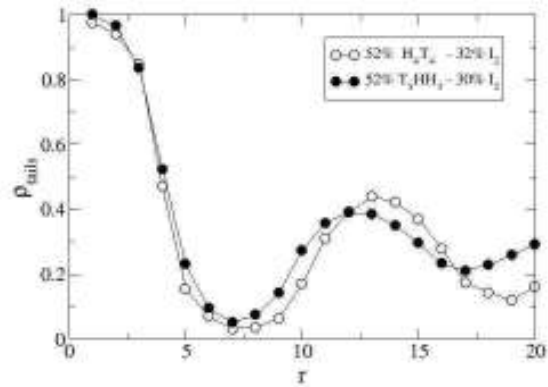

**Fig. 7.25**. Tail density profiles of systems containing two different surfactants.

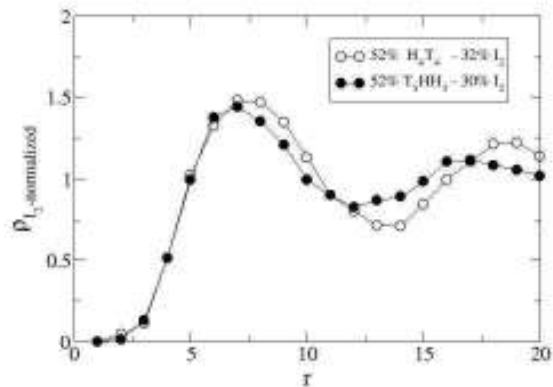

**Fig. 7.26**. Normalized density profiles of the pure silica precursor, $I_2$, for systems containing two different surfactants.

Such a result is in complete agreement with the experimental results presented by Stucky and coworkers [*Zhao et al.*, 1998], who synthesized mesoporous materials by substituting the ionic surfactants used by Beck in the synthesis of MCM-41 with block surfactants, and observed a significant increase in the wall thickness.

## 7.3    Many-Scale Simulations

One of the most challenging tasks towards understanding the properties of ordered mesoporous materials is to create as realistic as possible models, that can predict their behavior when used as adsorbents for gas separation or storage. The model we have selected to describe the self-assembly of such materials does not have the necessary features to satisfy this request, where a detailed representation of the silica molecules constituting the pore walls and of the functional organic groups is necessary. In fact, our aim was not to study the adsorption in the mesoporous structure, but the phase and aggregation behavior of amphiphilic self-assembly systems. The research works concentrating on the adsorption properties, usually apply atomistic models to cylindrical pores of variable roughness, in some cases obtained from coarse-grained simulations [*Coasne et al.*, 2006], or to perfect cylinders around which the silica layer can organize to form the mesopore [*Schumacher et al.*, 2006a].

Simulating the self-assembly by using an atomistic model would be too computationally demanding. A possible alternative is to use a coarse-grained model to simulate the self-assembly of the mesophase, and then an atomistic model for the condensation of silica and the adsorption in the resulting mesoporous material. We have designed a many scale approach to fill the gap between atomistic and coarse-grained models [*Prosenjak et al.*, 2007]. In particular, from the self-assembled hexagonal liquid crystal phases obtained by applying a coarse-grained model, we have isolated a single templating cylindrical micelle around which the condensation reaction of silica has been simulated by using the kinetic MC (kMC) scheme developed by Schumacher *et al.* [*Schumacher et al.*, 2006b]. The kMC method consists of an off-lattice simulation where an atomistic representation of the silicon and oxygen atoms leads to detailed models of the pores. These atomistic representations of the material are then used for adsorption studies using grand canonical Monte Carlo simulation.

In Figure 7.27 (a), a snapshot of the hexagonally ordered phase observed in the system $H_4T_4/I_2/S$ is given.

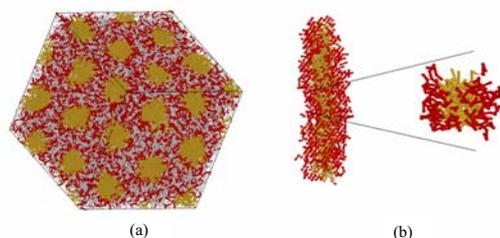

(a)                    (b)

**Fig. 7.27. (a)** Configuration of a $H_4T_4/I_2/S$ system, simulated for $90\times10^9$ MC steps at $T^*$=8.0 in a lattice box of size 40×40×40. **(b)** Cylindrical aggregate extracted from this configuration, where the part used as template in the kMC simulation is indicated. The surfactant tails are in yellow; the surfactant heads in red; and the inorganic precursor in gray. Volume fractions: $H_4T_4$ 52%, $I_2$ 32%.

From this configuration, a cylindrical aggregate, being the inner organic core of its corresponding mesopore, is randomly selected and one part is used as a template in the kMC simulation for the condensation of silica (Figure 7.27 (b)).

This selection was made by defining an axis connecting the center of a given cross section of the cylindrical aggregate with another cross section belonging to the same aggregate. All the cross sections sharing at least one tail segment were considered part of the cylinder, and the surfactant chains belonging to these cross sections were included in the cylindrical micelle. A part of the cylinder, being an appropriate sample for a realistic simulation of silica condensation, was used in the kMC simulation as template, represented as a continuous potential function (Figure 7.28).

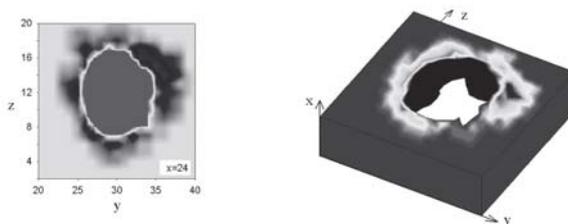

**Fig. 7.28.** Potential map of a cross section (left) belonging to a cylindrical sample (right) for the system $H_4T_4/I_2/S$ at $T^*$=8.0 in a lattice box of size 40×40×40. The central area represents the solvophobic core, and the brightness around it, depends on the concentration of surfactant head segments.

The potential map reported in Figure 7.28 has been calculated by considering the solvophobic core as a forbidden region for the silica precursor, whose interactions with the tails would be highly repulsive. In particular, the potential between the oxygen atoms of the silica precursor and the cylindrical aggregate depends on the distance between an oxygen atom and the edge of the aggregate: it is attractive near the aggregate and repulsive if the precursor penetrates into the aggregate. The silicon and oxygen atoms are represented explicitly, whereas hydrogen atoms are not taken into account directly, therefore the simulation does not distinguish between protonated and deprotonated forms of silanol groups. Covalent bonds between oxygen and silicon atoms are modeled as harmonic spring functions and a soft repulsive potential is applied between non-bonded atoms [*Prosenjak et al.*, 2007; *Schumacher et al.*, 2006b].

The kMC simulation is composed of three steps: (1) formation of the silica layer around the cylindrical aggregate; (2) aggregation of the replicas of the cylindrical aggregate into a hexagonal array; and (3) calcination. Figure 7.29 shows the snapshots referring to these steps when a linear diblock surfactant, $H_4T_4$, is used as structure directing agent (SDA) in the formation of the mesophase.

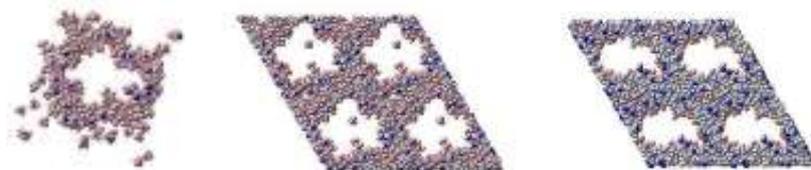

**Fig. 7.29.** Snapshots of the silica layer around a cylindrical micelle obtained from the self-assembling of $H_4T_4$, at the beginning of the simulation (left), during the aggregation step (center), and after the calcination (right). Silica atoms (blue), oxygen atoms (pink), and OH groups (gray) are represented [*Prosenjak et al.*, 2007].

The use of $H_4T_4$ as SDA yields a material with an exaggerated surface roughness. An other problem was caused by the length of the solvophilic block of the surfactant chains, which lead to a poorly defined silica layer formed around the aggregate during the first steps of the kMC simulation and to an excessive shrinkage during calcination. In particular, we observed a shrinkage almost five times bigger than that observed with a perfect cylindrical templating aggregate. Thus, we decided to modify the linear head with a branched head, being shorter and more compact than that of a diblock surfactant, as this would allow an easier control of the wall thickness to core size ratio.

A branched-head surfactant was chosen containing *10* tail segments and six head segments connected to a branched head: $T_{10}HH_6$ (Figure 7.30). At the reduced temperature $T^*$=12.0, this surfactant is able to self-assemble into ordered hexagonal phases, as observed in Figure 7.30, where a configuration of the binary system composed of 60% of $T_{10}HH_6$ and solvent, obtained after 80×10⁹ MC steps at $T^*$=12.0, is reported.

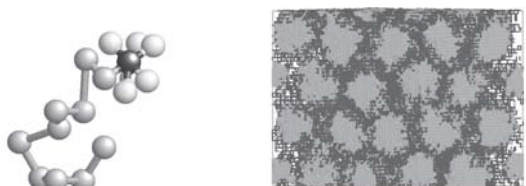

**Fig. 7.30. (Left)** Schematic representation of the branched-head surfactant $T_{10}HH_6$. The branched-head is black, the other heads are light gray, and the tail segments are dark gray. **(Right)** Configuration of a $T_{10}HH_6$/S system with 60% of surfactant, simulated for 80×10⁹ MC steps at $T^*$=12.0 in a lattice box of size 50×50×50. The surfactant tails are light and heads are dark shaded, respectively.

The tail density profile of $T_{10}HH_6$ in this binary system is shown in Figure 7.31 and compared with the tail density profile obtained with $H_4T_4$ at the same surfactant concentration, but at the reduced temperature $T^*$=8.0.

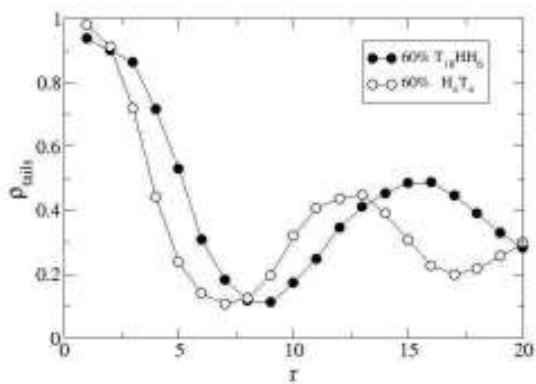

**Fig. 7.31**. Tails density profiles of systems containing $H_4T_4$ and $T_{10}HH_6$ at the reduced temperatures $T^*$=8.0 and $T^*$=12.0, respectively.

The comparison at the same temperature is not possible because: (1) the system $H_4T_4/S$ does not form ordered structures at $T^*$=12.0 (see Figure 5.6); and (2) simulating the system $T_{10}HH_6/S$ at lower temperatures would dramatically decrease the probability to accept the Monte Carlo movements, since the surfactant chain significantly longer than that of $H_4T_4$.

The minimum of the $T_{10}HH_6$ profile is shifted to larger distances from the center of the cylindrical aggregates, compared to the $H_4T_4$ profile. This means that the radius of the cylindrical aggregate obtained by the self-assembly of $T_{10}HH_6$ is larger than that obtained with $H_4T_4$, and this can give rise to mesoporous materials with bigger pores than those synthesized with $H_4T_4$. Also the peaks of the two profiles do not match, but to conclude something on the wall thickness of the resulting material, we should consider Figure 7.32, where the head density profiles are reported. The maximum in the head density distribution of $T_{10}HH_6$ seems to be narrower than that of $H_4T_4$, leading to the synthesis of a mesophase with thinner walls than those observed in the material assembled with $H_4T_4$.

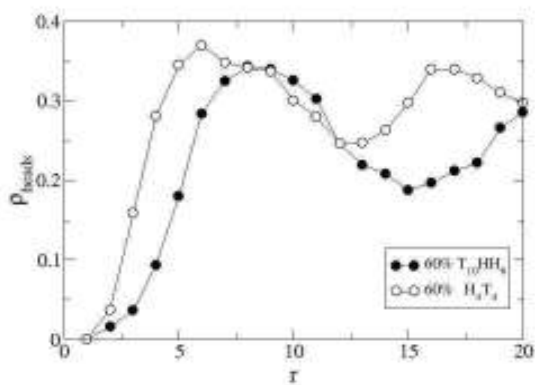

**Fig. 7.32**. Heads density profiles of systems containing $H_4T_4$ and $T_{10}HH_6$ at the reduced temperatures $T^*$=8.0 and $T^*$=12.0, respectively.

By repeating the same procedure as for the linear surfactant $H_4T_4$, we obtained the potential map of a cylindrical section (Figure 7.33), and a more realistic model for the design of mesoporous adsorbents, as reported in Figure 7.33.

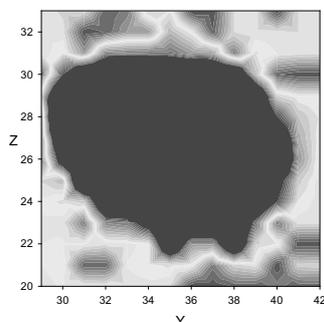

**Fig. 7.33.** Potential map of a cross section belonging to a cylindrical sample obtained with the branched-head surfactant $T_{10}HH_6$ at $T^*=12.0$ in a lattice box of size 50×50×50. The central area represents the solvophobic core, and the brightness around it, depends on the concentration of surfactant head segments.

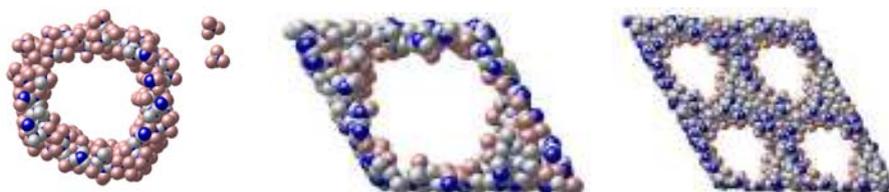

**Fig. 7.34.** Snapshots of the silica layer around a cylindrical micelle obtained from the self-assembling of $T_{10}HH_6$, at the beginning of the simulation (left), during the aggregation step (center), and after the calcination (right). Silica atoms (blue), oxygen atoms (pink), and OH groups (gray) are represented. (Courtesy of C. Prosenjak, unpublished results)

The isotherm of adsorption calculated at 263 K is given in Figure 7.34 and compared to the isotherm of adsorption obtained by assuming a perfect cylinder as templating aggregate.

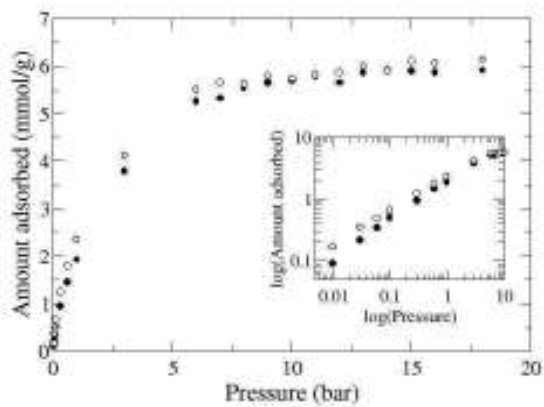

**Fig. 7.35.** Isotherms of adsorption of ethane at 263 K calculated by assuming a model self-assembled mesopore (open circles) and a perfect cylindrical mesopore (solid circles). (Courtesy of C. Prosenjak, unpublished results)

The agreement between the two curves is very good. The amount of ethane adsorbed is slightly higher when a coarse-grained self-assembled mesopore is considered, specially at low pressures. This is probably due to the pore walls which are not as smooth as those of the cylindrical mesopore. However, a deeper analysis comparing them with the experimental results is planned in order to detect to which extent a perfect cylindrical mesopore can be considered a valid approximation of a self-assembled mesopore.

**References Chapter 7**

Coasne, B., F. R. Hung, R. J. M. Pellenq, F. R. Siperstein, and K. E. Gubbins, Adsorption of sample gases in MCM-41 materials: The role of surface roughness, *Langmuir*, 22, 194-202, 2006.

Hoffmann, F., M. Cornelius, J. Morell, and M. Froba, Silica-based mesoporous organic-inorganic hybrid materials, *Angew Chem Int Ed Engl*, 45, 3216-3251, 2006.

Imperor-Clerc, M., P. Davidson, and A. Davidson, Existence of a microporous corona around the mesopores of silica-based SBA-15 materials templated by triblock copolymers, *Journal of the American Chemical Society*, 122, 11925-11933, 2000.

Israelachvili, J., *Intermolecular & Surface Forces*, 2nd ed., 450 pp., Academic Press, London, 1991.

Prosenjak, C., A. Patti, F. R. Siperstein, and N. A. Seaton, Modelling the synthesis of periodic mesoporous silicas, *Studies in Surface Science and Catalysis*, in press, 2007.

Schumacher, C., J. Gonzalez, M. Perez-Mendoza, P. A. Wright, and N. A. Seaton, Design of hybrid organic/inorganic adsorbents based on periodic mesoporous silica, *Industrial & Engineering Chemistry Research*, 45, 5586-5597, 2006a.

Schumacher, C., J. Gonzalez, P. A. Wright, and N. A. Seaton, Generation of atomistic models of periodic mesoporous silica by kinetic Monte Carlo simulation of the synthesis of the material, *Journal of Physical Chemistry B*, 110, 319-333, 2006b.

Zhao, D. Y., J. L. Feng, Q. S. Huo, N. Melosh, G. H. Fredrickson, B. F. Chmelka, and G. D. Stucky, Triblock copolymer syntheses of mesoporous silica with periodic 50 to 300 angstrom pores, *Science*, 279, 548-552, 1998.

*E quindi uscimmo a riveder le stelle.*
*Dante Alighieri*

# Conclusions

Lattice MC simulations in the *NVT* ensemble have been performed to study the phase and aggregation behavior of self-assembling amphiphilic systems composed of a diblock surfactant, an inorganic precursor (with the possibility of containing an organic functional group), and a solvent. These systems are of fundamental importance for the synthesis of ordered hybrid materials, which can be obtained thanks to the strategic role played by the surfactant, acting as structure directing agent, and to the strong interactions established by the surfactant heads with the inorganic precursor. The removal of the organic template by calcination or solvent extraction leads to the formation of a mesoporous ordered structure, whose features strictly depend on the nature of the structure directing agent, inorganic precursor, and synthesis conditions.

The coarse-grained model used in this work was able to highlight the most important tendencies of such ternary systems, being very useful to correctly understand their phase and aggregation behavior, and to furnish a proper tool as a guide for their design. Although this model cannot give information on an atomistic scale, it is very helpful to appreciate the periodicity of the mesophases formed, and to discover the necessary conditions for the synthesis of ordered structures. We have seen that, according to the surfactant and precursor concentration, these amphiphilic systems are able to phase separate in a solvent-rich (dilute) phase, where, in some cases, micelles are observed, and in a surfactant-rich (dense) phase, where ordered liquid crystal phases can form. The nature of the precursor, its solubility in the solvent and the presence of a more or less solvophilic organic group, affect the driving force for the phase separation, which mainly depends on the interactions established by the inorganic precursor with the surfactant heads and the solvent.

The phase separation is key to synthesize hybrid materials even when the global surfactant concentration in the system is very low. We observed that in binary surfactant/solvent systems, hexagonally ordered phases are obtained only at very high surfactant concentrations, usually above 50% by volume. In ternary systems, the presence of an inorganic precursor leads to a phase separation where the dense phase can contain enough surfactant for liquid crystals to be observed. This is why it is very important to know when phase separation occurs, and which are the conditions favoring it. In this context, the importance of ternary phase diagrams is clear, presenting a different immiscibility gap according to the inorganic precursor used. The phase diagrams obtained are characterized by a phase separation between a dilute phase, mainly occupied by the solvent, and a concentrated phase presenting a high content of the amphiphilic moiety and hybrid or pure inorganic particles. In systems with non-functionalized precursors, the phase separations can be clearly classified between associative or segregative phase separations. When hybrid precursors are used, this differentiation is not so clear, as an interplay of associative and segregative effects are observed.

The agreement observed in the ternary phase diagrams between MC simulations and QCT was very good when the system does not self-assemble into organized structures, and we observe qualitative agreement if ordered structures are present. In those cases where ordered aggregates are formed, the immiscibility gap calculated by the QCT was smaller than the one calculated by MC simulations, because this theory does not allow for the formation of aggregates or more complex liquid crystal phases, and leads to a higher content of solvent in the surfactant-reach phase.

In general, the driving force for the phase separation is the result of two main factors: (1) the strong attraction between surfactant heads and inorganic precursor (associative phase separation); and (2) the repulsion between the solvent and the inorganic precursor (segregative phase separation). Therefore, a proper choice of the precursor affects the concentration of surfactant in the dense phase and hence the possibility to observe ordered liquid crystal phases. If this concentration is high enough, one of the necessary conditions for the ordered phases to be formed is satisfied, the others being strictly connected to the nature of the precursor used.

As a matter of fact, the solubility of the precursor, the presence of a functional organic group, and the solvophilicity of such a group, are all parameters to take into account for a correct and exhaustive estimation of the driving force for the phase separation. By tuning these parameters, we can modify the highest surfactant concentration in the phases at equilibrium and their order. When terminal hybrid precursors are used, if the organic group is solvophobic, we observe that it is not possible to obtain ordered structures, regardless of the surfactant concentration, or the miscibility of the inorganic segment with the solvent. If the organic segment is solvophilic, we only observe the formation of hexagonal or lamellar phases when the inorganic segment is immiscible with the solvent. This could explain the difficulty in experiments in obtaining well ordered structures with terminal organosilicas, where the organic/silica ratio is high [*Hatton et al.*, 2005]. The precursor restricts the formation of ordered phases observed with purely siliceous precursors, or those with a smaller organic/silica ratio. As a matter of fact, the presence of the precursor, depending on its nature, can be viewed as a cosolvent or a cosurfactant that can change the aggregation number of the aggregates or modify the *solvent* properties, making the surfactant more soluble and less likely to form ordered structures.

By using a bridging hybrid precursor, it is generally easier to obtain ordered aggregates, especially when the organic group is solvophilic. With a solvophobic group, we observed the formation of spherical aggregates (not observed in the analogous terminal precursor) when the inorganic source is soluble in the solvent. In this case, it is interesting to note that the solvophobic organic group can penetrate more deeply into the micelle corona in order to stay closer to the solvophobic core and as far as possible from the interface with the solvent. Of course, the distribution of the precursor in the corona is the result of a compromise between the interactions established between the different beads.

Generally, when ordered phases are formed, their structural properties (density profile and radial distribution function) are mostly determined by the surfactant concentration, but the nature of the precursor can lead to interesting, although not radical, changes. When hybrid precursors are used, the distribution of inorganic and organic segments in the framework is usually homogeneous. Only in some cases a slightly larger concentration of organic segments is obtained near the solvophobic core when the organic group is solvophilic while the inorganic is assumed to be immiscible with the solvent (*I'H* and *I'HI'*) and when the bridging organic group is solvophobic and the inorganic source is soluble in the solvent (*ITI*).

This suggests that when hybrid precursors are used, it is likely that some of the organic groups will not be accessible to the adsorbed molecules as they are located in sections of the framework that do not have access to the pore network. This is especially evident when the inorganic precursor presents a partial solubility with the solvent and a solvophilic group in terminal position. In this case, the inorganic source penetrates the corona of the aggregate in order to stay away from the solvent, and creates a sort of barrier between the solvophobic tails and the solvophilic functional group. Presumably, an inorganic precursor being slightly solvophobic to prevent self-assembling and slightly solvophilic to prevent the functionalization of the pore walls, would be a better choice to observe a mesoporous material containing available functional organic groups. Only by carefully selecting the precursor solubility and the organic solvophilicity, is it possible to obtain a high concentration of accessible organic groups in the framework. Evidently, we have ignored many parameters that may play a role in the determination of the structure and properties of hybrid mesoporous materials, nevertheless, we consider that the models used capture the main properties of the systems studied.

Some of these properties, such as the distance between neighboring pores, the thickness of the pore walls, and the distribution of the precursor beads around the corona, have been analyzed by calculating the surfactant and precursor density profiles. We observed that the concentration of the surfactant and the inorganic precursor are the main parameters affecting the distance between the pores and thickness of the pore walls. Generally, when the surfactant concentration increases, the cylindrical aggregates get closer and the pore walls thinner, although such a tendency is limited by the content of inorganic precursor.

The pore size distribution is affected by the distance between neighboring aggregates and by the wall thickness. These are the two main factors affecting the diameters of the mesopores when the surfactant architecture is not changed. When we compared different surfactants, then some interesting aspects were observed. Not only the tail length affects the pore diameter (the longer the tail, the larger the pore), but also the head structure has been shown to have a significant importance. An interesting development of this last conclusion would be to analyze how the architecture of the surfactant head can affect the phase and aggregation behavior of ternary amphiphilic systems, and if hexagonally ordered phases are observed, the distribution of the pore size.

We have presented in this work a branched-head surfactant with a long solvophobic chain containing ten tail segments. Such a surfactant has been selected as structure directing agent for the self-assembly of cylindrical aggregates, used as templating structures for the formation, aggregation, and condensation of an atomistic representation of a silica precursor. The interest in finding a common point between a coarse-grained and an atomistic model, is justified by the possibility of improving the design of mesoporous materials for selective adsorption. The choice of a long-chain surfactant seems to solve the problem of the excessive roughness and shrinkage observed with a short-chain surfactant. The adsorption isotherms, calculated for ethane, are in very good agreement with the ones obtained with a perfect cylindrical aggregate. A detailed comparison with the experimental results is ongoing and will clarify to which extent a perfect cylindrical geometry can be assumed as a valid approximation of a self-assembled mesopore.

It is not straightforward to perform a quantitative comparison between the results presented here and experiments. The number of details not included in our study can be of fundamental importance to make a reasonable comparison. However, the tendencies observed by applying a very simple model have furnished a qualitative idea of the factors affecting (1) the phase separation of ternary amphiphilic systems, (2) the formation of ordered liquid crystal phases, (3) the organization and the accessibility of the functional organic groups, and (4) the structural properties of the porous framework, such as the wall thickness and pore size.

At $T^*$=8.0, we have observed phase separation in all the systems studied, but not all these systems gave rise to a concentrated phase with a surfactant content high enough to form liquid crystal phases. As a general trend, systems with bridging hybrid precursors can push the phase separation further than those containing terminal hybrid precursors, especially when the inorganic source is soluble in the solvent. As a matter of fact, experimentally it was observed that it is not easy to synthesize ordered materials by using amphiphilic solutions containing terminal hybrid precursors [*Hatton et al.*, 2005]. In general, a second, pure inorganic precursor with stabilizing properties is added, or the terminal precursor is grafted after the synthesis is completed [*Edler*, 2005; *Hatton et al.*, 2005; *Hoffmann et al.*, 2006]. The last method is quite expensive due to the high surfactant concentration required and to the necessity of a post-synthesis treatment.

### References

Edler, K. J., Current understanding of formation mechanisms in surfactant-templated materials, *Australian Journal of Chemistry*, 58, 627-643, 2005.

Hatton, B., K. Landskron, W. Whitnall, D. Perovic, and G. A. Ozin, Past, present, and future of periodic mesoporous organosilicas - The PMOs, *Accounts of Chemical Research*, 38, 305-312, 2005.

Hoffmann, F., M. Cornelius, J. Morell, and M. Fröba, Silica-Based Mesoporous Organic-Inorganic Hybrid Materials, *Angew. Chem. Int. Ed.*, 45, 3216-3251, 2006.